%% file: main.tex
\documentclass[
  aps,prab,            
  preprint,            
  superscriptaddress,  
  nofootinbib,
  floatfix,
  longbibliography]{revtex4-2}

\usepackage{graphicx}
\usepackage{float}             
\usepackage{url}
\usepackage{amsmath}
\usepackage{amssymb}
\usepackage{placeins}             
\usepackage{booktabs}             
\usepackage{siunitx}
\usepackage{adjustbox}
\usepackage[labelformat=simple]{subfig}   

\usepackage{chngcntr}
\counterwithin{figure}{section}               
\counterwithin{equation}{section}             
\counterwithin{table}{section}                

\begin{document}

\title{`White Paper' on a Novel FFA-Based CEBAF Upgrade to 22~GeV}

\author{S.~A.~Bogacz} \thanks{bogacz@jlab.org} 

\author{%
 R.~M.~Bodenstein, 
  K.~E.~Deitrick,
  B.~R.~Gamage,
  R.~Kazimi,
  D.~Khan, 
  E.~Nissen,
  S.~Ogur,
  Y.~Roblin,
  P.~Rossi
}
  \affiliation{Thomas Jefferson National Accelerator Facility, Newport News, VA, USA}
\author{%
  J.~S.~Berg,
  S.~J.~Brooks,
  D.~Trbojevic}
 \affiliation{ Brookhaven National Laboratory, Upton, NY, USA}
 \author{V.~Morozov}
 \affiliation{Oak Ridge National Laboratory, Oak Ridge, TN, USA} 

\author{\\ for the FFA@CEBAF Collaboration $^{\dagger}$}

\noaffiliation

\date{\today}

\begin{abstract}
We present a conceptual design for a cost-effective upgrade of the Continuous Electron Beam Accelerator Facility (CEBAF) to 22~GeV using non-scaling fixed-field alternating-gradient (FFA) arcs built from Halbach-style permanent magnets. Building on the eight-pass energy-recovery demonstration at the Cornell--BNL CBETA test accelerator, the design replaces the highest-energy recirculation arcs with a pair of FFA arcs that simultaneously transport six passes spanning a factor-of-two momentum range. We describe the machine layout; including the injector, recirculating linacs, spreaders and recombiners, FFA arcs, splitters, transition and extraction regions; together with beam-dynamics validation studies covering emittance growth, synchrotron-radiation-driven depolarization, and orbit correction, and we summarize permanent-magnet design, prototyping, and radiation-resiliency results. This white paper documents the accelerator physics underpinning a staged path to 22~GeV that preserves CEBAF's multi-hall, high-luminosity, polarized-beam capabilities.
\end{abstract}

\maketitle

\clearpage
\begin{center}
{\large\textbf{$^{\dagger}$ The FFA@CEBAF Collaboration}}\\[2ex]
R.~M.~Bodenstein, S.~A.~Bogacz, K.~E.~Deitrick, B.~R.~Gamage, R.~Kazimi,
D.~Khan, E.~Nissen, S.~Ogur, Y.~Roblin, R.~Ruber, T.~Satogata, N.~Sereno, A.~Seryi, V.~Ziemann \\[0.6ex]
{\itshape Thomas Jefferson National Accelerator Facility, Newport News, VA, USA}\\[2ex]
J.~S.~Berg, S.~J.~Brooks, P.~N'gotta, D.~Trbojevic\\[0.6ex]
{\itshape Brookhaven National Laboratory, Upton, NY, USA}\\[2ex]
V.~Morozov\\[0.6ex]
{\itshape Oak Ridge National Laboratory, Oak Ridge, TN, USA} \\[0.6ex]
G.~H.~Hoffstaetter\\[0.6ex]
{\itshape Cornell University, Ithaca, NY, USA}
\end{center}
\clearpage

\date{\today}

\maketitle

\tableofcontents

\newpage
\section{Project Overview}
\subsection{Preface}
The initial motivation for this study was provided during the Summer of 2020, shortly after a site selection for the EIC was finalized. The idea of using FFA style arcs was not entirely new to CEBAF. Two decades earlier, when the 12 GeV CEBAF upgrade was being developed, there was a brief exploratory study of this very concept~\cite{2004_tn} - adding a pair of FFA arcs to 'double' the energy from 6 GeV to 12 GeV. The conclusion back then was not favorable, because of complex design and large size of the required conventional FFA magnets. The feasibility of this scheme has radically changed with the advent of Halbach style permanent magnet technology pioneered by the CBETA Test Accelerator~\cite{hoffstaetter2017} over a decade later. With this new enabling technology in hand, we revisited this idea. The FFA@CEBAF Collaboration between JLAB, BNL and Cornell University was formed in November of 2020 and we launched a vigorous accelerator design effort, which continued for the next five years through regular weekly meetings. The conceptual design of 22 GeV CEBAF upgrade presented here is a result of this effort.     

\subsection{Scientific Background}
The Cornell–Brookhaven National Laboratory (BNL)–ERL Test Accelerator facility at Cornell~\cite{hoffstaetter2017} has demonstrated eight-pass recirculation of an electron beam with energy recovery (four accelerating beam passes and four decelerating beam passes). All eight beams are recirculated by single arcs of fixed field alternating (FFA) gradient magnets. This exciting new technology would enable a cost-effective method to double the energy of CEBAF, enabling significant new scientific opportunities–including new mass ranges for meson spectroscopy, enabling precision studies of the nucleon sea, providing precision data into the abiding mystery of nuclear anti-shadowing, and extending the kinematic range of nucleon imaging studies. Technical studies of the implementation of FFA (Fixed Field Alternating Gradient) technology at CEBAF are in progress. The community organized several workshops designed to underscore the physics reach that such an upgrade could enable. 
Further resources are required to continue accelerator science and technology to study the possibility of intense (polarized) positron beams, and to study the possibility of increasing the CEBAF beam energy in a cost-effective manner using an FFA (Fixed Field Alternating Gradient) configuration. 
This scheme aims at extending the energy reach of CEBAF up to 22 GeV within the existing tunnel. The proposed energy upgrade envisions increasing the number of recirculations, while using the existing CEBAF cavity system. The energy gain per pass remains unchanged, while the number of passes through the accelerating cavities is nearly doubled. A proposal was formulated to replace the highest-energy arcs with Fixed Field Alternating-gradient (FFA) arcs. The new pair of arcs would support simultaneous transport of additional six passes with energies spanning a factor of two, using the non-scaling FFA principle implemented with Halbach-derived permanent magnets - a novel magnet technology that significantly saves energy and lowers operating costs.

\FloatBarrier

\subsection{Project History}
\subsubsection{12 GeV CEBAF}
  The upgrade from 6 GeV to 12 GeV at CEBAF~\cite{12GeV_CEBAF} was driven by the need to probe the quark–gluon structure of matter with greater depth and precision than the original facility could reach. By the early 2000s, advances in superconducting RF cavity performance and emerging QCD questions—such as the origin of nucleon spin, 3D parton imaging, and confinement dynamics—made the case for a higher-energy machine compelling. The conceptual design was developed in 2003–2006, and the U.S. Department of Energy approved the project in 2008. Major construction and installation took place between 2009 and 2013, followed by first 12 GeV beam delivery in 2014 and full physics operations beginning in 2016.
The upgrade included ten new high-gradient C100 cryo-modules, raising the accelerating gradient from roughly 7.5 MV/m to 20 MV/m and doubling of CEBAF energy reach, while maintaining beam quality. The cryogenic plant capacity was doubled, dipole and quadrupole magnets were upgraded to support higher fields, and a new experimental area (Hall D) was added, reached via a half-pass recirculation arc for full-energy photon-beam experiments.
The optics redesign was a cornerstone of the 12 GeV project. This involved rebuilding the recirculation arcs with an isochronous lattice, providing matched beta functions and near-zero momentum compaction across the full energy range.  High-field dipoles (up to 1.5 T) and stronger quadrupoles were used to maintain tight transverse focusing and dispersion control.
 These optics preserved emittance and phase stability during multi-pass acceleration, keeping energy spread below $2\times10^{-4}$ even at 12 GeV. The lattice also allowed fine tuning of path length to synchronize arrival times through all recirculations, a key requirement for delivering beams simultaneously to multiple halls.
The 12 GeV CEBAF layout retains its twin-linac, five-and-a-half-pass design but now operates with digital RF feedback, higher injector energy, and four simultaneous experimental halls (A–D). These advances transformed CEBAF from a pioneering 6 GeV machine into a precision, high-luminosity facility at the frontier of modern nuclear physics—capable of performing electroweak measurements, mapping 3D nucleon structure, and exploring gluonic excitations that reveal the dynamics of confinement.
 A natural evolution is to build upon the existing CEBAF lattice to construct the 22 GeV FFA machine. This will be accomplished by keeping the same isochronous arc design for the first three passes. Magnets will be swapped or replaced accordingly to accommodate the higher beam momentum in the lower passes. 

\FloatBarrier

\subsubsection{CBETA as a proof-of-principle FFA}
\label{subsec:cbeta}

The Cornell-BNL ERL Test Accelerator (CBETA)~\cite{hoffstaetter2017} is a
superconducting energy recovery linac (ERL) that accelerates electrons to
150~MeV across four passes through a single main linac cryo-module (MLC), then
symmetrically decelerates them to recover their energy. Operated at Cornell University, it demonstrated 4-pass energy recovery in 2019,
establishing it as the world's first multipass superconducting
ERL~\cite{bartnik2020}. CBETA introduced two technologies new to ERLs: a single
fixed field alternating gradient (FFA) return arc serving all four beam
energies simultaneously, and Halbach-style permanent magnets as the arc
lattice elements~\cite{brooks2020}.

\paragraph{Machine Layout}

The machine comprises four subsystems. A 6~MeV photoinjector (design power
0.5~MW) feeds the MLC, which contains six 1.3~GHz niobium cavities imparting
$\pm36$~MeV per pass; alternating stiffened and unstiffened cavity designs are
powered by 5~kW and 10~kW amplifiers respectively~\cite{banerjee2018}.
Mechanical vibration isolation and an active microphonics compensation
loop~\cite{banerjee2019} reduced cavity detuning substantially.

The FFA return arc transports all four design energies (42, 78, 114, and
150~MeV) within a single beam pipe~\cite{berg2018}. Doublet focusing cells
whose defocusing magnets carry a dipole component perform the bending; 24
adiabatic transition cells on each side connect arc and straight sections.
Five types of Halbach permanent magnets are employed~\cite{brooks2020}, held
in temperature-regulated aluminium frames. After assembly, rotating-coil
measurements guided the insertion of iron correction wires, reducing residual
field errors to below 3~parts in~$10^{4}$~\cite{brooks2020}. Horizontal and
vertical dipole correctors surrounding every magnet provide simultaneous orbit
correction for all passes.

Individual splitter (SX) and recombiner (RX) beamlines for each energy
connect the MLC to the FFA arc and enforce the correct return phase,
momentum compaction $R_{56}$, and betatron match at the linac
entrance. Each line contains eight quadrupoles and three sliding path-length
joints.

\begin{figure}[htbp]
    \centering
\includegraphics[width=0.5\columnwidth]{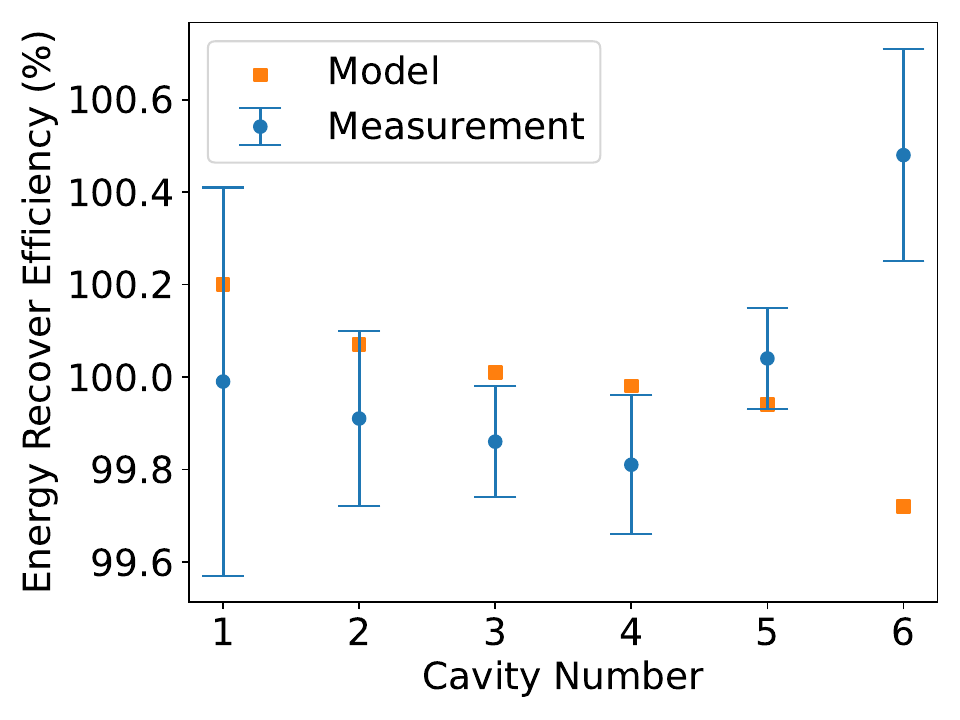}
    \caption{Measured per-cavity energy recovery efficiency in 1-pass operation,
    corrected for beam losses, compared with machine-model
    predictions~\cite{gulliford2021}.}
    \label{fig:efficiency}
\end{figure}

\paragraph{1-pass operation.}
In 1-pass mode the beam was injected at 6~MeV, accelerated to 42~MeV, and
decelerated back to 6~MeV~\cite{gulliford2021}. A peak current of nearly
70~$\mu$A was reached; higher currents were constrained by radiation limits
from an inadequately shielded beam dump and an inoperative fast-shutdown
system, not by ERL physics. Per-cavity energy recovery efficiency was
extracted by comparing RF power in acceleration-only and energy-recovery
configurations~\cite{gulliford2021}. Operating all cavities at a uniform
1$^{\circ}$ phase to suppress RF jitter sensitivity, the dump beam energy
was only $\sim$1~keV above the 6~MeV injection energy, confirming
high-efficiency energy recovery in each cavity (Fig.~\ref{fig:efficiency}).

\begin{figure}[htbp]
    \centering
 \includegraphics[width=0.5\columnwidth]{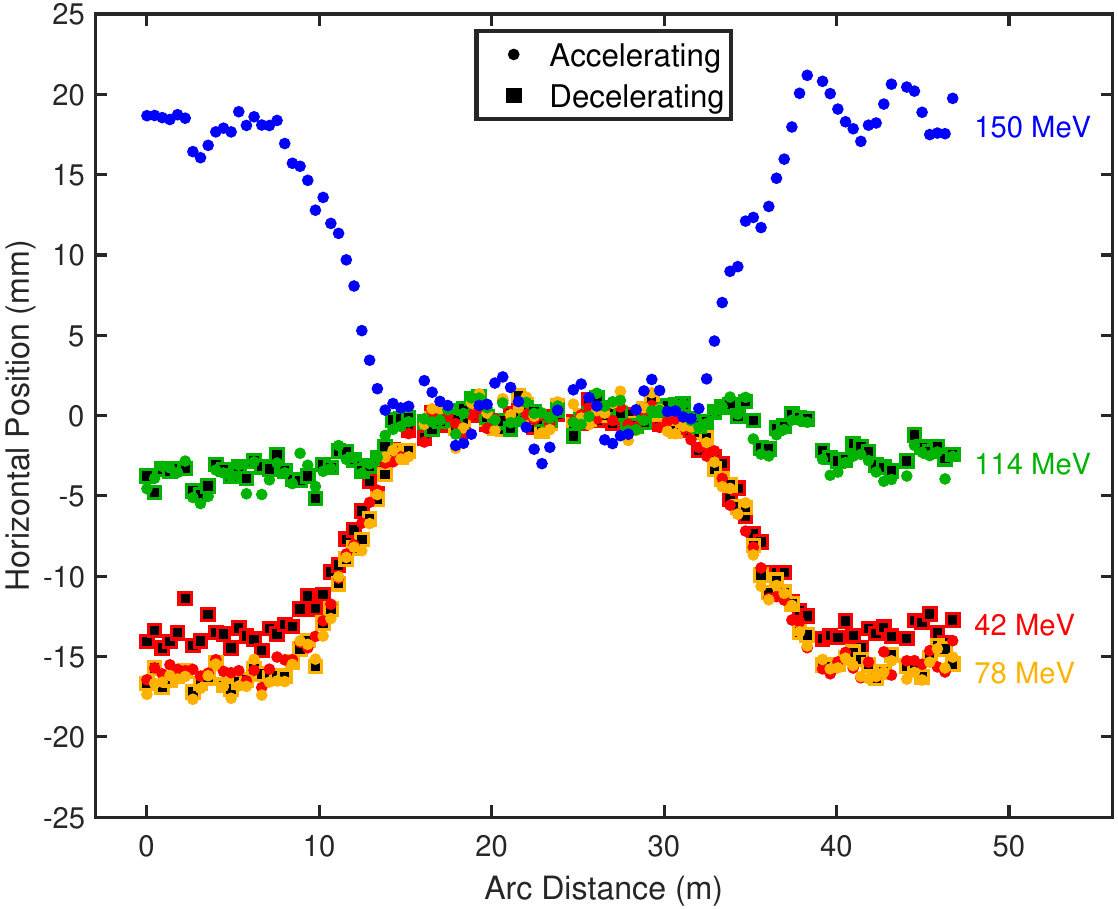}
    \caption{Corrected horizontal orbits in the FFA return arc during 4-pass
    operation~\cite{bartnik2020}.}
    \label{fig:orbits}
\end{figure}

\begin{figure}[htbp]
    \centering
\includegraphics[width=0.5\columnwidth]{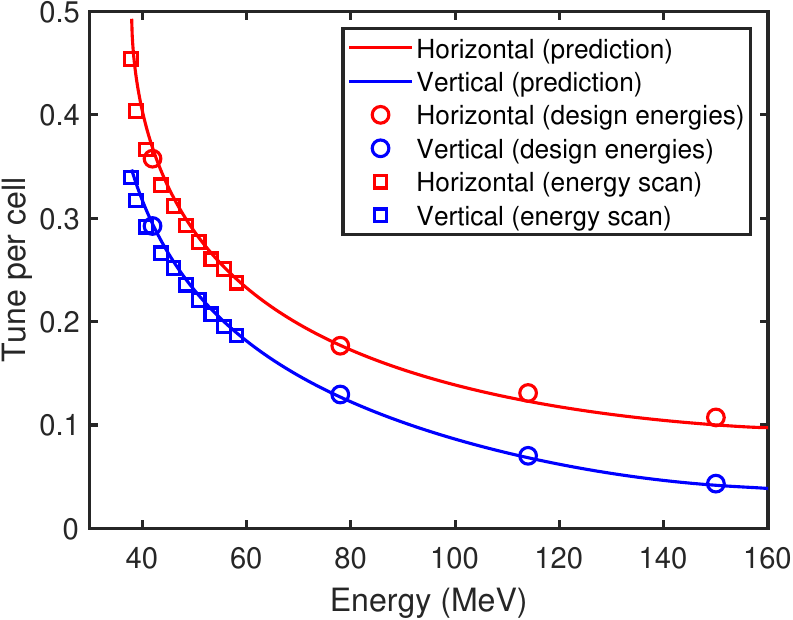}
 
    \caption{FFA arc betatron tunes per cell vs.\ beam energy, compared with
    model predictions~\cite{bartnik2020,gulliford2019}.}
    \label{fig:tunes}
\end{figure}

\paragraph{4-pass operation.}
Four-pass energy recovery was demonstrated at 1~nA beam current (5~pC bunch
charge)~\cite{bartnik2020}, requiring simultaneous closed-orbit correction at
all four energies in the shared FFA arc (Fig.~\ref{fig:orbits}). Betatron
tunes measured across a continuous energy range via a single-pass scan and at
the four design energies during 4-pass circulation agreed well with model
predictions based on magnet field maps~\cite{bartnik2020,gulliford2019}
(Fig.~\ref{fig:tunes}). Linac arrival phases matched design values to within
a few degrees, and precise path-length adjustment in the splitter lines was
found to be essential for sustaining all eight passes~\cite{bartnik2020}.

\paragraph{Challenges}

RF phase and voltage fluctuations caused frequent beam trips and forced
some cavities below design voltage~\cite{banerjee2019}. A slow drift in
measured RF amplitude, traced to an inconsistency in the control
loop~\cite{banerjee2022}, was managed operationally by retuning splitter
magnets. Beam-based alignment revealed a 3~mm vertical offset of the
MLC cryo-module relative to external BPMs~\cite{gulliford2019b}, which was
corrected by shimming. The compact splitter layout introduced magnetic
crosstalk between adjacent beamlines, hysteresis requiring degaussing cycles,
stiff path-length joints, and large chromaticity from heavily constrained
optics~\cite{brooks2022}. In 4-pass operation a significant current loss
occurred between the last two return passes, tentatively attributed to the
78~MeV recombiner line, though its origin was not fully
understood~\cite{bartnik2020}. At bunch charges above $\sim$2~pC, evidence of
microbunching was found~\cite{hoffstaetter2024}, and cathode-generated beam
halo was imaged in the injector~\cite{banerjee2020}; both effects could
limit high-current operation.

\paragraph{Future Prospects}

CBETA remains intact, and three directions are identified for resumed operation. Its ERL capabilities closely match the requirements of the electron cooler proposed for the Electron-Ion Collider (EIC) at BNL~\cite{gulliford2023,kayran2024}. The proposed CEBAF 22~GeV energy upgrade uses a splitter-plus-FFA design analogous to CBETA~\cite{khan2024},
and targeted CBETA studies could resolve outstanding splitter-line issues. Finally, adding a 150~MeV return line with an interaction point could convert CBETA into an inverse Compton scattering photon source capable of producing intense monochromatic photons above 100~keV~\cite{deitrick2021}.

\FloatBarrier

\section{Scientific Justification}
\subsection{Motivation}
For over two and a half decades, the Continuous Electron Beam Accelerator Facility (CEBAF) has provided the global scientific community with electron beams of unparalleled multi-GeV intensity and precision. A major turning point arrived in 2016 with the successful launch of the 12 GeV upgrade program, initiating a crucial new phase for the Laboratory.

This 12 GeV phase is now firmly established, already responsible for numerous significant experimental results that have been published \cite{Arrington2022}. Furthermore, a promising suite of experiments, officially approved by the Program Advisory Committee (PAC), is scheduled to run for at least the next decade.

In parallel with the successful execution of the 12 GeV scientific program, the CEBAF user community is strategically planning the facility's future. This planning focuses on the novel science that would become accessible through a potential, cost-effective upgrade to 22 GeV. This higher-energy capability would secure a rich and distinctive experimental nuclear physics program, successfully marrying CEBAF's prestigious history with a vibrant path forward and extending the facility's lifespan well beyond the 2030s. The compelling scientific case for the 22 GeV energy upgrade is built on two key strengths: exploiting the facility's unique, world-leading high-luminosity operations and leveraging existing or already planned Hall equipment.

The proposed upgrade would establish a new global benchmark for nuclear physics research focusing on clarifying the fundamental behavior of Quantum Chromo Dynamics (QCD), particularly in the valence quark-dominated region and the transition toward the sea quark region. By focusing on the ``luminosity frontier'' with its fixed-target configuration, coupled with large acceptance detectors and high-resolution spectrometers, CEBAF would continue to deliver unique insights into the nature of QCD and the emergence of hadron structure for future decades.

Crucially, even with this energy boost, CEBAF will still operate at orders of magnitude higher luminosity than the future Electron-Ion Collider (EIC). JLab’s current and planned capabilities would offer essential scientific opportunities that perfectly complement the EIC’s operational scope. In fact, while high-energy facilities like the EIC are essential for illuminating the perturbative regime of QCD and uncovering the fundamental role of gluons within nucleons and nuclei, a medium-energy electron accelerator operating at the luminosity frontier is critical to fully map the rich and extraordinary variety of non-perturbative effects that manifest in hadronic structure. Understanding these non-perturbative mechanisms is the linchpin for comprehending how hadrons and nuclei materialize from the underlying theory of QCD. In this non-perturbative domain, the underlying dynamics of quarks and gluons undergo a profound transformation. This process results in the appearance of effective degrees of freedom, a phenomenon known as emergent behavior. Although these new structures originate from the fundamental QCD fields, their experimental interpretation remains challenging, positioning strong interaction physics as a prime example of an emergent system. Ultimately, the combined capabilities of the EIC and JLab could provide the essential tools necessary to unravel the intricate mechanism by which QCD constructs hadronic matter.

\FloatBarrier

\subsection{New Scientific Opportunities}
This section briefly outlines a research program comprising several strategic objectives. All these efforts are unified by the common goal of investigating the complex, non-perturbative dynamics that govern hadron structure and the rich physical phenomena within these strongly interacting systems. For a detailed discussion of the extensive physics scope enabled by a 22 GeV CEBAF upgrade, see the initial White Paper \cite{Accardi2024}, and the most recent refinements detailed in \cite{Accardi2026}.

The development of the proposed 22 GeV physics program is strategically structured around three core pillars: \textbf{Uniqueness, Enrichment,} and \textbf{Complementarity}. While a sharp distinction between these categories is sometimes challenging, these pillars collectively ensure that the facility’s potential is leveraged to achieve both revolutionary scientific breakthroughs and crucial continuity with existing research efforts.

The \textbf{Uniqueness} pillar focuses on maximizing JLab's discovery potential by pursuing measurements exclusively enabled by 22 GeV energy. Key efforts include: \textit{Hadron Spectroscopy},  \textit{Proton Gluonic Structure}, and \textit{Precision Electroweak Tests}.

\textit{Hadron Spectroscopy}. Recent experimental discoveries in hadron spectroscopy have revealed numerous exotic states, such as tetraquarks and pentaquarks \cite{BESIII},\cite{Belle},\cite{Belle2011}, which challenge our current understanding of QCD confinement. While the existing 12 GeV program at JLab provides valuable insights into light-quark systems, it lacks the reach necessary to fully investigate the production and nature of these complex, heavy-quark exotic resonances. A 22 GeV energy upgrade is therefore essential to extend JLab’s reach into higher-lying resonance regions, enabling a comprehensive mapping of these poorly constrained states. This upgrade would leverage the complementary, high-precision capabilities of the CLAS12 \cite{Burkert} and GlueX \cite{GlueX} spectrometers in Hall B and Hall D respectively, to access a unique production environment essential for understanding the nature of charm-containing pentaquark and tetraquark candidates. By utilizing linearly polarized real and virtual photons, this program will provide a unique, powerful tool to disentangle production and decay dynamics that remain inaccessible through other experimental methods. 

\begin{figure}[!hbt]
    \centering
    \includegraphics[width=0.7\textwidth]{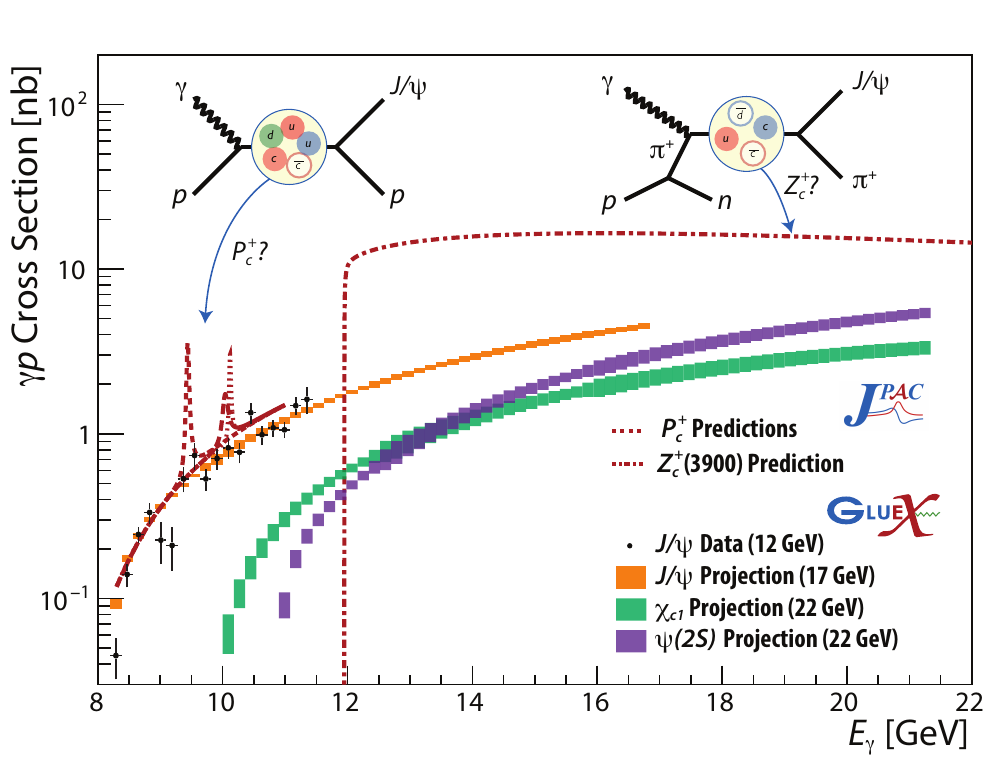}
    \captionsetup{justification=raggedright,singlelinecheck=false}
    \caption{Photoproduction cross sections of states containing $c\bar{c}$ as a function of photon beam energy. The points are GlueX data \cite{GlueX2023} The colored boxes are projections of statistical precision using the GlueX detector with different assumptions about the electron energy. The collection of dashed and dotted curves indicate how pentaquark Pc [\cite{Hiller} or tetraquark Zc \cite{Winney} candidates might appear.}
    \label{fig:spectroscopy}
\end{figure}

The last two decades have produced numerous discoveries of new particles in the charm and bottom sectors by experiments like BaBar, BESIII, and Belle at $e^{+} e^{-}$ machines, as well as LHCb at the LHC. Those experiments have identified numerous ``XYZ'' states, many of which defy conventional quark-model descriptions due to their anomalous charges, masses, or decay patterns. Extensive reviews of these new particles can be found in \cite{Olsen}, \cite{Labed}, \cite{Briceno}, \cite{Brambilla}. A central challenge in this field is that most of these candidates have only been observed in a single production environment, leading to uncertainty over whether they represent genuine hadronic resonances or merely kinematic ``triangle singularities.'' Distinguishing between these possibilities is essential to understanding the non-perturbative QCD dynamics of hadron formation. The proposed 22 GeV JLab upgrade will expand the kinematic reach required to produce XYZ states. By leveraging photoproduction, where the photon behaves as a virtual vector meson scattering off protons or pion clouds, JLab will enable two-to-two scattering studies. As these processes are free from the triangle singularities that often obscure results in $e^{+} e^{-}$ and B-decay experiments, this facility will offer the crucial, complementary data needed to elucidate the nature of these exotic states. Theoretical predictions for pentaquark ($P_c^+$) and tetraquark ($Z_c^+$) structures are shown inFigure~\ref{fig:spectroscopy}.

\textit{Proton Gluonic Structure}. 
Recent experimental results from JLab show the feasibility of extracting gluonic structure from near-threshold $J/\psi$ production \cite{Duran}. \cite{GlueX-Jpsi}, \cite{clas12-Jpsi}. The theoretical interpretation of these measurements remains an area of active research, raising several questions that cannot be fully resolved with current 12 GeV data and instead require a broader kinematic range. While more precise measurements of $J/\psi$ photo/electroproduction at 11 GeV are planned in Hall B, Hall C, Hall D, and with the future SoLID detector \cite{solid} which will provide important insights, the proposed 22 GeV JLab upgrade will be essential to fully realizing the potential of this program. The extended energy range will make it possible to measure the interplay of kinematical variables dependence over a range sufficient for separating the spin-0 and 2 contributions and testing/improving the proposed models of the reaction mechanism. The 22 GeV fixed-target energy covers exactly the region where the differences between different reaction models are maximal and can be distinguished by the data. Figure~\ref{fig:spectroscopy} illustrates the anticipated photoproduction cross sections for various charmonium states as a function of the photon beam energy, $E_\gamma$. The existing GlueX data at 12 GeV demonstrate the current kinematic reach, while the colored boxes represent projected statistical precision for higher energies. The JLab 22 GeV facility would be uniquely positioned to achieve unprecedentedly precise measurements of $J/\psi$ and higher-mass charmonium states, such as $\chi_{c1}$ and $\psi(2S)$, near their respective production thresholds.  

\textit{Precision Electroweak Tests}. A key opportunity enabled by the JLab 22 GeV upgrade is measuring the $\pi^0$ radiative decay width ($\Gamma(\pi^0 \to \gamma\gamma)$) to sub-percent precision by scattering off atomic electrons, a reaction that features a production threshold of 18 GeV. This measurement is critical for testing the chiral anomaly of QCD and resolving discrepancies with high-order calculations \cite{Adler}. While using traditional nuclear targets introduces complex backgrounds, such as strong production, nuclear interference, and incoherent processes that require careful angular fitting, switching to an electron target entirely eliminates these nuclear complications. Consequently, this clean experimental approach provides a pristine test of low-energy QCD dynamics and the internal structure of light pseudoscalar mesons.

The \textbf{Enrichment} and \textbf{Complementarity} strategies often merge, focusing on extending 12 GeV results and creating essential kinematic links to the future Electron-Ion Collider. Specifically, the 22 GeV upgrade would significantly expand the experimental phase space. ``Enrichment'' is achieved by extending measurements to higher momentum transfer $Q^2$. ``Complementarity'' is realized by establishing an essential kinematic bridge to the future EIC, ensuring that JLab 22 GeV data provides the necessary baseline and scaling information to fully exploit the EIC’s high-energy frontier. Given the vast reach of the physics program that will be simultaneously enriched and complemented by this upgrade, the following sections provide only a concise overview of the key research topics. For a comprehensive discussion of the scientific case, we refer the reader to the original white paper and its subsequent updates \cite{Accardi2024, Accardi2026}.

Access to higher $Q^2$ and larger hadronic transverse momenta significantly enhances our ability to study the momentum space tomography of nucleons and nuclei via \textit{Transverse Momentum Dependent (TMD)} parton distribution functions.
By integrating high luminosity with the precise detection of multiparticle final-state observables in multidimensional space, JLab will be uniquely equipped to isolate the genuine intrinsic transverse structure of hadrons within TMDs, while maintaining controlled systematics. Such capabilities are essential for interpreting measurements from both JLab and the EIC, and for achieving a comprehensive understanding of nucleon structure and hadronization. Furthermore, JLab maintains a fundamental role in the EIC era by performing precision measurements to separate the longitudinal $\sigma_L$ and transverse $\sigma_T$ photon contributions to the cross section, which are critical for investigating both semi-inclusive and exclusive processes.

The 22 GeV upgrade is essential for executing experiments involving \textit {elastic} and \textit{hard-exclusive processes}, as these measurements necessitate sufficient energy to reach the scaling and factorization regimes, high luminosity for low-rate processes and multivariable differential analysis, and superior detector resolution for accurate cross-section measurements. Key applications include the high-quality extraction of the D-term form factor of the QCD energy-momentum tensor and the internal ``pressure'' distribution of the proton \cite{Polyakov}, \cite{Burkert2}, \cite{Burkert3}. The facility will also enable fully differential 3D imaging of the nucleon through innovative processes like Double Deeply Virtual Compton Scattering (DDVCS) and exclusive diphoton production. Furthermore, it will facilitate the exploration of hadron structure via new exclusive channels such as $N\rightarrow N^*$ transition Generalized Parton Distributions (GPDs) and $N\rightarrow$ meson transition distribution amplitudes. Finally, the upgrade extends nucleon, pion, and resonance transition form factor measurements to momentum transfers of $Q^2\sim30$ GeV$^2$. This reach is critical for probing short-range structure, QCD interactions, and the mass emergence mechanism within the Dyson-Schwinger framework.

This upgrade offers critical insights for precise \textit{partonic structure} studies by filling the kinematic gap between the existing 12 GeV JLab program and the future EIC. With its increased energy range, the 22 GeV upgrade will enable precision measurements of the nucleon light sea in the intermediate to high-x region, which will validate theoretical predictions for intrinsic sea components and support Beyond Standard Model collider searches. Furthermore, the program will allow for the precise determination of nucleon helicity structure at large x and the extraction of the strong coupling constant to sub-percent levels. Finally, the  upgrade provides unique opportunities to investigate the internal structure of mesons in the intermediate to high-x range. 

The high-intensity 22 GeV beam will create an unprecedented opportunity for \textit{Nuclear Science} to significantly advance the understanding of QCD nuclear force dynamics at core distances. The program explores nuclear repulsion by conducting the first direct studies of nuclear Deep Inelastic Scattering structure at $x > 1.25$ and measuring deuteron structure at sub-femtometer distances via exclusive break-up reactions with missing momenta exceeding the GeV scale. It also provides unambiguous identification of three-nucleon short-range correlations within light-front nuclear structure by verifying new nuclear scaling at $x > 2$ and $Q^2 = 10\text{--}15 \text{ GeV}^2$ in inclusive e-A scattering. Furthermore, the program extends the range of medium modification studies into the antishadowing region through highly precise measurements using diverse targets and techniques, including tagging. Finally, it provides experimental proof of Color Transparency in the baryonic sector; and offering a vast, unprecedented kinematic reach for investigating hadronization within the nuclear medium.

The proposed 22 GeV program acts as a critical and synergistic bridge between the existing JLab 12 GeV program and the future EIC. It specifically targets essential aspects of hadron emergence that are inaccessible to the 12 GeV regime or are outside the kinematic reach of the EIC. Furthermore, a major benefit is the ability to conduct these sophisticated measurements using the existing, well-characterized JLab12 detectors, minimizing both cost and technical development risk.

\FloatBarrier

\section{Machine Layout}
\subsection{Existing CEBAF Infrastructure - Upgrades}
\subsubsection{Recirculating injector to 650 MeV}
Replacement of the current 123 MeV injector with a higher energy, 650 MeV, injector is needed to be compatible with the new multi-pass Linacs, which double the number of passes to 11 in the North Linac and to 10 passes in the South Linac. 
The resulting large number of Linac passes makes optical matching virtually impossible due to extremely high energy span ratio (1:175) at the beginning of the North Linac.
The injector system for the proposed 22 GeV CEBAF upgrade is being designed to deliver electrons from a novel, 1 mA highly polarized electron source \cite{bruker}. This next-generation RF gun is under development at Jefferson Lab through a Lab-Directed R\&D initiative and is being considered for both the 22 GeV CEBAF electron upgrade, referred to as FFA@CEBAF, and the 12 GeV CEBAF positron upgrade, known as Ce$^{+}$BAF~\cite{positron}. For FFA@CEBAF, the injector will deliver 650 MeV polarized electrons. In the Ce$^{+}$BAF configuration, the same injector system will be used to generate positrons by directing the electron beam onto a tungsten target \cite{tungsten-target} or liquid metal jet \cite{liquid-target}. The resulting positrons will be captured and accelerated in a modified Linac to reach energies up to 123 MeV.

\begin{figure}[!htb]
    \centering
    \includegraphics[width=1\linewidth]{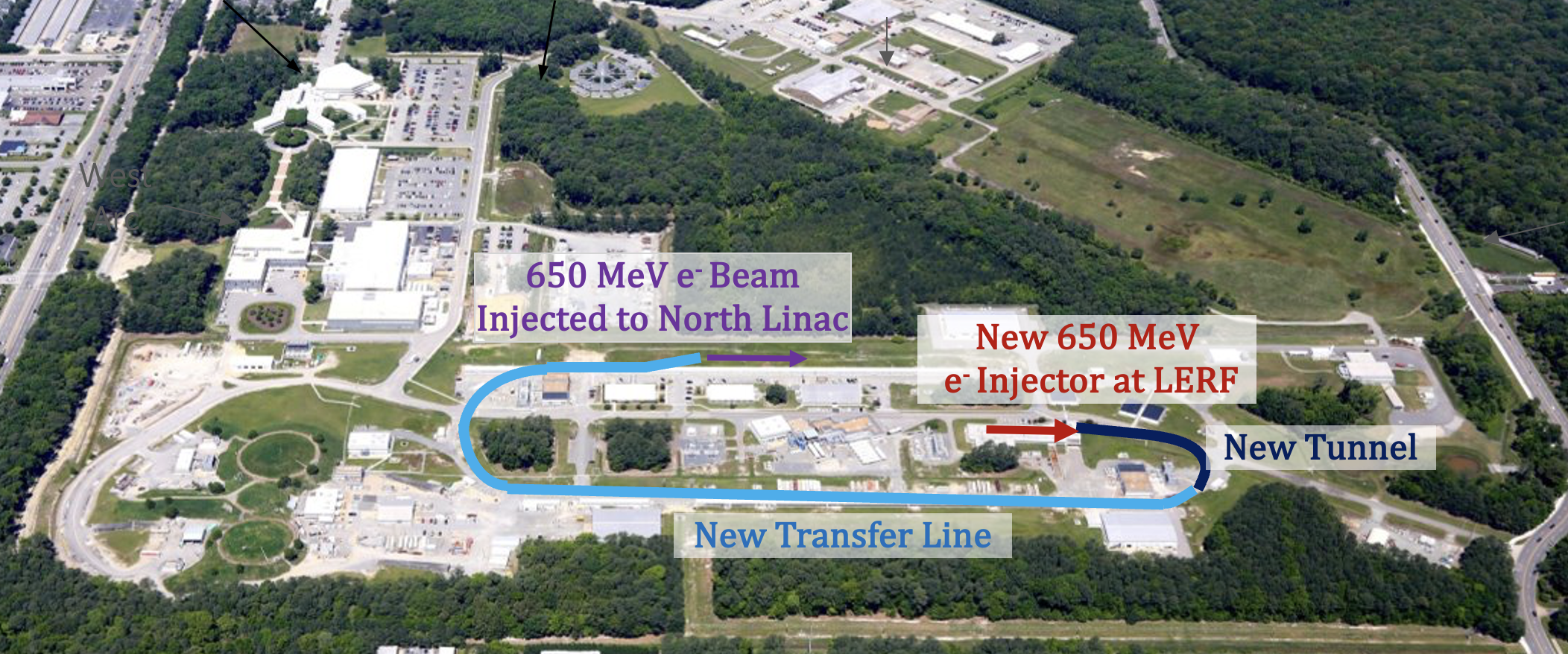}
    \caption{An overview of the FFA@CEBAF upgrade proposal. The injector would be housed in the LERF and new transfer line would be suspended from the existing CEBAF tunnel. The LERF and CEBAF would need to be connected via a new tunnel, shown with dark blue in the figure.}
    \label{fig:layout_CEBAF}
\end{figure}

Both CEBAF upgrade plans would share the 1 mA polarized electron source,  utilize the same superconducting Linac cavities as well as the same transfer line with bipolar dipole magnets. The transfer line would ultimately merge into CEBAF through the north Linac, following a configuration similar to the current injector layout. The new injectors are proposed to be housed in the existing Low Energy Recirculation Facility (LERF). A schematic of the CEBAF upgrade layout is shown in Figure~\ref{fig:layout_CEBAF}.

The new 650 MeV injector~\cite{bogacz_tn} is envisioned as a 3-pass recirculating racetrack based on the existing LERF infrastructure. Here, we present a compact injector design within the LERF vault, which would serve both the 22 GeV CEBAF and the positron programs, compatible with electron source needed to produce positrons for Ce+BAF.

\begin{figure}[!htb]
  \centering
  \includegraphics[width=\columnwidth]{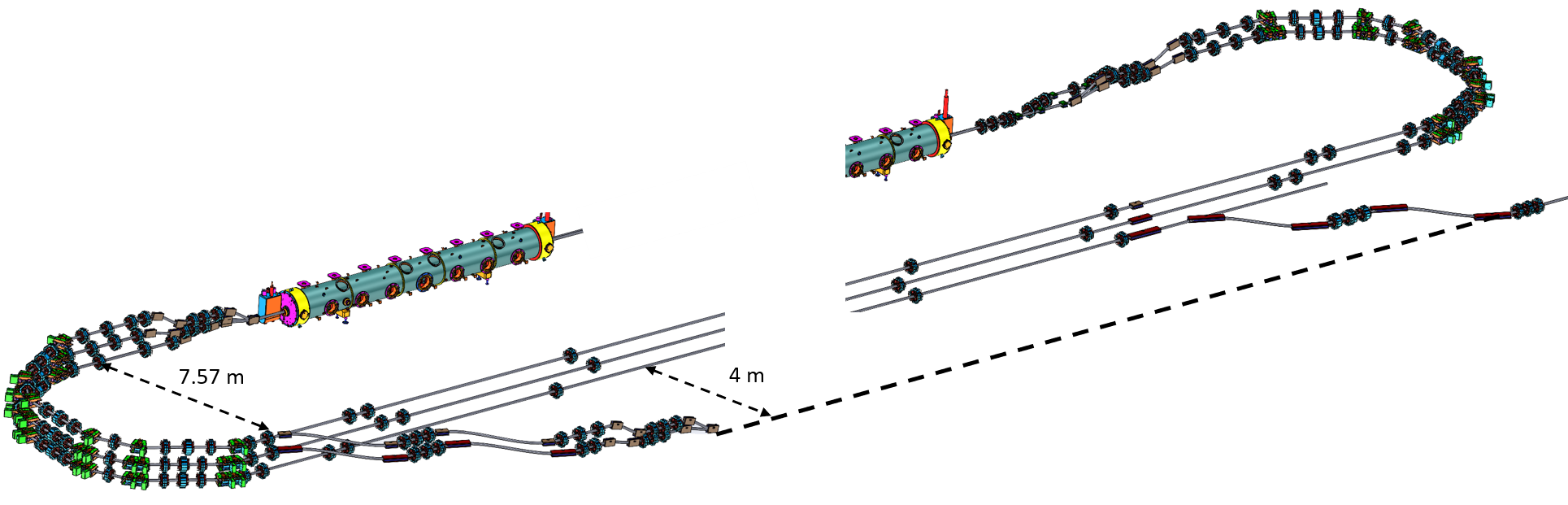}
  \caption{Engineering rendering of the 3-pass racetrack, featuring vertical
           stacks of arcs on each side (3 West\,+\,2 East), separated by
           spreaders and recombiners. The Linac, composed of three C-75
           cryo-modules, achieves a final energy of \SI{650}{MeV} in three
           passes (middle part cut out to enhance clarity).}
  \label{fig:racetrack_layout}
\end{figure}

\paragraph{Layout and lattice architecture}
The injector complex is arranged in a racetrack configuration hosting three C-75 cryo-modules, each containing eight 5-cell cavities operating at \SI{1497}{MHz}; all three cryo-modules are located in a single straight. Starting from an \SI{8}{MeV} injection energy, the final energy of \SI{650}{MeV} is reached in three recirculation passes, with each cryo-module providing a \SI{71.3}{MeV} energy boost. The beam is
injected into the racetrack via a three-bend isochronous merger. Five $180^\circ$ arcs provide recirculation: three on the West side and two on the East side of the racetrack. They are separated vertically by \SI{90}{cm} (\SI{45}{cm} + \SI{45}{cm}), with the lowest-energy arcs at the top of the stack. A vertical switch-yard is laid out as a 'three-tier-spreader', initiated by a single vertical bend common to all three passes. The lower-energy spreaders (1st and 2nd pass) are
configured as two-step ascents with dispersion suppression in between. The top-energy spreader (3rd pass) features a compact four-point-chicane achromat whose second magnet is shared with the 2nd-pass spreader. Recombiners on the East end mirror the spreaders. The overall layout is illustrated in Figure~\ref{fig:racetrack_layout}. All arcs employ the Flexible Momentum Compaction (FMC) lattice architecture.  The overall optics has been inspired by the PERLE at Orsay~\cite{perle_TDR}, \cite{perle_BD}. 

\paragraph{Merger}

\begin{figure}[!htb]
  \centering
  \includegraphics[width=\columnwidth]{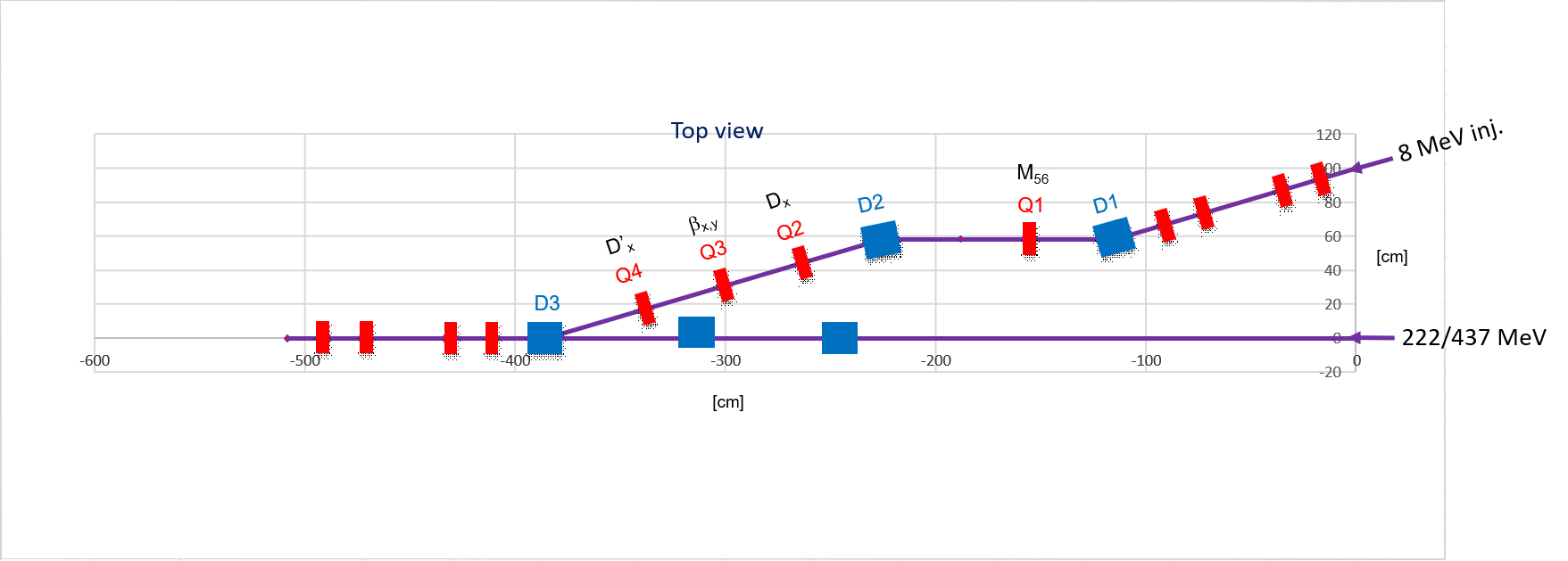}
  \caption{Top view of the 3-bend merger connected to the Linac straight,
           including a re-injection chicane that closes the orbit bump
           initiated by the last merger bend D3.}
  \label{fig:merger_layout}
\end{figure}

\begin{figure}[!htb]
  \centering
  \includegraphics[width=\columnwidth]{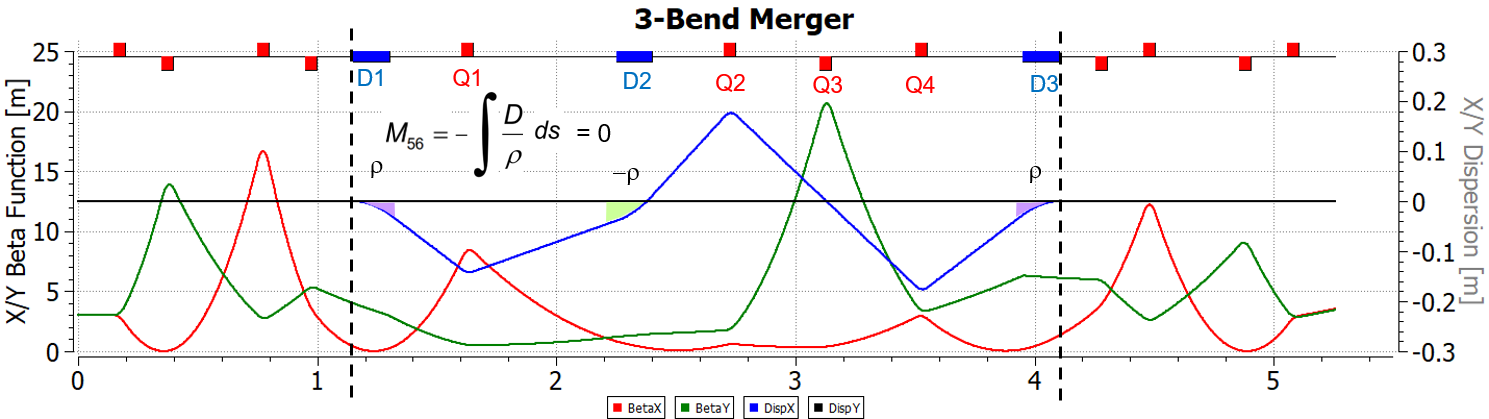}
  \caption{Optics of the 3-bend isochronous merger. Twiss functions at
           the boundary (dashed lines) are optimized for minimum betas
           inside the merger.}
  \label{fig:merger_optics}
\end{figure}

Injection at \SI{8}{MeV} is accomplished through a fixed-field,
three-bend merger whose last magnet is placed at the entrance of the
Linac (Fig.~\ref{fig:merger_layout}). The three-bend configuration was
chosen for its compactness while maintaining functional modularity. Two
quadruplet families flank the merger to provide optimized input/output
Twiss functions. The initial quad in the dispersive region, Q1, controls
momentum compaction; the subsequent pair Q2 and Q4 closes the achromat;
an additional defocusing quad Q3 at the zero-dispersion crossing
balances the beta functions strongly constrained in the horizontal plane.
The merger optics (Fig.~\ref{fig:merger_optics}) is tuned to the
isochronous condition, yet the architecture also supports non-zero
momentum compaction with orthogonal tunability of horizontal dispersion
and $M_{56}$.

\paragraph{Multi-pass Linac}

The last merger bend closes the orbit bump at the lowest injection energy
but would deflect higher-pass beams. To suppress the resulting
higher-pass bumps, a re-injection chicane is completed by placing two
opposing bends upstream of the last chicane magnet; these magnets are
thus invisible to the first-pass beam.

\begin{figure}[!htb]
  \centering
  \includegraphics[width=\columnwidth]{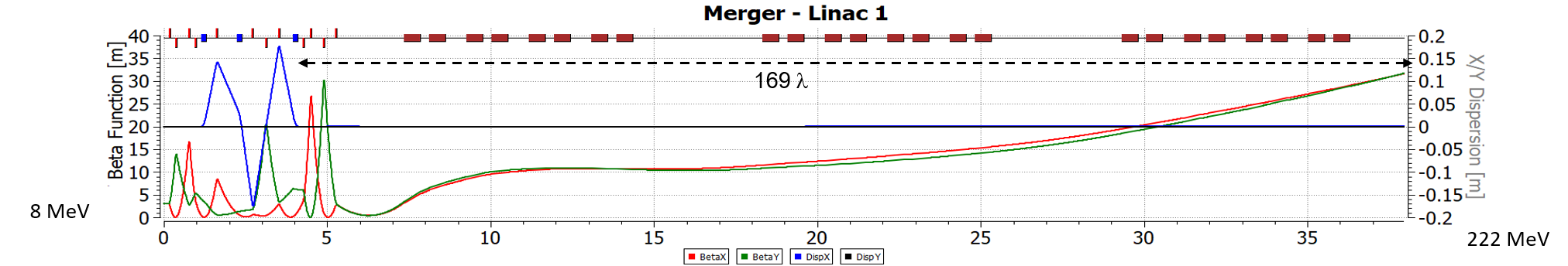}
  \caption{Linac configured with three C-75 cryo-modules. The 1st-pass
           injection optics, tunable via an initial quadrupole quadruplet,
           is dominated by strong cavity end-field focusing at the front
           end of the Linac.}
  \label{fig:Linac_1pass}
\end{figure}

\begin{figure}[!htb]
  \centering
  \includegraphics[width=\columnwidth]{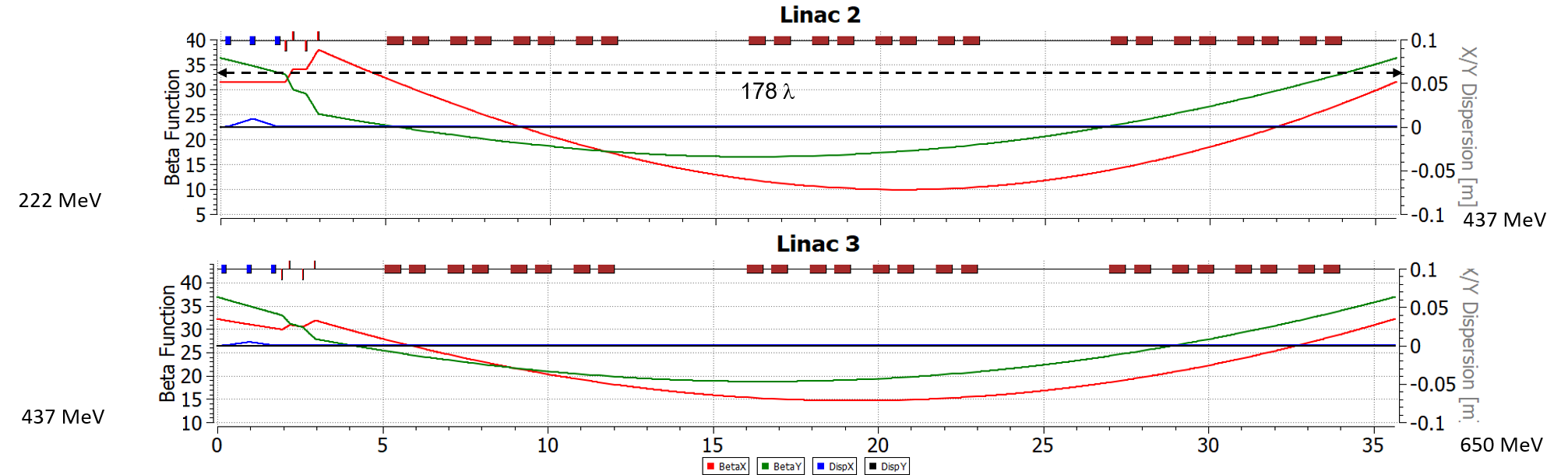}
  \caption{Drift-Linac optics for the 2nd and 3rd passes, featuring
           quasi-parabolic beta functions slightly perturbed by the
           quadrupole quadruplet.}
  \label{fig:Linac_higher}
\end{figure}

The multi-pass Linac optics is designed as a 'drift Linac' with no
quadrupoles between the cryo-modules. This choice yields quasi-parabolic
beta functions that are nearly identical for the 2nd and 3rd passes
(Fig.~\ref{fig:Linac_higher}); they can be symmetrized across the Linac,
with maximum beta functions at the Linac ends comparable to the Linac
length of approximately \SI{35}{m}. This simplifies the overall
racetrack architecture and allows nearly identical arcs on both ends of
the racetrack.

In contrast, the 1st-pass optics is dominated by cavity end-field
focusing (Fig.~\ref{fig:Linac_1pass}). A weak quadrupole quadruplet
is added to provide Twiss flexibility at injection while remaining nearly
transparent to the higher passes.

\paragraph{Recirculating arcs}

The spreaders immediately follow the Linac to separate beams of
different energies and route them to their respective arcs; recombiners
perform the inverse operation before the next Linac pass. Each spreader
begins with a vertical bending magnet common to all beams. The 1st-pass
beam steps up \SI{90}{cm} via a two-step vertical spreader, with three
appropriately placed quadrupoles suppressing vertical dispersion between
the two steps; this spreader uses four rectangular $30^\circ$ bends, the
first shared with all three passes. The 2nd-pass spreader rises by
\SI{45}{cm} in a similar two-step arrangement. The 3rd-pass is returned
to the Linac level by a three-point achromat chicane.

\begin{figure}[!htb]
  \centering
  \includegraphics[width=\columnwidth]{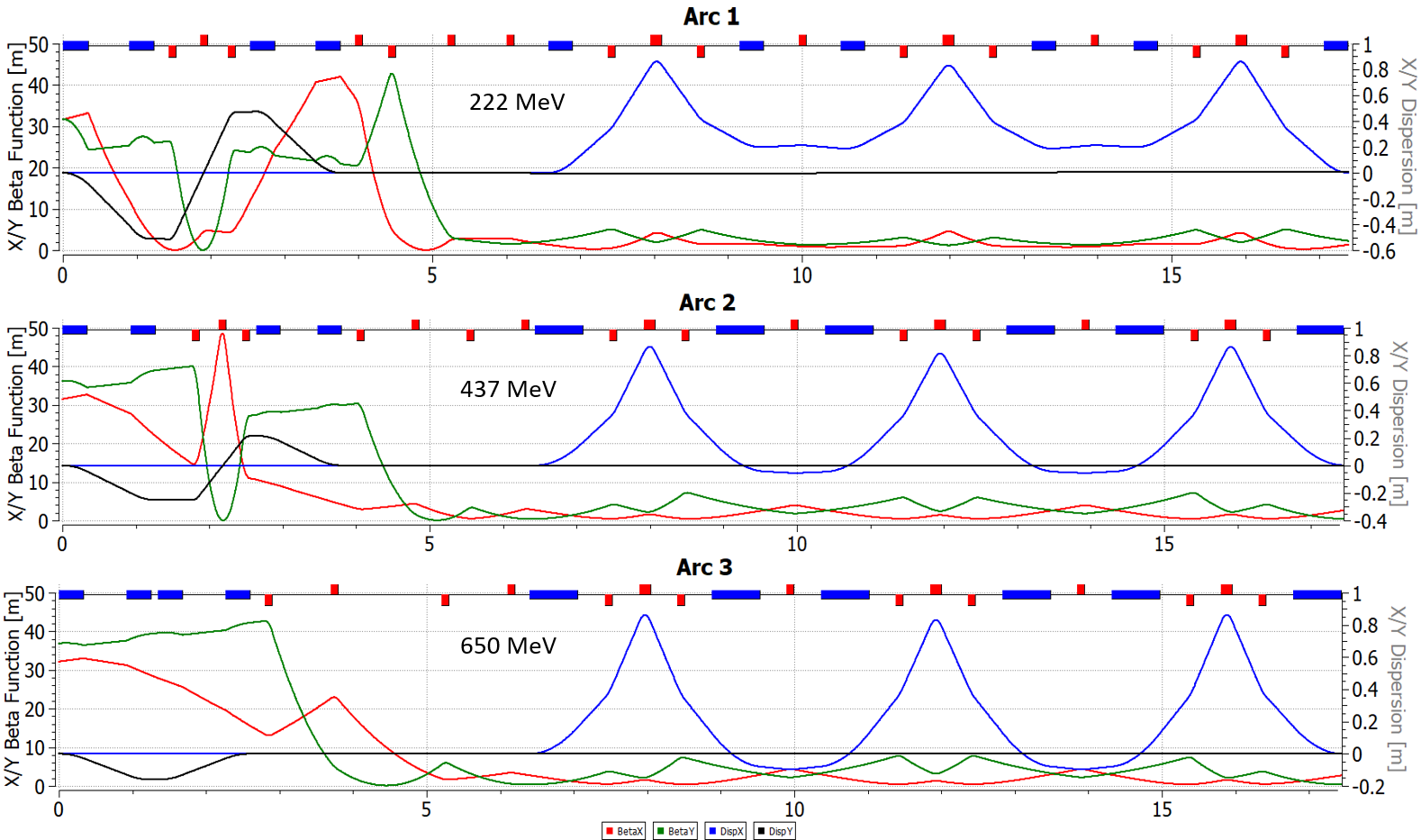}
  \caption{Optics architecture based on the FMC cell. The arcs are tuned
           to the isochronous condition (net-zero momentum compaction),
           balancing contributions from horizontal (arc) and vertical
           (spreader) bends.}
  \label{fig:fmc_arc}
\end{figure}

Four matching quads follow each spreader to bridge the Twiss functions
(two betas and two alphas) between the spreader and the $180^\circ$ arc.
By virtue of the mirror-symmetric higher-pass Linac optics, the arcs on
the opposite side of the racetrack are simply inverted versions of Arcs
1 and 2.

All five $180^\circ$ horizontal arcs employ FMC optics
(Fig.~\ref{fig:fmc_arc}) to allow independent adjustment of the momentum
compaction factor in each arc (needed for longitudinal phase-space
reshaping). Arc~1 uses six \SI{33}{cm} sector bends together with two
triplets and one singlet; Arcs 2 and 3 use double-length \SI{66}{cm}
sector bends. Consequently, the shorter and longer bends operate at
approximately \SI{1.1}{T} for Arcs 1 and 2, and at \SI{1.65}{T} for
Arc~3. The path length of each arc is chosen to be an integer number of
RF wavelengths.

\paragraph{Mirror-symmetric straights}

\begin{figure}[!htb]
  \centering
  \includegraphics[width=\columnwidth]{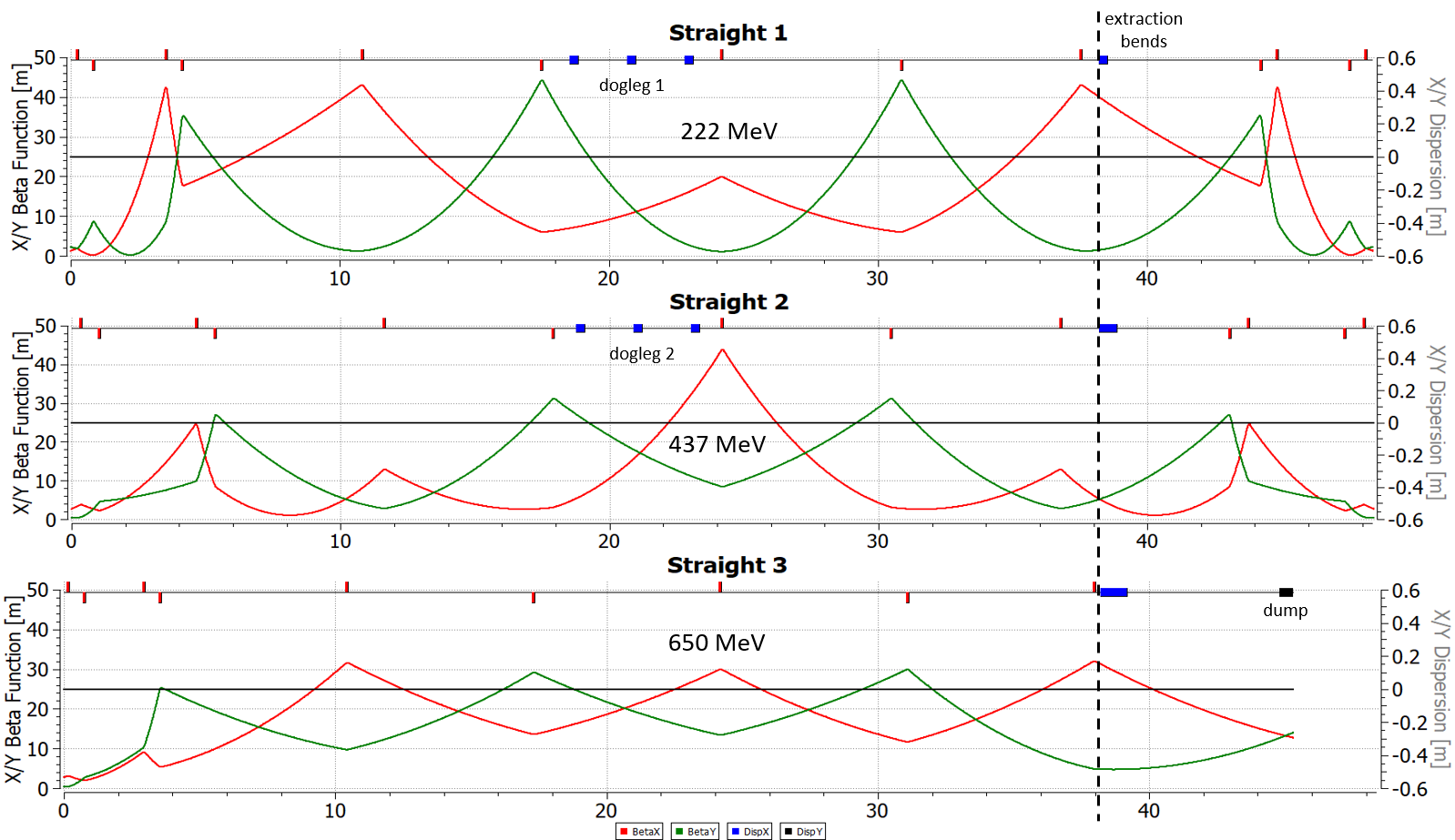}
  \caption{Optics of the mirror-symmetric straights closing the
           racetrack.}
  \label{fig:straights}
\end{figure}

The racetrack is completed with three mirror-symmetric straights
(Fig.~\ref{fig:straights}): Straight~1 bridges Arc~1 and its East-side
mirror; Straight~2 does the same for Arc~2; Straight~3 at top energy
terminates at an in-line dump (no Arc~3 exists on the East side). To
match the small betas and steep alphas of the arcs, each straight uses a
FODO-like structure flanked by two doublets. Lower-energy straights also
incorporate path-length correcting doglegs. Each straight contains a
single extraction dipole acting as a switch-yard.

\paragraph{Extraction lines}

\begin{figure}[!htb]
  \centering
  \includegraphics[width=\columnwidth]{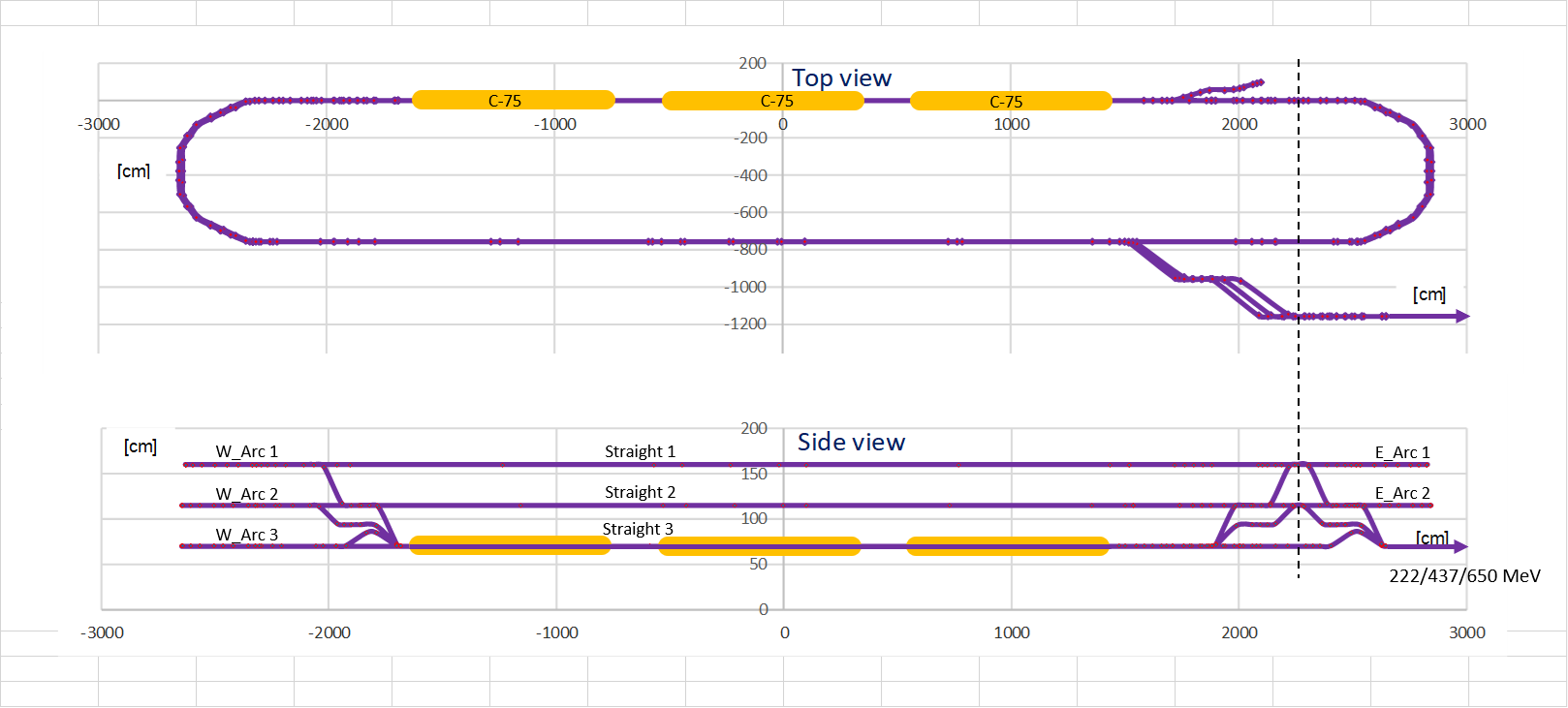}
  \caption{Layout of individual extraction lines featuring a \SI{4}{m}
           horizontal jog and a recombiner.}
  \label{fig:extraction_layout}
\end{figure}

This ``Universal Injector'' provides individual extraction lines for all
three passes (Fig.~\ref{fig:extraction_layout}). Each line begins with a
single bend in the corresponding straight and shares a two-step
horizontal jog layout, followed by a replica of the vertical
recombiner, which brings all three beams back to the ground-level Linac
plane (\SI{70}{cm} above the floor).

\begin{figure}[!htb]
  \centering
  \includegraphics[width=\columnwidth]{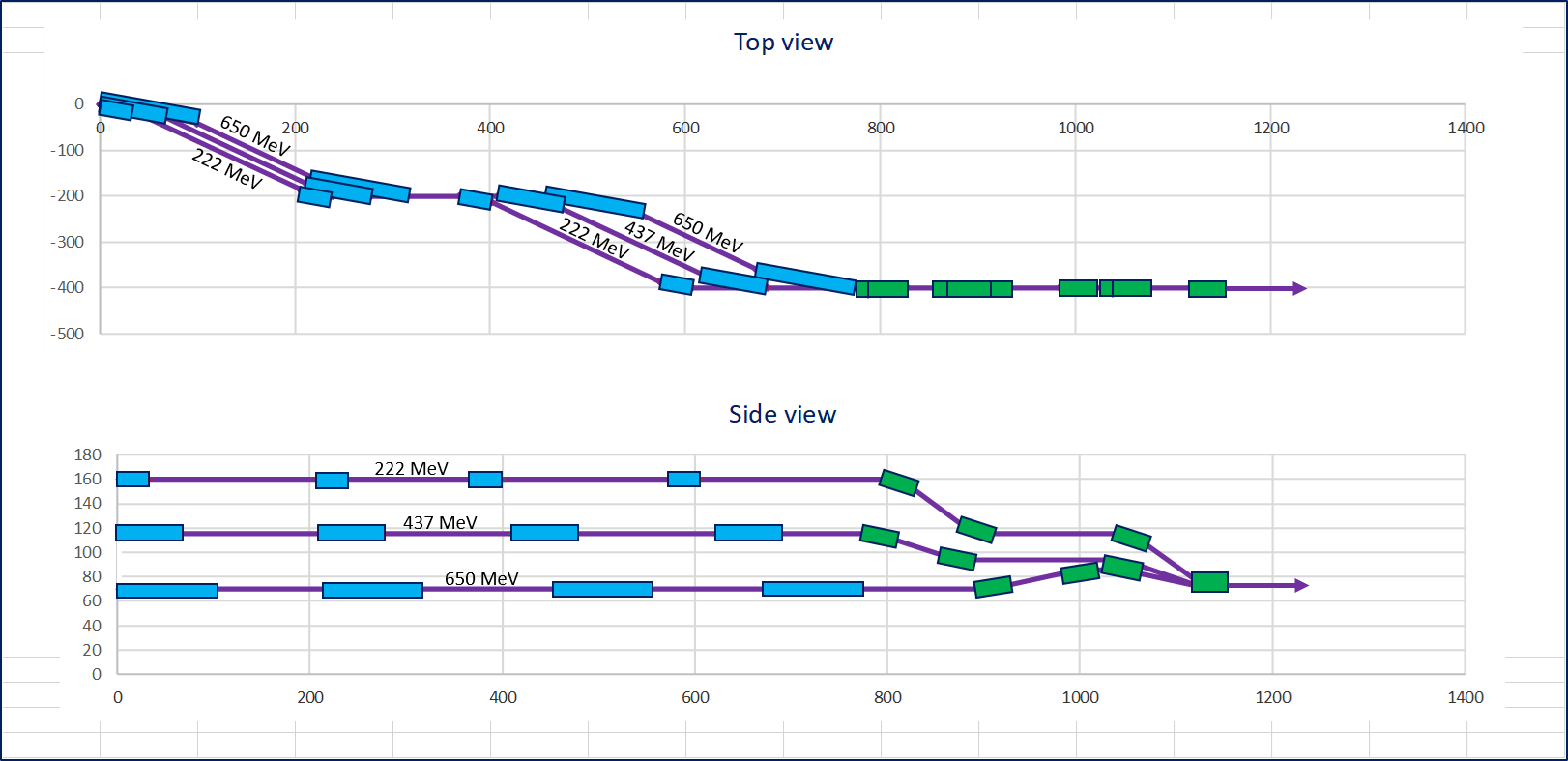}
  \caption{Individual extraction lines with \SI{33}{cm}, \SI{66}{cm},
           and \SI{99}{cm} rectangular bends (blue) for the three energy
           passes. The vertical recombiner replica uses \SI{33}{cm}
           rectangular bends (green).}
  \label{fig:extraction_magnets}
\end{figure}

\begin{figure}[!htb]
  \centering
  \includegraphics[width=\columnwidth]{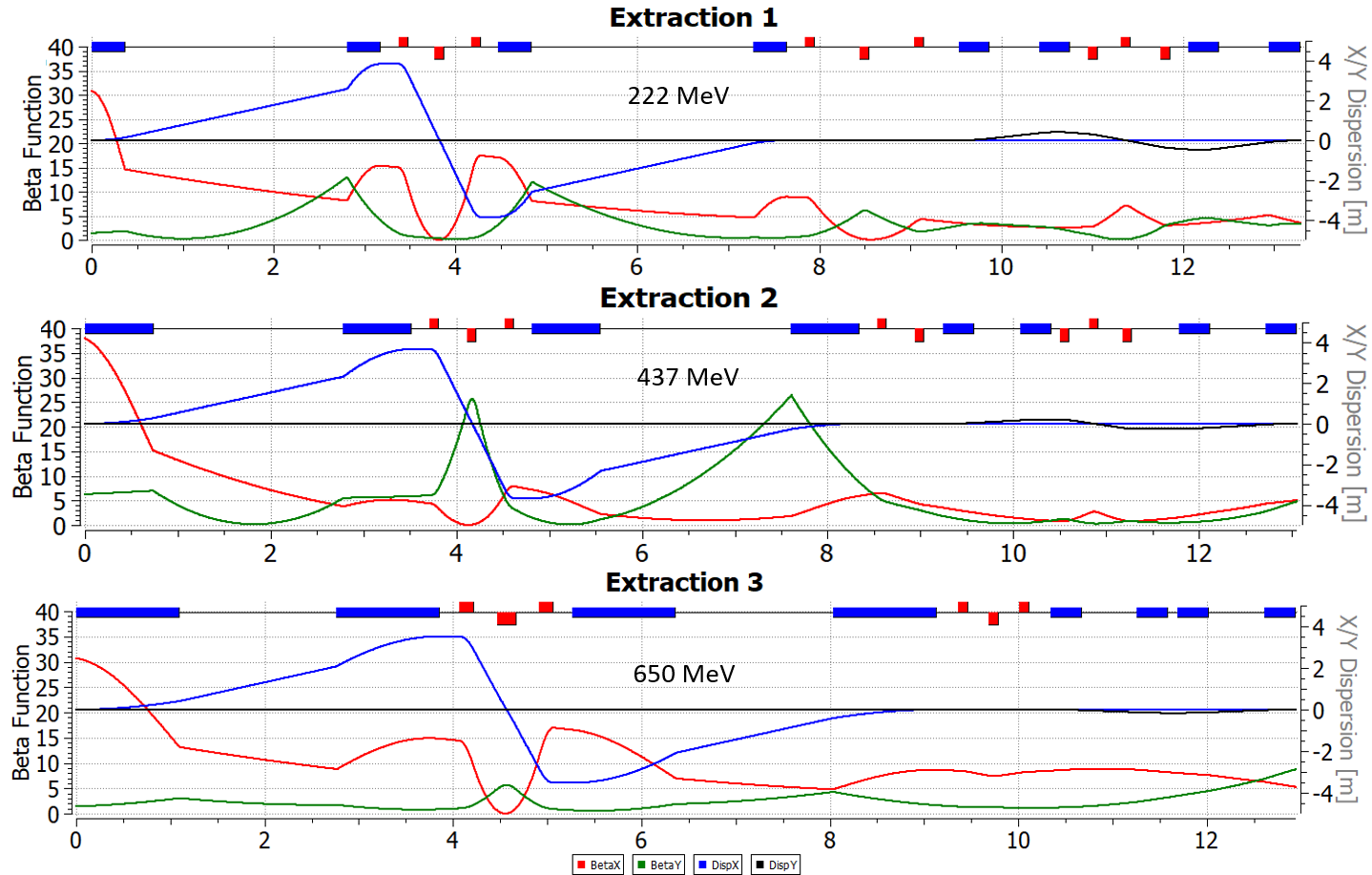}
  \caption{Optics of the individual extraction lines, featuring a 2-step
           horizontal jog achromat followed by a replica of the vertical
           recombiner, as in the arcs.}
  \label{fig:extraction_optics}
\end{figure}

Figures~\ref{fig:extraction_magnets} and \ref{fig:extraction_optics}
show detailed magnet layouts and optics of the extraction lines.
The 1st-, 2nd-, and 3rd-pass lines use \SI{33}{cm}, \SI{66}{cm}, and
\SI{99}{cm} rectangular bends, respectively, while the vertical
recombiner replica employs \SI{33}{cm} rectangular bends throughout.

In summary, a compact injector within the LERF vault would serve both the \SI{22}{GeV} CEBAF \cite{cebaf22gev} and the Ce$^+$BAF \cite{positron} programs. The injector complex is arranged in a racetrack configuration hosting three C-75 cryo-modules. Starting from an \SI{8}{MeV} merger, a final energy of \SI{650}{MeV} is reached in three recirculation passes.
The injector offers flexible extraction at 1st, 2nd pass
energies, as required for positron production and 3rd pass for the \SI{22}{GeV} CEBAF injector.
The multi-pass Linac optics is configured as a drift Linac with no quadrupoles between cryo-modules, yielding quasi-parabolic Twiss functions that are nearly identical for both higher passes. All five $180^\circ$ horizontal arcs employ FMC optics, enabling independent adjustment of the momentum compaction factor in each arc. The presented
arc optics architecture offers a high degree of modular functionality for momentum compaction management as well as orthogonal tunability of beta functions and dispersion.
A comprehensive beam dynamics validation program is underway, including the orbit correction scheme and start-to-end tracking with lattice misalignment and magnet errors.

\subsubsection{LERF to North Linac transfer line}

The LERF-to-North-Linac (NL) transport line is a shared infrastructure serving two distinct upgrade programs at CEBAF: the Ce\textsuperscript{+}BAF polarized positron and the FFA@CEBAF 22\,GeV energy upgrade. In both cases the line connects the Low Energy Recirculator Facility (LERF) building to the CEBAF North Linac, and its design is governed by the requirement not to interfere with the existing CEBAF electron beam delivery.

In the Ce\textsuperscript{+}BAF positron mode, a dedicated electron driver impinges on a high-$Z$ converter target at LERF, and the resulting positrons are captured and accelerated to 123\,MeV before being injected into the North Linac for further acceleration to 12\,GeV. Because a secondary positron beam is inherently diffuse, the optics of this line were originally optimized to accept and transport a beam with large transverse and longitudinal emittances at the relatively low injection energy of 123\,MeV \cite{salim_positron2026}. 

In the FFA@CEBAF electron mode, the same beamline transports a primary electron beam at 650\,MeV. Electrons are produced with substantially smaller transverse and longitudinal emittances, and the higher rigidity of the 650\,MeV beam eases many of the aperture and focusing constraints that govern the positron case. The two operational modes are made compatible through bipolar power supplies on all dipole and quadrupole magnets, allowing the magnetic fields—and therefore the beam optics—to be scaled between the two energies while reusing the same physical hardware. At the handover point into the North Linac the transport line is designed to match the required Twiss parameters independently for each mode, while simultaneously suppressing the horizontal and vertical dispersion functions and their derivatives to zero.

\begin{figure*}[!tbh]
    \centering
    \includegraphics*[width=\textwidth]{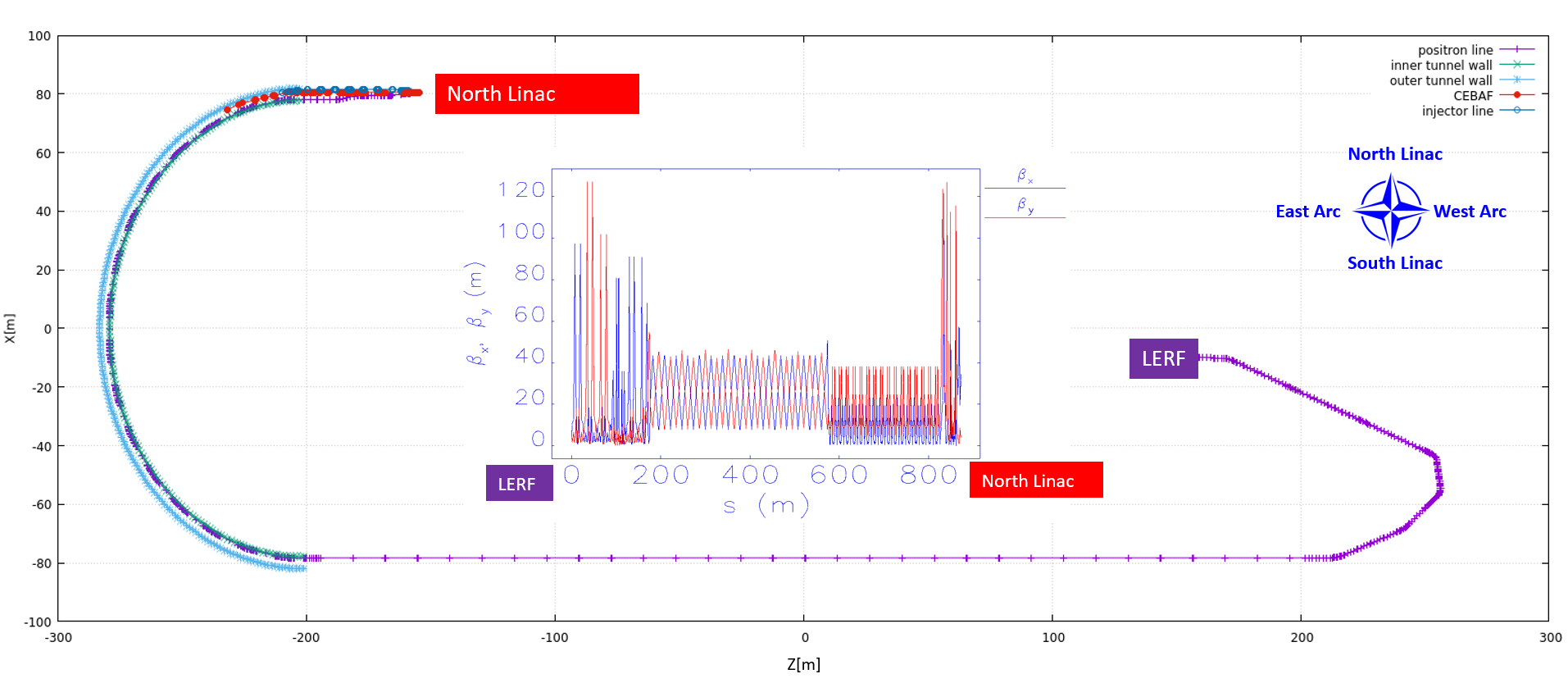}
    \caption{Layout of the shared LERF-to-North-Linac transport line shown in purple, with hyphens indicating accelerator element locations. Actual coordinates of the designed beamline and existing tunnel walls are shown in cyan and light green, respectively. The current CEBAF beamline is indicated in red. The inset displays the beam optics with the transverse $\beta$-functions.}
    \label{fig:transport_layout}
\end{figure*}

The layout, generated from the beam optics survey, is given in Figure~\ref{fig:transport_layout}.

\begin{figure*}[!htb]
    \centering
    \includegraphics[width=\textwidth]{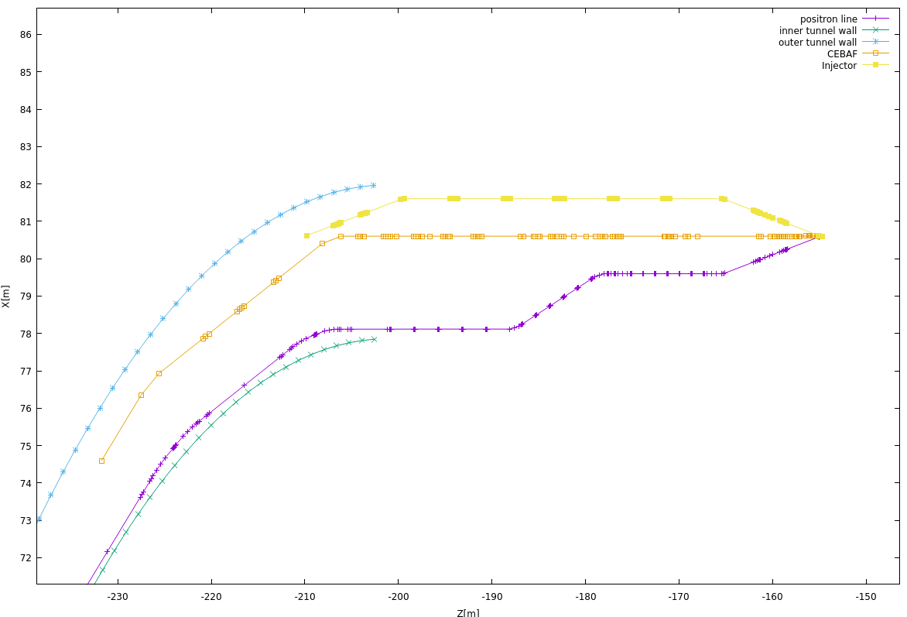}
    \caption{The merger section of the shared transport line (purple) with the existing CEBAF accelerator (orange) and the existing electron injector (yellow). Inner and outer tunnel walls are shown in green and cyan, respectively.}
    \label{fig:merger}
\end{figure*}

\begin{figure}[!htb]
    \centering
    \includegraphics[width=1\linewidth]{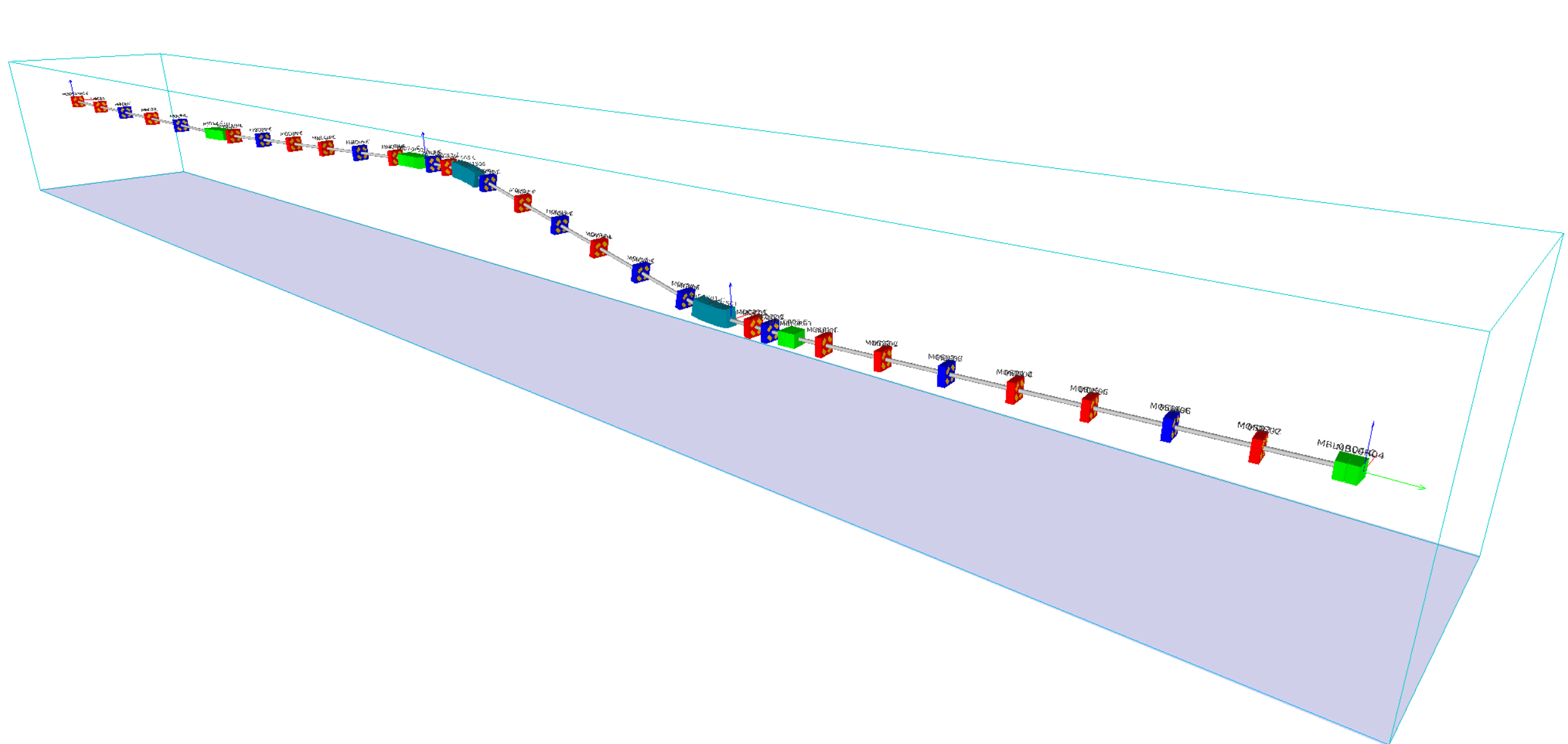}
    \caption{3-D view of the shared transport line descending from the CEBAF tunnel ceiling and bending toward the North Linac. Red and blue rectangles represent focusing and defocusing quadrupoles, cyan rectangles are vertical dipoles, and green rectangles are horizontal dipoles.}
    \label{fig:3D-merger}
\end{figure}

The beam optics design routes the transport beamline along the CEBAF tunnel ceiling, suspended from the inner wall opposite the accelerator on the outer wall. The line merges into the North Linac by mirroring the geometry of the last two dipoles of the operational CEBAF electron injector chicane. The zoomed view of the conjunction of the transport line and CEBAF is presented in Figure~\ref{fig:merger}. The line descends from the ceiling and joins the existing injector chicane at dipole \textit{MBL0R04}, maintaining sufficient clearance for tunnel access during cart operations. A 3-D optics plot of the merging section, generated with \textsc{Elegant}~\cite{elegant}, is shown in Figure~\ref{fig:3D-merger}.

The optics design of the shared transport line represents a practical compromise that links the LERF building, the existing tunnel, and the CEBAF North Linac without requiring tunnel enlargement or significant structural modifications, while providing efficient transport for both the large-emittance 123\,MeV positron beam and the lower-emittance 650\,MeV electron beam. To satisfy these constraints, the transverse beta functions are held below 130\,m and the horizontal dispersion within $\pm0.3$\,m throughout the line. Because the magnet excitations must be scalable between the two beam rigidities via bipolar power supplies, the lattice was designed so that a single family of optics solutions—differing only in magnet strengths—satisfies the Twiss matching and dispersion closure conditions at the North Linac for each mode independently.

The \textsc{Elegant} optics file is available in Refs.~\cite{github1,github2}. The following subsections describe each physical section of the line.

\paragraph{LERF to South Transport Line}

\begin{figure}[!htb]
    \centering
    \includegraphics[width=\linewidth]{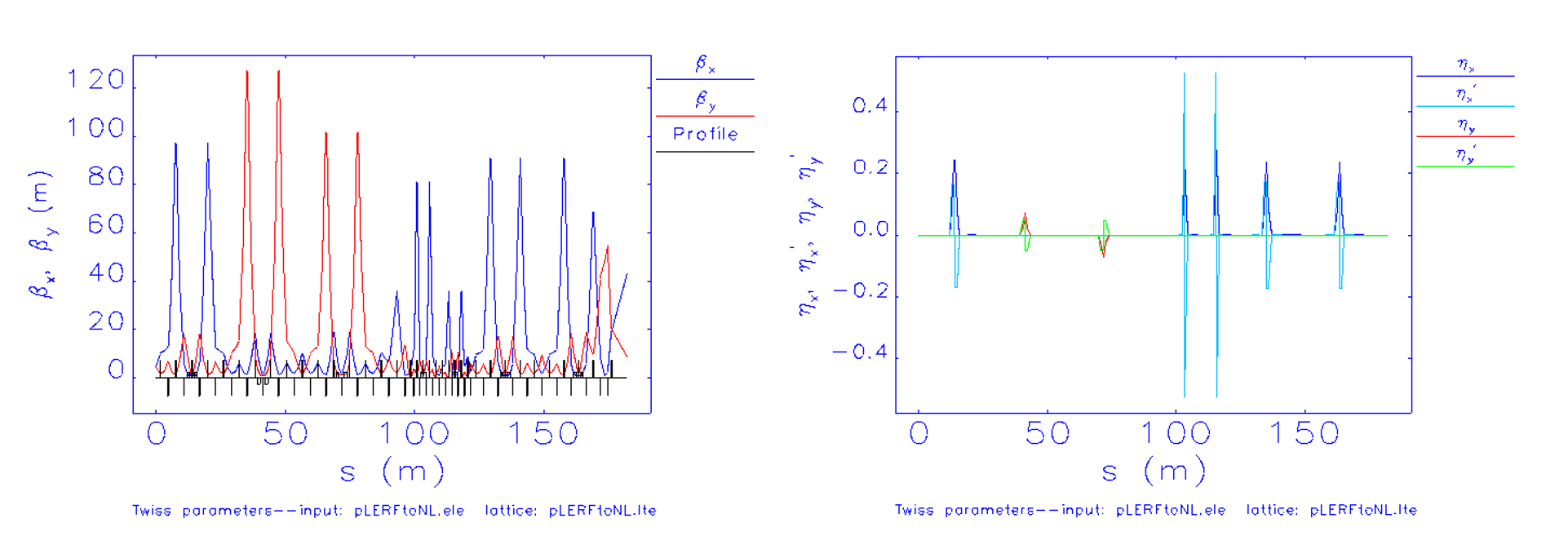}
    \caption{LERF to South Transfer Line optics. Beta functions are shown on the left; dispersion functions and their derivatives on the right. Optics maxima for the beta and dispersion functions are maintained throughout the transport line.}
    \label{fig:lerf_to_stl}
\end{figure}

The LERF enclosure sits at a higher elevation than the tunnel, so the beamline incorporates both horizontal and vertical bending elements in this initial segment. The resulting dispersions and their derivatives are properly matched and closed throughout, as illustrated in Figure~\ref{fig:lerf_to_stl}.

\paragraph{South Transport Line}

\begin{figure}[!htb]
    \centering
    \includegraphics[width=0.7\linewidth]{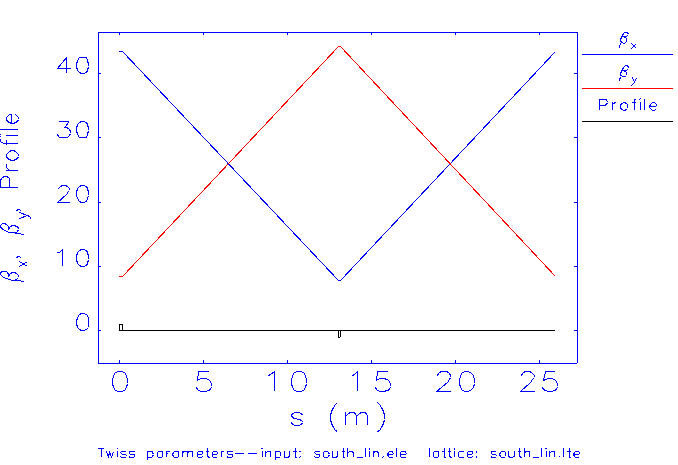}
    \caption{The FODO cell designed for the south transport line.}
    \label{fig:south_lin}
\end{figure}

Following the optics layout shown in Figure~\ref{fig:lerf_to_stl}, a FODO lattice was designed with constrained beta functions, as illustrated in Figure~\ref{fig:south_lin}. The design comprises 14 FODO cells, optimized to minimize the number of quadrupoles and thereby reduce overall project cost. Despite this optimization, the trace of the cell transfer matrix remains below 2, ensuring stable beam transport within the FODO structure for both the 123\,MeV positron and 650\,MeV electron operating points.

\paragraph{Double Bend Achromat for the West Transport Line}

After the south transport line, the beamline executes a 180-degree turn running parallel to the CEBAF west arcs (see Figure~\ref{fig:transport_layout}). To accommodate this geometry while maintaining compact $\beta$ and dispersion functions, the arc is implemented using Double-Bend Achromat (DBA) cells.

The transfer matrix between the two dipoles can be written in terms of Twiss parameters as
\[
M =
\begin{pmatrix}
\cos\mu + \alpha \sin\mu & \beta \sin\mu \\[6pt]
-\gamma \sin\mu & \cos\mu - \alpha \sin\mu
\end{pmatrix},
\]
where $\mu$ is the horizontal betatron phase advance and $\alpha, \beta, \gamma$ are the Twiss parameters with $\gamma = (1+\alpha^2)/\beta$.

\begin{figure}[!htb]
    \centering
    \includegraphics[width=0.7\linewidth]{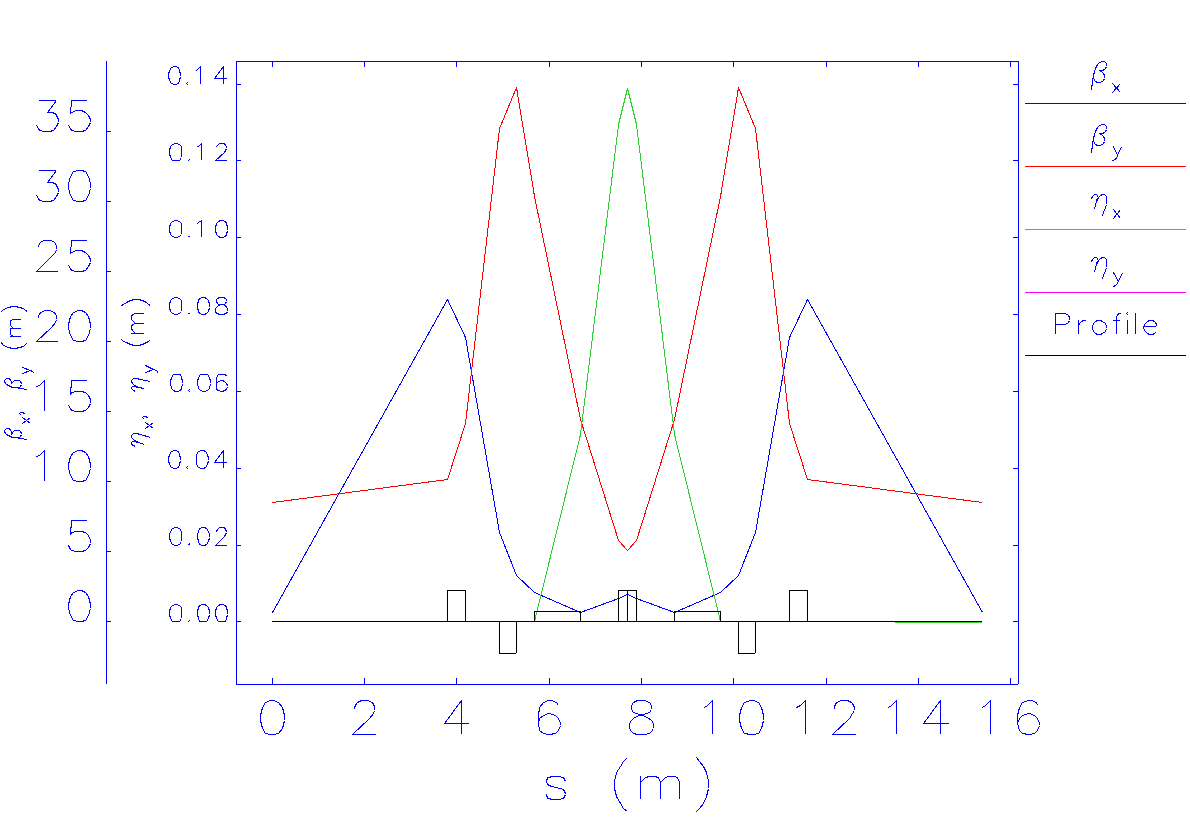}
    \caption{Double Bend Achromat cell designed for the $\pi$-turn in the transport line linking the south to the north of CEBAF.}
    \label{fig:DBA}
\end{figure}

The achromat condition requires
\[
m_{12} = 0, \qquad m_{22} = -1.
\]
From the matrix elements,
\[
m_{12} = \beta \sin\mu = 0 \quad \Rightarrow \quad \mu = n\pi,
\]
and
\[
m_{22} = \cos\mu - \alpha \sin\mu = \cos(n\pi) = (-1)^n = -1.
\]
Thus $n$ must be odd, giving the achromat condition $\mu = \pi \pmod{2\pi}$. At the solution point the transport matrix takes the form
\[
M =
\begin{pmatrix}
-1 & 0 \\[4pt]
m_{21} & -1
\end{pmatrix},
\]
which is the required form for horizontal dispersion cancellation in a DBA. Using \textsc{Elegant}, the horizontal transfer matrix between the centers of the two dipoles satisfies $m_{12} \approx 0$ and $m_{22} \approx -1$, corresponding to a horizontal phase advance $\mu_x \approx 3.135 \approx \pi$\,rad, confirming that the cell satisfies the standard DBA achromat condition. This condition is preserved under magnet strength scaling between the two operating energies, so the same cell geometry serves both the positron and electron modes. The individual DBA cell is illustrated in Figure~\ref{fig:DBA}.

\paragraph{Merging into the North Linac}

The West Transport Line terminates in a merger section leading into the North Linac. Suspended from the ceiling and running close to the tunnel's inner wall, the line is guided outward: the merger bends toward the outer wall while still suspended, then gradually descends to the Linac level. The final segment is designed to be symmetric with the last two dipoles of the existing injector chicane, ensuring sufficient separation to accommodate the new transport-line magnets while reusing the existing final chicane dipole (\textit{MBL0R04}), without requiring any physical modification to the CEBAF machine.

This symmetric geometry allows the beam—whether positrons at 123\,MeV or electrons at 650\,MeV—to be injected into the North Linac by scaling the magnetic field of \textit{MBLR04} rather than changing its mechanical configuration. To enable operation at both energies, \textit{MBLR04} is equipped with a bipolar power supply and rated for the higher 650\,MeV electron rigidity.

\begin{figure}[!htb]
    \centering
    \includegraphics[width=0.7\linewidth]{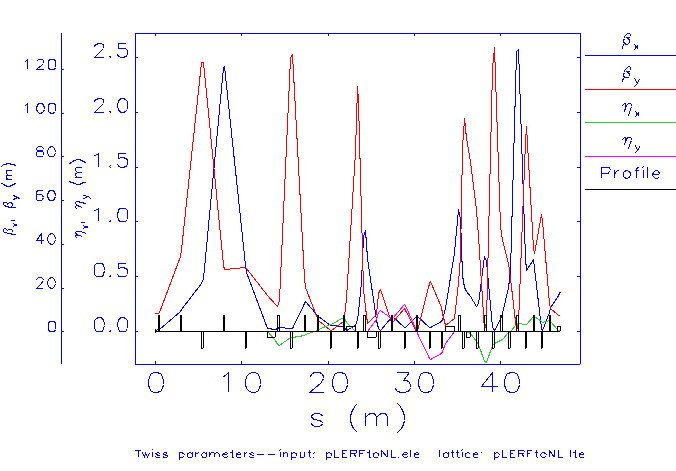}
    \caption{West Transfer Line to North Linac merger optics (123\,MeV positron mode shown). Non-zero horizontal and vertical dispersions arising from the beamline geometry are visible mid-section and are fully suppressed at the end of the line. The 650\,MeV electron mode uses the same lattice with strengths scaled to the higher rigidity.}
    \label{fig:Linac_merger_optics}
\end{figure}

Beyond dispersion closure, the merger must deliver a full Twiss match to the North Linac acceptance at both operating energies. The optics solutions for the two modes differ in magnet excitation but share the same physical apertures and element positions, so the hardware installed for Ce\textsuperscript{+}BAF positron injection serves equally as the FFA@CEBAF electron injection path. In both cases the horizontal and vertical dispersions and their derivatives are suppressed to zero at the North Linac handover point. The positron-mode merger optics are presented in Figure~\ref{fig:Linac_merger_optics}; the electron-mode solution follows the same lattice structure with strengths scaled to 650\,MeV.

The LERF-to-NL transport line is designed around two primary objectives: fitting within the existing tunnel infrastructure without interfering with CEBAF operations, and providing efficient beam transport across both operating modes. In the Ce\textsuperscript{+}BAF positron mode, the dominant challenge is the large transverse and longitudinal emittance of the secondary beam at 123\,MeV. In the FFA@CEBAF electron mode at 650\,MeV, the beam emittance is substantially smaller, but the higher magnetic rigidity places more stringent demands on magnet excitation and power supply range.

\begin{figure}[!htb]
    \centering
    \includegraphics[width=1\linewidth]{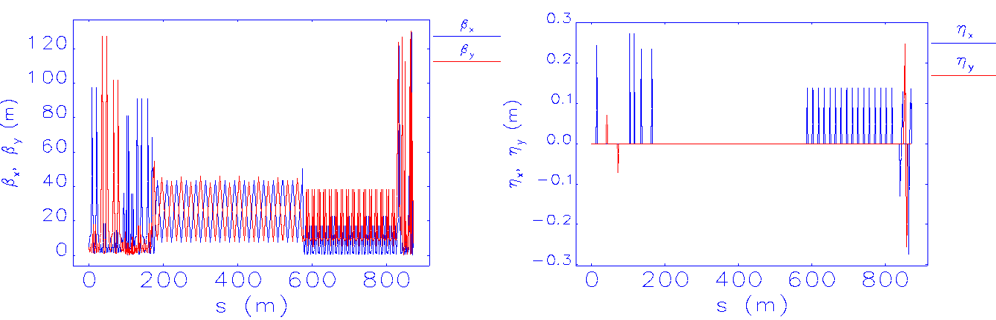}
    \caption{Beta and dispersion functions throughout the full LERF-to-North-Linac transport line.}
    \label{fig:overall_optics}
\end{figure}

Both challenges are addressed through a combination of FODO lattices in the straight sections and DBA cells in the arc, which together keep the beta and dispersion functions within the constraints established in Section~2. The complete optics for the LERF-to-NL transport line are illustrated in Figure~\ref{fig:overall_optics}.

\begin{figure}[!htb]
    \centering
    \includegraphics[width=0.6\linewidth]{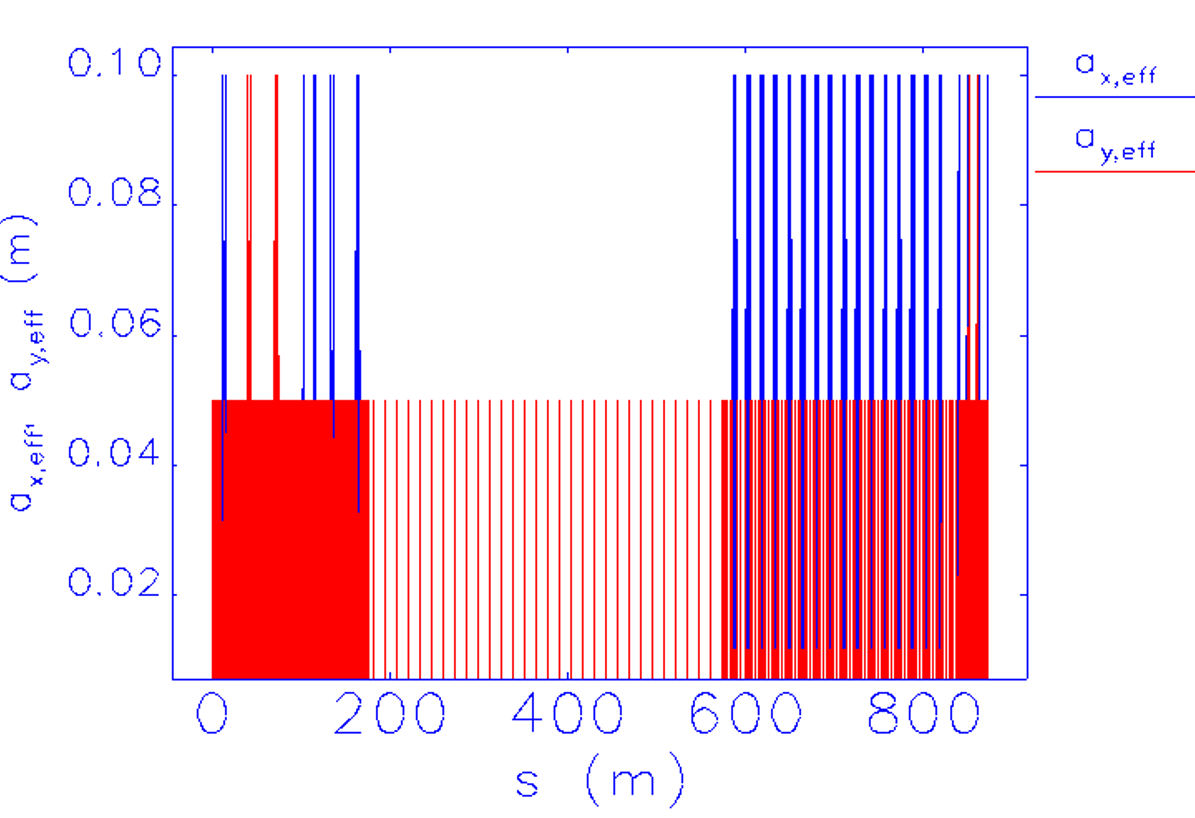}
    \caption{Aperture sizes of all LERF-to-NL magnets plotted throughout the transport line.}
    \label{fig:bores}
\end{figure}

Dipoles and quadrupoles are based on standard CEBAF sector bends and normal-conducting quadrupoles, with all magnets equipped with bipolar power supplies to support operation at 123\,MeV (positron mode) and 650\,MeV (electron mode). The sector dipoles have an effective magnetic length of approximately 1\,m; all quadrupoles are 15\,cm long. Magnet designs are documented in Refs.~\cite{benesch1,benesch2,benesch3}. The corresponding magnet bore sizes are summarized in Table~\ref{tab:bores} and illustrated in Figure~\ref{fig:bores}.

\begin{table}[!htb]
   \centering
   \caption{LERF-to-NL magnets and their aperture sizes.}
   \begin{ruledtabular}
\begin{tabular}{lcc}
       \textbf{Magnet Type} & \textbf{Horizontal Bore Radius [m]} & \textbf{Vertical Bore Radius [m]} \\
       \colrule
       Horizontal Bend & 0.1    & 0.0115 \\
       Vertical Bend   & 0.0115 & 0.1    \\
       Quadrupole      & 0.05   & 0.05   \\
   \end{tabular}
\end{ruledtabular}
   \label{tab:bores}
\end{table}

\begin{figure}[!htb]
    \centering
    \includegraphics[width=0.6\linewidth]{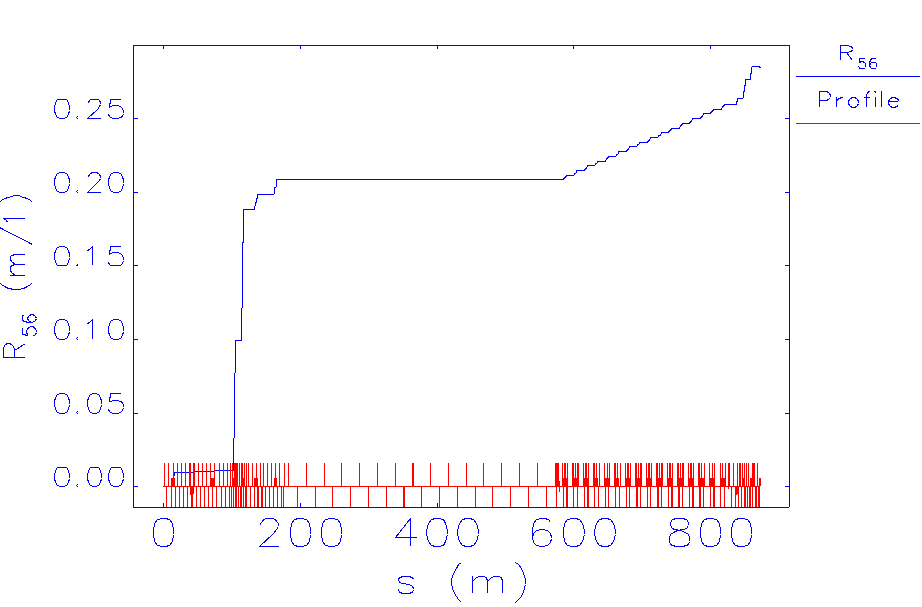}
    \caption{$R_{56}$ evolution through the LERF-to-NL transport line.}
    \label{fig:R56}
\end{figure}

In addition to the large transverse emittance inherent to the positron mode, the longitudinal emittance is also significant at 123\,MeV. Accordingly, the evolution of the $R_{56}$ element of the transport matrix along the beamline is shown in Figure~\ref{fig:R56}. At the injection point into the North Linac, $R_{56} = 0.284$\,m. Alternative optics solutions have been developed that yield smaller, and even negative, values of $R_{56}$ to enable bunch length compression where needed. For the 650\,MeV electron mode the longitudinal dynamics are less critical owing to the smaller energy spread and shorter bunch length of the primary beam, but the $R_{56}$ profile can be adjusted via the same bipolar magnet scaling. The final $R_{56}$ configuration for each mode will be determined as the project matures, guided by the longitudinal characteristics of the upstream positron beam~\cite{ushakov} and by CEBAF acceptance measurements~\cite{amy}.

\FloatBarrier

\subsubsection{North and South Linacs - Multi-pass optics architecture}
\begin{figure}[!htb]
  \centering
  \includegraphics[width=\columnwidth]{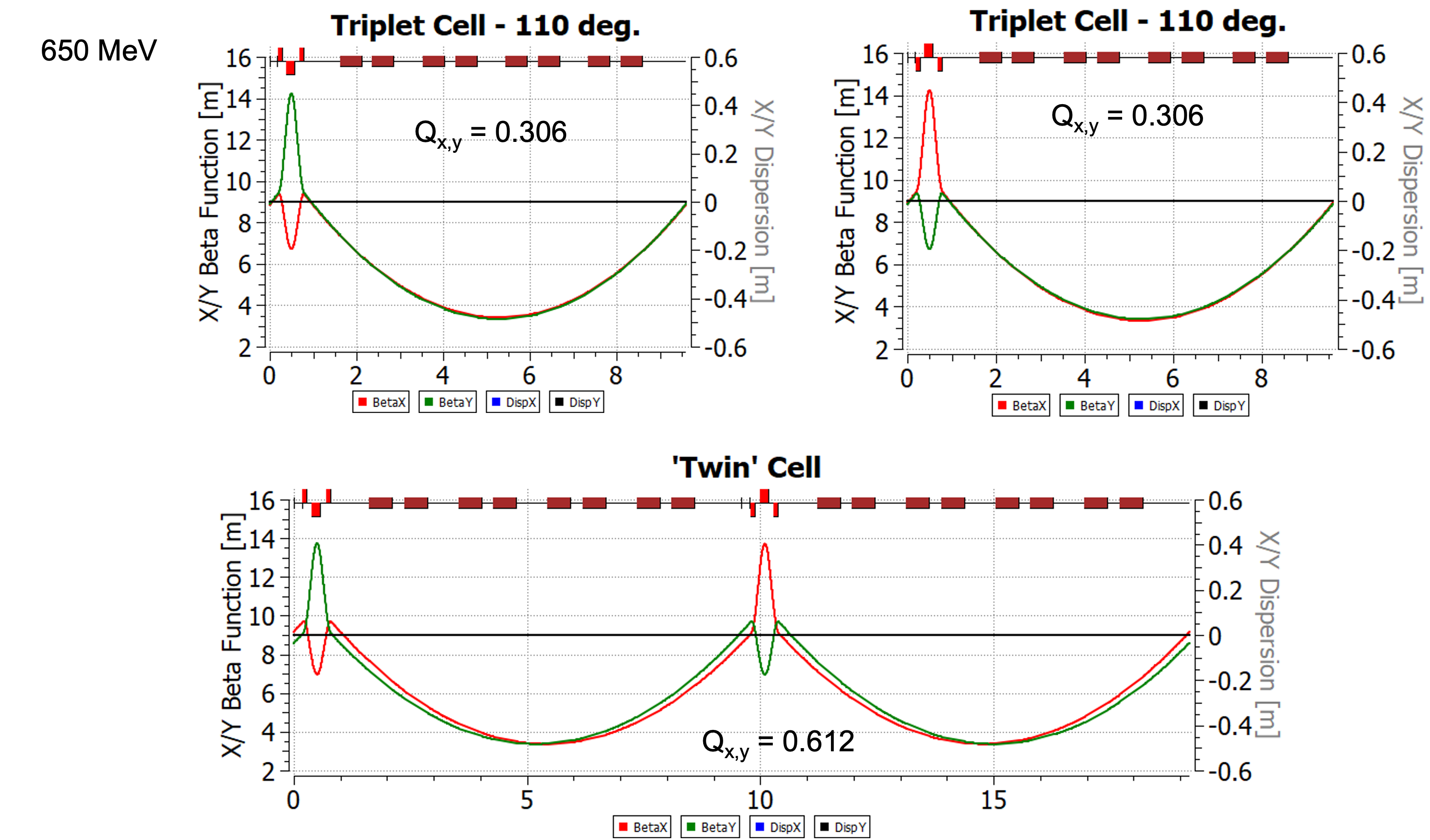}
  \caption{'Twin Cell' based on Strongly focusing alternating triplets }
  \label{fig:twin_cell}
\end{figure}

One of the challenges of the multi-pass Linac optics is to provide uniform focusing in a vast range of energies, using a fixed field lattice. 
The current CEBAF is configured with a 123 MeV injector feeding into a racetrack Recirculating Linear Accelerator (RLA) with a 1.1 GeV Linac on each side. Increasing number of Linac passes to 10-11 makes optical matching virtually impossible due to extremely high energy span ratio (1:175) at the beginning of the North Linac. The proposed new building block of the multi-pass Linac optics (to replace current FODO structure) are envisioned as a sequence of strongly focusing triplet cells (110 deg. phase advance per cell) with an alternating triplet polarity, as illustrated in Figures~\ref{fig:twin_cell} 

\begin{figure}[!htb]
  \centering
  \includegraphics[width=\columnwidth]{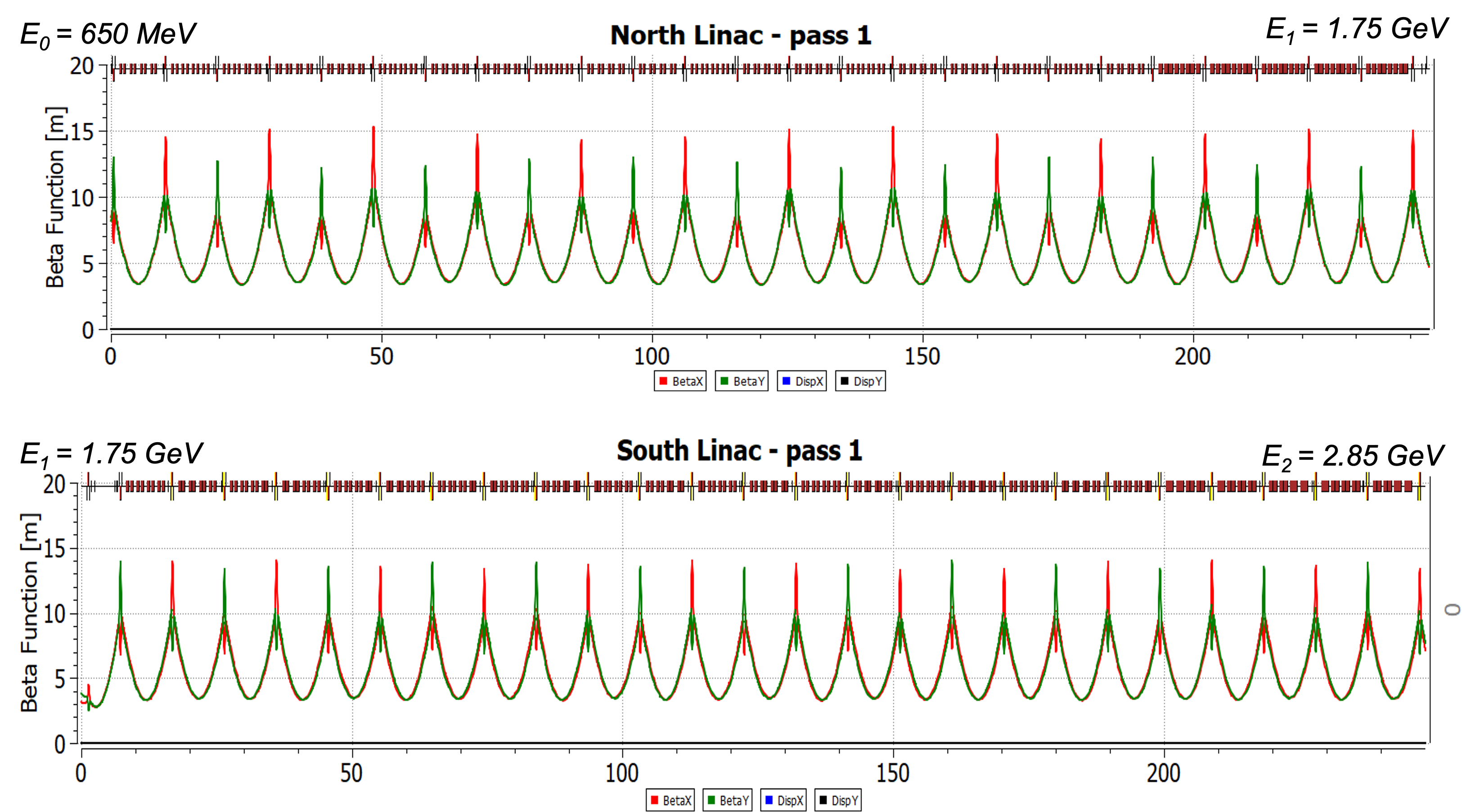}
  \caption{Periodic 1-st pass Linac optics based on 'Twin Cells' for both the North and South Linacs }
  \label{fig:NL_SL}
\end{figure}

This last feature greatly enhances transverse beam confinement and assures uniform phase advance evolution for higher Linac passes, as the alternating triplets ‘morph’ into singlets, so the Linac optics transitions into a FODO-like style, as the energy increases. Initial triplet cell at the Linac beginning is based on \SI{20}{T/m}
quads, with each triplet being flanked by two cryo-modules. Complete optics for both the NL and SL are illustrated in Figures~\ref{fig:NL_SL}

\begin{figure}[!htb]
  \centering
  \includegraphics[width=\columnwidth]{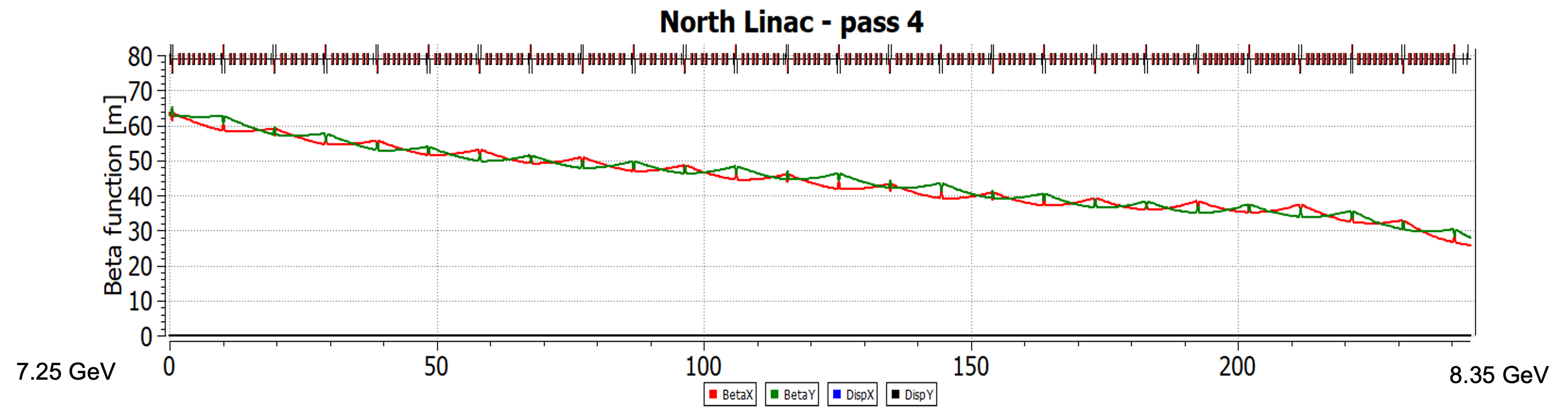}
  \caption{Quasi-periodic optics for 4-th pass in the North Linac }
  \label{fig:NL_4}
\end{figure}

Subsequent triplet strengths will scale with increasing momentum along the Linac, so that it allows one to maintain quasi-periodic optics for the lower four passes, going into conventional electromagnetic arcs. Figure~\ref{fig:NL_4} illustrates such quasi-periodic optics for the 4-th pass in the North Linac, features very low $\beta$ functions across the Linac

\begin{figure}[!htb]
  \centering
  \includegraphics[width=\columnwidth]{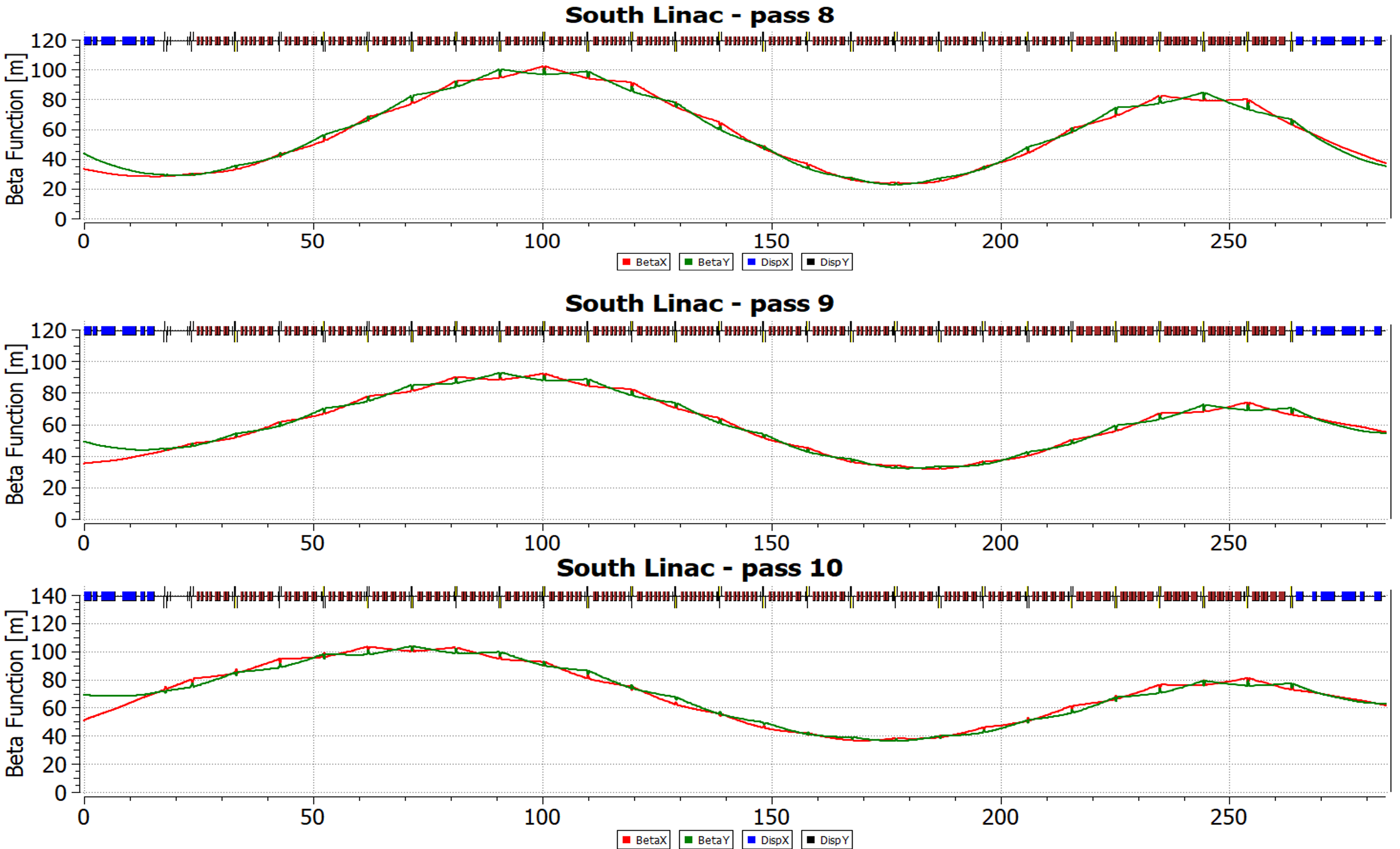}
  \caption{SL optics for passes 8, 9 and 10 featuring by-design $\beta$ beating nodes at the Linac ends. }
  \label{fig:SL_top}
\end{figure}

At higher passes (5-11), directed into the FFA arcs, the Linac optics depart by design from being periodic. Our design goal is to induce an optimized $\beta$ beat across the Linac to achieve reasonably small $\beta$’s at both Linac ends at the expense of an increase in the maximum $\beta$ inside the Linacs ($\beta$ beat creates ‘nodes’ at Linac ends). Optics for the final passes through the South Linac are illustrated in Figures~\ref{fig:SL_top}.

This feature greatly relaxed matching requirements into the FFA transition to arcs by significantly reducing $\beta$’s, and adjusting $\alpha$’s to nearly zero as favored by FFA matching. This style Linac focusing provides stable multi-pass optics compatible with much smaller beta functions in the FFA arcs and it is capable of covering energy ratio of 1:33. The feasibility of the above Linac design sets the minimum injection energy at 650 MeV. Therefore, it is proposed to replace the old 123 MeV injector with a 650 MeV, 3-pass recirculating, injector based on the existing LERF facility augmented by three C-75 cryo-modules. The beam is then transferred from the LERF vault through a dedicated fixed energy 650 MeV transport line and injected into the North Linac.

\subsubsection{Spreaders/Recombiners}
\label{Spreaders}

The proposed FFA@CEBAF energy upgrade, which aims to double the machine's final energy to approximately \SI{22}{GeV} \cite{Deitrick:IPAC23-MOPL182}, necessitates a significant redesign of the Spreader and recombiner systems. The Spreaders are the vertically-bending sections that separate the beams of different energies after each Linac pass and direct them into their respective recirculation arcs \cite{Chao:JLAB-TN-07-016}. Recombiners are beamlines which bring the separate passes back together in a mirror-symmetric manner to the Spreaders.

\paragraph{Motivation for Redesign}

The primary driver for the redesign is the change in injection energy, which increases from \SI{123}{MeV} to \SI{650}{MeV} \cite{Bodenstein:IPAC22-THPOST023,Deitrick:IPAC23-MOPL182}. This change alters the energy ratios of all subsequent passes as they enter the Spreaders, making the current optics solution unusable. Furthermore, magnets in the existing Spreaders must be upgraded to handle the higher beam energies.

Table \ref{tab:Spreader_Energies} shows the new nominal energies for all 20 passes entering the Northeast (odd passes) and Southwest (even passes) Spreaders. The first eight passes (1-8) are directed to traditional electromagnetic (EM) arcs, while the final 12 passes (9-20, shown in \textbf{BOLD}) are directed into the new Fixed-Field Alternating Gradient (FFA) arcs.
\begin{table}[!htb]
   \centering
   \caption{Energies Entering Each Spreader.}
   \begin{ruledtabular}
\begin{tabular}{cccc}
       \multicolumn{2}{c}{\textbf{Northeast}} & \multicolumn{2}{c}{\textbf{Southwest}} \\
       \cline{1-2} \cline{3-4}
       \textbf{Pass Number} & \textbf{Energy (\SI{}{GeV})} & \textbf{Pass Number} & \textbf{Energy (\SI{}{GeV})} \\
       \colrule
       1  & 1.750  & 2  & 2.850  \\
       3  & 3.950  & 4  & 5.050  \\
       5  & 6.150  & 6  & 7.250  \\
       7  & 8.350  & 8  & 9.450  \\
       \textbf{9}  & \textbf{10.550} & \textbf{10} & \textbf{11.650} \\
       \textbf{11} & \textbf{12.750} & \textbf{12} & \textbf{13.850} \\
       \textbf{13} & \textbf{14.950} & \textbf{14} & \textbf{16.050} \\
       \textbf{15} & \textbf{17.150} & \textbf{16} & \textbf{18.250} \\
       \textbf{17} & \textbf{19.350} & \textbf{18} & \textbf{20.450} \\
       \textbf{19} & \textbf{21.550} & \textbf{20} & \textbf{22.650} \\
   \end{tabular}
\end{ruledtabular}
   \label{tab:Spreader_Energies}
\end{table}
\FloatBarrier

\paragraph{New Spreader Design}

Upon entering a Spreader, all passes are in a common line. They are first separated vertically by a common dipole magnet. The existing EM passes (1-8) are sent into a two-step elevation change designed to cancel vertical dispersion.

The new design must accommodate the final six FFA passes (per Spreader) as well. Due to their higher beam rigidity, these passes are bent less by the common dipoles. They share a common septum with the final EM pass (pass 7 or 8). After being separated from the EM passes, the six FFA passes are sent through a series of reverse bends. This mirror-symmetric chicane brings the FFA passes back to the original Linac height and cancels their dispersion before they are recombined into the common FFA arc.

\begin{figure*}[!htb]
    \centering
    \includegraphics[width=\textwidth]{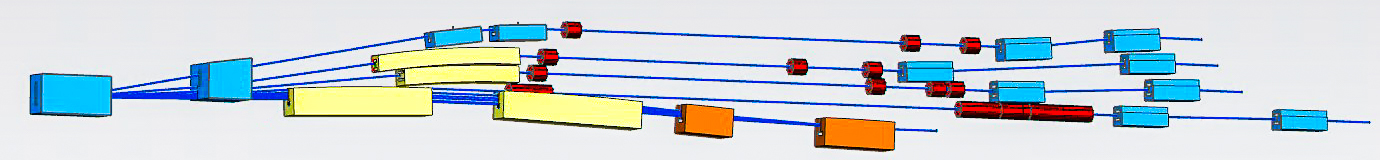}
    \caption{A 3D model of one version of the new Southwest Spreader. Ten passes enter from the left and are separated vertically. The EM passes are separated into independent arcs, while the FFA passes are recombined into an FFA arc. Blue, yellow, and orange blocks are dipoles or septa; red blocks are quadrupoles. Model by K. Tremblay \cite{Tremblay:PC}.}
    \label{fig:SW_Spreader}
\end{figure*}

\begin{figure}[!hbt]
    \centering
    \includegraphics[width=0.49\textwidth]{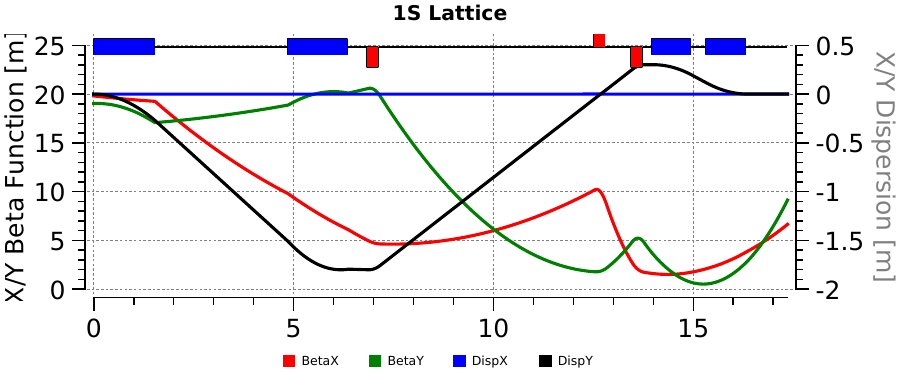}
    \includegraphics[width=0.49\textwidth]{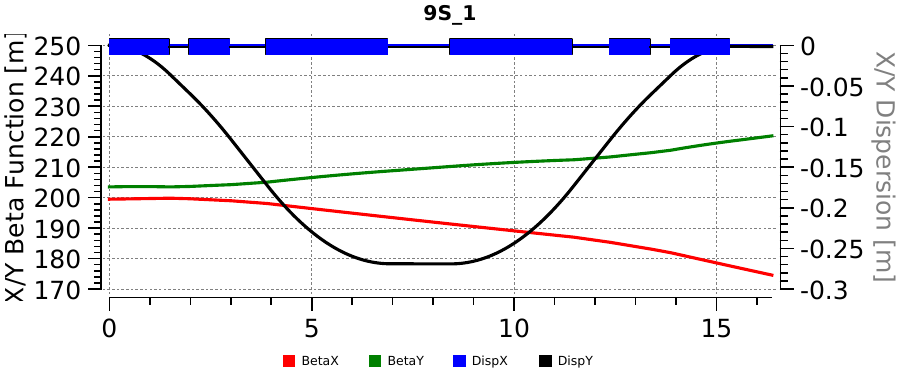}
    \includegraphics[width=0.49\textwidth]{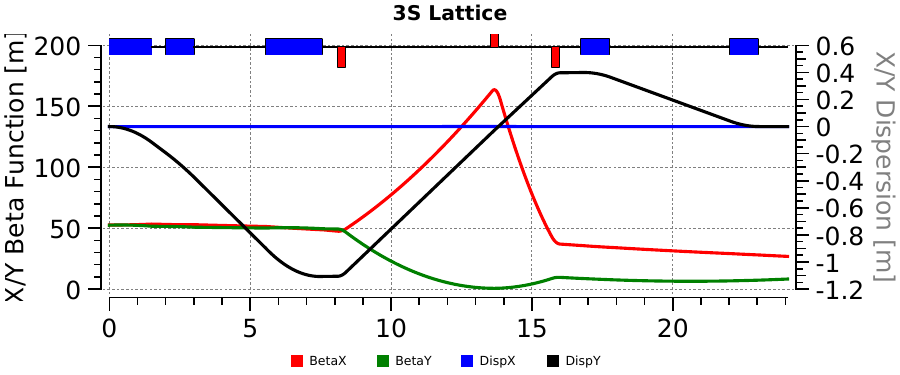}
    \includegraphics[width=0.49\textwidth]{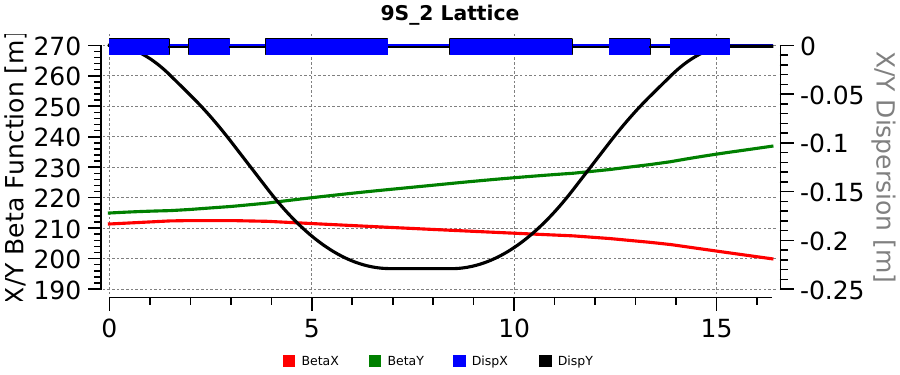}
    \includegraphics[width=0.49\textwidth]{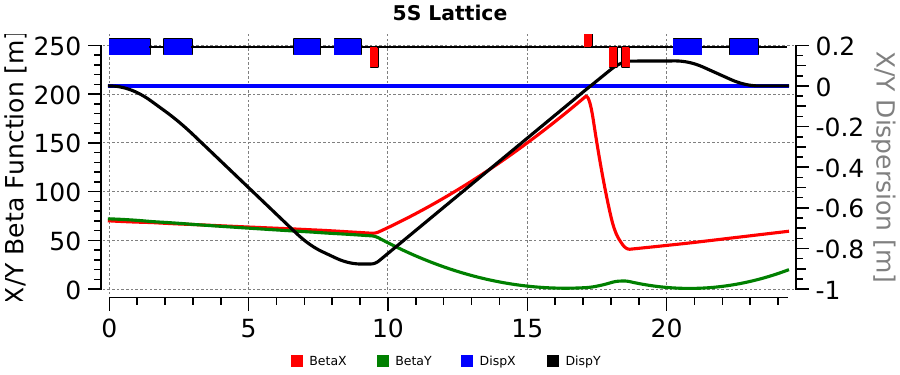}
    \includegraphics[width=0.49\textwidth]{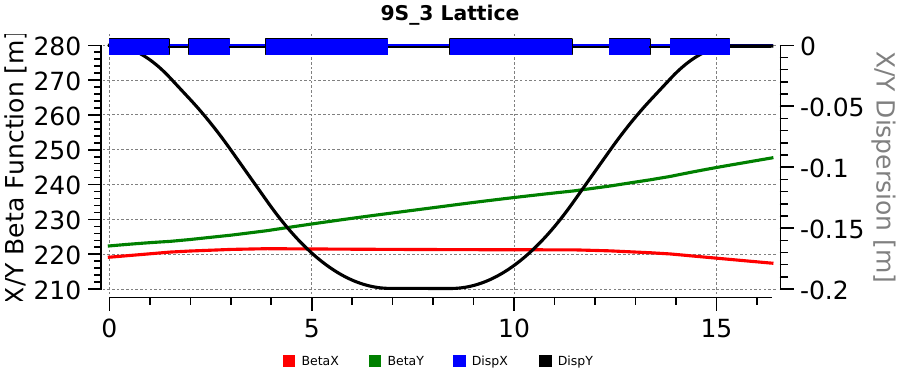}
    \includegraphics[width=0.49\textwidth]{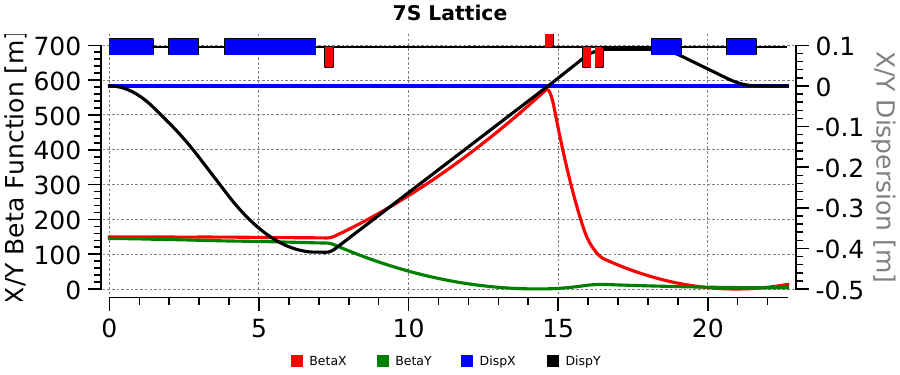}
    \includegraphics[width=0.49\textwidth]{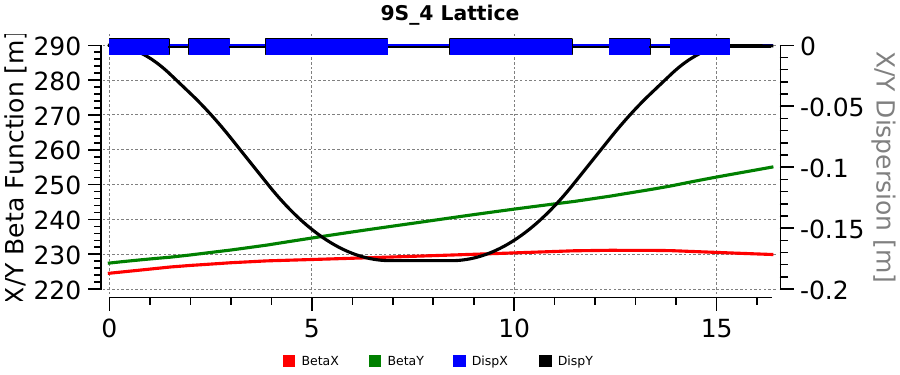}
    \includegraphics[width=0.49\textwidth]{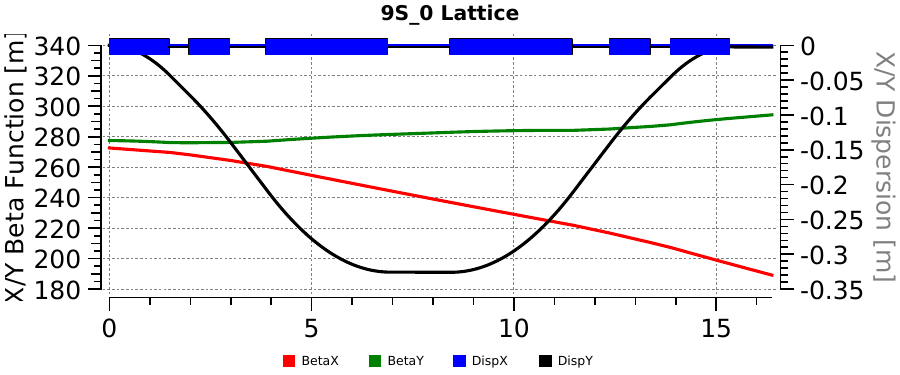}
    \includegraphics[width=0.49\textwidth]{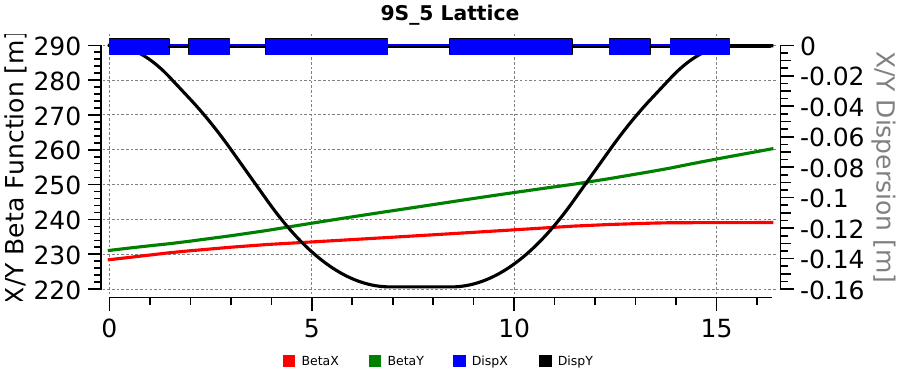}
    \caption{Lattice functions of the North East Spreader both for the lower energies and the FFA passes.}
    \label{fig:NE_SpreaderLat}
\end{figure}

\begin{figure}[!hbt]
    \centering
    \includegraphics[width=0.49\textwidth]{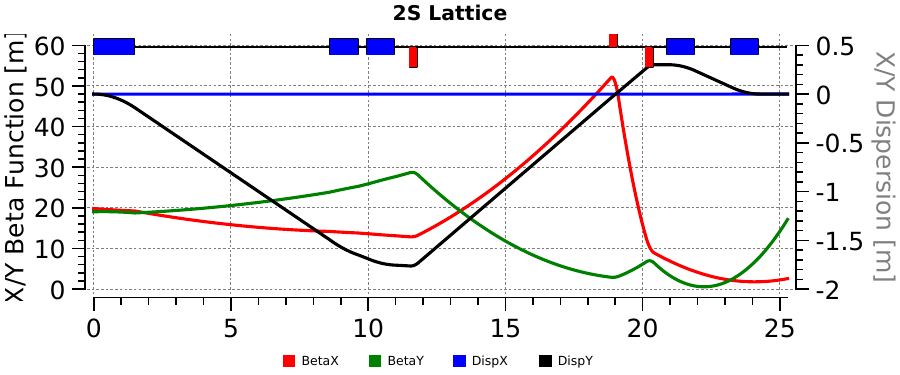}
    \includegraphics[width=0.49\textwidth]{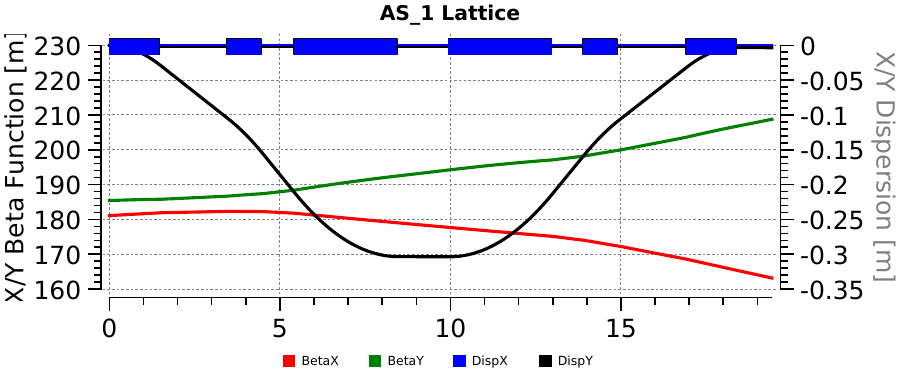}
    \includegraphics[width=0.49\textwidth]{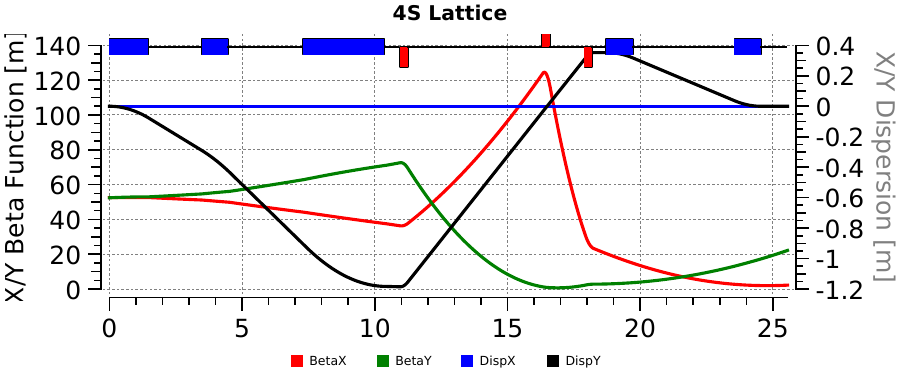}
    \includegraphics[width=0.49\textwidth]{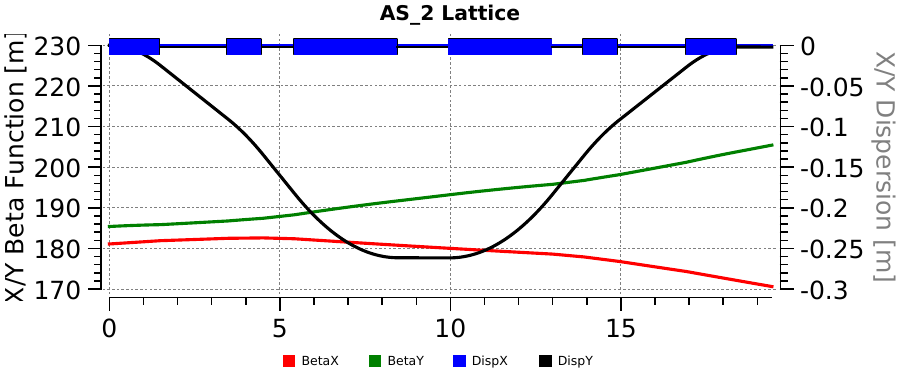}
    \includegraphics[width=0.49\textwidth]{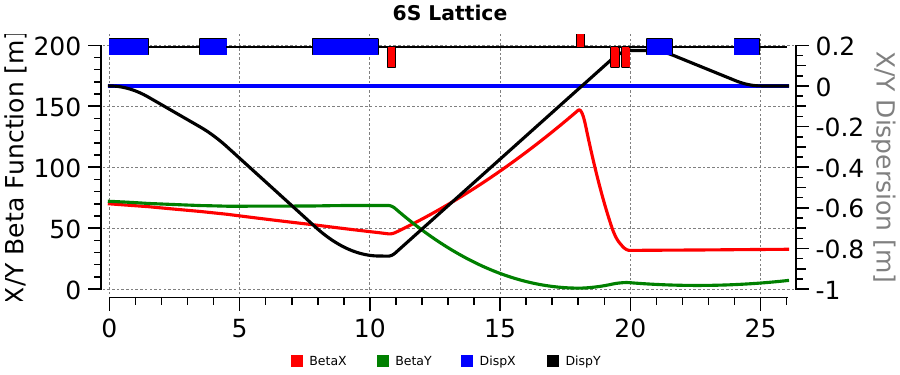}
    \includegraphics[width=0.49\textwidth]{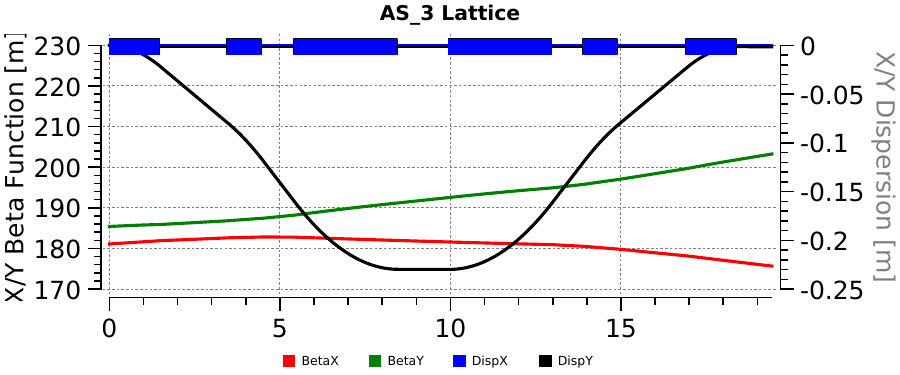}
    \includegraphics[width=0.49\textwidth]{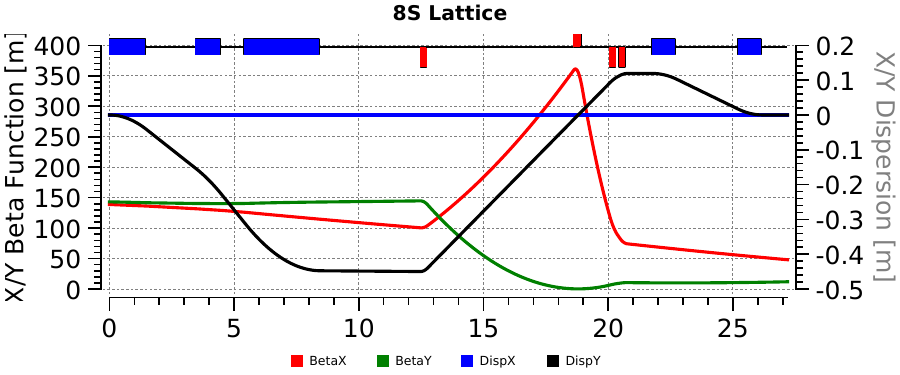}
    \includegraphics[width=0.49\textwidth]{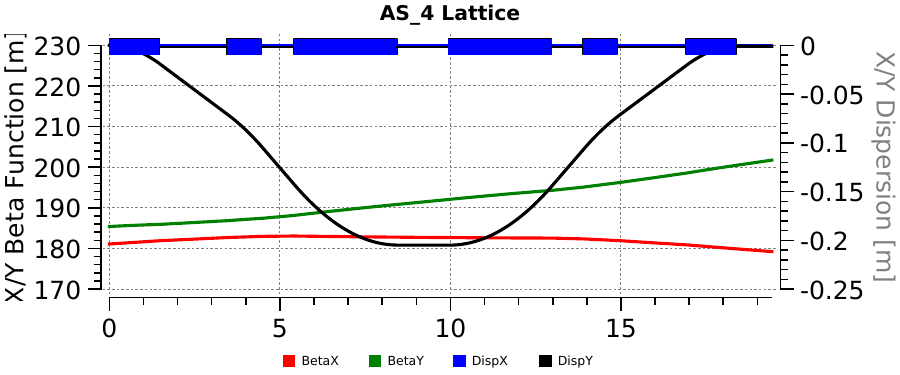}
    \includegraphics[width=0.49\textwidth]{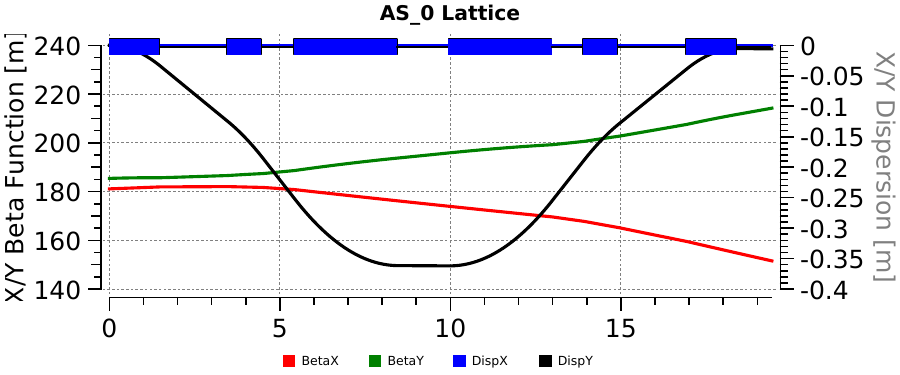}
    \includegraphics[width=0.49\textwidth]{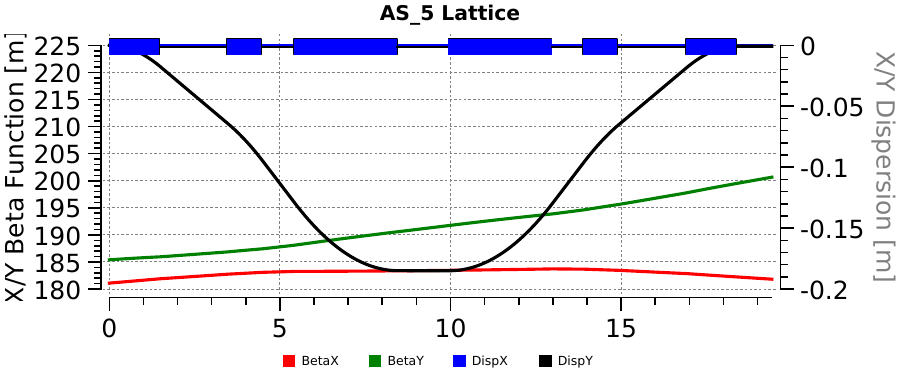}
    \caption{Lattice functions of the South West Spreader both for the lower energies and the FFA passes.}
    \label{fig:SW_SpreaderLat}
\end{figure}

Figure \ref{fig:SW_Spreader} shows a 3D model of the redesigned Southwest Spreader. This design is one of several versions, and it will continue to be refined as start-to-end simulations for the entire FFA@CEBAF upgrade evolve \cite{Coxe:IPAC23-MOPL177}. The lattice functions for the North East and South West spreaders are shown in Figures~\ref{fig:NE_SpreaderLat} and \ref{fig:SW_SpreaderLat}. These plots include the higher passes which are sent to the FFA line. The integrated magnet strength of the different spreader lines is shown in Tab. \ref{tab:spreadermagnets}. The higher passes all see the same field with only minor pathlength differences, so we only show arcs 9 and A.

It is important to note that this design does not properly accommodate the clearances required for many of the magnets. Some of the quadrupoles are likely overlapping dipoles and septa. Small changes to address these engineering concerns will need to be performed.

\begin{table}
    \centering
    \caption{This table shows the integrated magnet strengths in Gauss-Centimeters for the different passes in the vertical spreader. The higher, FFA, passes go through the same magnets as 9S and AS with only minor pathlength differences.}
    \label{tab:spreadermagnets}    
    \scriptsize
    \begin{tabular}{clr|clr}
        Section & Magnet & BDL (Gcm) & Section & Magnet & BDL (Gcm)\\
        \hline \hline
1S	& MAQ1S01 & 2082100	& 2S & MXR2S01 & 1792300 \\
 &	MAI1S03	& -2082100	& &	MXH2S02	& -896130 \\
 &	MQB1S01	& -10807 & & MXY2S03	& -896130\\
 &	MQB1S02	& 45000	 & &MQB2S01	& -13913\\
 &	MQB1S03	& -65684 & &MQB2S02 & 51000\\
 &	MAI1S04	& 1278000	 & & MQB2S03 & -78264\\
 &	MAI1S06	& -1278000 & & MXK2S05 & 1215300\\
 &			& & &	MXK2S06	& -1215300\\
 \hline
 3S	& MAQ1S01 &2046400	&	4S & MXR2S01	& 1785000\\
 & MXQ3S02 & 652740	& & MXT4S02 & 994870\\
&	MAI3S03	& -2699200 & & MXV4S03 & -2779900	\\
&	MQB3S01	& -23695 & & MQB4S01 & -32221 \\
&	MQB3S02	& 45002	& & MQB4S02 & 75000\\
&	MQB3S03	& -65276	& & MQB4S03	& -108150\\	
&	MAI3S05	& 1003700	& & MXF4S05	& 1234800 \\
&	MAI3S06	& -1003700 & & MXF4S06 & -1234800\\
\hline
5S & MAQ1S01 & 2041600	 & 6S& MXR2S01 & 1783300\\	
&	MXQ3S02	& 646440 & &	MXT4S02 & 990070\\
&	MXN5S03	& -1353400 & & MXU6S03 & -2773300	\\
&	MXW5S04	& -1334600	& & MQB6S01 & -33819	\\
&	MQB5S01	& -26446	& & MQB6S02	& 105000	\\
&	MQB5S02	& 120000	& & MQB6S03	& -86329	\\
&	MQB5S03	& -106050	& & MQB6S04	& -88307	\\
&	MQB5S04	& -105950	& & MXB6S05	& 1260200	\\
&	MXD5S05	& 1261900	& & MXB6S06	& -1260200	\\
&	MXD5S06	& -1261900	& & 	& \\	
\hline
7S & MAQ1S01 &	2040000	& 8S & MXR2S01 &	1782600 \\
&	MXQ3S02	 & 644450	&	& MXT4S02	& 988230\\
&	MXN7S03	& -2684500	&	& MZA8S03 &	-2770800\\
&	MQB7S01	& -38750	&	& MQB8S01 &	-51831\\
&	MQB7S02	& 135000	&	& MQB8S02 &	135130\\
&	MQB7S03	& -97863	&	& MQB8S03 &	-104600\\
&	MQB7S04	& -98572	&	& MQB8S04 &	-105050\\
&	MXD7S05	& 999850	&	& MXE8S05 &	1096100\\
&	MXD7S06	& -999850	&	& MXE8S06 &	-1096100\\
\hline
9S &	MAQ1S01 &	2039400 &	AS & MXR2S01 & 1782300 \\
&	MXQ3S02	& 643580 &	&	MXT4S02	& 987330\\
&	MXN7S03	& -2682900 &	&	MZA8S03	& -2769600\\
&	MXN9S04	& -2682900	&	& MZA10S04	& -2769600\\
&	MXS9S05	& 643580 &	&	MXS10S05	& 987330\\
&	MXS9S06	& 2039400 &	&	MXS10S06	& 1782300\\
    \end{tabular}

\end{table}

\FloatBarrier

\subsubsection{Electromagnetic recirculating arcs, Arcs 1-8 }
The EM arcs consist of three substructures: the spreader, arc proper, and recombiner. The spreader is a system of vertical dipole and quadrupole magnets that extract the beam from the preceding Linac and inject it to its respective arc. The arc proper is an array of horizontal dipole magnets that bends the beam a full $180^{\circ}$ into the proceeding Linac. The recombiner (the mirror image of the spreader) is responsible for injecting the beam back into the proceeding Linac. The combined operation of these three substructures must be achromatic ($\eta_{x,y} \sim 0$ and $\eta'_{x,y} \sim 0$) and isochronous ($R_{56}\sim 0$).

The upgraded injection energy will require two major changes in the EM arcs. The first will be to completely redesign the existing spreader/recombiner beamlines in each arc with ones that can accommodate the increased energy \cite{ryan2023}. The second is to discard the lowest energy arcs (Arc 1 and Arc 2) and physically move the remaining arc proper sections ``up-by-one'' (Arc 3 moves up to Arc 5's position, Arc 4 moves up to Arc 6's position, etc.) in an effort to keep the dipole bending magnets within their magnetic field threshold \cite{jay2023,ryan2023}. 

The energy upgrade will increase the beam's rigidity and thus, require larger magnetic field strengths, $B_{new}$, to maintain the optics. The field strengths required are found by the ratio of new and old beam momentum,

\begin{align}\label{eq:eq1}
|B_{new}| = \frac{p_{new}}{e \rho_o} = \frac{p_{o} (p_{new}/p_o)}{e \rho_o} = B_o \chi,
\end{align}

\noindent where $e$ is the electron charge, $p_o$ is the old momentum, $\rho_o$ is the old bend radius, and $\chi=p_{new}/p_o$ ($\chi \sim 1.48$). Using Eq.~(\ref{eq:eq1}), we obtain a set of $B_{new}$ to compare with the $B_{old}$ and the maximum allowed field strength (see Table~\ref{tab:table1}) \cite{ced,jay2023}. The new field strength requirements exceeds the maximum capable field strength for all arc dipoles and will be sufficient to bend the beam $180^{\circ}$. 

\begin{table*}[!htb]
   \centering
   \caption{ARC proper dipole strengths for the old (12~GeV) and new (22~GeV) beamlines.}
   \begin{ruledtabular}
\begin{tabular}{ccccccc}
       \textbf{ARC} & \textbf{Type} & \textbf{Quantity} & \textbf{$L_B$ [m]} & \textbf{$B_{\text{old}}$ [T]} & \textbf{$B_{\text{new}}$ [T]} & \textbf{$B_{\text{max}}$ [T]} \\
       \colrule
       1 & JD & 32 & 1 & 0.79 & 1.14 & 0.84 \\	
       2 & JI & 32 & 2 & 0.75 & 0.93 & 0.80 \\
       3 & JD & 32 & 1 & 1.11 & 1.29 & 1.17 \\
       4 & JC & 32 & 2 & 0.73 & 0.83 & 0.77 \\ 
       5 & JC & 32 & 2 & 0.91 & 1.01 & 0.96 \\
       6 & JC & 32 & 2 & 1.09 & 1.19 & 1.14 \\
       7 & JA & 32 & 3 & 0.85 & 0.91 & 0.89 \\    
       8 & JA & 32 & 3 & 0.96 & 1.03 & 1.01 \\
   \end{tabular}
\end{ruledtabular}
   \label{tab:table1}
\end{table*}

To alleviate the magnetic field requirement on the bending magnets, a solution of discarding the lowest energy arcs and moving each subsequent arc ``up-by-one'' (the spreaders/recombiners will remain unmoved) is being adopted. Under this regime, the magnetic field of each magnet is within spec.

With the aforementioned changes each arc will need to have its optics tuned in two main areas. The first is to make the entire beamline isochronous ($R_{56}=0$) to maintain the proper phasing between particles and the acceleration fields in the RF cavities. The second is to match the Twiss parameters ($\beta, \alpha$) from the spreader into the arc proper and from the arc proper to the recombiner sections. 

The $R_{56}$ is defined as

\begin{align}\label{eq:eq2}
R_{56} = \int \left({ \frac{ \eta_{x} }{\rho_{x}} + \frac{ \eta_{y} }{\rho_{y}} }\right) ds,
\end{align}

\noindent where $\eta_{x,y}$ is the first-order dispersion function, $\rho_{x,y}$ is the bending radius and $ds$ is the accelerator coordinate. The spreaders/recombiners will have negative $R_{56}$ (see Table~\ref{tab:table1}) and so, the arc proper will need to have equal and opposite $R_{56}$. There are four periodic cells contained in the arc proper. Therefore, a single cell's $R_{56}$ must equal $-1/2$ of the spreader's $R_{56}$ in order for the entire beamline to be isochronous. 

\begin{figure}[htb]
 \centering
  \includegraphics[width=0.6\linewidth]{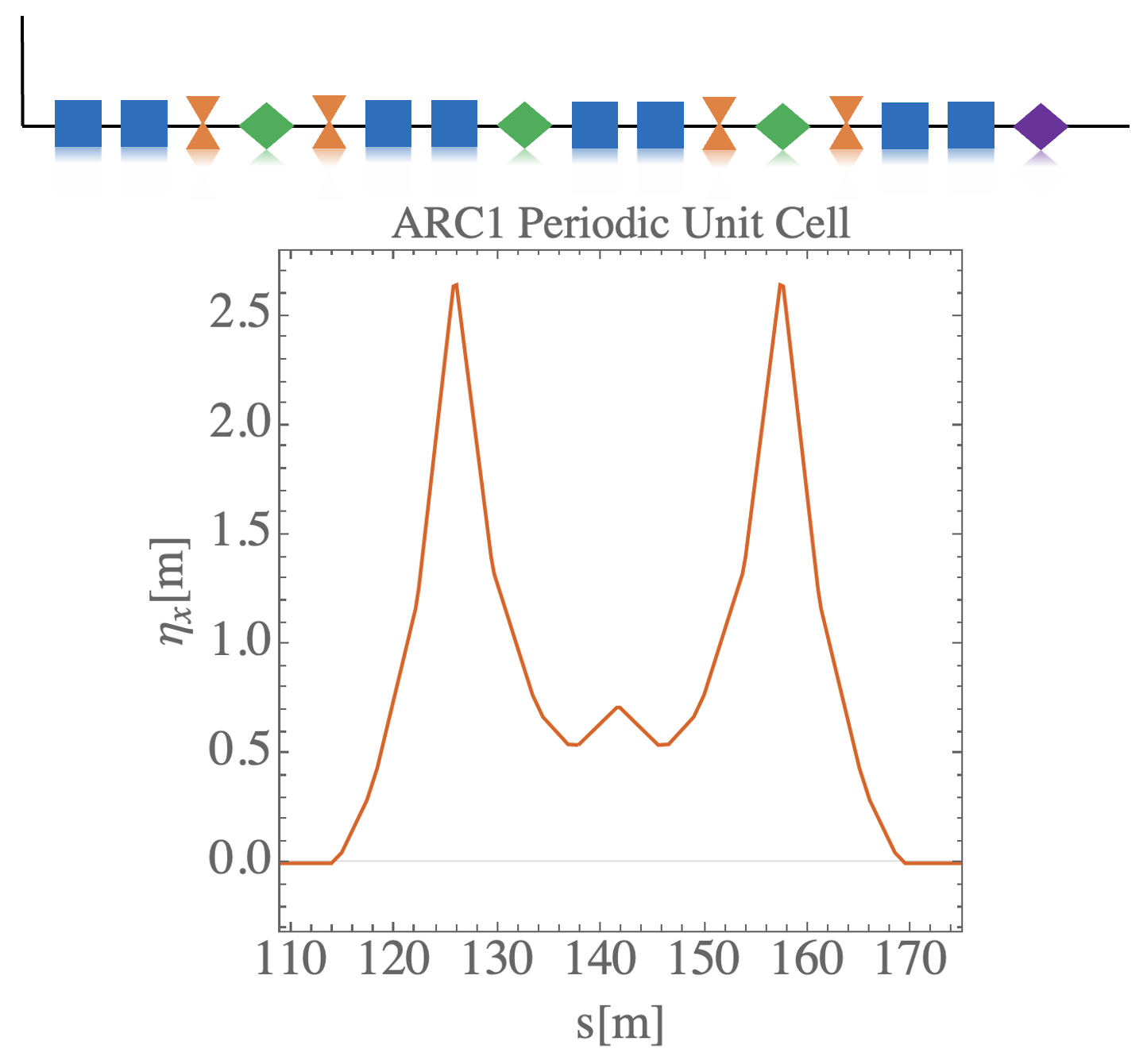}
  \caption{\label{fig:fig3} 4-fold symmetry periodic lattice structure in ARC proper. Top: Cartoon of the periodic cell (bends are squares, focusing and defocusing quadrupoles are diamonds and hourglasses, respectively). The purple diamond is a free parameter to vary the periodic $\beta$ and $\alpha$. Bottom: The $\eta_x$ function in the ARC periodic cell. Each cell is achromatic.}
\end{figure}

The arc proper beamline consists of a periodic array of bending magnets and quadrupoles (see Figure~\ref{fig:fig3}). The only free parameters in the cell are the quadrupole strengths, $K_1$ (the bends are constrained by the $180^{\circ}$ bend requirement). The quadrupoles enclosed within the bends (orange and green quadrupoles in Figure~\ref{fig:fig3}) will affect the peaks in the horizontal dispersion function and can tune the $R_{56}$ while ensuring the cell is achromatic. The lone quadrupole outside the bends (purple quadrupole in Figure~\ref{fig:fig3}) has no impact on the dispersion and thus can be varied to tune the cell's periodic Twiss functions.

After determining the periodic Twiss functions, the matching section quadrupoles are varied to match into and out of the arc proper. The matching section after the spreader but before the arc proper is used to match into the periodic arc proper Twiss functions. The quadrupole strengths are optimized to maintain the match into the arc proper but also minimize the beta functions. The quadrupole strengths are optimized to minimize the beta functions in both the recombiner and proceeding Linac section. The west arcs have an added complication of matching into the reinjector and Linac beamlines, both of which have static optics.

\FloatBarrier
\clearpage
\subsection{Novel Machine Additions}
\subsubsection{Fixed Field Alternating Gradient arcs}\label{-ffa-arcs}

\paragraph{Lattice constraints}
The fixed-field (FFA) arcs of the CEBAF energy upgrade \cite{Deitrick:IPAC23-MOPL182,VasilyIPAC24} will replace the final electromagnetic arcs of the current 12\,GeV CEBAF but allow a wide energy range to be transmitted.  For the east arc studied in this paper, this range will be approximately 10.5--21\,GeV.  Option A in Table~\ref{tab:latticerules} is the baseline as of December 2022, which is a linear field FFA.  All lattices in this study have the simple structure \textbf{BD~o~BF~o} where \textbf{o} is a drift space.

\begin{table}[!htb]
   \centering
   \caption{Lattice Option Design Rules}
   \begin{ruledtabular}
\begin{tabular}{lccccccc}
       \textbf{Option} & \multicolumn{2}{c}{\textbf{Energy (GeV)}} & \multicolumn{2}{c}{\textbf{Cell tune (cycles)}} & \textbf{Max. Dipole} & \textbf{Gradient} & \textbf{Sextupole} \\
        & \textbf{Min.} & \textbf{Max.} & \textbf{Min.} & \textbf{Max.} & \textbf{(T)} & \textbf{(T/m)} & \textbf{(T/m$^\mathbf2$)} \\
       \colrule
A & 10.494 & 21.014 & 0.0363* & 0.3943* & 1.2815* & 43.44* & 0 \\
B & 10.494 & 21.014 & 0.035 & 0.4 & 2 & 100 & 0 \\
C & 10.494 & 21.014 & 0.035 & 0.4 & 2 & 100 & 2000 \\
D & 9 & 21 & 0.04 & 0.39 & 2 & 100 & 400 \\
E & 9 & 21 & 0.05 & 0.32 & 2 & 100 & 400 \\
   \end{tabular}
\end{ruledtabular}
\footnotesize{* Point design values, rather than optimization limits.}
   \label{tab:latticerules}
\end{table}

Lines B--E of Table~\ref{tab:latticerules} give constraints for coupled lattice--magnet optimizations that otherwise try to reduce permanent magnet volume.  Option B is a re-optimization for minimum magnet volume while keeping the same energy and tune ranges.  Option C allows a (potentially large) amount of sextupole to be added to the magnets, this flexibility potentially reducing magnet size further, while otherwise preserving the conditions of B.  Options D and E attempt to use the additional flexibility provided by the sextupole to give an extended lower energy range, which is important for fine-tuning the facility beam energy, or dealing with other situations where the SRF Linac energy gain is reduced and the beam enters the FFA arcs at below nominal energy.  Option E tries a narrower cell tune range avoiding the $Q=1/3$ resonance that could be excited by sextupoles.

Lattice cells must also conform to the 80.6\,m CEBAF tunnel radius.  This is enforced by setting magnet bend angles $\theta_m=(L_\mathrm{cell}/80.6$\,m$)L_m/(L_\mathrm{BF}+L_\mathrm{BD})$ where $L_m$ is the magnet length, which is allowed to vary.  The drift spaces are set to a constant 9\,cm in optimizations B--E.

\newcommand{\subsubsubsection}[1]{\textit{#1}\\}
\subsubsubsection{Magnet alignment details}
Long permanent magnets are generally built in rectangular sections rather than sectors.  The magnet bend angle $\theta$ is therefore implemented (in the layout reference line) as two corners of angle $\theta/2$ at either end of the magnet, with straight lines in between.  A soft-edged fringe field proportional to $\frac12+\frac12\tanh(z/(2.5\,\mathrm{cm}))$ is used at both ends of all magnets.  To deal with orbit curvature within the magnets, they are split into six longitudinal sections, each individually centered on the range of beam orbits.

\paragraph{Optimization method}
Candidate lattice cells were tracked with the \texttt{Muon1} code~\cite{Muon1} and optimised with its built-in genetic algorithm.  The optimiser started from random designs with no manually-set starting point, so the following scoring ranges were used to guide it towards viable designs.
\begin{enumerate}
\item If the first energy is unstable or has an unacceptable tune, the cell is scored by how far $\cos \phi$ (calculated from the trace of the transfer matrix) deviates from the desired tune range limits, where $\phi$ is the phase advance.
\item If the first energy has correct tunes but later ones do not, the cell is scored by the percentage of the FFA energy range that is acceptable.
\item If all energies have correct tunes, the \texttt{HalbachArea} code~\cite{HalbachArea} is called to attempt magnet designs.  If the magnet design fails, the cell is scored by peak field in the bore of the accelerator, with lower being better.
\item If all energies have correct tunes and magnets exist, the cell is scored by the average magnet area $(\sum_{\mathrm{Elements\ }e}A_e L_e)/\sum_e L_e$ weighted by length through the cell, with lower being better.
\end{enumerate}

\paragraph{Resulting lattice}
The geometries of the optimised lattice cells are given in Table~\ref{tab:latticegeom} and the magnetic fields are given in Table~\ref{tab:latticefields}.

\begin{table*}[!htb]
   \centering
   \caption{Lattice Geometries}
   \begin{ruledtabular}
\begin{tabular}{lccccccc}
       \textbf{Option} & \multicolumn{4}{c}{\textbf{Lengths (m)}} & \multicolumn{3}{c}{\textbf{Angles (mrad)}} \\
        & \textbf{BD} & \textbf{BF} & \textbf{Drifts} & \textbf{Cell} & \textbf{BD} & \textbf{BF} & \textbf{Cell}\\
       \colrule
A & 1.2448 & 1.6731 & 0.1162 & 3.1504 & -7.11 & -31.98 & -39.09 \\
B & 1.1832 & 1.1892 & 0.09 & 2.5524 & 15.79 & 15.87 & 31.67 \\
C & 1.5195 & 1.4505 & 0.09 & 3.1500 & 20.00 & 19.09 & 39.08 \\
D & 1.4625 & 1.9760 & 0.09 & 3.6184 & 19.09 & 25.80 & 44.89 \\
E & 1.3814 & 1.7898 & 0.09 & 3.3512 & 18.11 & 23.47 & 41.58 \\
   \end{tabular}
\end{ruledtabular}
   \label{tab:latticegeom}
\end{table*}

\begin{table*}[!hbt]
   \centering
   \caption{Lattice Magnetic Field Specifications}
   \begin{ruledtabular}
\begin{tabular}{lcccccc}
       \textbf{Option} & \multicolumn{2}{c}{\textbf{Dipole (T)}} & \multicolumn{2}{c}{\textbf{Gradient (T/m)}} & \multicolumn{2}{c}{\textbf{Sextupole (T/m$^\mathbf2$)}} \\
        & \textbf{BD} & \textbf{BF} & \textbf{BD} & \textbf{BF} & \textbf{BD} & \textbf{BF} \\
       \colrule
A & -0.3828 & -1.2815 & 43.44 & -41.13 & 0 & 0 \\
B & 0.8629 & 0.8629 & 55.155 & -69.369 & 0 & 0 \\
C & 0.9590 & 0.9590 & 59.960 & -89.189 & -1411.41 & 974.97 \\
D & 0.8228 & 0.8228 & 45.345 & -48.549 & -400 & 339.94 \\
E & 0.8148 & 0.8148 & 47.548 & -50.951 & -400 & 351.95 \\
   \end{tabular}
\end{ruledtabular}
   \label{tab:latticefields}
\end{table*}

\begin{table*}[!hbt]
   \centering
   \caption{Lattice Results and Figures of Merit}
   \begin{ruledtabular}
\begin{tabular}{lccccc}
       \textbf{Option} & \multicolumn{2}{c}{\textbf{Cell tune (cycles)}} & \textbf{$|\mathbf B|_\mathrm{max}$} & \textbf{Orbit excursion} & \textbf{Path length change} \\
         & \textbf{min.} & \textbf{max.} & \textbf{(T)} & \textbf{(mm)} & \textbf{(mm)} \\
       \colrule
A & 0.0363 & 0.3943 & 1.5346 & 44.968 & 1.233 \\
B & 0.0357 & 0.3994 & 1.6140 & 28.607 & 0.525 \\
C & 0.0352 & 0.3993 & 1.4922 & 23.602 & 0.344 \\
D & 0.0426 & 0.3898 & 1.4689 & 41.739 & 0.916 \\
E & 0.0500 & 0.3194 & 1.5438 & 42.966 & 0.910 \\
   \end{tabular}
\end{ruledtabular}
   \label{tab:latticeresults}
\end{table*}

Table~\ref{tab:latticeresults} shows some performance statistics for the optimised cells.  Option C has the smallest average magnet area, which is also the optimiser's figure of merit: 48\% below the baseline A.  This makes sense as option C was the least constrained.

\begin{figure}[!htb]
   \centering
   \includegraphics*[width=0.7\columnwidth]{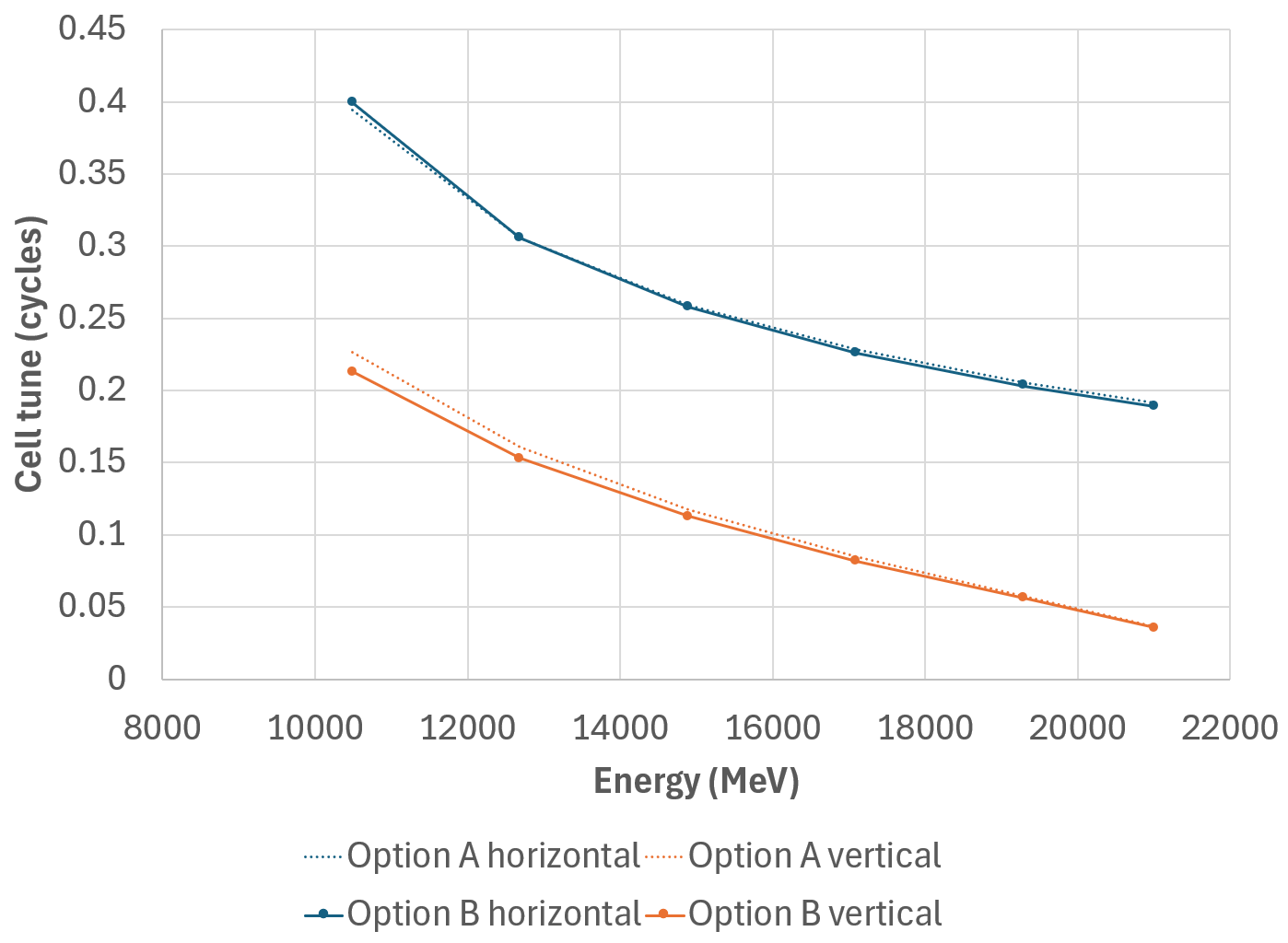}
   \caption{Cell tunes of the optimised linear field option B compared to the baseline option A.}
   \label{fig:compareAB}
\end{figure}

Figure~\ref{fig:compareAB} shows the cell tunes as a function of energies within the range.  Option B has nearly identical tunes to the baseline A, with its 36\% reduction in orbit excursion and 11\% reduction in magnet area mainly coming from shortening the cell length from 3.15 to 2.55\,m.

\begin{figure}[!htb]
   \centering
   \includegraphics*[width=0.7\columnwidth]{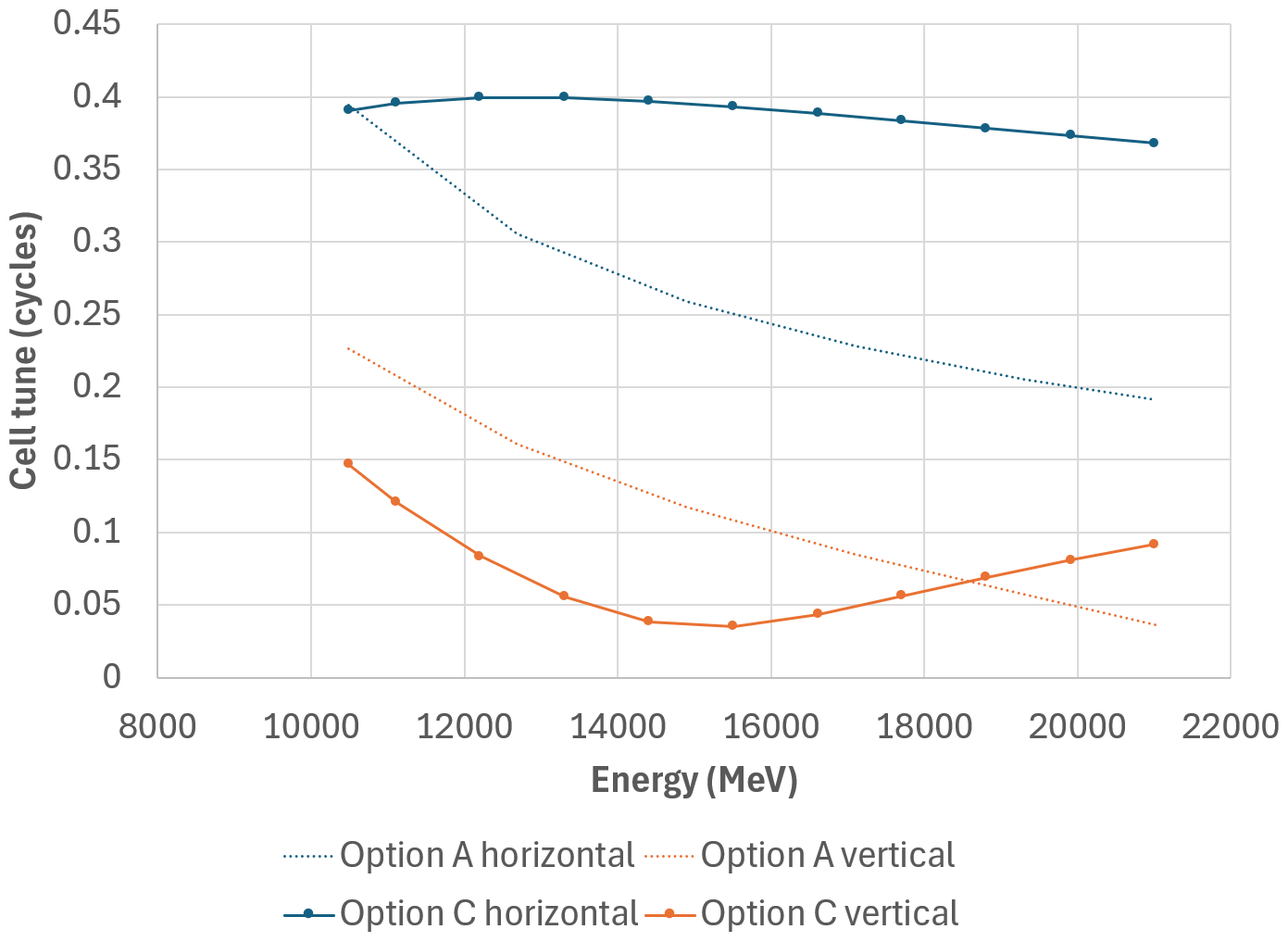}
   \caption{Cell tunes of the optimised lattice with sextupole (option C) compared to the baseline option A.}
   \label{fig:compareAC}
\end{figure}

Figure~\ref{fig:compareAC} shows that the tune dependence changes radically when a large amount of sextupole is allowed.  This may be a problem for the FFA orbit correction scheme, which relies on the multiple beams being linearly independent by having different tunes.  This is the reason for the 400\,T/m$^2$ sextupole limit used in optimizations D and E.

\begin{figure}[!htb]
   \centering
   \includegraphics*[width=0.7\columnwidth]{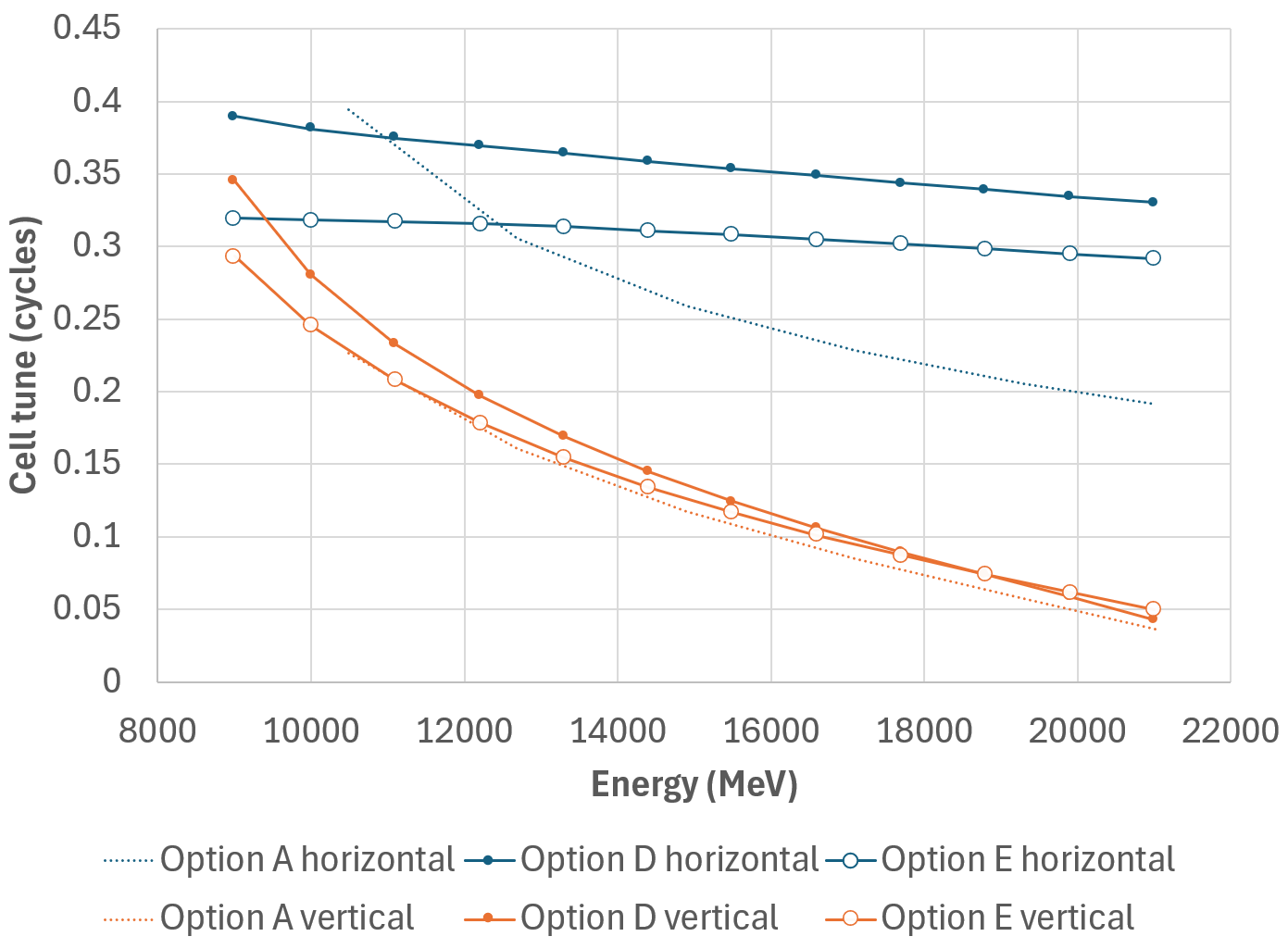}
   \caption{Cell tunes of the extended energy range sextupole lattices (D and E) compared to the baseline option A.}
   \label{fig:compareADE}
\end{figure}

Figure~\ref{fig:compareADE} shows that a smaller amount of sextupole has a less drastic impact on the tunes.  The attempt at keeping the tunes below 1/3 has had the side effect of making option E's horizontal tunes nearly constant, which is bad for orbit correction.  However, option D looks more reasonable.
\FloatBarrier
\clearpage
\subsubsection{TOF correction Splitters}
\label{subsubsec:splitter_designs}
 
As part of the FFA@CEBAF energy upgrade, Khan \textit{et al.}~\cite{Khan2024splitter} present two exploratory horizontal splitter beamline designs for routing six recirculating beams, with energies of approximately 11, 13, 16, 18, 20, and 22~GeV, into the East FFA arc. We summarize this work in this subsection~\cite{Khan2024splitter}. The splitters must bridge the beam from the North Linac to the East FFA arc, matching Twiss parameters ($\beta_{x,y}$, $\alpha_{x,y}$), $R_{56}$, time-of-flight, transverse offset, and dispersion at the arc entrance. The CBETA multi-pass splitter bend system~\cite{bartnik2020} serves as the guiding design precedent.
 
\paragraph{Design constraints}
 
The primary challenge is spatial: the total allowable transverse footprint within the existing CEBAF tunnel is approximately 3~m, with only $\sim$1.6~m available to the outer tunnel wall and $\sim$1.4~m to the inner wall. A total longitudinal extent of $\sim$97~m is available. These constraints are compounded by the high beam rigidity in the 10--22~GeV energy range, which demands large bending and focusing magnets. The designs employ dipole magnets (3~m length, 0.5~m aperture, $B_{\mathrm{max}} = 1.8$~T) and quadrupole magnets (0.8~m length, 0.3~m aperture, gradient up to 54~T/m), with each beamline requiring a minimum of eight quadrupoles for optics control~\cite{Khan2024splitter}. All simulation work was performed using the \textsc{elegant} code~\cite{Borland2000}.
 
\paragraph{Symmetric and asymmetric Splitter configurations}
 
\begin{figure}[ht]
    \centering
    \includegraphics[width=0.48\textwidth]{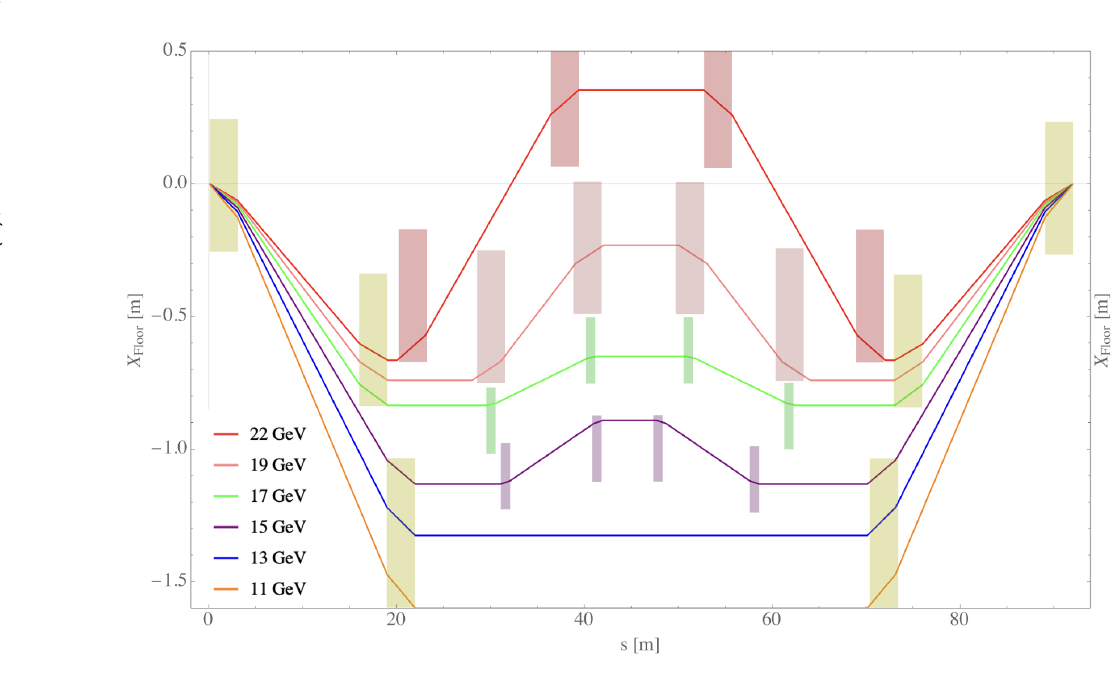}
    \hfill
    \includegraphics[width=0.48\textwidth]{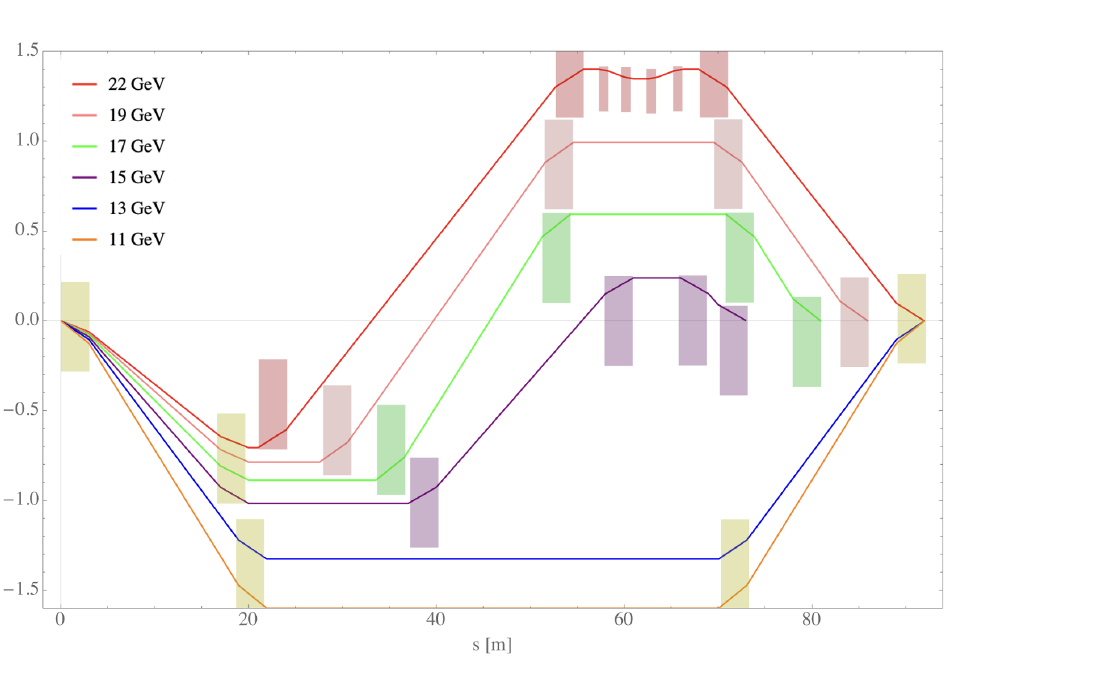}
    \caption{Horizontal beam trajectories for the symmetric (left) and asymmetric (right) splitter designs for the North-East corner of the FFA@CEBAF upgrade. Only dipole magnets are shown; quadrupole magnets are omitted for clarity. Figure reproduced from~\cite{Khan2024splitter}.}
    \label{fig:splitter_layouts}
\end{figure}

Two distinct topologies were developed (see Figure~\ref{fig:splitter_layouts}):
 
\begin{itemize}
    \item \textbf{Symmetric Splitter:} Designed with operational simplicity as the primary objective. Magnets can be powered in series, and higher-order dispersion terms vanish by symmetry~\cite{Stulle2004}. The four highest-energy beamlines employ nested chicanes to achieve the required transverse separation.
 
    \item \textbf{Asymmetric Splitter:} Optimized to reduce synchrotron radiation effects by minimizing the magnet count per beamline. The extended drift distances between opposite-polarity chicane sections preserve space for additional focusing elements or beam extraction. Both designs share a common beamline geometry for the two lowest-energy passes.
\end{itemize}
 
\paragraph{Optics matching and synchrotron radiation}
 
\begin{figure}[ht]
    \centering
    \includegraphics[width=\textwidth]{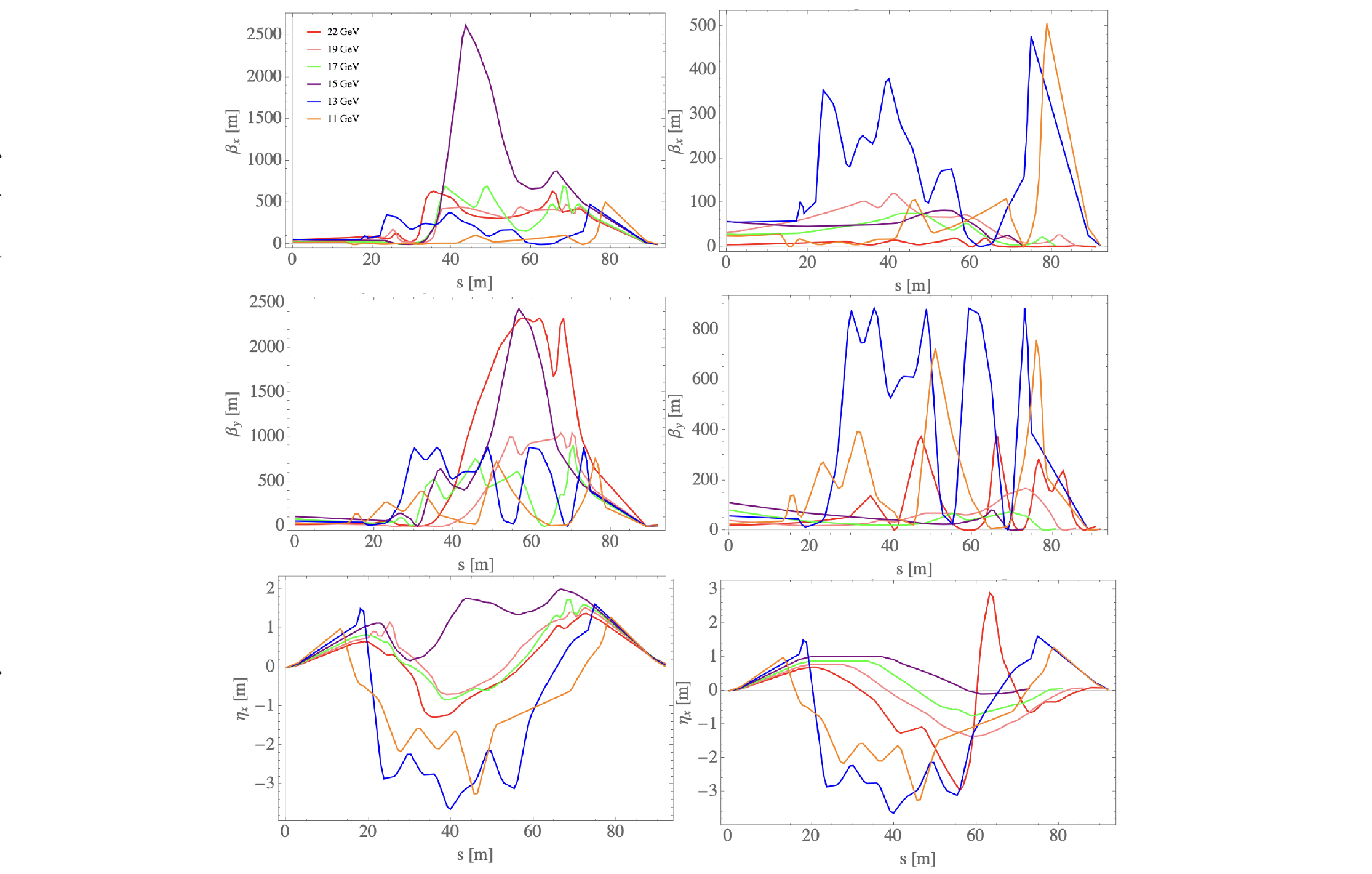}
    \caption{Optical functions ($\beta_x$, $\beta_y$, $\eta_x$) along the symmetric (left column) and asymmetric (right column) splitters for all six energy passes. Figure reproduced from~\cite{Khan2024splitter}.}
    \label{fig:optical_functions}
\end{figure}

Seven exit parameters per beamline, $\beta_x$, $\alpha_x$, $\beta_y$, $\alpha_y$, $R_{56}$, $\eta_x$, and $\eta'_x$, were matched using the native simplex optimization routine in \textsc{elegant}, excluding the transverse position $X$ and angle $X'$. Entrance and exit matching targets are summarized in Tables~\ref{tab:entrance_optics} and~\ref{tab:exit_optics}, respectively. Matching is achieved at the exit of each beamline, though the vertical beta functions tend to grow substantially in the interior of the splitter lines due to coupling between the transverse planes and between $R_{56}$ and the Twiss parameters (see Figure~\ref{fig:optical_functions}).
 
\begin{table}[!hbt]
   \centering
   \caption{Estimated entrance optics for the North-East splitters~\cite{Khan2024splitter}.}
   \begin{ruledtabular}
\begin{tabular}{ccccc}
       \textbf{FFA Pass} & \textbf{$\beta_x$ (m)} & \textbf{$\alpha_x$} & \textbf{$\beta_y$ (m)} & \textbf{$\alpha_y$} \\
       \colrule
1 & 25.5 & 0.12    & 29.5  & $-0.16$ \\
2 & 57.5 & 0.09    & 59.7  & 0.51    \\
3 & 58.2 & 0.49    & 111.1 & 1.16    \\
4 & 29.6 & 0.25    & 82.8  & 1.50    \\
5 & 33.1 & $-0.46$ & 41.6  & 1.00    \\
6 & 53.5 & $-0.82$ & 22.8  & 0.09    \\
   \end{tabular}
\end{ruledtabular}
   \label{tab:entrance_optics}
\end{table}

\begin{table*}[!hbt]
   \centering
   \caption{Estimated exit matching parameters for the North-East splitters~\cite{Khan2024splitter}.}
   \begin{ruledtabular}
\begin{tabular}{ccccccccc}
       \textbf{Pass} & \textbf{$E$} & \textbf{$R_{56}$} & \textbf{$\beta_x$} & \textbf{$\alpha_x$} & \textbf{$\beta_y$} & \textbf{$\alpha_y$} & \textbf{$\eta_x$} & \textbf{$\eta'_x$} \\
        & \textbf{(GeV)} & \textbf{(mm)} & \textbf{(m)} & & \textbf{(m)} & & \textbf{(cm)} & \textbf{($10^{-3}$)} \\
       \colrule
1 & 10.6 & $+71$  & 4.2 & 3.0 & 6.5  & $-3.2$ & 2.7 & $-2.0$ \\
2 & 12.8 & $-45$  & 3.0 & 1.8 & 6.5  & $-3.0$ & 4.6 & $-2.6$ \\
3 & 15.0 & $-128$ & 2.7 & 1.5 & 7.0  & $-3.2$ & 6.1 & $-3.1$ \\
4 & 17.2 & $-192$ & 2.6 & 1.4 & 8.0  & $-3.6$ & 7.3 & $-3.4$ \\
5 & 19.4 & $-242$ & 2.5 & 1.3 & 10.1 & $-4.5$ & 8.4 & $-3.5$ \\
6 & 21.6 & $-281$ & 2.5 & 1.2 & 16.8 & $-7.5$ & 9.3 & $-3.5$ \\
   \end{tabular}
\end{ruledtabular}
   \label{tab:exit_optics}
\end{table*}
 
Synchrotron radiation is an anticipated concern given the 10--22~GeV energy range. The normalized emittance growth and relative energy spread induced in a bending magnet scale as $\gamma^6/\rho^2$ and $\gamma^5/\rho^2$, respectively, motivating designs that maximize the magnet packing factor to reduce effective bend radii per unit length~\cite{Khan2024splitter}.
 
\paragraph{Conclusions and outlook}
 
Both splitter topologies satisfy the geometric layout requirements and achieve the seven target exit optics parameters for each energy pass. Key areas for future development include refining quadrupole placement to decouple the $R_{56}$ and Twiss knobs, adjusting the FFA cell match point to relax optics requirements, and exploring the use of off-axis quadrupoles for bending in the highest-energy lines~\cite{Khan2024splitter}. The use of small permanent magnets analogous to those in the FFA arc cells is also under consideration. The ultimate objective is a complete conceptual design suitable for start-to-end simulations integrated with the broader FFA@CEBAF lattice.

\FloatBarrier

\subsubsection{FFA arc-to-Linac Transition}
\input{Transition.tex}

\subsubsection{Multi-hall extraction scheme}
\label{sec:extraction}
\begin{figure}[!htb]
    \centering
    \includegraphics[width=0.75\linewidth]{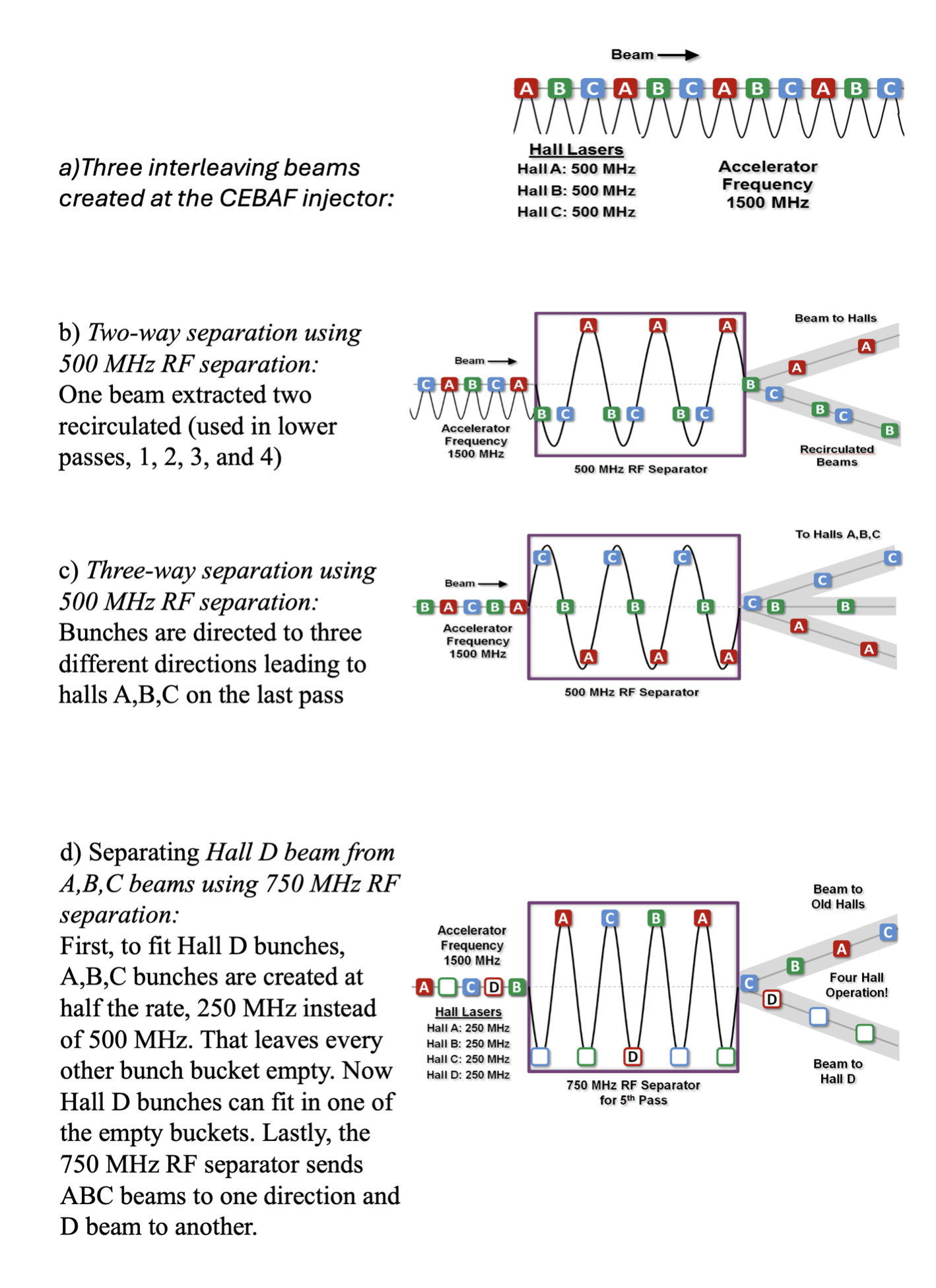}
    \caption{CEBAF RF separation.}
    \label{fig:separators}
\end{figure}

The present CEBAF beam extraction system using RF separators is functioning with high reliability \cite{Kazimi2013}. Therefore, the FFA@CEBAF would inherit and got inspired from RF separator based extraction systems, which is shown in~Fig.~\ref{fig:separators}.

Let us begin with the beam delivery system for Experimental Halls A, B, and C. At the injector stage, three interleaved electron beams—designated A, B, and C—are generated using three independent lasers illuminating the photocathode of the electron gun. Each laser produces a beam with a 500 MHz bunch repetition rate. Combined, these beams form a single composite beam with an effective bunch structure of 1500 MHz (see Figure~\ref{fig:separators}a).

This composite beam is then accelerated through the main accelerator. At the final stage, it is directed through an RF separator system, which consists of resonant cavities imparting transverse dipole kicks~\cite{Kazimi1992} and operates at 500 MHz. By appropriately adjusting the RF phase, various beam extraction schemes can be realized: for instance, isolating a single 500 MHz beam (Fig.\ref{fig:separators}b) or distributing all three beams into separate beamlines (Fig.\ref{fig:separators}c).

To incorporate Hall D into the beam delivery system, the injector’s bunch frequency configuration was modified to support a fourth laser. A 750 MHz RF separator was subsequently introduced to enable the extraction of the D beam \cite{Kazimi2013-2}. The resulting “four-hall operation” mode, illustrated in Figure~\ref{fig:separators}d, is detailed in \cite{Kazimi2013}.

\begin{figure} [!htb]
    \centering
    \includegraphics[width=0.9\linewidth]{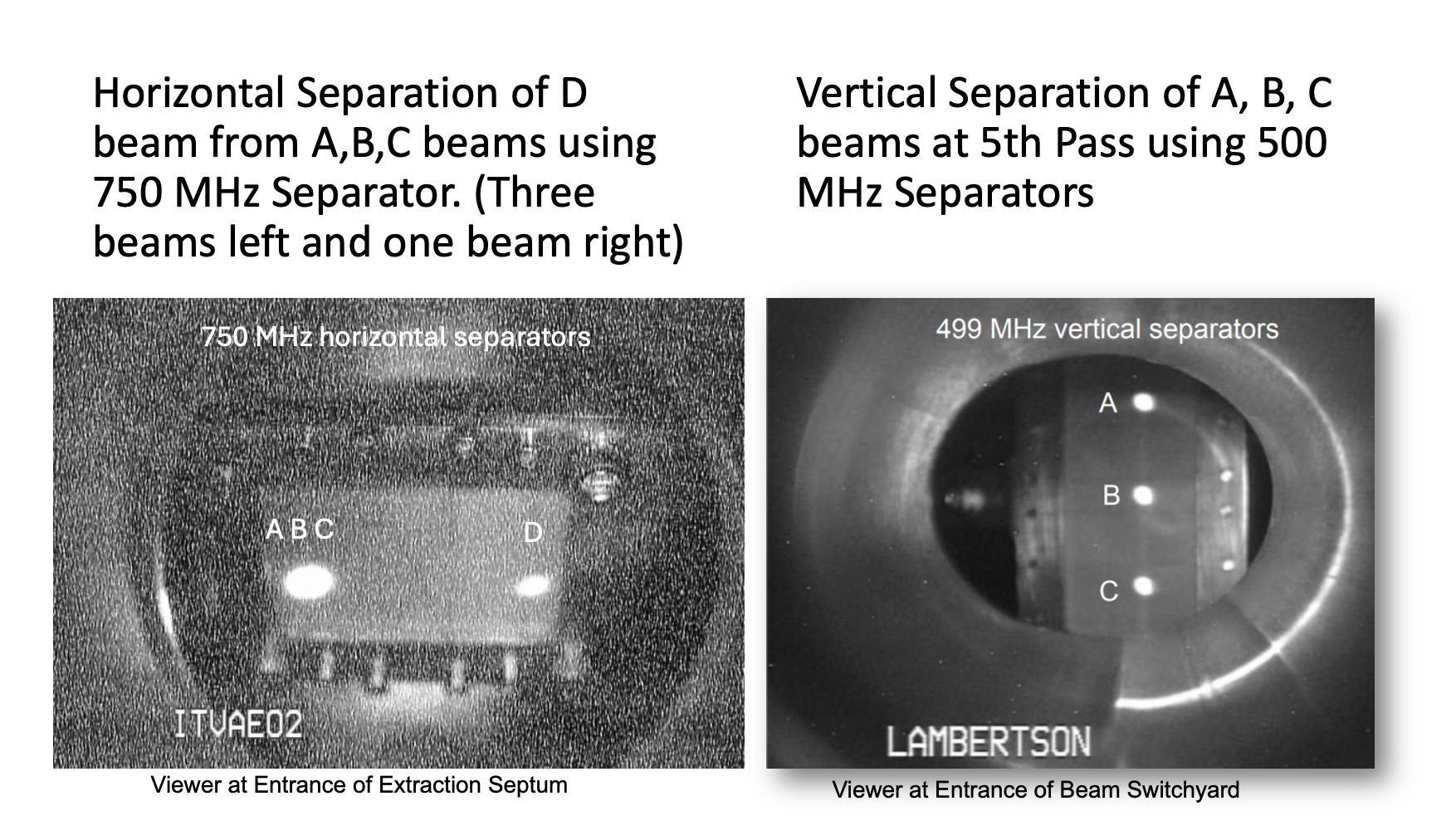}
    \caption{Screen shots from CEBAF operation.}
    \label{fig:cebaf4beams}
\end{figure}

Today, all of these beam extraction patterns are actively used in the 12 GeV CEBAF accelerator. For instance, Figure\ref{fig:cebaf4beams} presents a screenshot from CEBAF operations: initially showing the separation of the D beam from the combined ABC beams (left side), followed by the subsequent separation of A, B, and C beams into their respective experimental halls. These operational modes correspond directly to the configurations depicted in Figure\ref{fig:separators}d and Figure\ref{fig:separators}c.

\begin{figure} [!htb]
    \centering
    \includegraphics[width=0.7\linewidth]{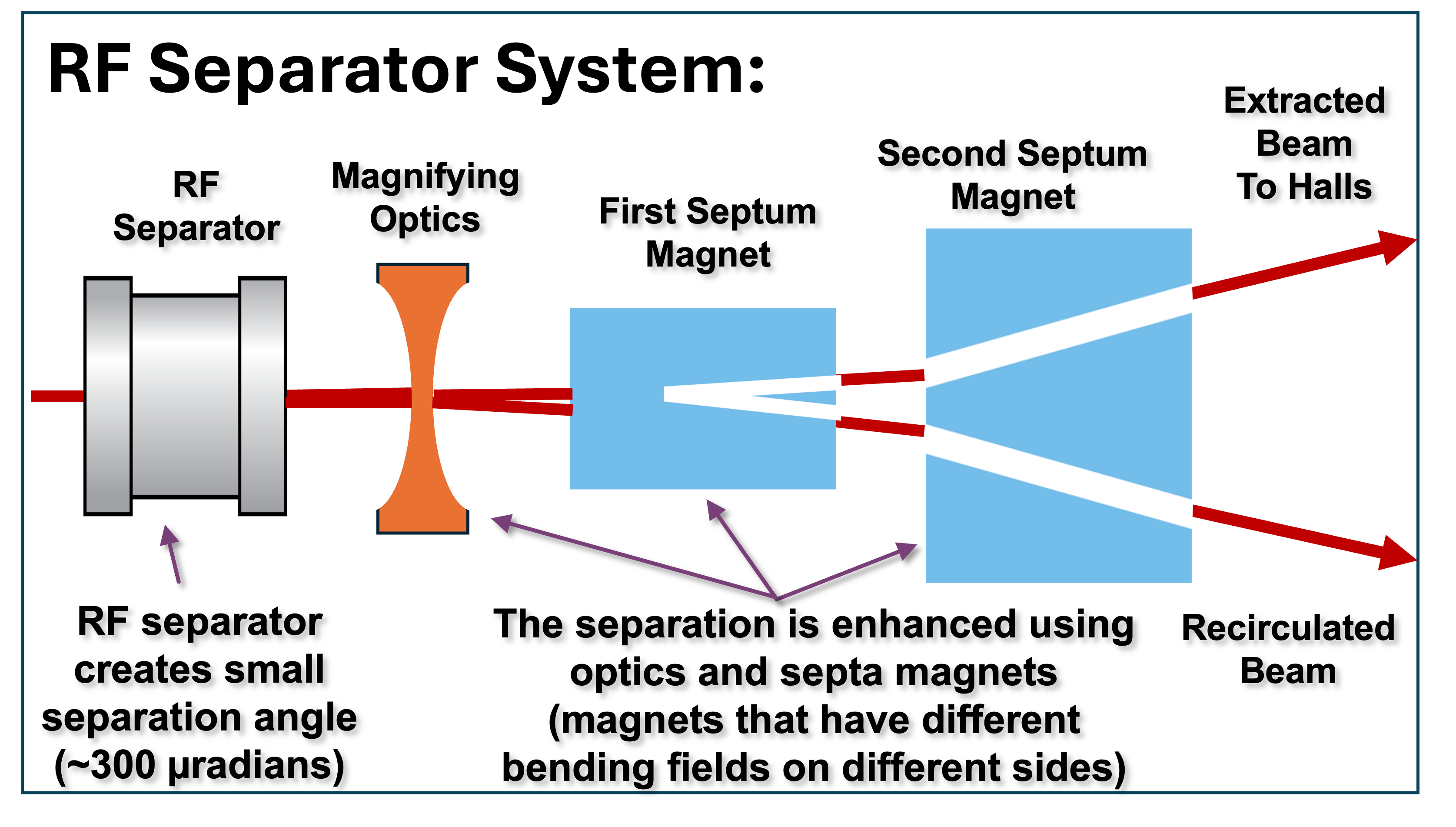}
    \caption{Basics of RF extraction.}
    \label{fig:rfextract}
\end{figure}

It is important to note that the transverse kick provided by the RF separator is relatively small—approximately 300 $\mu$rad in angular separation. However, this initial deflection is significantly amplified downstream by the beamline optics, which include magnetic lenses and carefully designed geometry. Only after the beams are sufficiently separated are septum magnets employed to complete the extraction process, as presented in Figure~\ref{fig:rfextract}.

All RF separation patterns discussed so far share a common principle: a subharmonic RF kicker operates on a single-energy beam, selectively deflecting certain bunches in one direction, and the rest tin another direction. In contrast, the Fixed Field Alternating Gradient (FFA) arcs simultaneously transport beams of different energies—corresponding to different passes—within the same beamline. This makes it impractical to implement individual separation systems for each pass.

\begin{figure} [!htb]
    \centering
    \includegraphics[width=0.7\linewidth]{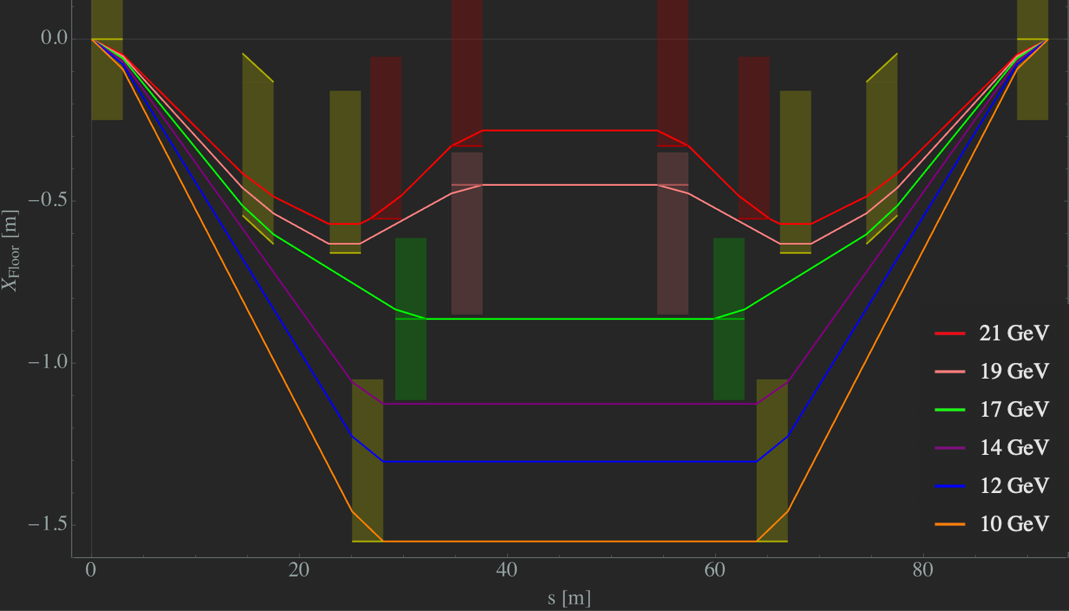}
    \caption{FFA@CEBAF Splitter Optics. The green lines represent six different energy passes through the splitter region, traversing dipole magnets (blue), extraction magnets (orange), and quadrupoles (red), each indicated as colored blocks. Note that the plot axes are not to scale: the vertical (Y) axis spans from –2 m to +2 m, while the horizontal (X) axis extends from 0 to 100 m.}
    \label{fig:splitterDonish}
\end{figure}

The only region where beams from different passes exhibit spatial separation is within the Splitter sections, as shown in Figure\ref{fig:splitterDonish}. However, even in these locations, installing RF separator cavities for each pass line is not feasible—both due to spatial constraints and the high associated cost. The central challenge, therefore, is to develop a method for extracting specific bunches from a particular pass without interfering with the continued recirculation of other bunches through the accelerator.

As discussed earlier (see Figure~\ref{fig:rfextract}), the angular deflection imparted by an RF separator or kicker cavity alone is insufficient for complete beam extraction. For an RF-based extraction system to function effectively, the initial kick must be followed by defocusing optics, such as magnetic lenses, to amplify the transverse angle. Only after sufficient separation is achieved can septum magnets be employed to complete the extraction process. This principle suggests a possible solution to the challenge posed by the FFA beamlines.

While installing dedicated RF separators for each FFA pass line is impractical, a viable alternative involves using \textbf{two strategically placed RF kickers} that act on all FFA passes simultaneously. The proposed configuration is as follows:

\begin{itemize}
    \item Place one RF kicker upstream and another downstream of the splitter section.
    
    \item Given that the current splitter design separates beam passes in the horizontal plane, the RF kickers should provide vertical kicks, deflecting the beam out of the splitter plane (e.g., downward).
    
    \item Design the optics of the splitter lines such that all passes have either even or odd multiples of $\pi$ phase advance (i.e., $2n\pi$ or $(2n+1)\pi$) ensuring that the vertical kicks from the two RF kickers cancel out for beams that are not to be extracted.
    
    \item For non-extracted passes, configure the splitter optics to suppress the effect of the RF kicks, preventing angular amplification.
    
    \item For the pass intended for extraction, set the optics to amplify the vertical separation introduced by the RF kickers. Additionally, activate the septum magnets only in the corresponding splitter line to complete the extraction.
\end{itemize}

This approach enables selective bunch extraction from a specific energy pass without interfering with the transport of other passes. It provides a practical path forward for implementing RF-based extraction within the spatial and optical constraints of the FFA beamlines.

\begin{figure}[!htb]
    \centering
    \includegraphics[width=0.6\linewidth]{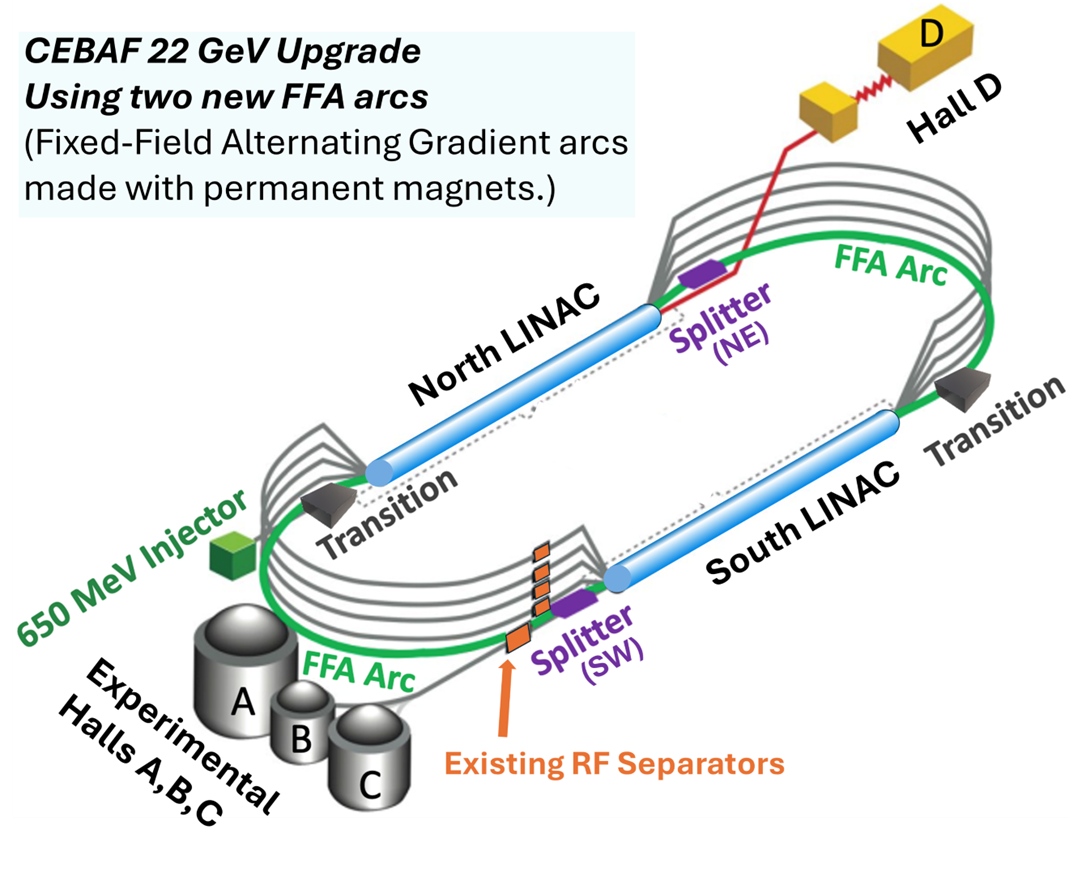}
    \caption{CEBAF 22 GeV upgrade.}
    \label{fig:cebafoverview}
\end{figure}

The proposed solution can accommodate different RF separation patterns; however, it is most practical to use the existing 750 MHz separation system currently employed to separate the Hall D beam from the ABC beams. For example, by applying this approach with 750 MHz RF kickers at the southwest (SW) splitter Figure~\ref{fig:cebafoverview}, the ABC beams can be separated from the D beam at any FFA pass, similar to the operation shown in Figure~\ref{fig:separators}c or Figure~\ref{fig:cebaf4beams}. In this configuration, the D beam continues its recirculation.

Meanwhile, the extracted ABC beams follow the established procedure used today to reach their respective experimental halls: first passing through a 500 MHz RF separator phased for three-way separation (see Figure~\ref{fig:separators}c or Figure~\ref{fig:cebaf4beams}), then being directed to Halls A, B, and C.

The D beam continues to recirculate to higher passes. Since the ABC beams have already been extracted, only the D beam remains at these higher passes and can be extracted magnetically in the northeast (NE) splitter lines, see Figure~\ref{fig:cebafoverview} without the need for additional RF separation.

In the example described, RF separation is employed only at the SW splitter, resulting in early extraction of the ABC beams and allowing Hall D to receive the beam at higher passes. To reverse this arrangement—so that the ABC halls receive beam at higher passes than Hall D—it would be feasible to add RF separation to the NE splitter, extracting the D beam before the ABC beams.

\FloatBarrier

\subsubsection{Orbit and optics correction scheme}
\label{ch:correction}
The following section summarizes the technical development of the orbit and error correction analysis detailed in the dissertation ``Error and Correction Analysis for the FFA@CEBAF Energy Upgrade'' by Alexander M. Coxe \cite{Coxe2024Dissertation}.

\paragraph{Overview and simulation framework}
The proposed FFA@CEBAF energy upgrade requires a novel correction protocol capable of simultaneously managing six electron beams at different energies (passes) within a single, shared FFA return arc. The successful transport of higher-energy passes is contingent on the stability of the lower-energy passes, and each pass may have different initial injection errors.

A proof-of-concept correction protocol was developed and validated using the Bmad accelerator simulation toolkit. The simulated lattice consists of 75 BF-O-BD-O (bending-focusing, drift, bending-defocusing, drift) cells. This lattice features 100 corrector magnets and 100 Beam Position Monitors (BPMs) \cite{Coxe2024Dissertation}.

Extensive error studies were performed to establish lattice tolerances. The studies concluded that the FFA lattice is highly resilient to the expected 1\% magnetic field and gradient errors. However, the lattice is highly sensitive to hardware misalignments, which were found to be the primary driver of beam quality degradation and beam loss \cite{Coxe2024Dissertation}.

\paragraph{Multi-pass SVD correction algorithm}
The core of the correction scheme is a multi-pass, iterative Singular Value Decomposition (SVD) algorithm. This protocol is designed to address the multi-pass challenge by correcting each of the six beams sequentially, ordered from the lowest energy to the highest \cite{Coxe2024Dissertation}.

The technical implementation follows these steps:
\begin{itemize}
    \item \textbf{Iterative matrix construction}: The algorithm builds a response matrix for each pass, incorporating the data from all preceding (lower-energy) passes. For a system with 100 correctors and 101 diagnostic points (including the exit), the response matrix for Pass 1 is $100 \times 101$, while the matrix for Pass 6 grows to $100 \times 606$ \cite{Coxe2024Dissertation}.
    \item \textbf{Exit condition weighting}: To ``lock in'' the correction for a pass, the algorithm multiplies the figure-of-merit weight on that pass's exit BPM by a factor of 100 after it is corrected. This ensures that optimizing the trajectory for higher-energy beams does not disturb the already-corrected orbits of the lower-energy beams \cite{Coxe2024Dissertation}.
    \item \textbf{Sequential field correction}: Within each energy pass, the algorithm first applies corrections to the quadrupole correctors using Twiss ($\alpha, \beta$) function data. It then sets the dipole correctors using orbit excursion and momentum data \cite{Coxe2024Dissertation}.
\end{itemize}

\paragraph{Algorithm performance and key findings}
The multi-pass SVD algorithm was found to be an ``unmitigated success'' for the error types and correction capability considered \cite{Coxe2024Dissertation}.

\begin{figure}[htb!]
    \centering
    \includegraphics[width=0.75\textwidth]{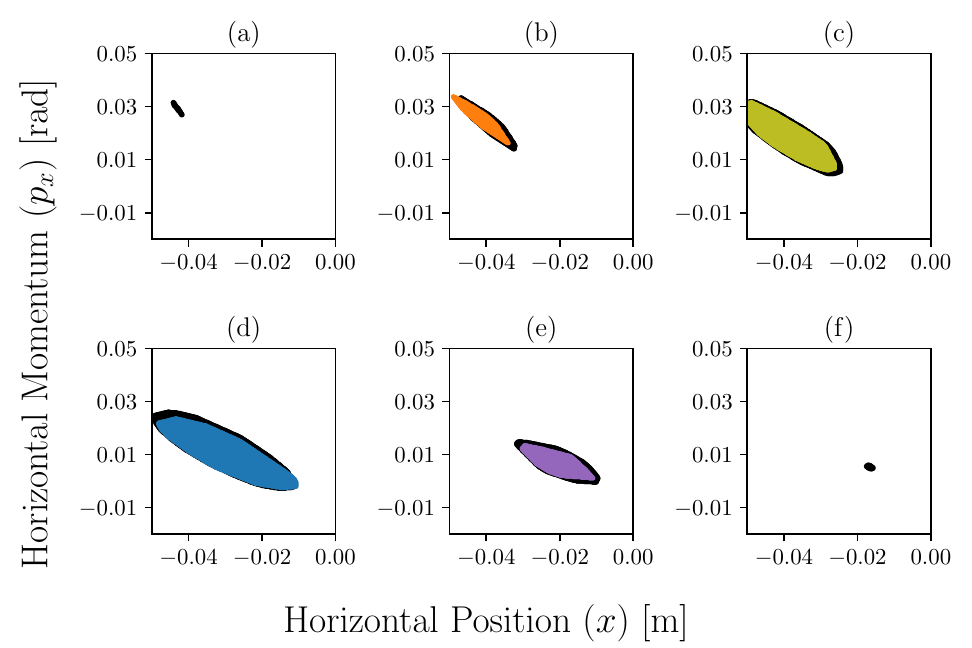}
    \caption[East FFA Arc Acceptance with Misalignments]{East Arc acceptance aperture with alignment errors. The colored areas show the small uncorrected aperture, while the black-outlined areas show the significantly larger aperture recovered by the correction algorithm. In panels (a) and (f), the uncorrected beam was completely lost, and the algorithm successfully recovered a usable aperture (black dot) \protect\cite{Coxe2024Dissertation}.}
    \label{fig:east_survey_ap}
\end{figure}

\begin{figure}[htb!]
    \centering
    \subfloat[Exit positions ($x, p_x$) \label{fig:exit_ap}]{%
        \includegraphics[width=0.75\textwidth]{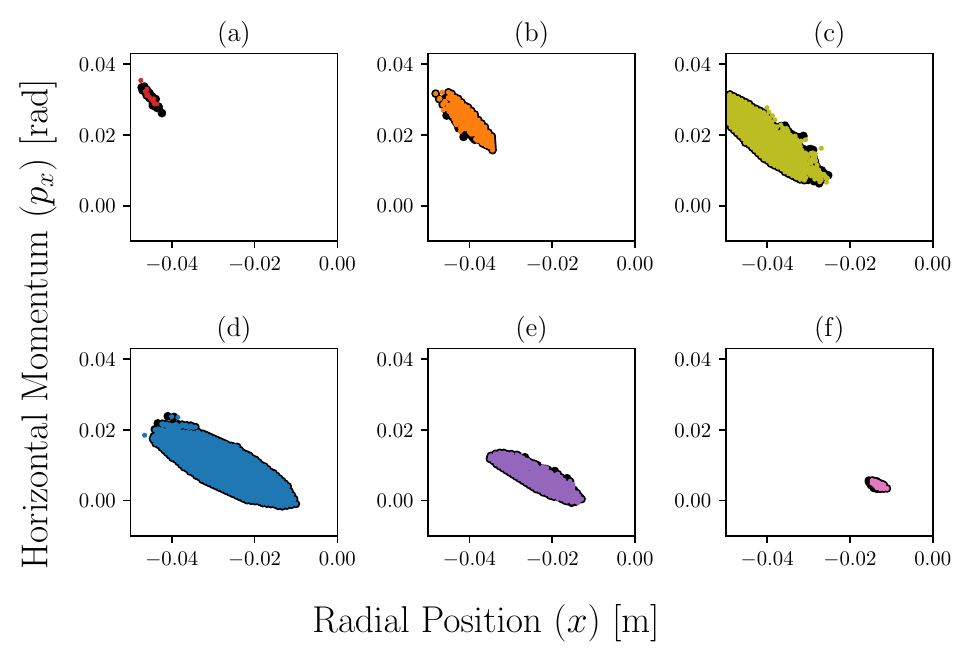}
    }
    \vfill
    \subfloat[Exit optics ($\beta, \alpha$) \label{fig:exit_optics}]{%
        \includegraphics[width=0.75\textwidth]{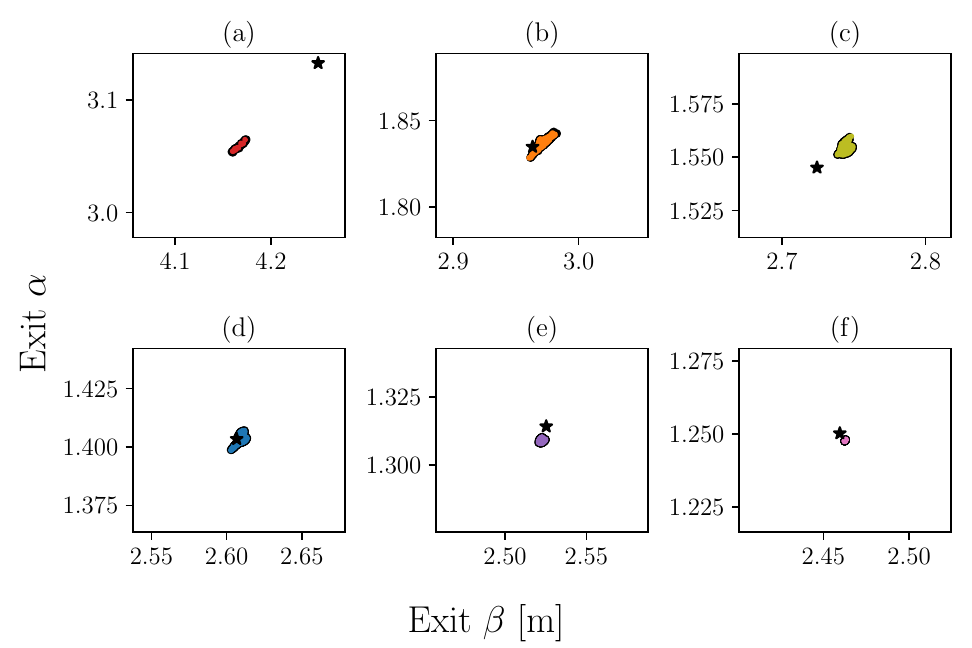}
    }
    \caption{Comparison of East Arc exit conditions for six passes. (a) The phase space shows significant overlap \protect\cite{Coxe2024Dissertation}. (b) The optical parameters are clearly distinct and clustered by energy \protect\cite{Coxe2024Dissertation}.}
    \label{fig:exit_comparison}
\end{figure}

\begin{itemize}
    \item \textbf{Beam recovery}: The most significant finding is the algorithm's ability to ``recover lost beam''. In simulations with severe alignment errors that resulted in 0\% beam transmission for the lowest and highest energy passes, the correction protocol successfully restored a stable, usable acceptance aperture \cite{Coxe2024Dissertation}. This effect is demonstrated in Figure~\ref{fig:east_survey_ap}.
    
    \item \textbf{Recombination insights}: The analysis provided a critical insight for the arc-to-Linac recombination section.
    \begin{itemize}
        \item At the arc exit, the horizontal phase space ($x, p_x$) of the six passes has a large overlap, making it ``untenable'' to separate the beams by their position or momentum. (See Figure~\ref{fig:exit_ap}).
        \item Conversely, the exit Twiss parameters ($\alpha, \beta$) are highly differentiated and ``tightly clustered'' by energy, showing ``no overlap''. (See Figure~\ref{fig:exit_optics}).
    \end{itemize}
It is therefore recommended that the transition from the FFA arcs to the Linacs make use of these differentiated optical parameters, or the related phase advance, to selectively manipulate the beams for matching \cite{Coxe2024Dissertation}.
\end{itemize}

\paragraph{Feasibility of neural network (ML/NN) corrections}
A feasibility study was conducted to determine if a machine learning model could perform corrections, with the goal of creating a future orbit-lock system for Continuous Wave (CW) operation \cite{Coxe2024Dissertation}.

\begin{figure}[htb!]
    \centering
    \includegraphics[width=0.7\textwidth]{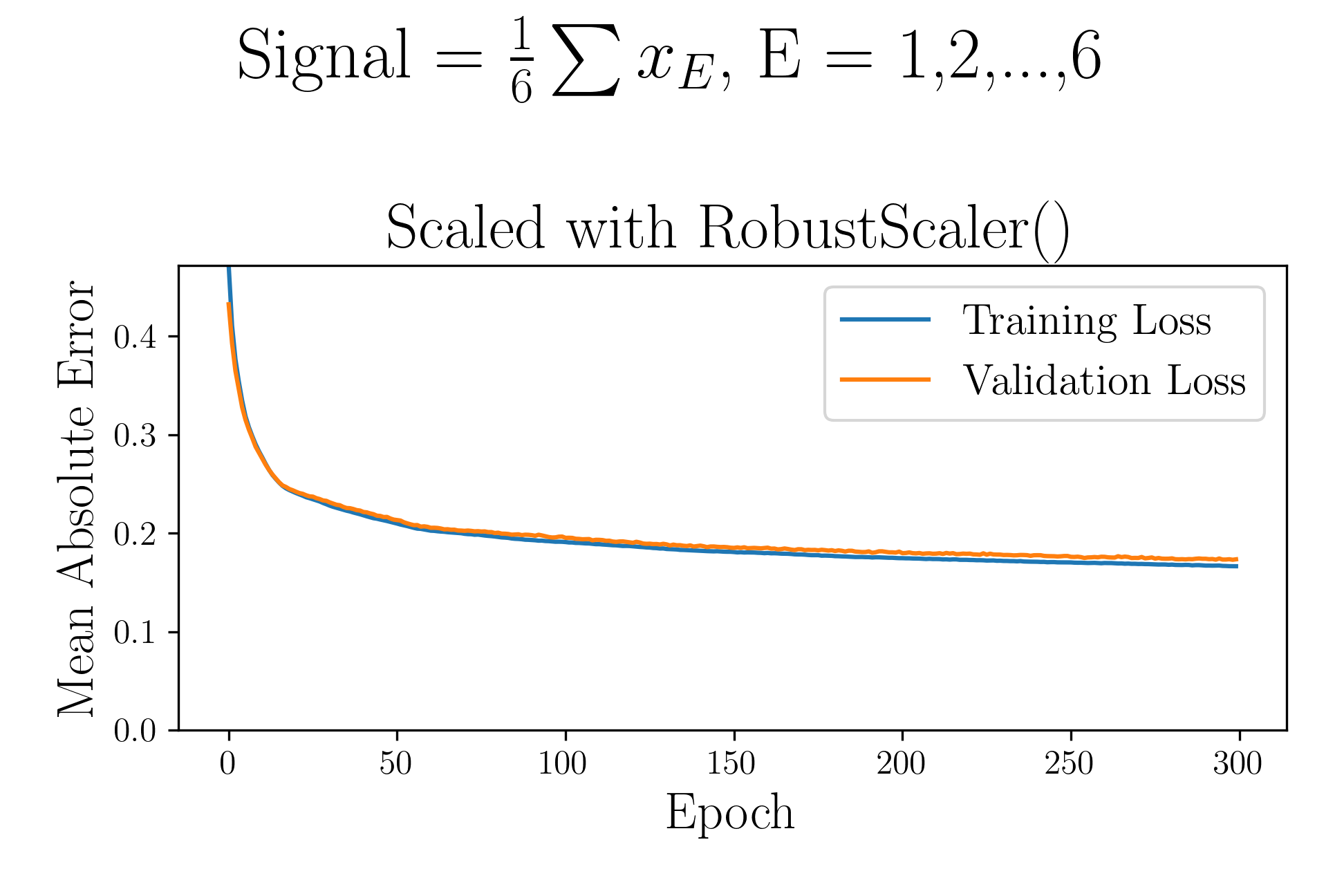}
    \caption[NN Training and Validation Loss]{Training and validation loss curves for a neural network trained on the mean of all six passes, scaled with a RobustScaler. The validation loss (orange) closely tracks the training loss (blue), demonstrating the network is successfully generalizing the correction problem \protect\cite{Coxe2024Dissertation}.}
    \label{fig:nn_loss}
\end{figure}

\begin{itemize}
    \item \textbf{Challenge}: In CW mode, BPMs cannot distinguish between the six passes and will read only a single, ``combination of all beams as one signal''. The SVD algorithm, which relies on pass-by-pass data, cannot be used \cite{Coxe2024Dissertation}.
    \item \textbf{Approach}: A feed-forward neural network was trained on simulation data. To simulate the CW BPM signal, the input to the network was a linear combination (e.g., the mean or sum) of all six passes at each BPM. The network was trained to output the 100 corresponding corrector strengths \cite{Coxe2024Dissertation}.
    \item \textbf{Results}: While the networks developed are ``not yet able'' to produce high-quality corrections, the training showed highly promising results. As shown in Figure~\ref{fig:nn_loss}, the training and validation loss curves track each other closely, indicating the network is successfully learning and ``generalizing'' the underlying physics of the correction problem rather than just ``memorizing'' the training data \cite{Coxe2024Dissertation}.
\end{itemize}

This work establishes the viability of using neural networks for this application. Future work, including tuning hyperparameters and, most importantly, increasing the 13,000-example training set, is expected to improve performance to an operational level \cite{Coxe2024Dissertation}.

In summary, this work successfully developed and demonstrated a viable correction protocol for the FFA@CEBAF arcs. The iterative, multi-pass SVD algorithm is shown to be a ``successful baseline'' for this non-scaling, linear transport lattice. It is capable of preserving beam quality and, critically, ``recovering lost beam'' in the presence of significant, realistic alignment errors. Furthermore, this research established the initial feasibility of using a neural network, trained on a composite BPM signal, as a future path toward an automated orbit-lock system for CW operations \cite{Coxe2024Dissertation}.

This work provides several key recommendations for the collaboration based on these findings. First, it is strongly recommended that the arc-to-Linac recombination design use the differentiated Twiss parameters ($\alpha$ and $\beta$) to separate the six beams, as their exit positions in phase space are shown to have untenable overlap. Second, while this protocol is successful, it is recommended that its parameters (e.g., corrector/BPM placement and SVD weights) be fully optimized using a multi-objective genetic algorithm once the lattice design is finalized. The author also recommends that the neural network approach be pursued, with the next steps being to tune hyperparameters and significantly increase the size of the 13,000-example training dataset \cite{Coxe2024Dissertation}.

\FloatBarrier

\section{Design Validation Beam Dynamics Studies}
\subsection{Synchrotron Radiation Effects}
Staying within the CEBAF footprint, while transporting high energy beams (10-22 GeV) poses significant impact of synchrotron radiation on single-particle beam dynamics, resulting in substantial emittance dilution. This calls for special mitigation measures to preserve beam quality. One of them is to increase the bend radius at the arc dipoles, which has been implemented by large packing factor in the FFA arcs (increased to about 92\%). 
Furthermore, the optics were designed to ease individual adjustment of momentum compaction and the horizontal emittance  dispersion, $\mathcal{H}$ (the curly H function), in each arc to suppress adverse effects of the synchrotron radiation on beam quality: dilution of the transverse and the longitudinal emittance due to quantum excitations. Both effects, along with the synchrotron radiation energy loss, will be addressed in the following section.
\subsubsection{Emittance growth budget}
For any high energy transport line, beam degradation due to synchrotron radiation needs to be examined. For the designs presented, this was done by using radiation integrals to calculate beam loss, increase in energy spread, and increase in emittance. The relevant radiation integrals for a transfer line are given below:

\begin{gather}
    L_2 = \int g^2\gamma_0^4ds \\
    L_3 = \int g^3\gamma_0^7ds \\
    L_{5x} = \int g^3\mathcal{H}_x\gamma_0^6ds,
\end{gather}
where $g = 1/\rho$, with $\rho$ being the radius of curvature, $\gamma_0$ being the relativistic factor, and $\mathcal{H}_x = \gamma_x\eta_x^2 + 2\alpha_x\eta_x\eta^{\prime}_x + \beta_x\eta^{\prime 2}_x$. The energy loss, increase in energy spread, and increase in normalized emittance are then given by:

\begin{gather}
    U_0 = \frac{2}{3}r_cmc^2L_2 \\
    \sigma_E^2 = \frac{4}{3}C_qr_c(mc^2)^2L_3 \\
    \epsilon_{nx} = \frac{2}{3}C_qr_cL_{5x}
\end{gather}
where $r_c$ is the classical electron radius, $m$ is the mass of the electron, $c$ is the speed of light, and $C_q = \frac{55}{32\sqrt{3}}\frac{\hbar}{mc}$~\cite{sandsPHYSICSELECTRONSTORAGE1970, helmEvaluationSynchrotronRadiation1973, jowettIntroductoryStatisticalMechanics1987}. 

For this design, there are four sections in which such effects need to be considered, the electromagnetic arcs, the spreaders/recombiners, the FFA arcs, and the splitters. The electromagnetic arcs are not included here, though due to the lower beam energy, they are unlikely to substantially impact beam quality to the halls at the highest energies. The beam degradation due to synchrotron radiation is shown for the FFA arcs in Table~\ref{TB:FFA_SR} and the spreaders/recombiners in Table~\ref{TB:SP_REC}.
\begin{table}[!hbt]
   \centering
   \caption{The beam energy, accumulated energy loss, accumulated normalized horizontal emittance growth, and accumulated energy spread due to the FFA arcs for the upper passes. The beam energy takes into account the energy loss of preceding passes.}
   \begin{ruledtabular}
\begin{tabular}{ccccc}
       \textbf{Pass} & \textbf{Beam} & \textbf{Accumulated} & \textbf{Accumulated norm.} & \textbf{Accumulated} \\
       \textbf{number} & \textbf{energy} & \textbf{energy loss} & \textbf{horiz. emit. growth} & \textbf{energy spread} \\
        & \textbf{(GeV)} & \textbf{(MeV)} & \textbf{(mm-mrad)} & \textbf{($10^{-4}$)} \\
       \colrule
5    & 11.6 & 53   & 0.03 & 2.6 \\
6    & 13.7 & 189  & 0.34 & 4.7 \\
7    & 15.7 & 352  & 1.8  & 6.2 \\
8    & 17.7 & 529  & 5.4  & 7.3 \\
9    & 19.7 & 712  & 11   & 7.9 \\
10   & 21.7 & 920  & 21   & 8.3 \\
10.5 & 22.7 & 1050 & 27   & 8.5 \\
   \end{tabular}
\end{ruledtabular}
   \label{TB:FFA_SR}
\end{table}

\begin{table}[!hbt]
   \centering
   \caption{The beam energy, accumulated energy loss, accumulated normalized vertical emittance growth, and accumulated energy spread due to the spreaders and recombiners for the upper passes. The beam energy does not take into account the energy loss of preceding passes.}
   \begin{ruledtabular}
\begin{tabular}{ccccc}
       \textbf{Pass} & \textbf{Beam} & \textbf{Accumulated} & \textbf{Accumulated norm.} & \textbf{Accumulated} \\
       \textbf{number} & \textbf{energy} & \textbf{energy loss} & \textbf{vert. emit. growth} & \textbf{energy spread} \\
        & \textbf{(GeV)} & \textbf{(MeV)} & \textbf{(mm-mrad)} & \textbf{($10^{-4}$)} \\
       \colrule
5  & 11.7 & 8   & 5  & 1.4 \\
6  & 13.9 & 19  & 18 & 1.6 \\
7  & 16.1 & 33  & 31 & 1.8 \\
8  & 18.3 & 52  & 45 & 2.1 \\
9  & 20.5 & 76  & 61 & 2.4 \\
10 & 22.7 & 105 & 79 & 2.7 \\
   \end{tabular}
\end{ruledtabular}
   \label{TB:SP_REC}
\end{table}

From the tables, it becomes clear that while degradation does occur, an electron beam of sufficient quality can likely be delivered to the halls, provided the degradation in the splitters is manageable and the FFA transition sections do not have substantially different radiation integrals from the FFA arcs. However, the splitters are likely to contribute substantially to the horizontal emittance growth, as there is significant bending and a larger $\mathcal{H}_x$ than in the FFA arcs; taken in combination, this could be a larger source of horizontal emittance growth than the arcs, despite providing less bending. This is similar to how the spreaders and recombiners cause more vertical emittance growth, despite less energy loss and increased spread, than the FFA arcs. 
\\
In summary, the lower limit for emittance dilution is set by the normalized emittance of about $100~mm\cdot mrad$ with a relative energy spread of $1.4 \cdot 10^{-3}$. Further recirculation beyond $22~GeV$ is limited by large, $1.2~GeV$ per electron, energy loss due to synchrotron radiation, which slightly exceeds the energy gain per Linac. This sets the limit of reasonable number of recirculations.

\subsubsection{‘Upper bound’ beam specs set by the users}
In consultation with the 22 GeV user community~\cite{Accardi2026}, a comparison between acceptable beam specs for different Halls at higher passes (5-10) and the projected from the above emittance dilution budget beam specs was made.  The results are summarized in Table~\ref{tab:pass-params}.

\begin{table}[!htb]
   \centering
   \caption{Normalized emittance and relative energy spread at each pass for Halls A/B/C and Hall D. Starred quantities are the ‘upper bound’ values set by the users.}
   \begin{ruledtabular}
\begin{tabular}{ccccccc}
       \textbf{Pass} & \textbf{Hall} & \textbf{$E$ [GeV]} & \textbf{$\varepsilon_N$ [m\,rad]} & \textbf{$\varepsilon_N^{*}$ [m\,rad]} & \textbf{$\sigma_{\Delta E/E}$} & \textbf{$\sigma^{*}_{\Delta E/E}$} \\
       \colrule
       5  & A/B/C & 10.5 & 3.3E-05 & 1.2E-04 & 3.1E-04 & 8.6E-03 \\
          & D     & 11.5 & 3.3E-05 & 3.3E-05 & 3.5E-04 & 4.3E-03 \\
       6  & A/B/C & 12.6 & 3.4E-05 & 1.3E-04 & 3.9E-04 & 7.2E-03 \\
          & D     & 13.7 & 3.5E-05 & 3.5E-05 & 4.6E-04 & 3.6E-03 \\
       7  & A/B/C & 14.8 & 3.6E-05 & 1.4E-04 & 5.3E-04 & 6.2E-03 \\
          & D     & 15.9 & 3.9E-05 & 3.9E-05 & 6.2E-04 & 3.1E-03 \\
       8  & A/B/C & 17.0 & 4.3E-05 & 1.8E-04 & 7.2E-04 & 5.6E-03 \\
          & D     & 18.1 & 4.8E-05 & 4.8E-05 & 8.4E-04 & 2.8E-03 \\
       9  & A/B/C & 19.2 & 5.6E-05 & 2.5E-04 & 9.7E-04 & 5.0E-03 \\
          & D     & 20.3 & 6.6E-05 & 6.6E-05 & 1.1E-03 & 2.5E-03 \\
       10 & A/B/C & 21.4 & 8.1E-05 & 3.7E-04 & 1.2E-03 & 4.4E-03 \\
          & D     & 22.4 & 1.0E-04 & 1.0E-04 & 1.4E-03 & 2.2E-03 \\
   \end{tabular}
\end{ruledtabular}
   \label{tab:pass-params}
\end{table}

Hall D is expected to have the most stringent beam quality and it serves as the baseline of operational requirements for 22 GeV CEBAF. The immediate goal is to support standard ``GlueX physics'' operations, with the coherent peak positioned at a comparable fraction of the endpoint energy to $12~GeV$ but shifted for $22~GeV$.
Furthermore the collimator diameter was reduced from $5~mm$ to $2.5~mm$ for improved beam quality.
 Resulting beam centroid stability considerations specified as ±$100~microns$ RMS were incorporated into the virtual spot size estimates presented in Table~\ref{tab:pass-params}.

\FloatBarrier

\FloatBarrier
\subsection{Tracking with Field Maps}
As part of the design validation for the FFA@CEBAF upgrade, tracking with field maps will be performed using the Bmad simulation framework to accurately model and evaluate beam transport through the proposed nonscaling FFA recirculation arcs. Unlike conventional lattice models based on linear or higher-order transfer maps, the field-map–based approach will enable direct particle tracking through the actual three-dimensional magnetic field distributions of the Halbach-style permanent magnets. This approach will be essential for capturing nonlinear effects inherent to the FFA design, particularly as the beam energy will span nearly a factor of two across multiple passes.

Field maps will be generated from detailed magnetostatic models of the Halbach arc elements and subsequently imported into Bmad using the standard 3D field map element interface. High-resolution tracking will be carried out for a representative range of energies covering all anticipated recirculation passes, ensuring accurate modeling of off-momentum particle dynamics. Each trajectory will be integrated through the full length of the recirculation arc using adaptive step-size Runge–Kutta integration to maintain symplectic accuracy and minimize numerical noise. Particular emphasis will be placed on preserving transverse emittance, characterizing chromatic response, and evaluating tune footprints across the energy range.

To validate lattice matching and assess longitudinal beam quality, the evolution of Twiss parameters, dispersion functions, and path-length dependence ($R_{56}$ and higher-order time-of-flight terms) will be extracted from the tracked particle ensemble. Comparisons between linear optics models and field-map tracking results will provide a direct measure of model fidelity and highlight the impact of intrinsic nonlinearities in the FFA magnets on beam stability. Phase-space distributions at the arc entrance and exit will be analyzed to ensure proper matching into the downstream Linac sections and to verify tolerance to alignment and field errors. These studies will demonstrate that field-map–based tracking in Bmad will provide a high-fidelity tool for validating the optical design and ensuring robust beam transport performance of the FFA arcs under realistic operating conditions.

\FloatBarrier
\subsection{Refinement of Initial 2D Permanent Magnet Design}
\begin{figure*}[!htb]
   \centering
   \includegraphics*[width=\textwidth]{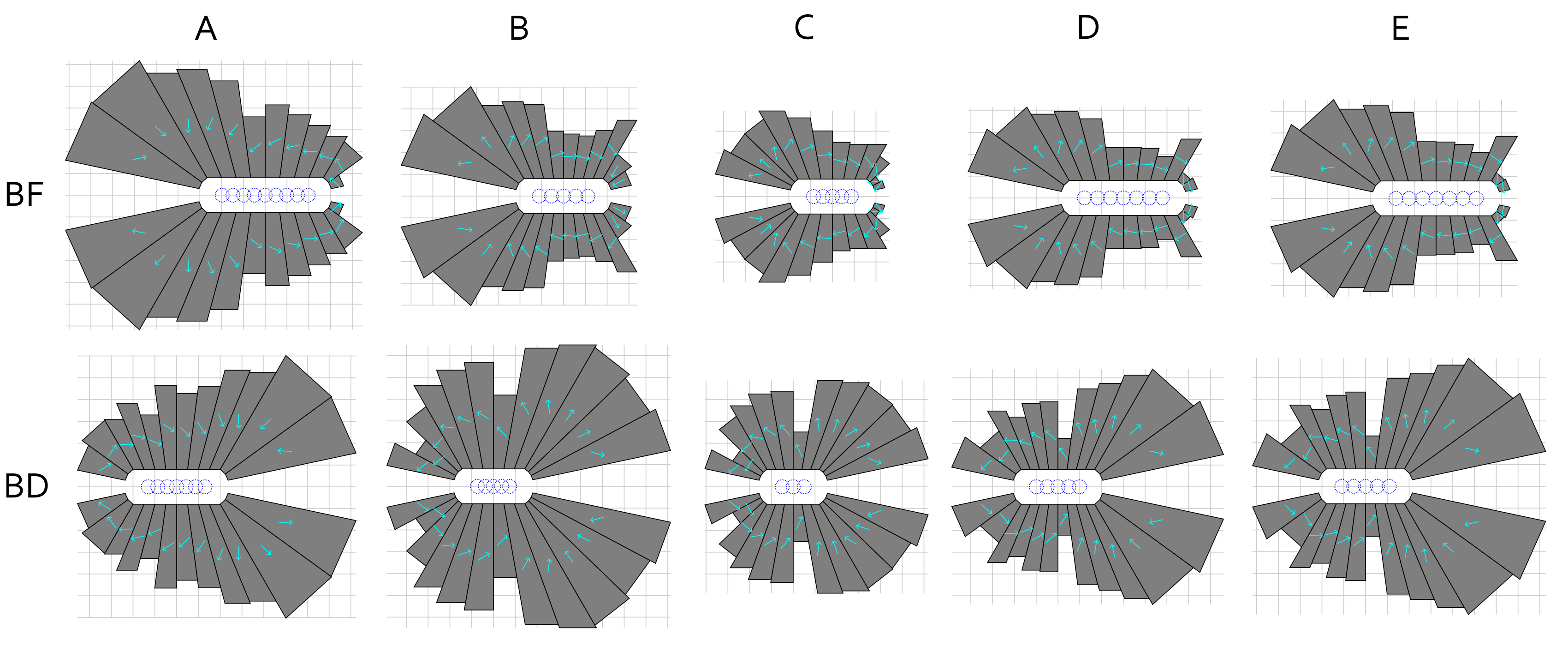}
   \caption{Cross-sections of both permanent magnets for all five lattice options.  Arrows indicate magnetization direction, grid spacing is 1\,cm.}
   \label{fig:magnetscollage}
\end{figure*}

When 2D permanent magnet designs are generated with the \texttt{HalbachArea} tool~\cite{HalbachArea}, they use the design rules given in Table~\ref{tab:magnetrules}, which ensure there is an open midplane to allow synchrotron radiation to escape, as well as allowing enough vertical room for the beam pipe.  Cross sections of the resulting magnets are shown in Figure~\ref{fig:magnetscollage}.

\begin{table}[!htb]
   \centering
   \caption{Permanent Magnet Design Rules}
   \begin{ruledtabular}
\begin{tabular}{lc}
       \textbf{Parameter} & \textbf{Value} \\
       \colrule
Number of wedges & 24 (12 per side) \\
Midplane angular gap & $\pm$12$^\circ$ \\
Vertical aperture & $\pm$8\,mm \\
Minimum midplane gap & $\pm$3\,mm \\
Material & NdFeB \\
Grade & N42EH \\
$B_r$ (real) & 1.28--1.33\,T \\
$B_{r1}$ (for $\mu_r=1$ model) & 1.248\,T \\
$\mu_0H_{cJ}$ & 2.9\,T \\
   \end{tabular}
\end{ruledtabular}
   \label{tab:magnetrules}
\end{table}

The magnet geometry is made of multiple wedges around the aperture, as used in CBETA~\cite{CBETAmagnets} and adapted in geometry for the CEBAF upgrade use case, including a recently constructed prototype~\cite{MagnetIPAC23}.  The material selected is one of the highest field grades that still admits reasonably good radiation resistance judged from its $H_{cJ}$ value, although this is also being tested experimentally at CEBAF~\cite{RyanRadiation}.  

\begin{table*}[!htb]
   \centering
   \caption{Magnet Figures of Merit for Lattice Options}
   \begin{ruledtabular}
\begin{tabular}{lcccccc}
       \textbf{Option} & \textbf{$|\mathbf B|_\mathrm{max}$} & \textbf{Orbit excursion} & \multicolumn{3}{c}{\textbf{Magnet areas (cm$^\mathrm2)$}} \\
         & \textbf{(T)} & \textbf{(mm)} & \textbf{Average} & \textbf{BD} & \textbf{BF} \\
       \colrule
A & 1.5346 & 44.968 & 84.69 & 87.43 & 94.41 \\
B & 1.6140 & 28.607 & 75.75 & 104.56 & 58.54 \\
C & 1.4922 & 23.602 & 44.29 & 59.32 & 34.04 \\
D & 1.4689 & 41.739 & 54.38 & 72.18 & 46.16 \\
E & 1.5438 & 42.966 & 64.24 & 86.07 & 53.86 \\
   \end{tabular}
\end{ruledtabular}
   \label{tab:magnetfoms}
\end{table*}

The resulting figures of merit for the lattice options defined in section \ref{-ffa-arcs} are shown in Table \ref{tab:magnetfoms}.  The automatically-produced design geometries for all magnets in options A--E are shown in Figure~\ref{fig:magnetscollage}.

\FloatBarrier
\newpage
\subsection{Depolarization Effects due to Synchrotron Radiation}

The FFA arcs of the 22\,GeV CEBAF upgrade operate in a regime where
synchrotron radiation (SR) is not merely a perturbative correction but the
dominant effect shaping both the transverse orbital dynamics and the spin
transport. At the sixth-pass energy of $21.55$\,GeV, quantum fluctuations
in the photon emission drive stochastic energy oscillations whose amplitude
far exceeds any classical envelope estimate, and the resulting correlated
excursions in energy and transverse orbit propagate through the non-closed
FFA optics to modulate the local spin-precession rate along each particle's
trajectory. A reliable prediction of the polarization delivered to the
experimental halls therefore requires multi-particle spin--orbit tracking
with SR included self-consistently, rather than single-particle or
envelope-level estimates. We have implemented such tracking in
{\sc Bmad}~\cite{ref:bmad} on the baseline 75-cell FFA FODO
arc~\cite{ref:lattice}, pursuing two objectives simultaneously: quantifying
SR-induced depolarization along a $21.55$\,GeV arc pass, and extracting the
betatron tune shift driven by the same radiation-induced energy oscillations
from the identical tracked ensemble.

\subsubsection{Simulation setup}

A sample of $N=10^{4}$ macroparticles is drawn from a 6-D Gaussian
distribution matched to the FFA emittance budget and propagated through
one complete 180\textdegree{} arc pass. The SR energy loss at this energy
is ${\sim}182$\,MeV per pass, as computed from {\sc Bmad}'s radiation
integrals for the arc proper; contributions from spreaders, recombiners,
and other beamline sections will be included in future start-to-end
runs~\cite{FFAStatus}. Non-linear longitudinal path-length effects are
retained to third order,
\begin{equation}
  \Delta L(\delta)=R_{56}\,\delta+T_{566}\,\delta^{2}+U_{5666}\,\delta^{3},
\end{equation}
which is necessary to faithfully reproduce the energy-oscillation amplitude
when SR is active. Each macroparticle carries a unit spin vector; two
simulation modes are compared. Those modes have SR loss and stochastic oscillations
\emph{disabled} (SR OFF) and \emph{enabled} (SR ON). Because
longitudinally polarized beam is required at the experimental halls, the injection spin orientation used here
is chosen to be illustrative ($s_x=-0.9$ at injection, giving
$s_z\simeq+0.9$ at the arc exit).

\begin{figure}[!ht]
\centering
\includegraphics[width=\columnwidth]{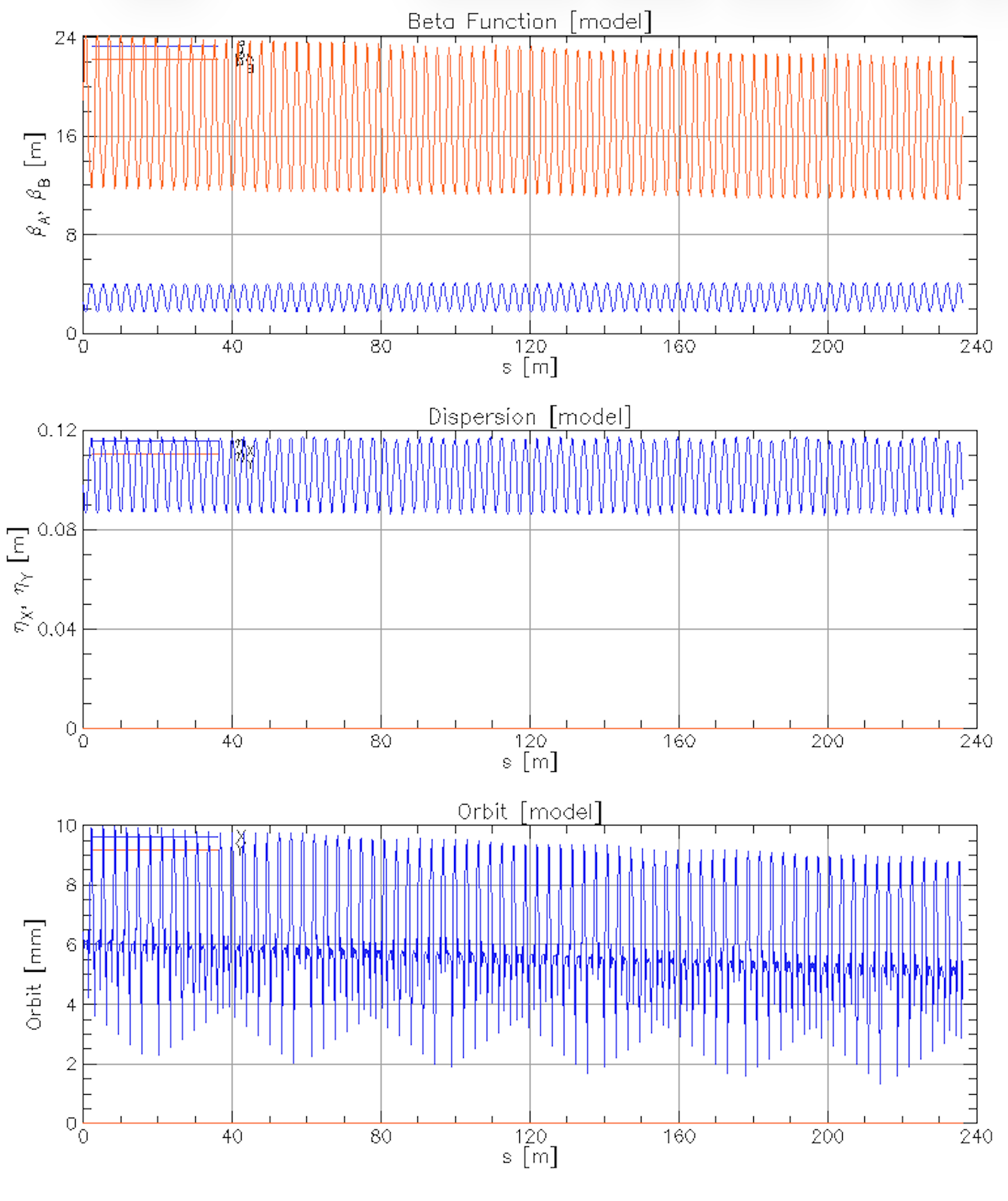}
\caption{Periodic optics of the 21.55\,GeV FFA East arc with synchrotron radiation. Top:
$\beta_{x,y}$. Middle:~dispersion. Bottom: orbit in the presence of SR.}
\label{fig:optics}
\end{figure}

\begin{figure}[!ht]
\centering
\includegraphics[width=\columnwidth]{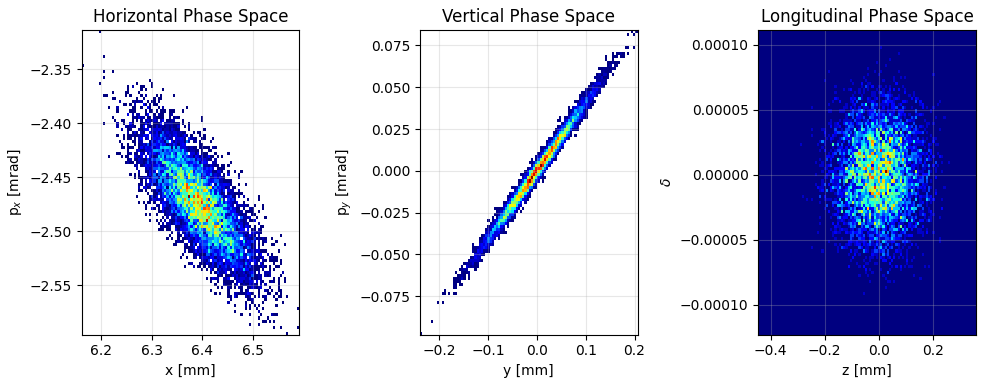}
\caption{Initial 6-D phase-space distribution at the arc entrance:
horizontal $(x,p_x)$, vertical $(y,p_y)$, longitudinal $(z,\delta)$.}
\label{fig:initbeam}
\end{figure}

The periodic Twiss functions, dispersion, and closed orbit of the
$21.55$\,GeV FFA arc with SR ON are shown in Figure~\ref{fig:optics}.
The optics are intentionally non-closed: the horizontal dispersion
$\eta_x\approx9$--$11$\,cm and the off-axis closed orbit of
${\sim}5$--$8$\,mm allow all six passes to share the same permanent-magnet
channel~\cite{ref:bogacz,FFAdemo,benesch1,benesch3}. The injected
phase-space distribution (Fig.~\ref{fig:initbeam}) consequently carries a
finite horizontal centroid and slope inherited from the open-geometry
dispersion, while the vertical plane is nominally matched and the
longitudinal plane is uncorrelated.

\subsubsection{Spin transport and depolarization}

Spin motion through the arc is governed by the
Thomas--Bargmann--Michel--Telegdi (Thomas--BMT) equation~\cite{Montague1984},
\begin{equation}
  \frac{d\vec{S}}{dt} = \vec{\Omega} \times \vec{S},
\end{equation}
where, for a purely magnetic lattice ($\vec{E}=0$), the precession vector is
\begin{equation}
  \vec{\Omega}
  = -\frac{e}{m\gamma}
  \left[
    (a\gamma + 1)\vec{B}_{\perp}
    + (a + 1)\vec{B}_{\parallel}
  \right],
  \label{eq:Omega}
\end{equation}
with $a=(g-2)/2=1.159\,652\times10^{-3}$ the anomalous magnetic moment and
$\vec{B}_{\perp}$, $\vec{B}_{\parallel}$ the field components perpendicular
and parallel to the particle velocity, in the form used internally by
{\sc Bmad} for purely magnetic elements. At $E=21.55$\,GeV the Lorentz
factor is $\gamma\approx4.220\times10^{4}$, giving a spin tune
\begin{equation}
  \nu_s = a\gamma \approx 48.9
\end{equation}
and a total spin precession per 180\textdegree{} pass of
\begin{equation}
  \theta_s = \pi\nu_s = \pi a\gamma \approx 154~\mathrm{rad}
  \quad(\text{nearly }24.4\text{ full }2\pi\text{ rotations}).
\end{equation}
Expressed along the arc path length $s$~\cite{Montague1984},
\begin{equation}
  \frac{d\vec{S}}{ds} = \frac{1}{v}\,\vec{\Omega}\times\vec{S},
\end{equation}
with $\vec{\Omega}$ from Eq.~(\ref{eq:Omega}); radial excursions driven by
the non-zero dispersion of the FFA cells modulate the local spin-precession
rate particle-by-particle. For CEBAF-style recirculation, the cumulative
spin evolution is the sum of arc-by-arc Thomas--BMT precessions;
{\sc Bmad} performs the tracking pass-by-pass.

The polarization magnitude of the ensemble is defined as
\begin{equation}
  P = \sqrt{\langle S_x\rangle^{2}+\langle S_y\rangle^{2}+\langle S_z\rangle^{2}},
\end{equation}
and the SR-induced change is $\Delta P = P_{\mathrm{final}}-P_{\mathrm{initial}}$.

\begin{figure}[!ht]
\centering
\includegraphics[width=\columnwidth]{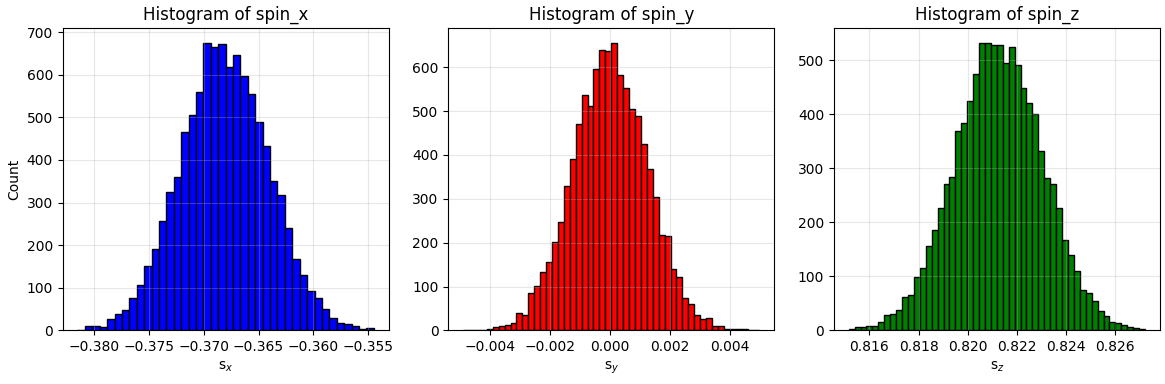}
\caption{Final spin distribution at the arc exit, SR OFF.
$\langle s_z\rangle=0.82127\pm0.00002$, $\sigma_{s_z}=1.8\times10^{-3}$,
$\Delta|S|=1.15\times10^{-5}$.}
\label{fig:spinNoSR}
\end{figure}

With SR disabled, the beam exits the arc in the configuration shown in
Fig.~\ref{fig:spinNoSR}: the mean longitudinal spin component is
$\langle s_z\rangle=0.82127\pm0.00002$ with a spread of
$\sigma_{s_z}=1.8\times10^{-3}$ and a polarization-magnitude change of only
$\Delta|S|=1.15\times10^{-5}$, confirming that deterministic spin transport
through the FFA lattice is inherently benign.

\begin{figure}[!ht]
\centering
\includegraphics[width=0.9\columnwidth]{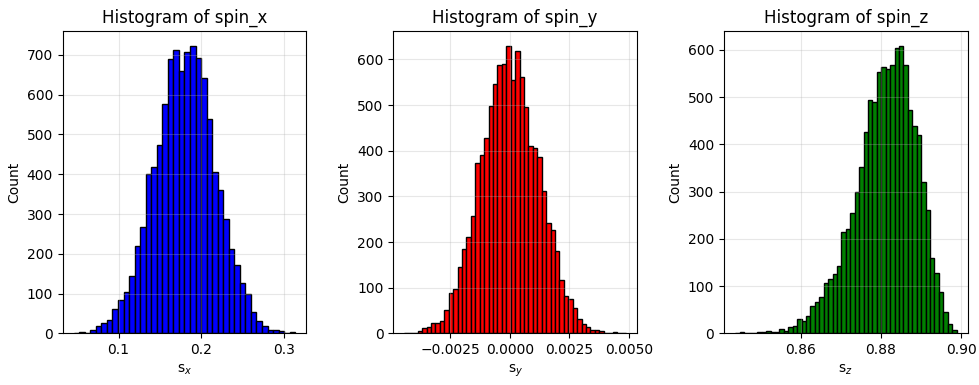}
\caption{Final spin distribution at the arc exit, SR ON.
$\langle s_z\rangle = 0.88125\pm0.00007$, $\sigma_{s_z}=7.0\times10^{-3}$,
$\Delta|S|=7.8\times10^{-4}$. Although $|S|$ barely changes, the spin
orientation is rotated significantly relative to the SR-OFF case.}
\label{fig:spinSR}
\end{figure}

Enabling SR loss and stochastic oscillations substantially alters the
picture (Fig.~\ref{fig:spinSR}). The mean longitudinal spin shifts to
$\langle s_z\rangle=0.88125\pm0.00007$, the spread grows to
$\sigma_{s_z}=7.0\times10^{-3}$, and the polarization-magnitude change
rises to $\Delta|S|=7.8\times10^{-4}$. This magnitude change remains well
within the $1\sigma$ resolution of the Compton, M\o{}ller, and Mott
polarimeters operated at CEBAF~\cite{ref:hallC} at all CEBAF operational
energies. However, the coherent rotation of the mean spin---a shift
$\Delta s_z\sim6\times10^{-2}$ relative to the SR-OFF case would manifest
as a measurable systematic offset if left uncorrected. The physical origin
is clear: SR-induced stochastic energy kicks, amplified by the non-zero
$\eta_x$ and the higher-order momentum compaction coefficients $T_{566}$ and
$U_{5666}$, cause each particle to traverse a slightly different local spin
tune $a\gamma$ along its trajectory, rotating the ensemble-averaged spin
vector.

\FloatBarrier

\subsubsection{Betatron tune shift induced by radiation-driven energy oscillations}

The same tracked ensemble provides a self-consistent measurement of the
horizontal betatron tune via a normalized phase-space estimator. The
horizontal betatron tune $Q_x$ counts the transverse oscillations per
lattice section and is related to the phase advance $\Delta\mu_x$ in
normalized phase space by $Q_x=\Delta\mu_x/(2\pi)$.
Courant--Snyder normalized coordinates are defined as
\begin{equation}
  X=\frac{x}{\sqrt{\beta_x}}, \qquad
  P_X=\sqrt{\beta_x}\,p_x+\frac{\alpha_x}{\sqrt{\beta_x}}\,x,
\end{equation}
and the entrance and exit distributions are cast in complex form,
$z_i=X_i+iP_{X,i}$ and $z_f=X_f+iP_{X,f}$. After subtracting the
centroid, $z'_{i,j}=z_{i,j}-\langle z_i\rangle$ and
$z'_{f,j}=z_{f,j}-\langle z_f\rangle$, the ensemble-averaged phase advance
follows from the argument of the complex inner product,
\begin{equation}
  \Delta\mu_x = \arg\!\left(\sum_j z'^{\,*}_{i,j}\,z'_{f,j}\right), \qquad
  Q_x = \frac{\Delta\mu_x}{2\pi}.
  \label{eq:tune}
\end{equation}
In the linear approximation, normalized-space transport reduces to a pure
rotation,
\begin{equation}
  \begin{pmatrix}X_f\\P_{X,f}\end{pmatrix}
  = R(\mu)\begin{pmatrix}X_i\\P_{X,i}\end{pmatrix}, \quad
  R(\mu)=\begin{pmatrix}\cos\mu & \sin\mu\\ -\sin\mu & \cos\mu\end{pmatrix},
\end{equation}
with $\mu=2\pi Q_x$, so Eq.~(\ref{eq:tune}) extracts this rotation angle
directly from the tracked ensemble without requiring explicit matrix
reconstruction.

With SR off, the estimator recovers the nominal arc tune
$(Q_x,Q_y)=(14.0787,2.3222)$ to better than $10^{-3}$. With SR on,
{\sc Bmad}'s optics module reports $(Q_x,Q_y)=(14.1272,2.4193)$,
corresponding to shifts $\Delta Q_x\simeq+0.05$ and
$\Delta Q_y\simeq+0.10$. These arise from the interplay between the
SR-driven energy oscillations and the chromaticity of the open-geometry FFA
cell. The tracking ensemble further reveals a $0.8$\,mm orbit shift and
$0.3$\,mrad angular deflection in the horizontal plane, and a significant
modification of the Twiss parameters,
\[
  \beta_x: 2.495\to3.830~\mathrm{m},\quad
  \alpha_x: 1.192\to1.344,\quad
  \gamma_x: 0.970\to0.925,\quad
  \varepsilon_x: 1.264\to2.213~\mathrm{nm},
\]
indicating a residual mismatch at the arc--Linac interface that must be
absorbed by the downstream matching section.

\begin{figure}[!ht]
\centering
\includegraphics[width=0.9\columnwidth]{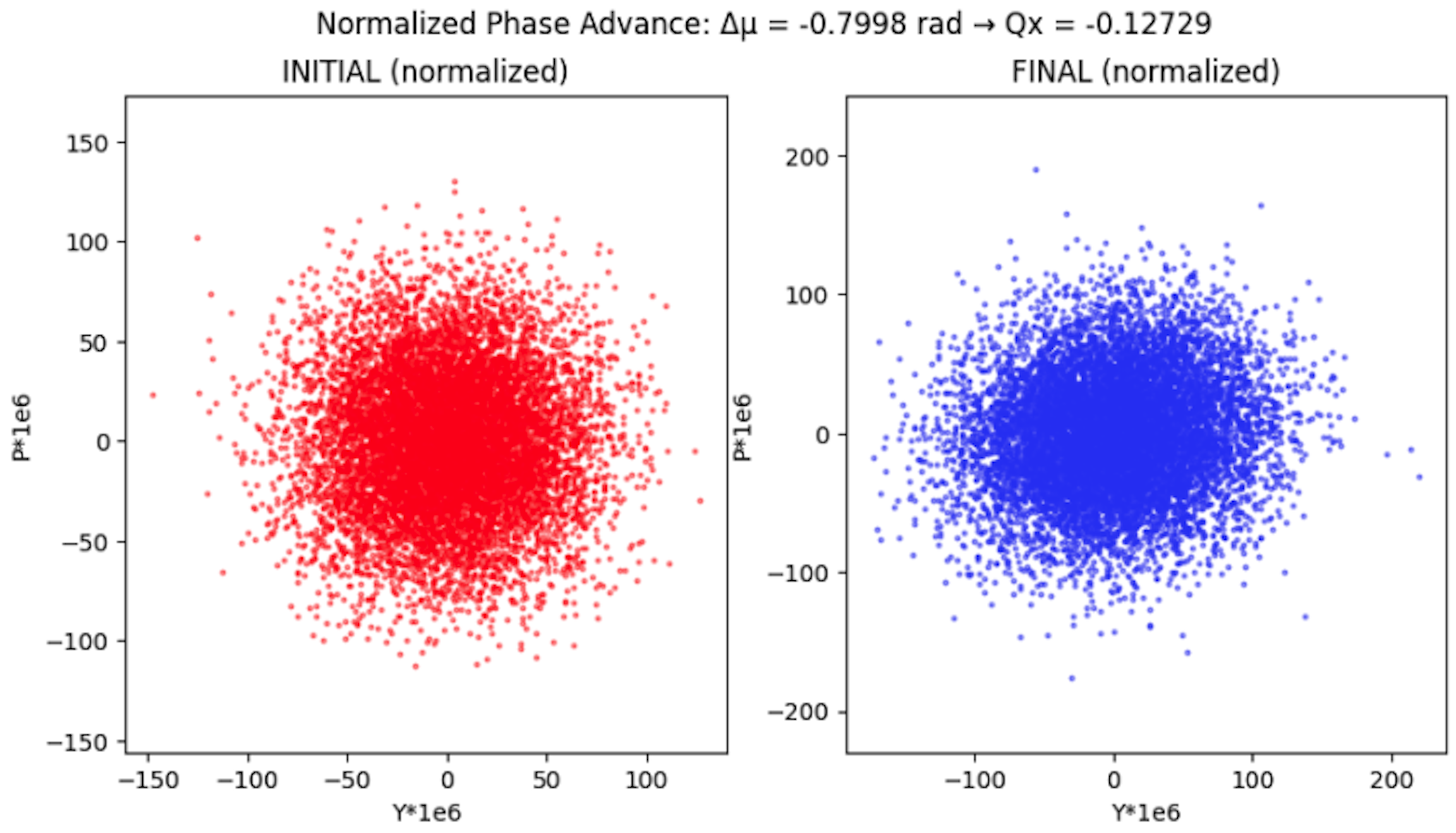}
\caption{Normalized horizontal phase-space distribution at the entrance
(red) and exit (blue) of the 21.55\,GeV FFA arc with SR ON. The
centroid-subtracted complex inner product yields the per-arc phase
advance $\Delta\mu = -0.7998$\,rad ($Q_x = -0.12729$ mod 1).}
\label{fig:tuneNorm}
\end{figure}

The normalized horizontal phase-space distributions at the arc entrance and
exit are displayed in Figure~\ref{fig:tuneNorm}. Applying
Eq.~(\ref{eq:tune}) to the centroid-subtracted ensemble yields a phase
advance of $\Delta\mu=-0.7998$\,rad, equivalent to $Q_x=-0.12729$\,(mod~1)
per arc, in agreement with the {\sc Bmad}-predicted tune shift. For
reference, the un-normalized (raw coordinate) fit gives
$\Delta\mu_{\mathrm{raw}}=-0.5365$\,rad ($Q_x=-0.0853$), a ${\sim}30\%$
discrepancy attributable entirely to the change in $\beta_x$ and $\alpha_x$
between the two measurement planes; the normalized estimator removes this
bias by construction.

\FloatBarrier
\section{Alternatives}

\subsection{FMC vs FODO FFA Arc Optics}
\label{subsec:fmc-vs-fodo}

Two optics architectures have been considered for the fixed-field arcs of the
22~GeV upgrade. The first is the Flexible Momentum Compaction (FMC) cell, which is similar to the
already employed CEBAF's electromagnetic arcs and in the 650~MeV
injector racetrack, which offers orthogonal control of the momentum-compaction
factor and near-isochronous transport in each arc. The second is a compact
FODO-type doublet cell, \textbf{BD~o~BF~o}, built from combined-function
Halbach permanent magnets in direct lineage from the CBETA return
loop~\cite{brooks2020}. The FODO cell is the present working baseline for the
FFA arcs: its short cell length maximizes the dipole packing factor, the key
lever for suppressing synchrotron-radiation (SR) induced emittance
dilution, and it minimizes magnet aperture and material volume. Its principal
drawback is that the momentum compaction is fixed by the magnet geometry and
varies with energy, so the residual time-of-flight differences between passes
must be absorbed by the splitter lines. This subsection summarizes the status
of the FODO West-Arc lattice, its extension to a wider energy band through a
sextupole field component, the cross-code benchmarking of its optics, and the
motivation for recovering FMC-like momentum-compaction control in a future
iteration.

\subsubsection{Synchrotron-radiation-optimized FODO lattice and its extended-range variant}
\label{subsubsec:fodo-lattice}

The West Arc comprises $N_{\mathrm{cell}}=98$ identical FFA cells of length
$L_{\mathrm{cell}}=2.440255$~m, each built from a focusing (BF) and a
defocusing (BD) combined-function magnet. The SR-optimized linear design
transports the six nominal FFA pass energies from $11.65$ to $22.65$~GeV, with
a total SR loss in the West Arc, summed over its five recirculated passes, of
$525.8$~MeV. To recover margin at the low-energy end of the band, which is important
for Linac energy flexibility, cavity de-rating scenarios, and polarization
optimization with the Wien filter was used; the lattice was subsequently extended down
to $9.45$~GeV by adding a sextupole component to the linear combined-function
gradient of each arc magnet. The sextupole term controls the energy dependence
of the cell tunes and of the momentum-compaction (time-of-flight) function,
both of which would otherwise drift out of the acceptable working range at the
edges of the enlarged band. The parameters of both variants are compared in
Table~\ref{tab:fodo-cell-comparison}; the peak fields of the extended-range
design remain within the permanent-magnet limits established by the prototype
program. The price of the wider acceptance is a larger SR loss of
$644.2$~MeV over the five West-Arc passes, which must be folded into the
pass-to-pass energy bookkeeping.

\begin{table}[!htb]
\caption{West-Arc FFA FODO cell parameters: the synchrotron-radiation-optimized
linear design ($11.65$--$22.65$~GeV) versus the extended-range design with
sextupole components ($9.45$--$22.65$~GeV). Sextupole strengths are quoted as
integrated components normalized to the beam rigidity,
$K_{2}L=\mathrm{SEX}/B\rho$.}
\label{tab:fodo-cell-comparison}
\begin{ruledtabular}
\begin{tabular}{lcc}
Parameter & SR-optimized & Extended range \\
\colrule
Energy range [GeV]                        & $11.65$--$22.65$ & $9.45$--$22.65$ \\
Number of cells $N_{\mathrm{cell}}$       & $98$             & $98$ \\
Cell length $L_{\mathrm{cell}}$ [m]       & $2.440$          & $2.440$ \\
\colrule
\multicolumn{3}{l}{\textit{Focusing magnet (BF)}}\\
Gradient $G_{F}$ [T/m]                    & $-71.606$        & $-96.440$ \\
Length $L_{QF}$ [m]                       & $1.400$          & $1.400$ \\
Bend angle $\theta_{F}$ [rad]             & $-0.025$         & $-0.025$ \\
Central field $B_{F}$ [T]                 & $-1.210$         & $-1.210$ \\
Peak field $B_{F,\mathrm{max}}$ [T]       & $-1.640$         & $-1.557$ \\
\colrule
\multicolumn{3}{l}{\textit{Defocusing magnet (BD)}}\\
Gradient $G_{D}$ [T/m]                    & $89.643$         & $108.224$ \\
Length $L_{BD}$ [m]                       & $0.880$          & $0.880$ \\
Bend angle $\theta_{D}$ [rad]             & $-0.006$         & $-0.006$ \\
Central field $B_{D}$ [T]                 & $-0.428$         & $-0.428$ \\
Peak field $B_{D,\mathrm{max}}$ [T]       & $-1.631$         & $-1.618$ \\
\colrule
\multicolumn{3}{l}{\textit{Sextupole components}}\\
Focusing $K_{2}L$ (SF)                    & ---              & $+1176.0/B\rho$ \\
Defocusing $K_{2}L$ (SD)                  & ---              & $-1911.1/B\rho$ \\
\colrule
SR loss, West Arc (five passes) [MeV]     & $525.810$        & $644.158$ \\
\end{tabular}
\end{ruledtabular}
\end{table}

\begin{figure}[!htb]
\centering
\includegraphics[width=0.8\columnwidth]{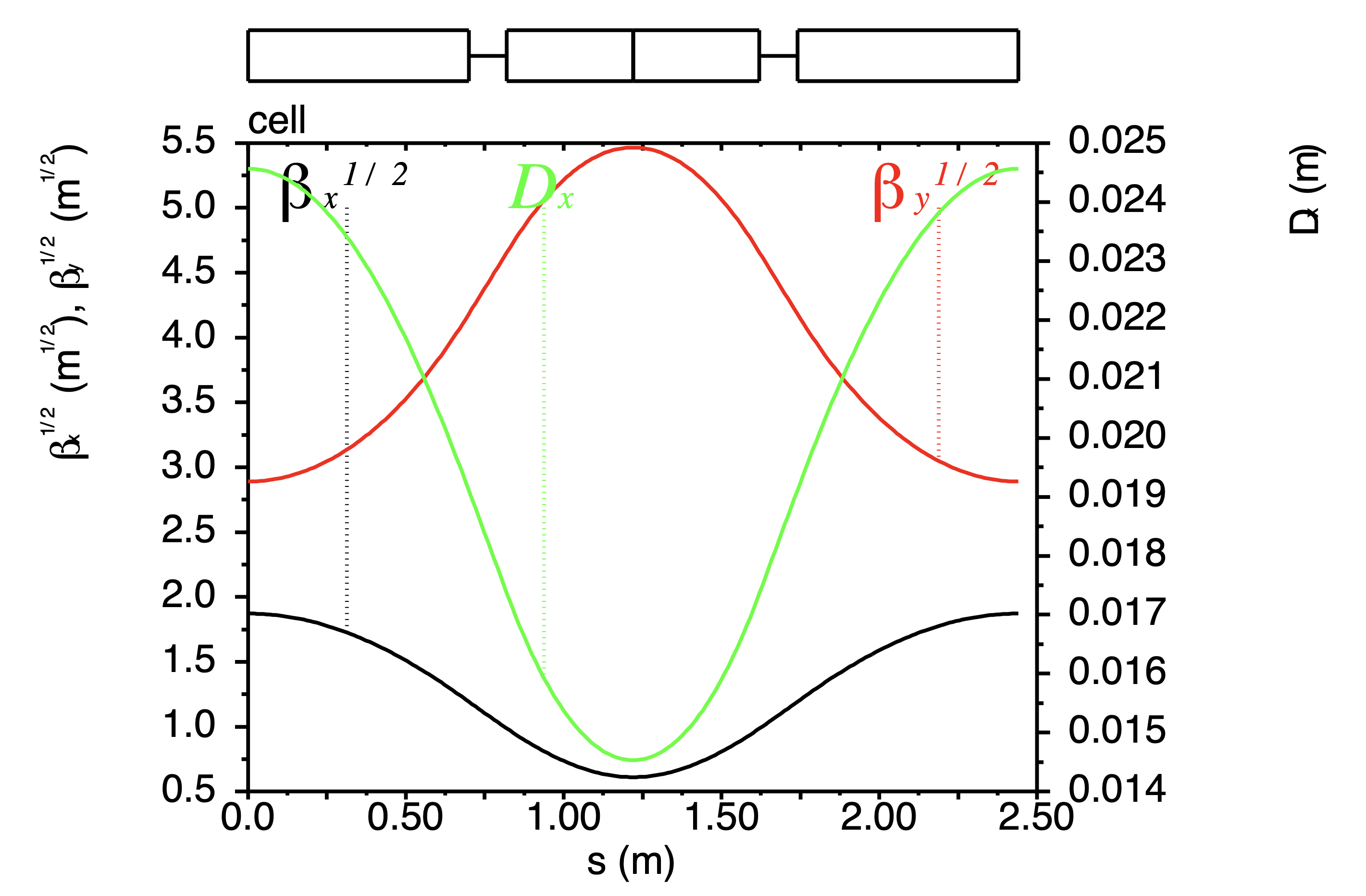}
\caption{Square-root betatron functions $\sqrt{\beta_x}$ (red),
$\sqrt{\beta_y}$ (black) and horizontal dispersion $D_x$ (green) along one
West-Arc FFA FODO cell of the extended-range design, computed with
\textsc{madx-ptc}. The magnet layout of the cell is drawn above the plot.}
\label{fig:fodo-cell-optics}
\end{figure}

The linear optics of a single cell of the extended-range design, obtained with
\textsc{madx-ptc}, are shown in Figure~\ref{fig:fodo-cell-optics}: the
square-root betatron functions remain bounded below
$\sim\!5.5~\mathrm{m}^{1/2}$ and the horizontal dispersion peaks at only a few
centimetres, as required for the compact good-field region of the
permanent-magnet aperture.

\subsubsection{Betatron tunes and cross-code benchmarking}
\label{subsubsec:fodo-tunes}

\begin{figure}[]
\centering
\includegraphics[width=0.8\columnwidth]{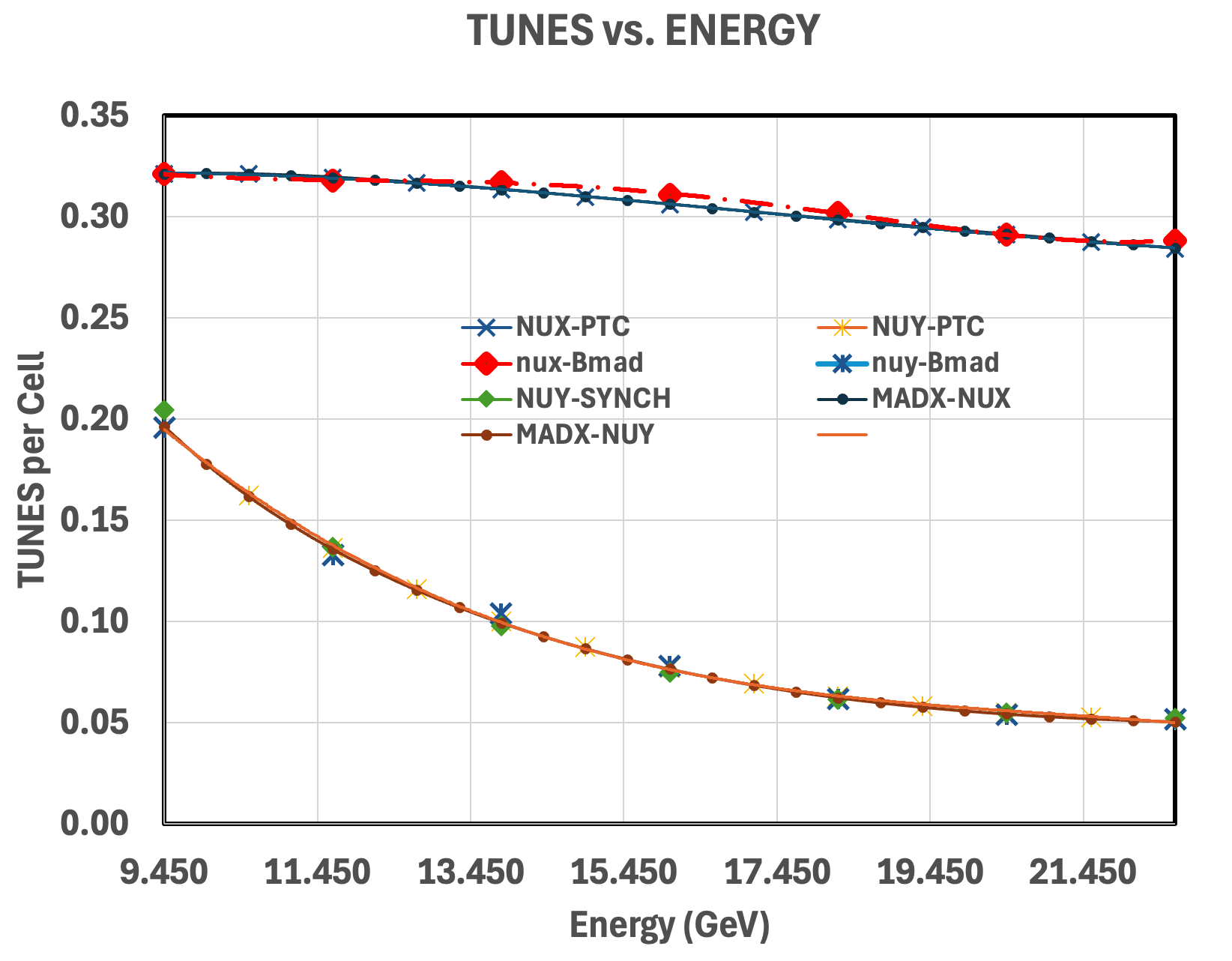}
\caption{Per-cell horizontal and vertical betatron tunes versus energy for the
West-Arc FODO lattice, computed with \textsc{ptc}, \textsc{madx-ptc},
\textsc{bmad}, and \textsc{synch}. The four codes overlap to within plotting
accuracy.}
\label{fig:fodo-tunes-codes}
\end{figure}

Because the arc is a fixed-field channel, the closed orbit, betatron tunes,
and transport optics all vary with beam energy. The per-cell tunes were
tracked across the full band with four independent codes, \textsc{ptc},
\textsc{madx-ptc}, \textsc{bmad}, and \textsc{synch}. As shown in
Fig.~\ref{fig:fodo-tunes-codes}, the codes agree closely. The horizontal tune
varies gently from $\nu_x/\mathrm{cell}\approx0.32$ at the low-energy end to
$\approx0.285$ at $22.65$~GeV, while the vertical tune falls more steeply from
$\nu_y/\mathrm{cell}\approx0.20$ at $9.45$~GeV to $\approx0.05$ at the top
energy. The sextupole component keeps this tune footprint away from low-order
structure resonances over all recirculated passes.

\begin{figure}[]
\centering
\includegraphics[width=0.8\columnwidth]{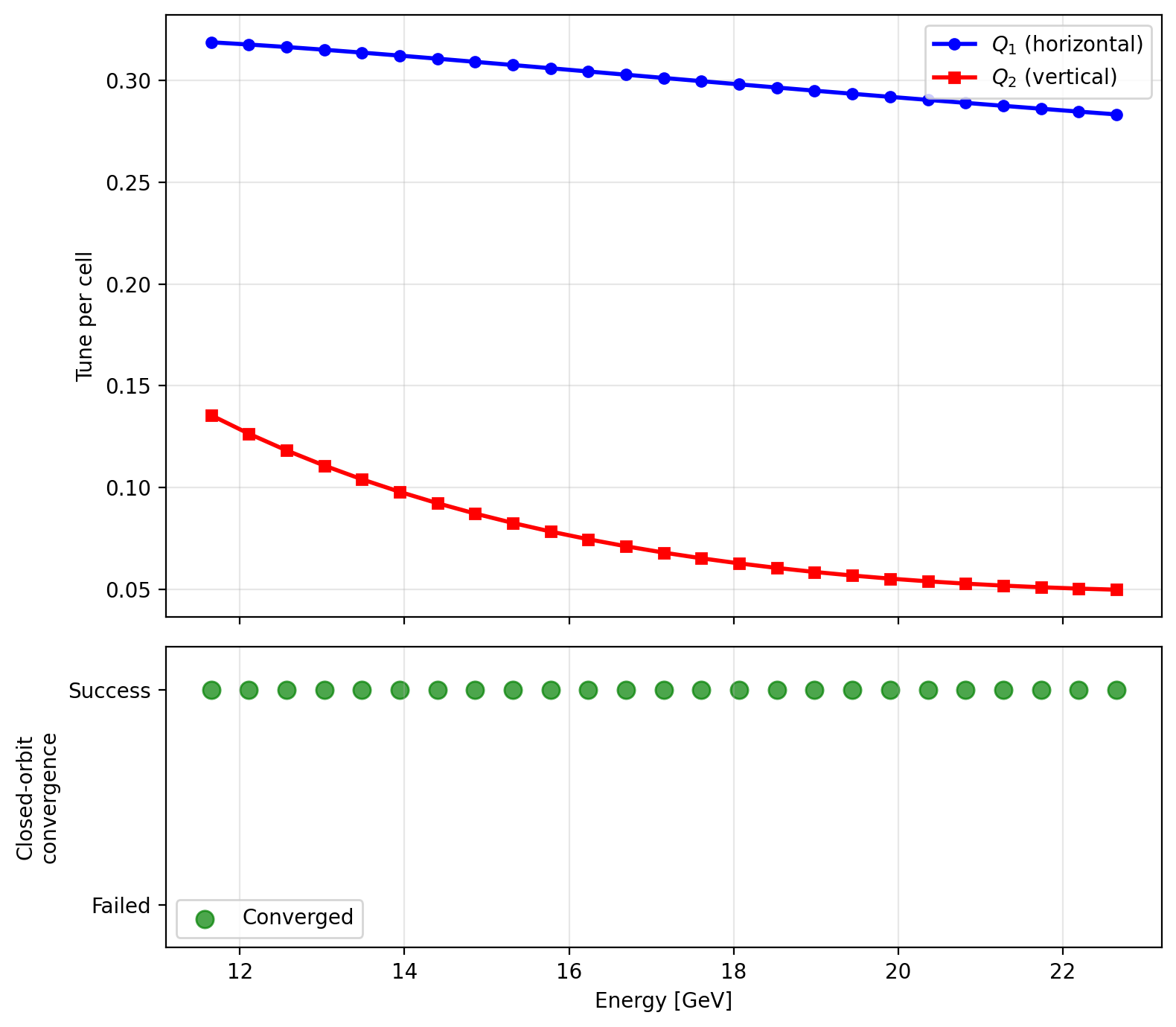}
\caption{Independently regenerated \textsc{madx-ptc} evaluation of the
West-Arc FODO lattice. Top: per-cell horizontal ($Q_1$) and vertical ($Q_2$)
tunes versus energy. Bottom: closed-orbit convergence status, confirming
successful convergence at all 25 sampled energies across the
$11.65$--$22.65$~GeV band.}
\label{fig:ptc-tunes-convergence}
\end{figure}

To verify the reproducibility of these results, the arc cell was
independently rebuilt from the design parameters in a \textsc{mad-x} model
driven through the \texttt{cpymad} interface, with slightly retuned gradient
and sextupole settings, and re-scanned in energy using \textsc{ptc}
closed-orbit calculations, each momentum point evaluated in a fresh
\textsc{mad-x} instance. The closed-orbit search converged at all 25 sampled
energies between $11.65$ and $22.65$~GeV, and the resulting per-cell
tunes, $Q_1$ decreasing smoothly from $0.318$ to $0.283$ and $Q_2$ from
$0.135$ to $0.050$, reproduce the multi-code results of
Fig.~\ref{fig:fodo-tunes-codes}. Both the tune scan and the convergence status
are summarized in Figure~\ref{fig:ptc-tunes-convergence}.

\subsubsection{Closed orbits, momentum compaction, and time of flight}
\label{subsubsec:fodo-m56}

\begin{figure}
\centering
\includegraphics[width=.7\columnwidth]{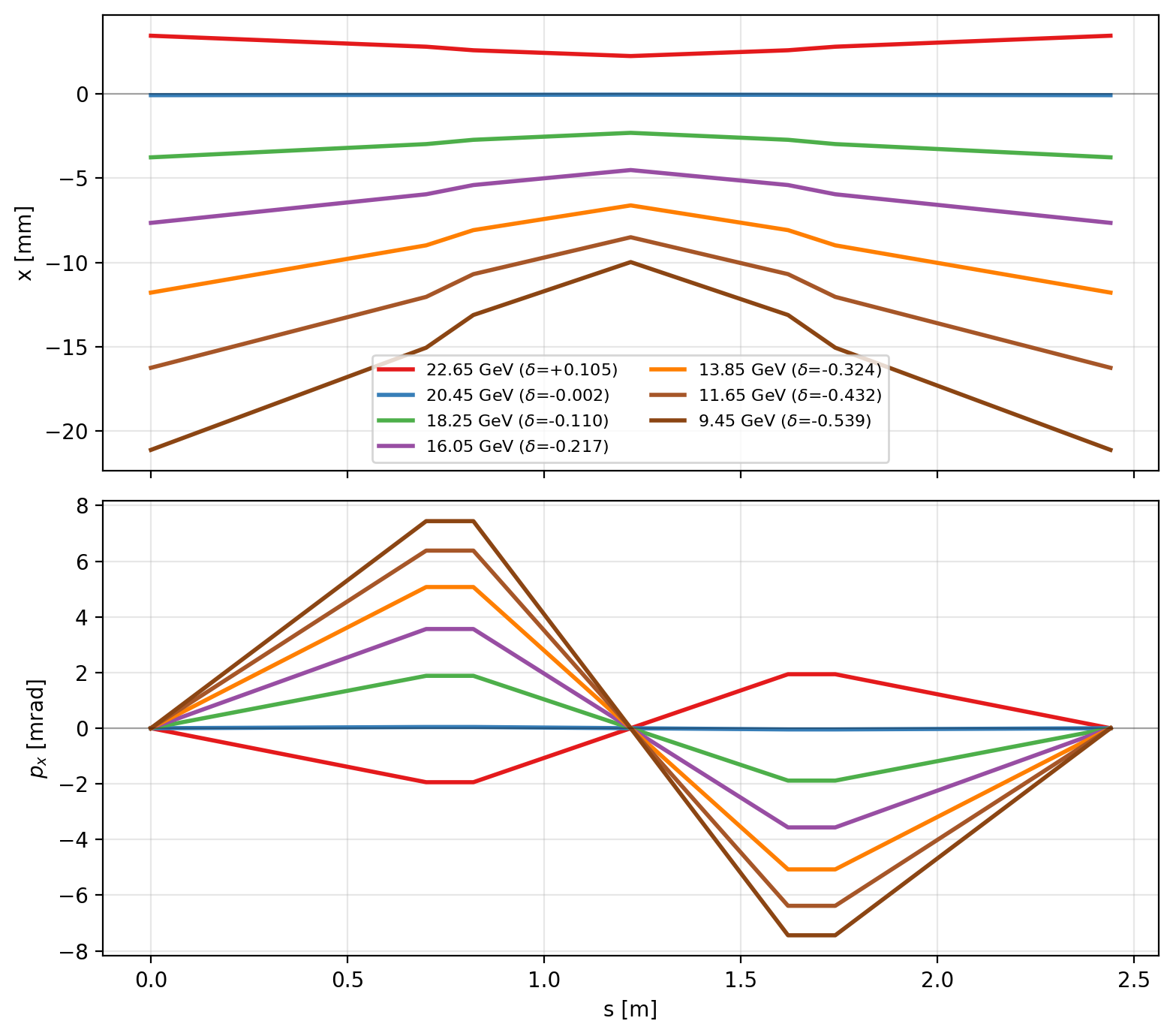}
\caption{Horizontal closed-orbit position (top) and angle (bottom) along one
West-Arc FODO cell of the regenerated \textsc{madx-ptc} model, for the seven
pass energies spanning $9.45$--$22.65$~GeV, quoted as momentum offsets
relative to the $20.50$~GeV reference momentum.}
\label{fig:ptc-closed-orbit}
\end{figure}

\begin{figure}
\centering
\includegraphics[width=0.7\columnwidth]{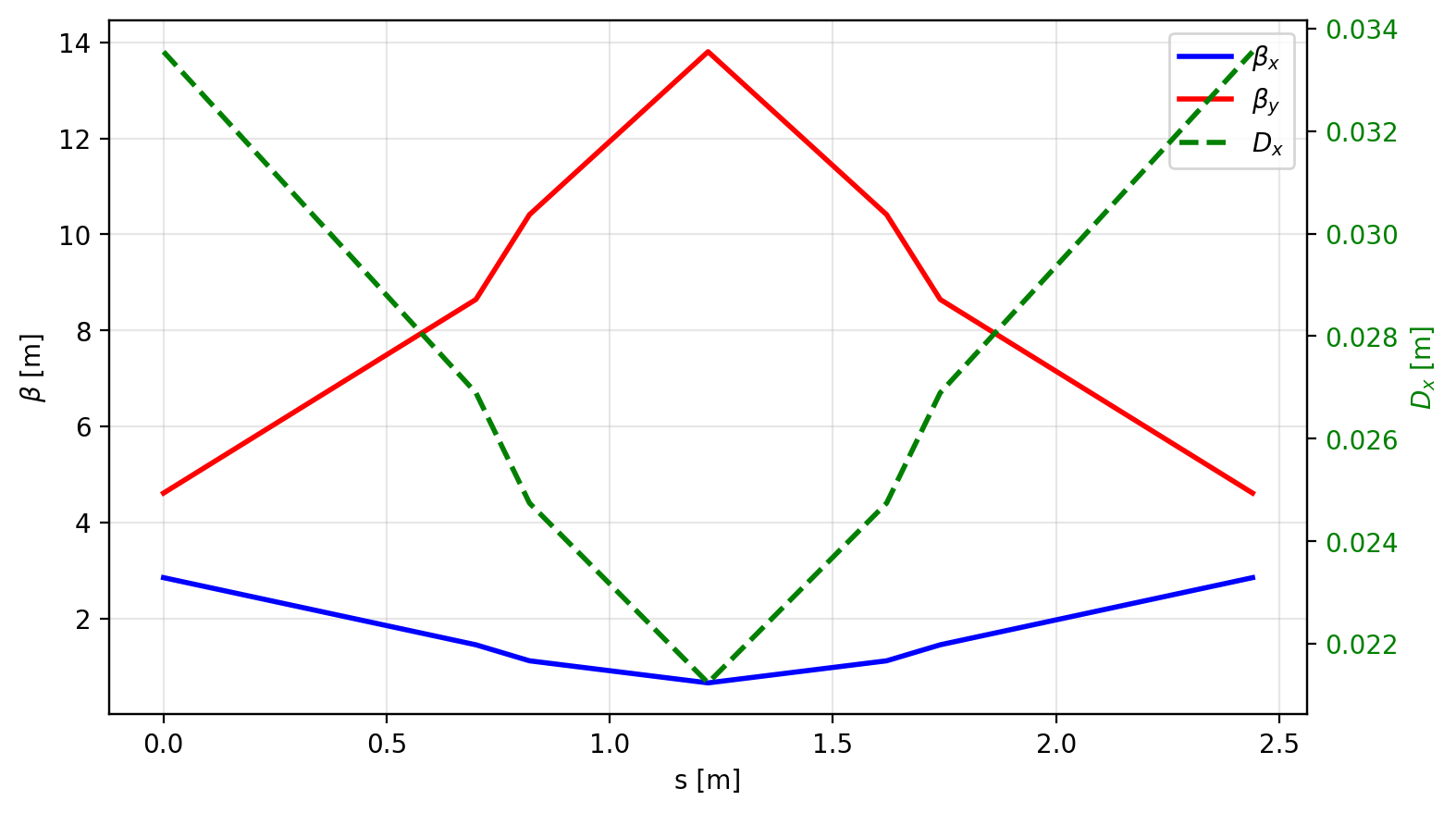}
\caption{Betatron functions and horizontal dispersion along one West-Arc FODO
cell at the $20.45$~GeV reference energy, from the regenerated
\textsc{madx-ptc} model.}
\label{fig:ptc-cell-optics}
\end{figure}

\begin{figure}[!htb]
\centering
\includegraphics[width=0.7\columnwidth]{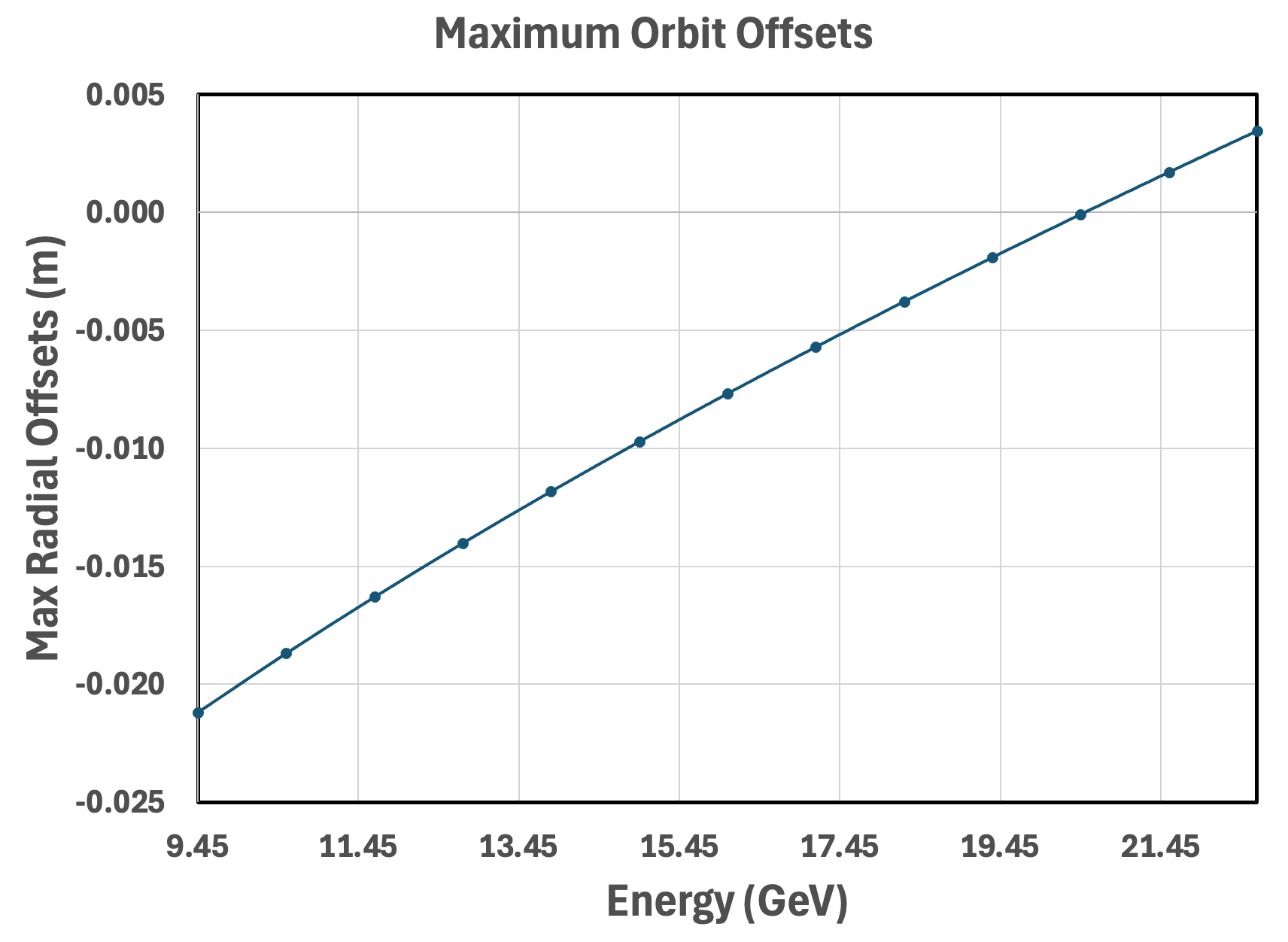}
\caption{Maximum radial closed-orbit offset in the West Arc versus energy for
the extended $9.45$--$22.65$~GeV FODO lattice.}
\label{fig:fodo-max-offsets}
\end{figure}

Figure~\ref{fig:ptc-closed-orbit} shows the \textsc{ptc} closed orbits of a
single cell of the regenerated model at the seven energies spanning
$9.45$--$22.65$~GeV, expressed as momentum offsets $\delta=\Delta p/p$
relative to the $20.50$~GeV reference momentum. The radial closed-orbit
excursion spans from $-21.1$~mm at $9.45$~GeV ($\delta=-0.539$) to $+3.4$~mm
at $22.65$~GeV ($\delta=+0.105$), with the reference (zero-offset) orbit
crossing near $20.45$~GeV; the corresponding orbit angles remain within
$\pm7.5$~mrad. This $\sim\!24$~mm total radial swing, consistent with the
maximum-orbit-offset scan of Figure~\ref{fig:fodo-max-offsets}, sets the
good-field aperture requirement for the arc magnets. The betatron functions
and dispersion at the $20.45$~GeV reference energy are shown in
Fig.~\ref{fig:ptc-cell-optics}: the beta functions stay below $14$~m and the
dispersion remains at the few-centimetre level ($2$--$3.4$~cm) throughout the
cell.

\begin{figure}[!htb]
\centering
\includegraphics[width=\columnwidth]{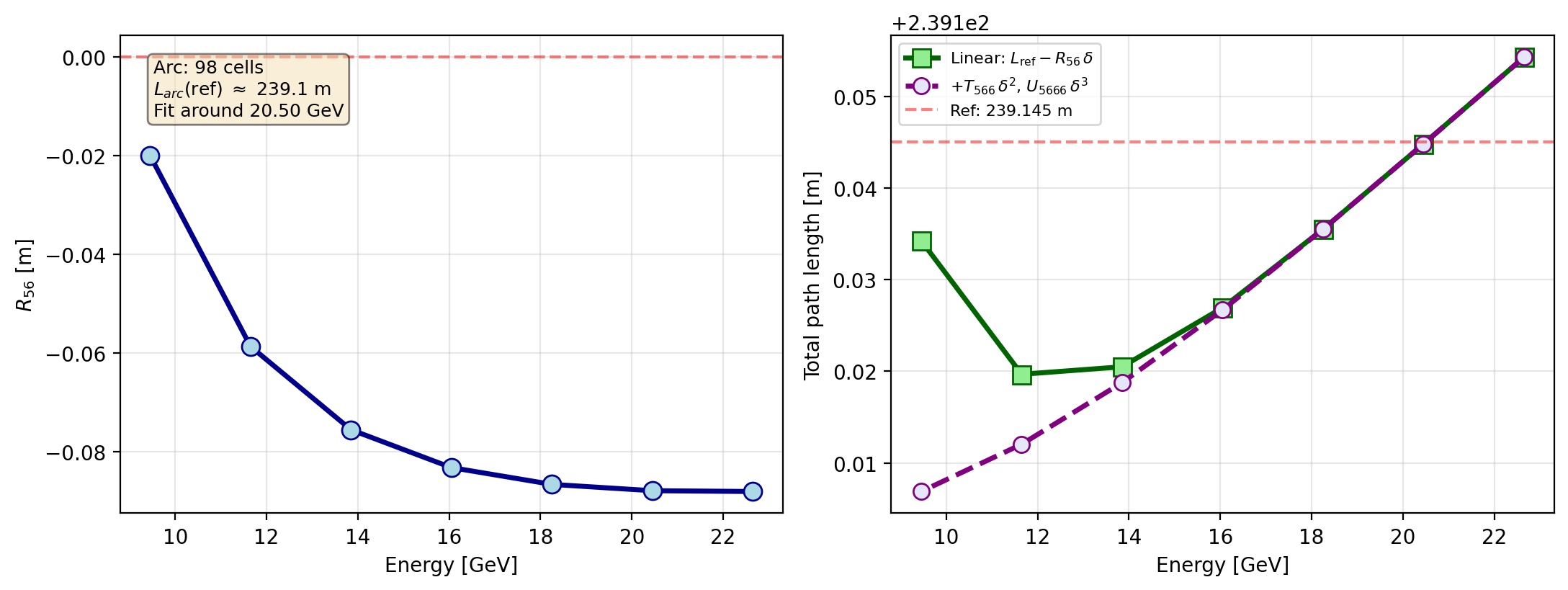}
\caption{Longitudinal properties of the 98-cell West Arc
($L_{\mathrm{arc}}\approx239.1$~m) from the regenerated \textsc{madx-ptc}
model. Left: per-arc $R_{56}$ versus energy, relative to the $20.50$~GeV
reference. Right: total path length versus energy, comparing the linear
estimate $L_{\mathrm{ref}}-R_{56}\,\delta$ with the higher-order expansion
including $T_{566}$ and $U_{5666}$.}
\label{fig:ptc-r56-path}
\end{figure}

\begin{figure}[!htb]
\centering
\includegraphics[width=0.7\columnwidth]{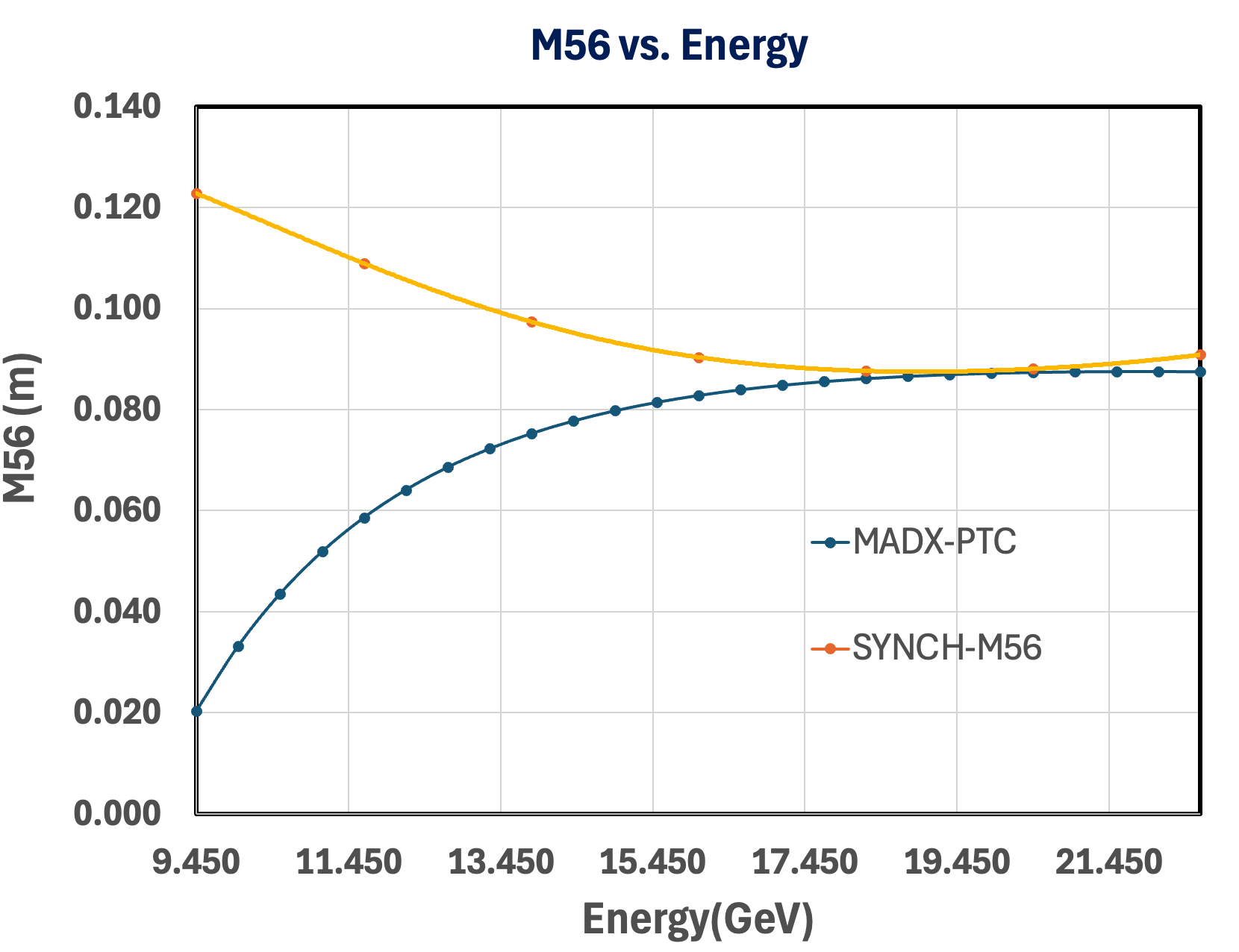}
\caption{Momentum-compaction element $M_{56}$ of the West Arc versus energy,
from \textsc{madx-ptc} and \textsc{synch}. The codes agree above
${\sim}17$~GeV but diverge at the low-energy end of the band.}
\label{fig:fodo-m56-codes}
\end{figure}

The energy dependence of the longitudinal transport is characterized by the
per-arc $M_{56}$ element (equivalently, the pass-to-pass time-of-flight
difference). Figure~\ref{fig:fodo-m56-codes} compares $M_{56}(E)$ obtained
from \textsc{madx-ptc} and from \textsc{synch}. The two codes converge to a
common value of $|M_{56}|\approx0.088$~m above ${\sim}17$~GeV, but disagree
markedly below it: \textsc{synch} predicts the magnitude rising toward
${\sim}0.12$~m at the low-energy end, whereas \textsc{madx-ptc} predicts it
falling toward ${\sim}0.02$~m. The regenerated model reproduces this
\textsc{ptc} behavior, as shown in Figure~\ref{fig:ptc-r56-path}: the per-arc
$R_{56}$ decreases in magnitude from $0.088$~m at the top energy to
$0.020$~m at $9.45$~GeV. Although, sharing the \textsc{ptc}
formalism, it cannot by itself arbitrate the discrepancy with
\textsc{synch}. A third-order fit of the path-length dependence
$\Delta L(\delta)=-R_{56}\delta-\tfrac{1}{2}T_{566}\delta^{2}
-\tfrac{1}{6}U_{5666}\delta^{3}$ around the $20.50$~GeV reference yields
effective arc coefficients $R_{56}=-0.0880$~m, $T_{566}=-0.0126$~m, and
$U_{5666}=+0.286$~m for the reference arc length
$L_{\mathrm{arc}}=239.145$~m. The total path length varies by
${\sim}48$~mm across the band, and the higher-order terms are essential to
reproduce it at the low-energy edge, where the linear estimate visibly
fails; a direct consequence of the large off-momentum orbit excursions
there. Definitive resolution of the code-to-code discrepancy remains an
outstanding item and is required before the splitter time-of-flight
corrections can be finalized.

\subsubsection{FMC-like momentum-compaction control}
\label{subsubsec:fmc-outlook}

The comparison above frames the trade-off between the two architectures. The
FODO FFA cell delivers the compactness, high packing factor, and small magnet
volume that make the permanent-magnet arcs affordable and SR-tolerant, and its
optics are now validated by four independent codes. However, its
momentum compaction is an uncontrolled by-product of the fixed magnet
geometry: the residual $|M_{56}|\approx0.09$~m per arc, together with its
strong energy dependence and higher-order terms, must be compensated
pass-by-pass in the splitter lines, complicating their design. An FMC-based
arc, by contrast, would allow the momentum compaction of each pass to be tuned
toward the isochronous condition, as was done in the electromagnetic arcs and the
injector racetrack. This would come at the cost of a longer, less densely packed, cell with
larger orbit excursions and magnet apertures.

The planned lattice improvements therefore aim to combine the strengths of
both approaches: resolving the remaining \textsc{bmad} compatibility and
$M_{56}$ questions; adjusting the drift spaces to satisfy vacuum and
instrumentation requirements; keeping the design within the maximum
permanent-magnet field limits; reinstating flexible-momentum-compaction
features in the arc optics to minimize $M_{56}$; completing the arc-to-straight
matching sections; and developing a splitter design based on triplet FFA cells
analogous to the arc-to-straight match.

\FloatBarrier

\subsection{Alternative 5-pass FFA Scenario with a Path to 22 GeV}

\begin{figure}[!htb]
  \centering
  \includegraphics[width=0.8\columnwidth]{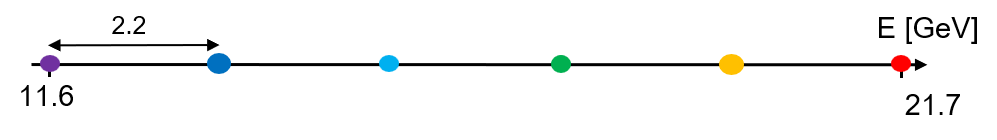}
  \caption{Six orbits of different pass energies within energy acceptance from about 11 GeV to 22 GeV.}
  \label{fig:alt1}
\end{figure}

One may envision a more conservative, staged scenario to upgrade energy to 22 GeV. Initially starting with 1.1 GeV per Linac and adding 5 rather than 6 new FFA passes, the top energy of about 20 GeV will be reached. Then through cryo-module upgrades (replacing lower gradient cryo-modules with C100) one would increase energy per Linac to 1.21 GeV, to bring up the top energy to 22 GeV. The FFA permanent magnets, as presently designed, are at the limit of their aperture and strength for accommodating six orbits of different energy within energy acceptance from about 11 GeV to 22 GeV (West FFA Arc). As a result, the requirement on the Linac energy margin is about $1\%$, as illustrated in Figure~\ref{fig:alt1}. This is very tight and would not allow the operational flexibility we currently have to reduce the energy slightly. If, e.g., several cavities de-rated, or one of the cryo-modules bypassed. Furthermore, additional Linac energy flexibility is required to optimize polarization to multiple halls in combination with the Wien filter; with a fixed Linac energy, only one hall can get the desired polarized beam. 
Adding a sextupole component to permanent magnets may extend the energy acceptance, which is still under study.

To alleviate these shortcomings, while staying with the current linear magnet design, as described above, we envision the following alternative 22 GeV upgrade scenario, summarized in two steps:

\begin{itemize}
\item \textbf{Step I} - Starting with 4 passes (electromagnetic arcs) followed by 5 (rather than 6) FFA passes and assuming 1.1 GeV/Linac, the final energy drops to 20 GeV. Assuming only 5 FFA passes significantly enhances the Linac energy flexibility, as illustrated in Figure~\ref{fig:alt2}. 
\begin{figure}[!htb]
  \centering
  \includegraphics[width=0.9\columnwidth]{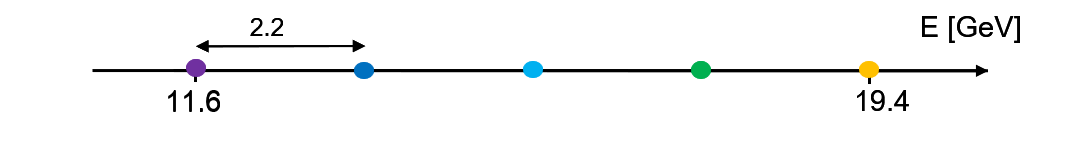}
  \caption{Only five orbits of different pass energies within the same energy acceptance from about 11 GeV to 22 GeV.}
  \label{fig:alt2}
\end{figure}
\item \textbf{Step II} - Then, subsequent upgrade of Linac energy to 1.21 GeV/Linac (cryo-module upgrades) would restore the final energy to 22 GeV (in 4 conventional and 5 FFA passes). Increasing Linac energy from 1.1 to 1.21 GeV boosts the lowest pass energy by 1.1 GeV ($5\times0.22$ GeV) and brings the final energy back to about 22 GeV with adequate Linac energy flexibility margin, as illustrated in Figure~\ref{fig:alt3}.

\begin{figure}[!htb]
  \centering
  \includegraphics[width=0.9\columnwidth]{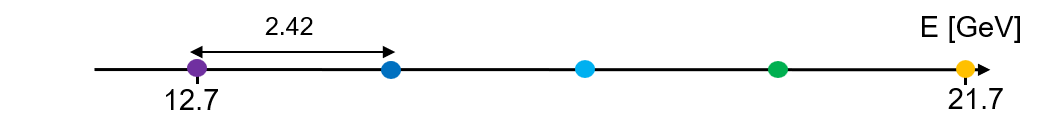}
  \caption{Diagram of updated Linac energy range.}
  \label{fig:alt3}
\end{figure}

\end{itemize}

This approach would assure adequate energy flexibility for the Linacs (like we have now), to slightly reduce energy, if, e.g., several cavities de-rated, or one of the cryo-modules bypassed. Furthermore, Linac energy flexibility is essential to optimize polarization to multiple halls in combination with the Wien filter; with a fixed Linac energy, only one hall can get polarized beam.

\FloatBarrier

\FloatBarrier
\section{Permanent Magnets}
\subsection{Magnet Design and Prototyping}

The CEBAF energy upgrade will require magnets with high fields to bend electron beams of up to 22\,GeV in the 80.6\,m radius tunnel.  A peak field in excess of 1.5\,T, together with a large gradient of 40\,T/m or more, are used in its fixed-field arc lattice to bend multiple recirculation energies in a single pipe.  Additionally, the magnet must have an open midplane to allow synchrotron radiation to be absorbed by a cooling channel.

A short 45\,mm section of NdFeB prototype has been designed and built as part of permanent magnet R\&D at BNL.  This satisfies all the above requirements and has had its integrated field tuned to better than 1 part in $10^3$.  This tuning process uses a technique with iron rods, together with measurements at a new compact field-mapping stand that is accurate to 1 part in $10^4$.

\subsubsection{Specification}
The requirements for the prototype permanent magnet are given in Table \ref{tab:spec}.  The combined-function field, gradient and good field region were taken from the lattice of the fixed-field (FFA) arcs for the CEBAF energy upgrade \cite{Bodenstein:IPAC22-THPOST023}.  This is evolving and has changed since last reported \cite{BrooksIPAC22} and continued to change after the prototype magnet was ordered.  As of early 2023, only one FFA loop is required in the upgrade, consisting of two slightly different 180$^\circ$ arcs each with two distinct types of permanent magnet.  The parameter ranges required in this most up-to-date lattice as of writing are:
\begin{itemize}
    \item dipole $-1.2815 \le B(0) \le -0.3828$\,T,
    \item gradient $41.13 \le |B'| \le 53.79$\,T/m and 
    \item peak field $1.515 \le B_\mathrm{max} \le 1.573$\,T,
\end{itemize}
which are still well represented by the prototype.  Designs of similar geometry exist for all four magnets in the new lattice.

\begin{table}[!htb]
   \centering
   \caption{Prototype Magnet Specification}
   \begin{ruledtabular}
\begin{tabular}{lcc}
       \textbf{Parameter} & \textbf{Value} & \textbf{Unit} \\
       \colrule
Dipole ($B(0)$) & $-0.9512$ & T \\
Gradient ($B'$) & 55.54 & T/m \\
$x$ good field region (GFR) & $\pm$10.5 & mm \\
$B_\mathrm{max}$ in GFR & ($-$)1.536 & T \\
       \colrule
Magnet length & 45 & mm \\
Vertical aperture (GFR) & $\pm$7.5 & mm \\
Minimum midplane gap & $\pm$3 & mm \\
       \colrule
Material & NdFeB & \\
Grade & N42EH & \\
$B_r$ & 1.28--1.33 & T \\
$\mu_0H_{cJ}$ & 2.9 & T \\
   \end{tabular}
\end{ruledtabular}
   \label{tab:spec}
\end{table}

The prototype magnet was built as a small 45\,mm length section in order to test the integrated field per unit length.  In reality, many sections will be clamped together longitudinally to make the full magnet, which will be roughly 1--2\,m long.  Initially, 60\,mm sections were asked for but there was a Neodymium ore price spike in early 2022, together with pandemic-related supply chain issues, making the material particularly expensive to obtain at that point.

The vertical full aperture of 15\,mm is about the minimum required to fit a beam pipe and any iron field tuning rods, while still having space for beam.

Neodymium Iron Boron (NdFeB) material was selected to give the high fields required to bend the beam in the fixed radius CEBAF tunnel, while also selecting the high-coercivity (high temperature) grade N42EH \cite{AllStarNeoGrades}, which sacrifices some of this field strength for improved radiation resistance and general resistance to demagnetization: it can withstand 2.9\,T of reverse field before losing all its magnetization.

\subsubsection{Design and features}
\begin{figure}[!htb]
   \centering
   \includegraphics*[width=0.7\columnwidth]{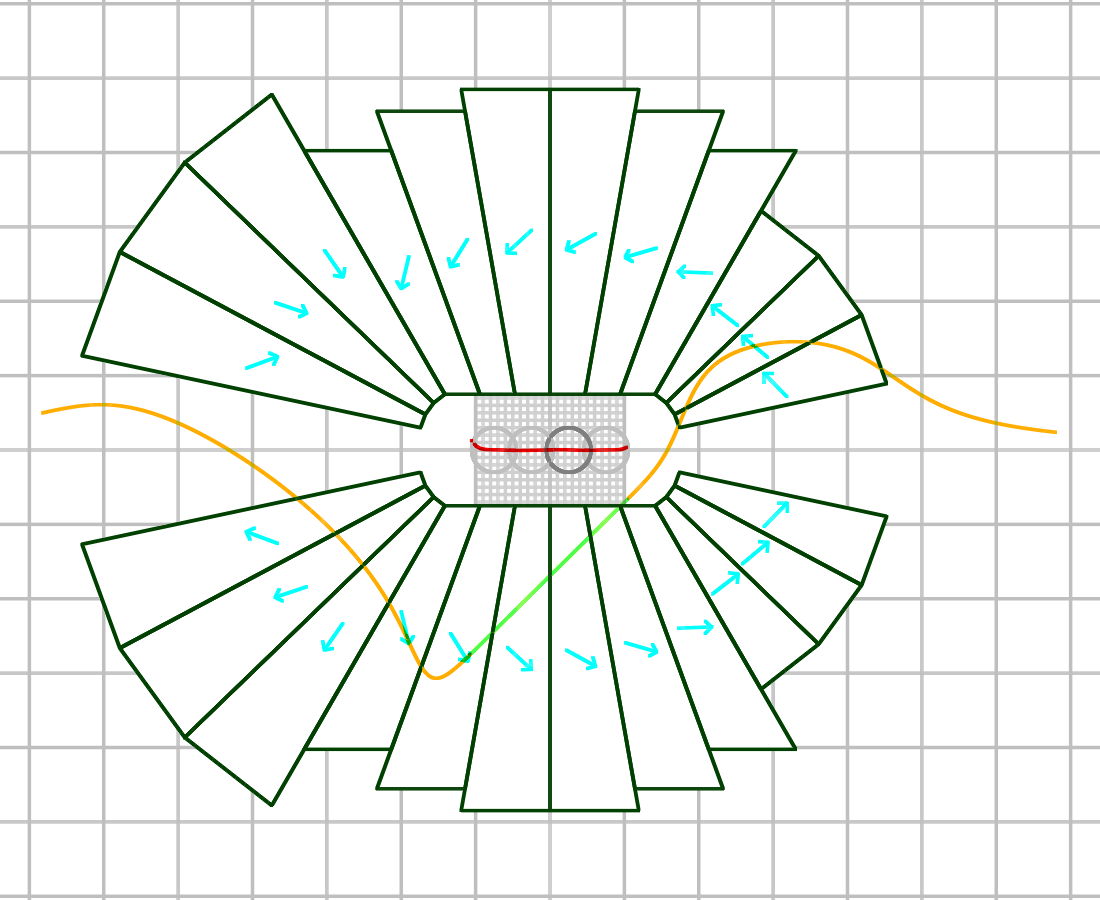}
   \caption{2D permanent magnet design (cm grid).  Midplane field is graphed in orange, with good field region in green.}
   \label{fig:design}
\end{figure}

The cross-section of the magnet is shown in Figure~\ref{fig:design} and has area 70.1\,cm$^2$.  It consists of 24 wedge-shaped pieces of NdFeB with geometry given in Table \ref{tab:geom}.  The outer wedges are symmetrical, while the top and bottom wedges have two edges parallel to the horizontal axis.  The wedge shape prevents the pieces from falling into the aperture under magnetic forces, Some of the wedges here are quite narrow, so future magnets will bind them in place with epoxy.

\begin{table}[!hbt]
   \centering
   \caption{Permanent Magnet Geometry}
   \begin{ruledtabular}
\begin{tabular}{lcc}
\textbf{Parameter} & \textbf{Value} \\
       \colrule
Number of wedges & 24 (3+6+3 per side) \\
Midplane angular gap & $\pm$12$^\circ$ \\
Wedge opening angles & 16$^\circ$/10$^\circ$/16$^\circ$ \\ 
   \end{tabular}
\end{ruledtabular}
   \label{tab:geom}
\end{table}

Each wedge has a different magnetization direction, although when comparing the magnetization vectors of symmetrical upper and lower blocks, the horizontal component is inverted while the vertical stays the same.  The field on the midplane good field region, calculated by a 2D simulation, differs by no more than 6.8\,Gauss, or $4.4\times10^{-4}$ of the peak field, from the ideal value.

The orientation of the magnetization in each wedge, as well as the wedge height, were determined by multi-parameter optimization similar to that used in the HalbachArea tool \cite{HalbachArea}.  The figure of merit was RMS field error in the midplane good field region.

The magnet has an open midplane to let synchrotron radiation emitted by the horizontally-bent high-energy electrons escape, but unlike the designs in \cite{BrooksIPAC22}, the midplane gap now has an opening angle.  This helps greatly with the mechanical strength of the vacuum chamber, as only a small distance of the unsupported span has to be of minimum thickness, with the material thickening as the midplane gap grows.

The central beam aperture is oval rather than circular, as this significantly reduces the amount of permanent magnet material required for a given horizontal orbit excursion \cite{ModifiedHalbach}.

\begin{figure}[!htb]
   \centering
   \includegraphics*[width=0.7\columnwidth]{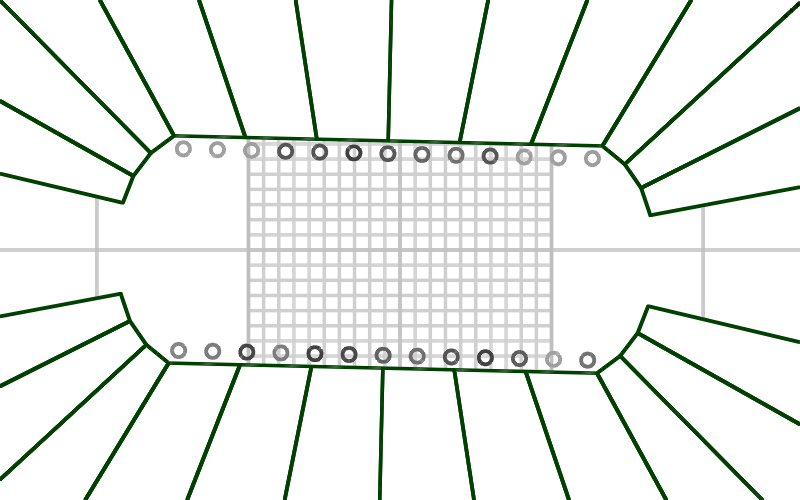}
   \caption{Locations of field tuning rods within the aperture.}
   \label{fig:designtuning}
\end{figure}

\paragraph{Field tuning method}
Given an initial measurement of the magnet's field, small iron rods can be placed just inside the aperture to cancel field errors, following the method in \cite{CBETAmagnets} adapted from CBETA \cite{CBETA-PRL,CBETA-DR} and miniaturised here.  This was adapted for a smaller aperture magnet by using rods of 0.89\,mm (35\,mil) diameter, reduced from 2.03\,mm (80\,mil), and adapted for the oval aperture by placing the rods above and below the midplane, as shown in Figure~\ref{fig:designtuning}.  The aperture appears tilted as the software also fits for magnet orientation on the measurement stand.  It was found that 26 rods placed at $y=\pm6.66$\,mm and spaced by $\Delta x=2.25$\,mm out to $|x|_\mathrm{max}=13.5$\,mm were sufficient to give good error correction in simulations.

\subsubsection{Construction}
\begin{figure}[!htb]
   \centering
   \includegraphics*[width=\columnwidth]{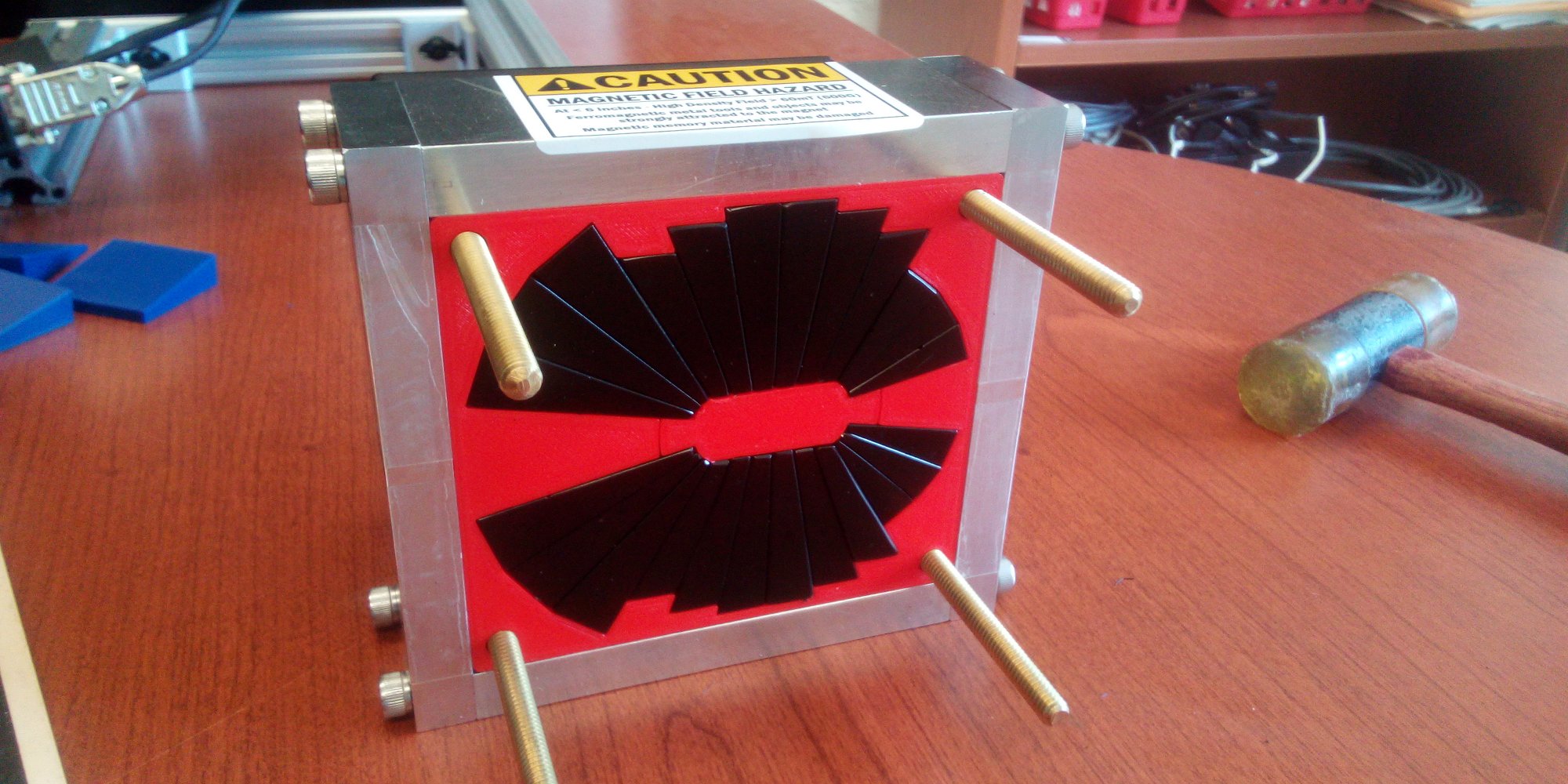}
   \caption{Magnet before central plastic plug is removed.}
   \label{fig:magnetfilled}
\end{figure}

\begin{figure}[!htb]
   \centering
   \includegraphics*[width=\columnwidth]{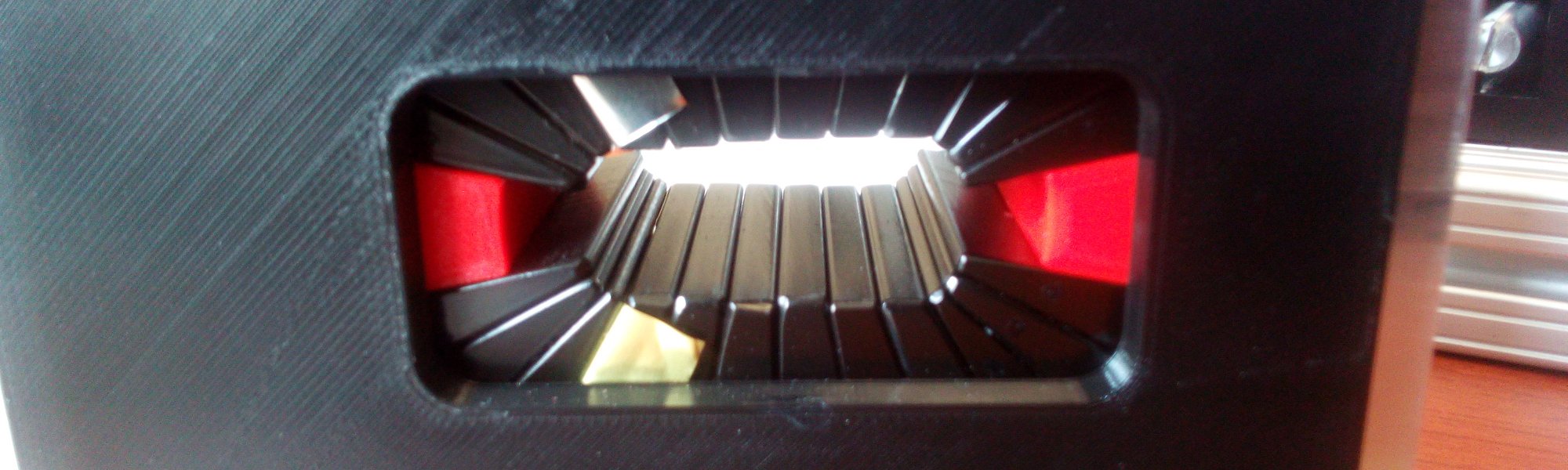}
   \caption{Aperture of magnet after central plug removed.}
   \label{fig:magnetaperture}
\end{figure}

This prototype magnet was made by placing the NdFeB wedges within a 3D printed PLA mould, contained by a rectangular aluminium frame, as shown in Figures~\ref{fig:magnetfilled} and \ref{fig:magnetaperture}.

\begin{figure}[!htb]
   \centering
   \includegraphics*[width=0.7\columnwidth]{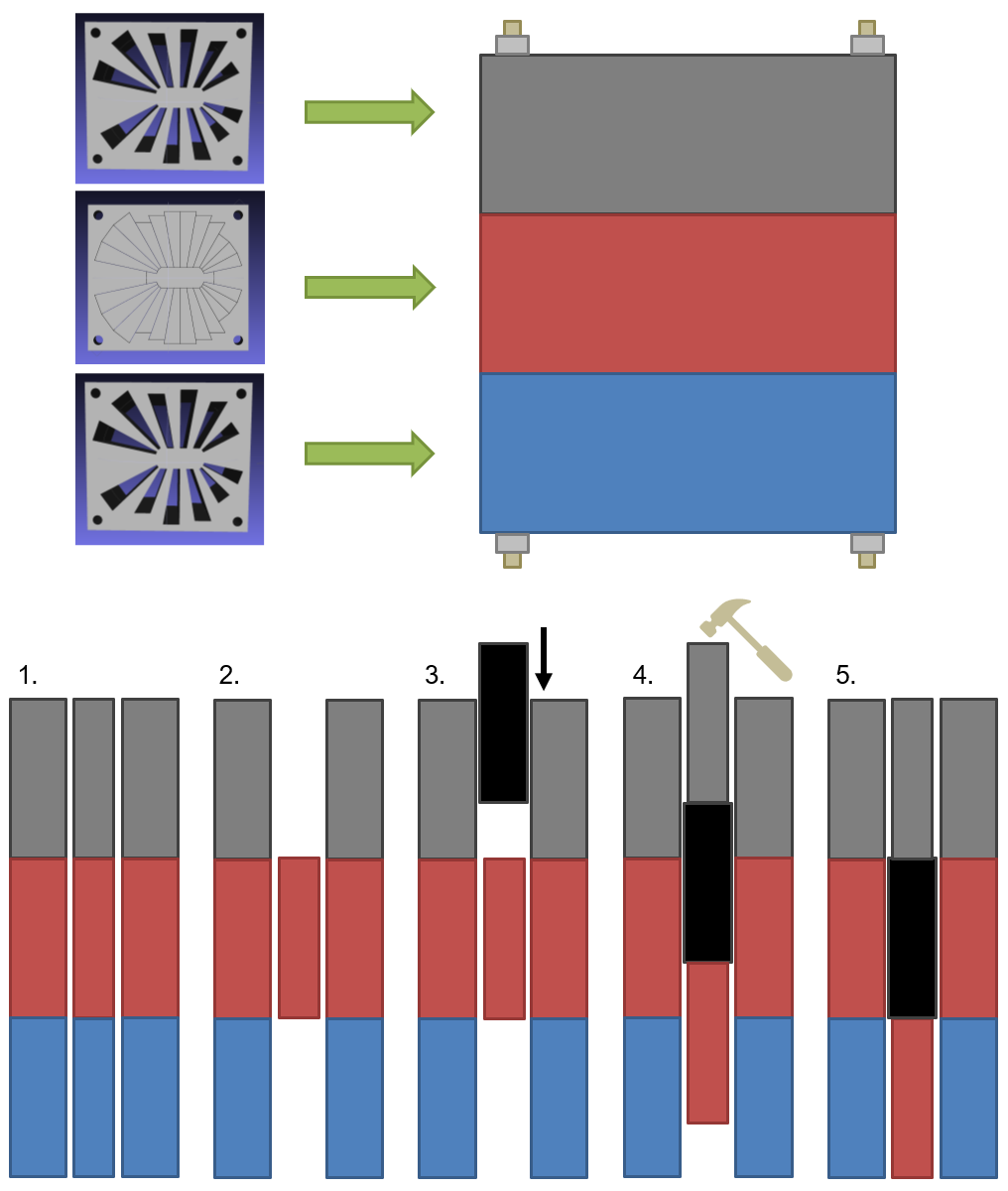}
   \caption{Assembly tooling and magnet insertion process.}
   \label{fig:assembly}
\end{figure}

The forces and torques between permanent magnet wedges are too large for assembly purely by hand, so channels to guide the magnets were 3D printed and attached to the mould, as shown in Figure~\ref{fig:assembly}.  The mould was initially filled with plastic dummy wedges, which were replaced one-at-a-time by magnets.  This was done by sliding the magnet down the channel and pushing out the dummy wedge, which also serves as the longitudinal end-stop.

\begin{figure}[!htb]
   \centering
   \includegraphics*[width=0.9\columnwidth]{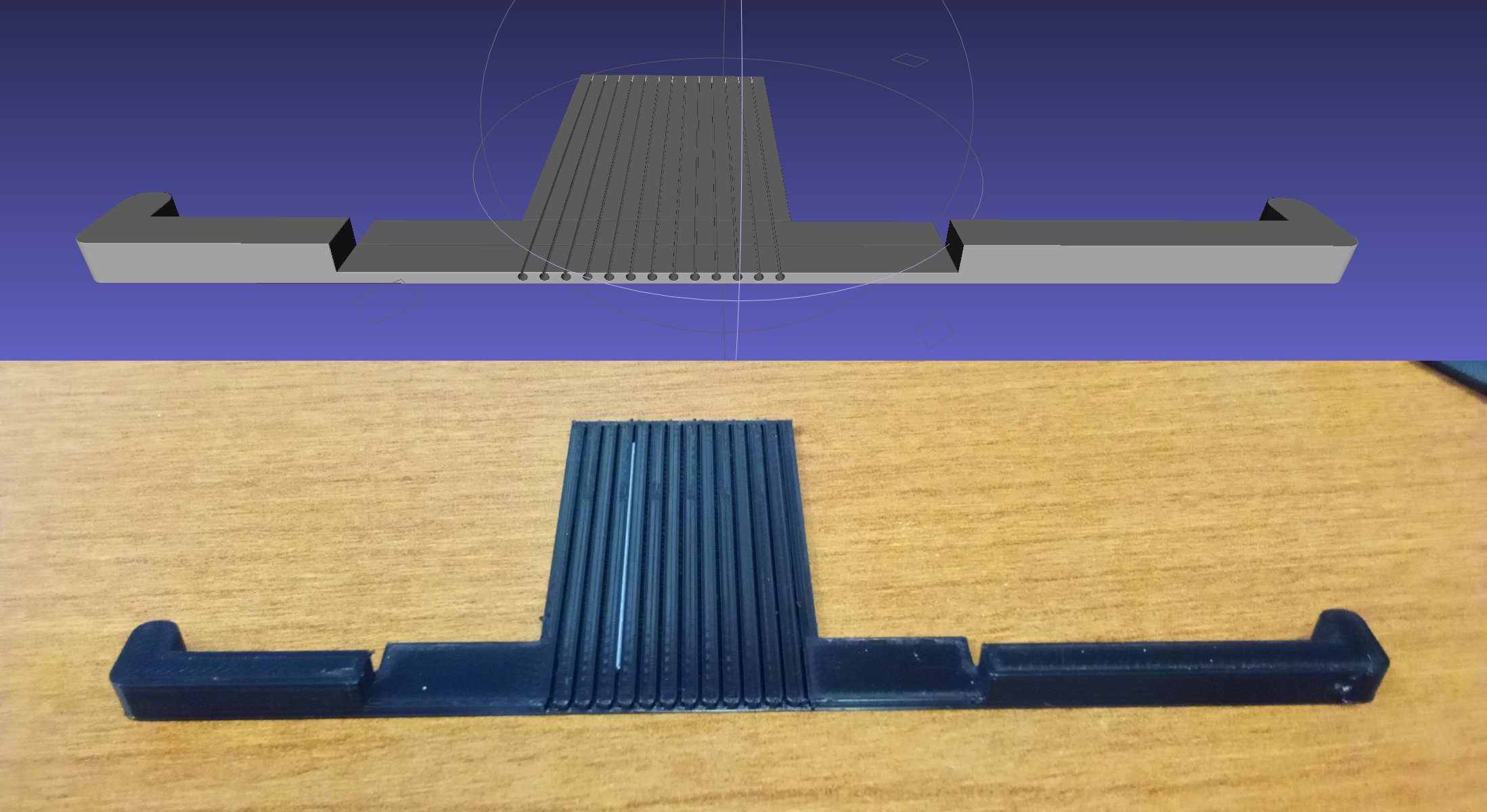}
   \caption{Tuning rod holder design and 3D printed part.}
   \label{fig:rodholder}
\end{figure}

\paragraph{Field tuning rod holders}
Plastic holders with channels to align the iron tuning rods within the magnet were 3D printed, as shown in Figure~\ref{fig:rodholder}.

\subsubsection{Field measurements}
\begin{figure}[!htb]
   \centering
   \includegraphics*[width=0.9\columnwidth]{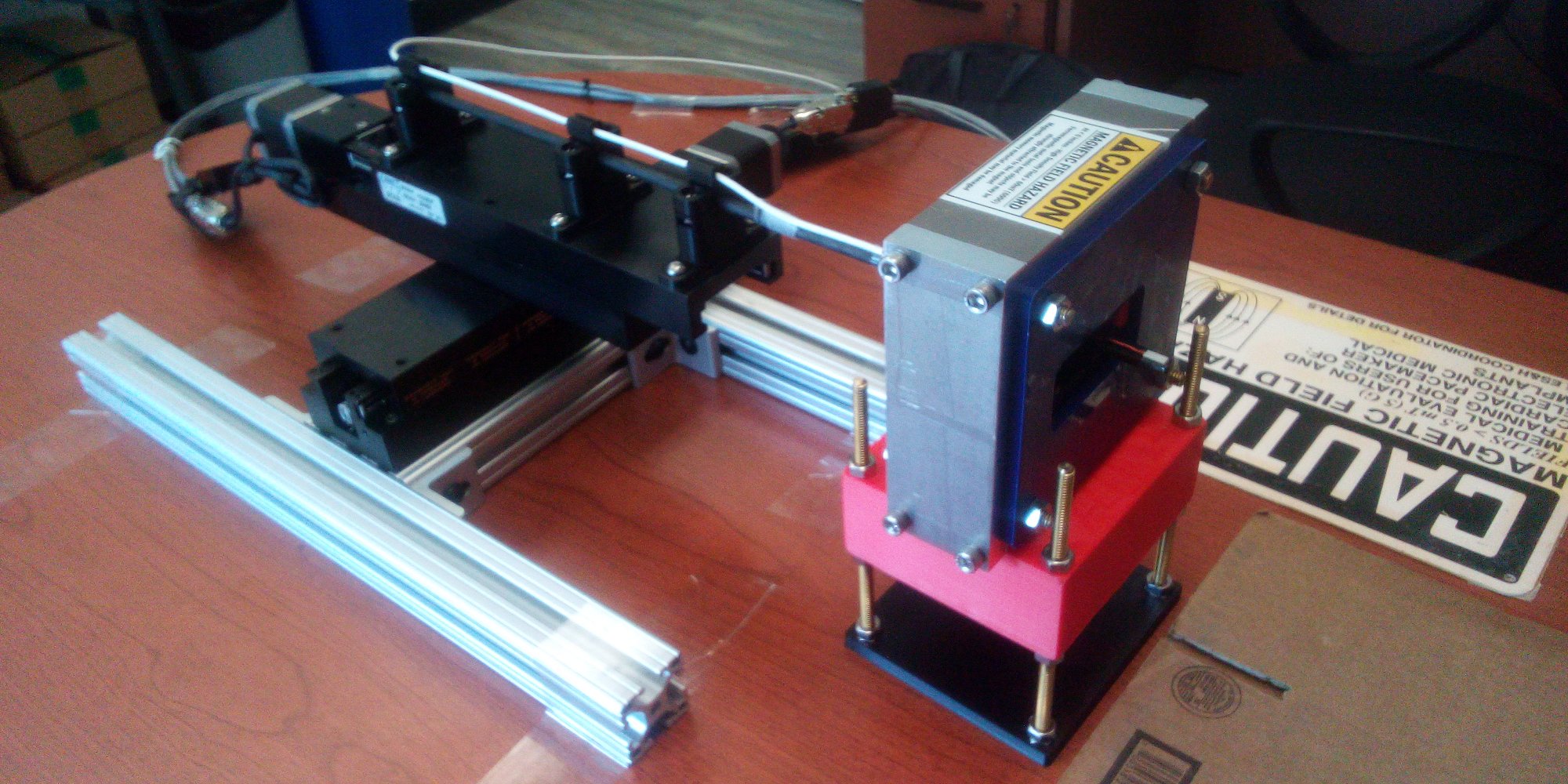}
   \caption{Field mapping the magnet midplane with a three-axis Hall probe moving transversely and longitudinally.}
   \label{fig:fieldmapper}
\end{figure}

Field mapping was done using a Senis 3MH6 Teslameter with a three-axis Hall probe that has $\pm0.01$\% accuracy.  Two stacked linear stages, seen in Figure~\ref{fig:fieldmapper}, scanned transversely and longitudinally in the magnet.  The accuracy of these stages was determined to be 1--2\,$\mu$m by cross-checking against the magnet field when the same point was approached from different directions.  All of this leads to a capability of measuring the magnetic field at the $10^{-4}$ level.

The linear stages have a full range of 200\,mm but in this case $x=\pm12$\,mm was scanned transversely in 0.5\,mm steps and the full range was scanned longitudinally in $2.5$\,mm steps to give the integrated field.

\begin{figure}[!htb]
   \centering
   \includegraphics*[width=0.8\columnwidth]{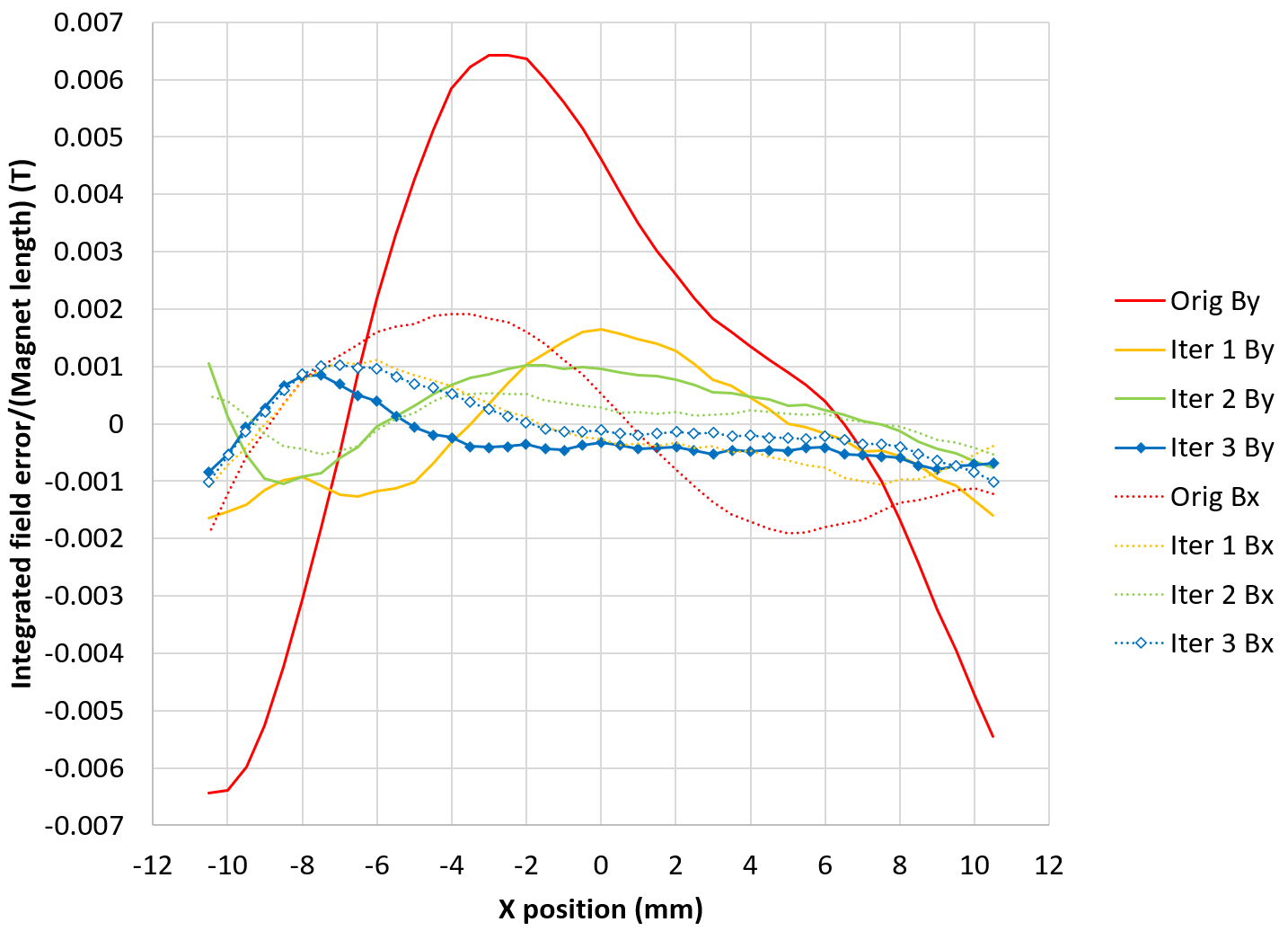}
   \caption{Integrated field error as a function of transverse position in the magnet, for successive field tuning iterations.}
   \label{fig:fielderrors}
\end{figure}

\begin{table}[!htb]
   \centering
   \caption{Measured integrated magnetic field errors in the good field region, in units of $10^{-4}$ of the maximum field.}
   \begin{ruledtabular}
\begin{tabular}{lccc}
\textbf{Tuning} & \textbf{$B_y$ max} & \textbf{$B_x$ max} & \textbf{$\mathbf B$ vector} \\
\textbf{iteration}  & \textbf{error} & \textbf{error} & \textbf{RMS error} \\
       \colrule
Original & 41.44 & 12.36 & 27.44 \\
1 & 21.03 & 7.06 & 17.30 \\
2 & 11.22 & 3.41 & 9.20 \\
3 & 7.88 & 6.56 & 5.62 \\
   \end{tabular}
\end{ruledtabular}
   \label{tab:relerrors}
\end{table}

Integrated field measurements of the magnet and three successive field tuning iterations are shown in Figure~\ref{fig:fielderrors} and Table \ref{tab:relerrors}.  These are given as deviations from the ideal linear integrated field, as otherwise the differences would be too small to see on a plot.  They are also presented as a function of transverse position, since multipole harmonics on a circle are no longer easy to define for an oval aperture magnet.  Only the $x=\pm10.5$\,mm good field region is shown in Figure~\ref{fig:fielderrors} and used for error statistics.

The field improves to better than 1 part in $10^3$ in both $B_y$ and $B_x$ error components.  This is roughly the field quality criterion used for accepting the CBETA magnets \cite{CBETAmagnets}.  It is likely that three tuning iterations are needed because of how short this R\&D magnet is, as the longer magnets from CBETA typically converged in one or two iterations.  Field contributions from the tuning rods in a short magnet will be less well-approximated by the 2D model.

The main source of the original magnet's large $B_y$ error contribution ($4.1\times10^{-3}$ relative) could be wedges `falling' in to different heights around the aperture under magnetic forces, as can be seen by looking closely at Figure~\ref{fig:magnetaperture}.  Top-bottom symmetric effects, such as this, produce $B_y$ errors, whereas $B_x$ (skew) errors come from asymmetric effects, which were smaller here.

\subsubsection{Conclusion}
The prototype permanent magnet demonstrates:
\begin{itemize}
\item High combined-function field levels up to 1.536\,T;
\item Zero energy consumption;
\item Good linearity with relative errors $<10^{-3}$;
\item An open midplane for synchrotron radiation emission, with opening angle for a stronger vacuum chamber;
\item An oval aperture for lower material use and cost.
\end{itemize}
These features are key for the 22 GeV CEBAF energy upgrade \cite{Bodenstein:IPAC22-THPOST023}, as well as advanced light source lattices that use high-gradient magnets \cite{CBII} and compact fixed-field hadron therapy gantries using permanent magnets \cite{DejanGantry}.

\subsection{Permanent Magnet Radiation Resiliency}
\label{ch:detailed_resiliency}

This section summarizes the experimental program that quantifies radiation-induced demagnetization of the permanent magnet materials proposed for the FFA@CEBAF energy upgrade. Results are drawn from the first full exposure campaign, reported in the NAPAC'25 proceedings \cite{Bodenstein:NAPAC2025}, the laboratory-versus-tunnel measurement comparison \cite{Bodenstein:TN25069}, and the comprehensive campaign summary \cite{Bodenstein:TN26055}. All values quoted here are corrected to a common 20$^\circ$C reference temperature.

\subsubsection{Experimental approach}

The FFA arcs proposed for the energy upgrade rely on permanent magnets, which are susceptible to radiation-induced demagnetization. A Laboratory Directed Research and Development (LDRD) project was established to quantify that susceptibility under realistic operating conditions, in the tunnel during normal machine operation, rather than in a dedicated irradiation facility.

\paragraph{Material selection}
Four commercial grades were selected to span the two dominant material families for accelerator applications:
\begin{itemize}
    \item \textbf{NdFeB N42EH}: the baseline grade specified for the Brookhaven prototype FFA magnet design.
    \item \textbf{NdFeB N52SH}: a higher-remanence, lower-coercivity alternative for comparison.
    \item \textbf{SmCo33H and SmCo35}: historically more resistant to radiation and thermal effects, included as a stability benchmark.
\end{itemize}

\paragraph{Scale and the differential measurement}
Sixty sample plates carrying 540 individual magnets were installed at eight tunnel regions: the four recirculating arcs, both Linacs, and two labyrinths. The primary quantitative dataset comes from 30 tunnel plates and nine laboratory control plates, each carrying one sample of all four grades in randomized slot positions.

That co-location is central to the analysis. Because all four materials on a plate are measured in the same session on the same instrument, the difference between the NdFeB and SmCo results cancels the dominant Helmholtz coil gain drift ($\pm 0.124\%$) by construction. This intra-plate differential, rather than the absolute change of any single material, is the primary observable, and it is what makes detection at the sub-percent level credible.

\begin{figure}[!htb]
    \centering
    \includegraphics[width=0.25\textwidth]{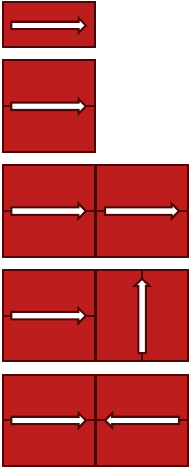}
    \caption{Magnet sample alignments for reverse-flux studies. The white arrow indicates the direction of magnetization. From top: a single sample, then the Delta, Alpha, Gamma, and Beta configurations. Varying the alignment simulates the differing magnetic loads experienced by wedges in a Halbach array.}
    \label{fig:alignments}
\end{figure}

\paragraph{Simulating the Halbach environment}
Magnets in the proposed FFA arcs will be arranged in Halbach configurations, where strong reverse fields act on individual wedges and can make the material more vulnerable to radiation damage. To reproduce those conditions without building full prototypes, the study uses pair assemblies in four orientations (Figure~\ref{fig:alignments}): Alpha (aligned), Beta (antiparallel), Gamma (90$^\circ$), and Delta (a single pair). Varying the alignment varies the demagnetizing field the samples experience.

\paragraph{Metrology}
A Magnetic Instrumentation Helmholtz coil integrates the total magnetic moment of each sample and provides the primary measurement. A Senis 3MH6 Teslameter records point field vectors ($B_x, B_y, B_z$) and is being developed as an independent cross-check. One constraint should be noted: the Beta (antiparallel) assemblies generate significant quadrupole content, which conflicts with the coil's linear field integration. Those configurations are therefore excluded from the quantitative Helmholtz analysis, and an alternative approach for them remains in development.

\paragraph{Dosimetry}
Two passive channels are co-located with every plate. Optically Stimulated Luminescence (OSL) area monitors record photon, beta, thermal neutron, and fast neutron exposure, and high-range optichromic rods record integrated gamma exposure. The OSL monitors saturate in the highest-dose regions, so integrated gamma values are taken from the rods, which do not saturate at these levels. Plates were sited near the laboratory's Neutron Dose Rate Meters with Extended Capabilities (NDX) where practical. Integrated gamma exposures across the deployment span roughly 0.3~Gy in the labyrinths to 23~kGy at the North Linac.

\paragraph{Temperature correction}
All results are corrected to 20$^\circ$C using material-specific coefficients (NdFeB $-0.10$ to $-0.11\%/^\circ$C; SmCo $-0.04\%/^\circ$C). This correction is not cosmetic. A dedicated comparison of laboratory measurements ($\sim$22--24$^\circ$C) against tunnel enclosure measurements ($>$30$^\circ$C) found a mean offset near $-0.5\%$ in uncorrected readings, with NdFeB shifting several times more than SmCo \cite{Bodenstein:TN25069}. Reversible thermal response of this size is comparable to the effect being measured, and temperature remains the dominant systematic on the differential, contributing $\pm 0.033\%$.

\begin{figure}[!htb]
    \centering
    \subfloat[Samples on a tripod in the North Linac. \label{fig:nl_setup}]{%
        \includegraphics[width=0.55\textwidth]{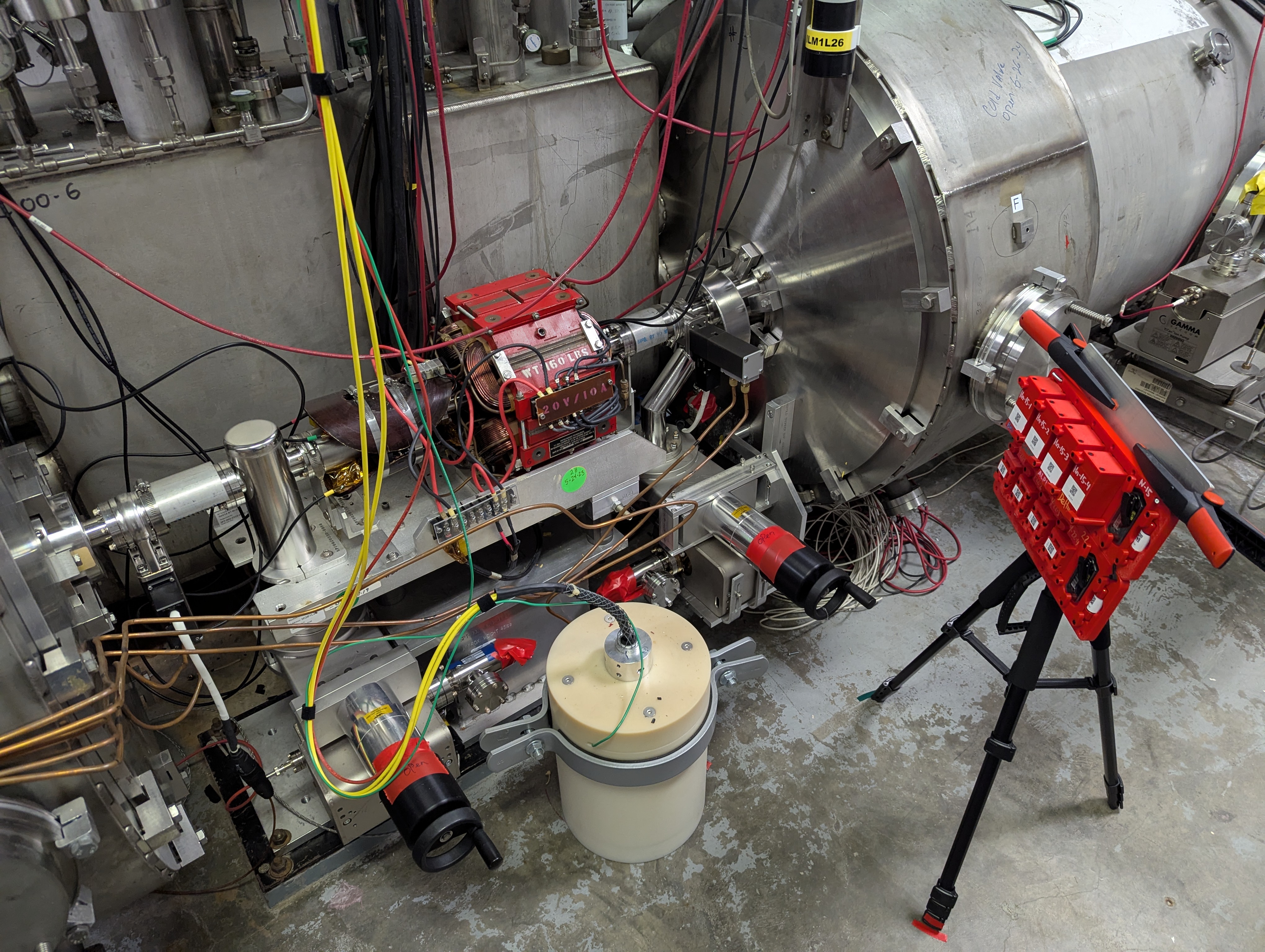}
    }
    \hfill
    \subfloat[Southeast corner installation. \label{fig:se_setup}]{%
        \includegraphics[width=0.30\textwidth]{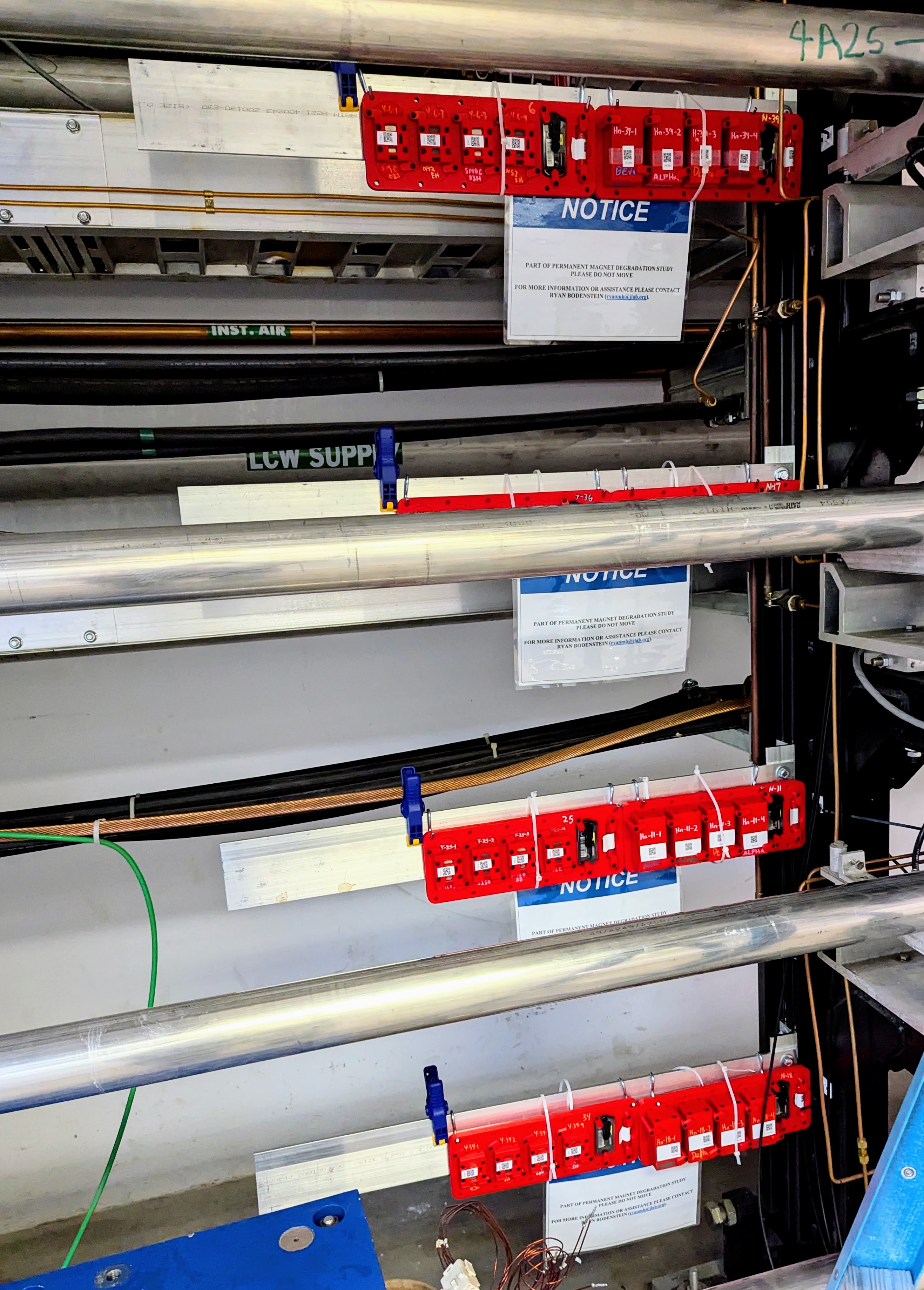}
    }
    \caption{Sample placement examples. (a) A tripod holding a single sample tray and a pair assembly tray, placed near an NDX detector in the North Linac. (b) Sample stacks installed on the outboard side of the recirculating arcs in the Southeast corner.}
    \label{fig:setup_photos}
\end{figure}

\subsubsection{First campaign results}

The samples were installed in January 2025 and retrieved in January 2026, approximately twelve months in the tunnel, of which roughly six months included beam operation at 2.12~GeV per pass.

\begin{figure}[!htb]
    \centering
    \includegraphics[width=0.72\textwidth]{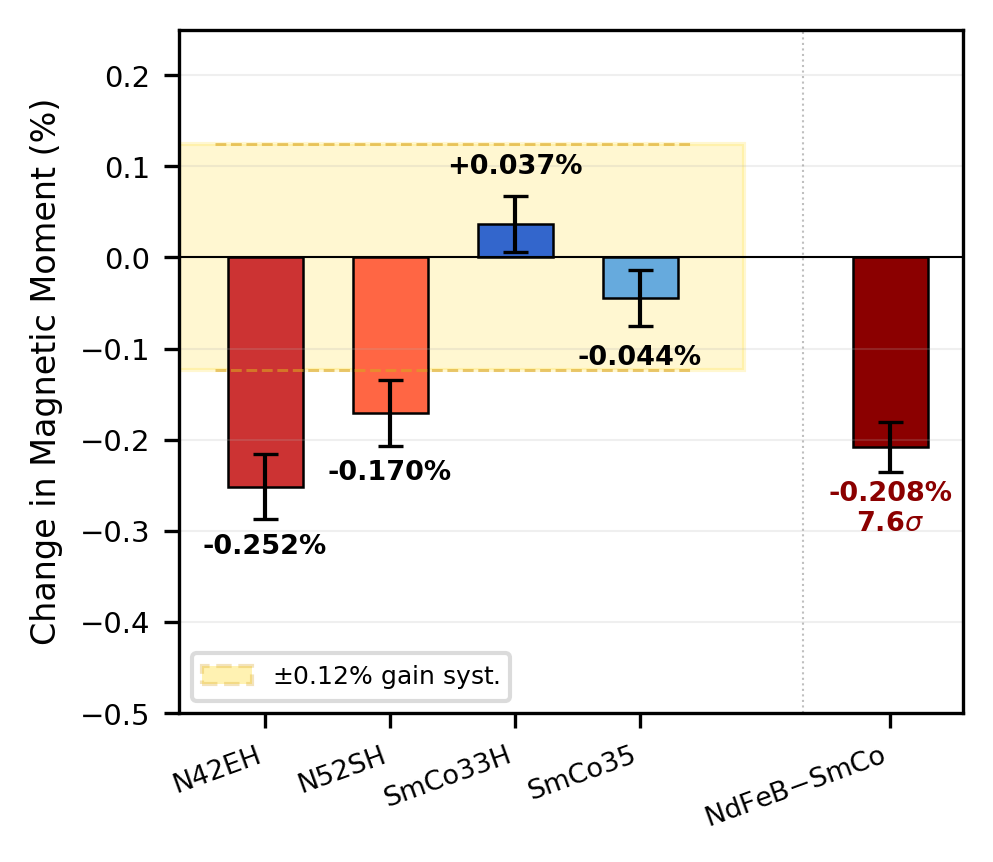}
    \caption{Mean percentage change in magnetic moment by material grade (30 tunnel plates, temperature-corrected to 20$^\circ$C). Error bars are $\pm 1$ SEM. The rightmost bar is the gain-immune intra-plate NdFeB$-$SmCo differential. The shaded band indicates the $\pm 0.12\%$ Helmholtz coil gain systematic, which cancels in the differential.}
    \label{fig:materials}
\end{figure}

Both NdFeB grades lost magnetic moment, while both SmCo grades are consistent with zero change (Figure~\ref{fig:materials}): N42EH $-0.252 \pm 0.036\%$, N52SH $-0.170 \pm 0.036\%$, SmCo33H $+0.037 \pm 0.031\%$, and SmCo35 $-0.044 \pm 0.031\%$. The NdFeB-minus-SmCo differential across the 30 tunnel plates is
\[
\delta = -0.208\% \pm 0.028\%~\text{(stat)} \pm 0.036\%~\text{(syst)},
\]
corresponding to 7.6$\sigma$ statistical and 4.6$\sigma$ combined significance. The nine laboratory control plates give $-0.007\% \pm 0.038\%$ (0.2$\sigma$), indicating that the measurement chain introduces no material-dependent bias, and the tunnel-minus-laboratory excess is independently significant at 4.3$\sigma$.

\paragraph{Radiation type and mechanism}
Correlating the differential against the co-located dosimetry separates the radiation types. Integrated gamma exposure shows no significant correlation (Spearman $\rho = 0.21$, $p = 0.27$) across nearly five orders of magnitude in dose, while neutron exposure does correlate ($\rho = 0.39$, $p = 0.03$). The damage therefore tracks neutron exposure rather than ionizing dose, which is consistent with a displacement-driven mechanism and with the damage hierarchy reported in the irradiation literature. Total integrated dose alone is not a sufficient predictor: the North Linac plates received the highest exposures but show a smaller mean change ($-0.099\%$) than the arc plates ($-0.263\%$), so local position within a sample stack appears to matter at least as much as regional dose.

\paragraph{Grade dependence}
One result runs counter to expectation. N42EH degrades more than N52SH despite its higher intrinsic coercivity. The two grades differ in heavy rare-earth content, notably dysprosium, which has a large thermal neutron capture cross-section. Whether dysprosium-bearing capture reactions account for the inversion is under investigation, and a quantitative treatment will be reported separately. SmCo's radiation hardness makes it attractive on this metric alone, but samarium-cobalt carries separate operational considerations, including radiological activation, which are treated in a companion note \cite{Bodenstein:TN26054}.

\begin{figure}[!htb]
    \centering
    \includegraphics[trim=0cm 1cm 0cm 0cm, clip, width=\textwidth]{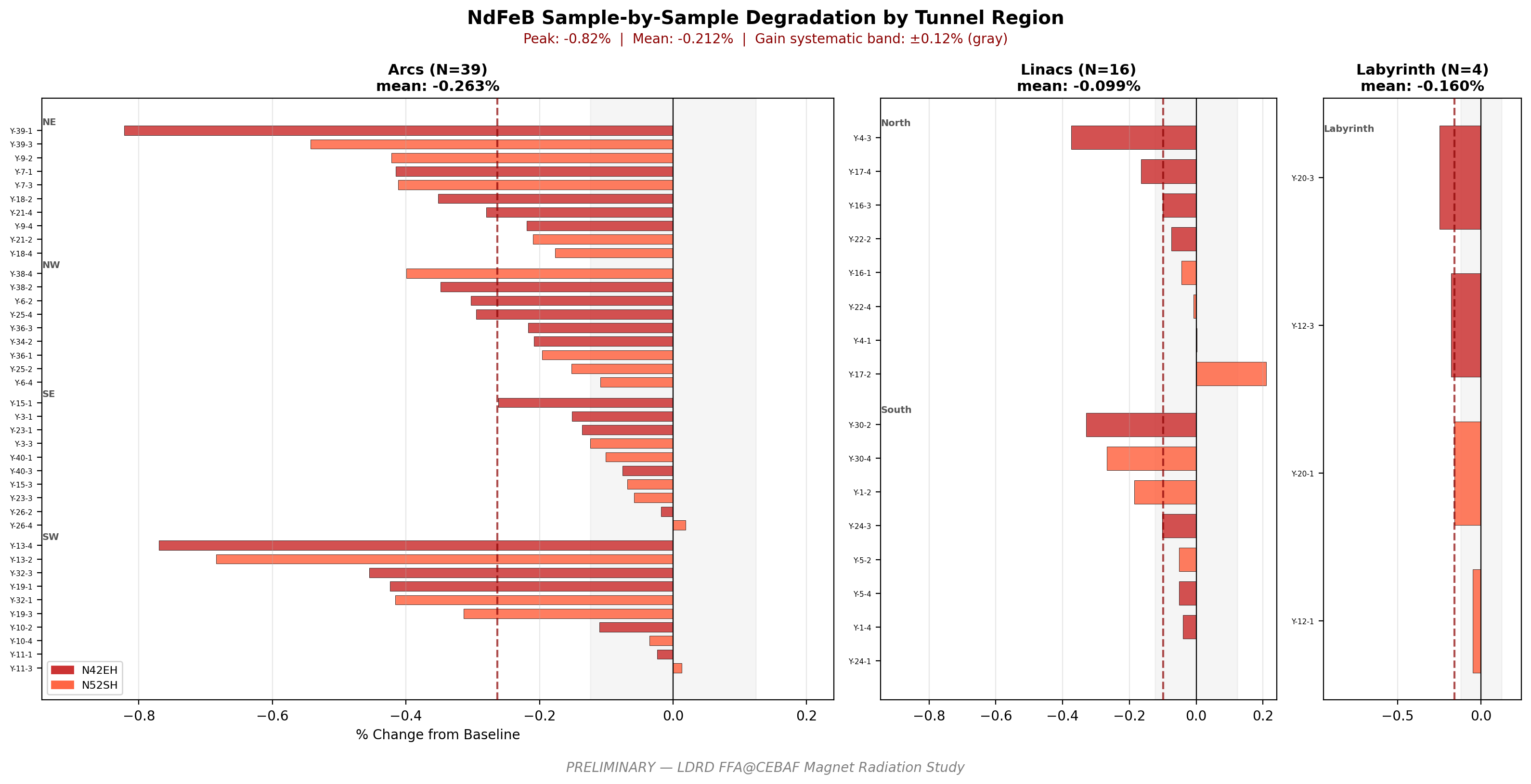}
    \caption{Sample-by-sample NdFeB degradation sorted by magnitude and grouped by tunnel region. The mean is $0.21\%$ and the largest single change is $0.82\%$. The gray band indicates the $\pm 0.12\%$ gain systematic.}
    \label{fig:waterfall}
\end{figure}

\paragraph{Non-uniformity}
The response is not uniform across samples. The mean NdFeB change is $0.21\%$, but individual samples reach $0.82\%$ (Figure~\ref{fig:waterfall}), a spread of roughly a factor of four among nominally identical samples, in some cases within the same tunnel region.

\subsubsection{Persistence of the observed change}

The samples were remeasured twice during the second campaign, once after a roughly three-month beam-off shutdown and again during subsequent lower-energy running. The differentials were $-0.235\% \pm 0.026\%$ and $-0.242\% \pm 0.033\%$, compared with a first-campaign endpoint of $-0.212\% \pm 0.029\%$. There was no recovery during the shutdown; the differential instead continued to deepen slightly. This indicates that the loss is permanent on the timescales examined rather than a reversible thermal or short-term relaxation effect, which is the relevant case for magnets expected to remain in the tunnel for the operating life of the machine.

\subsubsection{Implications for Halbach array field quality}

In a Halbach array, each wedge has a distinct easy-axis orientation, permeance coefficient, and local demagnetizing field, and wedges at different positions within a magnet will not receive identical exposure. Wedges within a single magnet are therefore expected to degrade at somewhat different rates. The sample-level spread already observed (Figure~\ref{fig:waterfall}) shows that non-uniform response is present even among nominally identical samples in comparable environments.

If the magnetization vectors of the constituent wedges evolve asymmetrically, the good-field region of the magnet will shift and deform. Unlike a uniform reduction, which appears principally as a dipole error, asymmetric evolution can introduce higher-order terms such as sextupole and octupole content that the planned Panofsky corrective quadrupoles are not designed to remove. With more than 70 cells per arc, perturbations that are individually small may accumulate through the lattice if they share a systematic pattern, for example if inner-arc wedges consistently degrade more than outer-arc wedges.

The configuration-dependent part of this question is not yet resolved experimentally. The pair assemblies were designed to isolate the effect of reverse-field geometry, but the differences between configurations are currently smaller than the measurement resolution, and the Beta configuration is not measurable by the primary method. Field-harmonic correction schemes for permanent magnet devices are under development elsewhere, including radial magic finger correctors for permanent magnet quadrupoles \cite{Brookbank:MEDSI2025}, which could in principle compensate for some radiation-induced harmonic content. Determining whether the degradation levels measured here are acceptable for the FFA design would require mapping a measured degradation profile onto individual wedges, computing the resulting multipole content, and tracking the result through the lattice. Wedge-level tolerances for the upgrade have not yet been established.

\subsubsection{Status}

These results are preliminary and represent the first of multiple planned exposure campaigns. Exposure to date corresponds to roughly six months of beam operation at 2.12~GeV per pass, a modest fraction of the machine's projected operating life at upgrade energies. Within that limited exposure the effect is small in absolute terms, well under one percent, but it is statistically unambiguous, material-dependent, correlated with neutron rather than gamma exposure, and persistent. A second exposure campaign is underway, with refined dose correlation, expanded configuration measurements, and a blind reanalysis of the frozen pipeline; those results will be reported separately. The questions most relevant to the upgrade are how the observed rates extrapolate across the full operating period and what wedge-level field-quality tolerance the FFA optics can absorb.

\section{Summary - Outlook}
The presented scheme aims to extend the energy reach of CEBAF up to 22 GeV within the existing tunnel. Proposed energy upgrade envisions increasing the number of recirculations, while using the existing CEBAF cavity system. The energy gain per pass remains unchanged, while the number of passes through the accelerating cavities is nearly doubled. A proposal was formulated to replace the highest-energy arcs with FFA arcs. The new pair of arcs would support simultaneous transport of additional six passes with energies spanning a factor of two, using the non-scaling FFA principle implemented with Halbach-derived permanent magnets - a novel magnet technology that significantly saves energy and lowers operating costs. The recirculating FFA arcs are configured with compact (3 m) FODO cells based on (3 m) FODO cells based on permanent multi-function Halbach magnets with dominant dipole and quadrupole fields. The optics have been designed for closely spaced orbits (4 cm spread) and low betas (a few meters) resulting from very strong focusing, reducing the horizontal dispersion function from meters in conventional separate functions arcs down to a few cm in the FFA arc. The arc optics were optimized to facilitate individual adjustment of momentum compaction and the horizontal emittance dispersion, $\mathcal{H}$ (to suppress adverse effects of the synchrotron radiation on beam quality). A novel concept of an extraction system and beam delivery to the experimental halls has been explored. Variations of the extraction system based on RF separators, capable of four-halls-operation have been studied based on the requirements of the experimental physics users. Individual hall lines are currently under investigation. Improvements to the magnetic septa are expected to be required. 

The proposed 22 GeV program acts as a critical and synergistic bridge between the existing JLab 12 GeV program and the future EIC. It specifically targets essential aspects of hadron emergence that are inaccessible to the 12 GeV regime or are outside the kinematic reach of the EIC. Furthermore, a major benefit is the ability to conduct these sophisticated measurements using the existing, well-characterized JLab12 detectors, minimizing both cost and technical development risk. This would  enable significant new scientific opportunities–including new mass ranges for meson spectroscopy, enabling precision studies of the nucleon sea, providing precision data into the abiding mystery of nuclear anti-shadowing, and extending the kinematic range of nucleon imaging studies. 

To conclude, significant progress has been made in the design of the energy upgrade for CEBAF using FFA beam transport. Over the last five years, we have settled on a solid design concept, developed more detailed optics solutions for various machine sections, and optimized the overall design, as simulations were performed. While the full accelerator design is not yet completed, we have summarizing current state of concept in this White Paper.

\section{Acknowledgments}
This material is based upon work supported by the U.S. Department of Energy, Office of
Science, Office of Nuclear Physics under Contract No. 89243126CSC000213 and authored in part by UT Battelle, LLC, under Contract No. DE-AC05-00OR22725.

Artificial intelligence tools were used in the preparation of this manuscript. Specifically, AI assistance was employed for editing and formatting purposes, including improving clarity, grammar, and structural consistency. Additionally, AI was used to help synthesize and organize content from existing literature into select sections of the paper. All AI-assisted content was reviewed, verified, and edited by the author(s), who take full responsibility for the accuracy, originality, and integrity of the final manuscript.
\newpage


\end{document}

%% file: Transition.tex
The transition beam line connecting the FFA arcs to the CEBAF linacs must
satisfy several matching conditions simultaneously for all six high-energy
passes while operating under tight spatial constraints. To keep the access
to the CEBAF tunnel as it is, the transition is planned to be a single
beamline.
 
\begin{figure}[htbp]
    \centering
    \includegraphics[width=\linewidth]{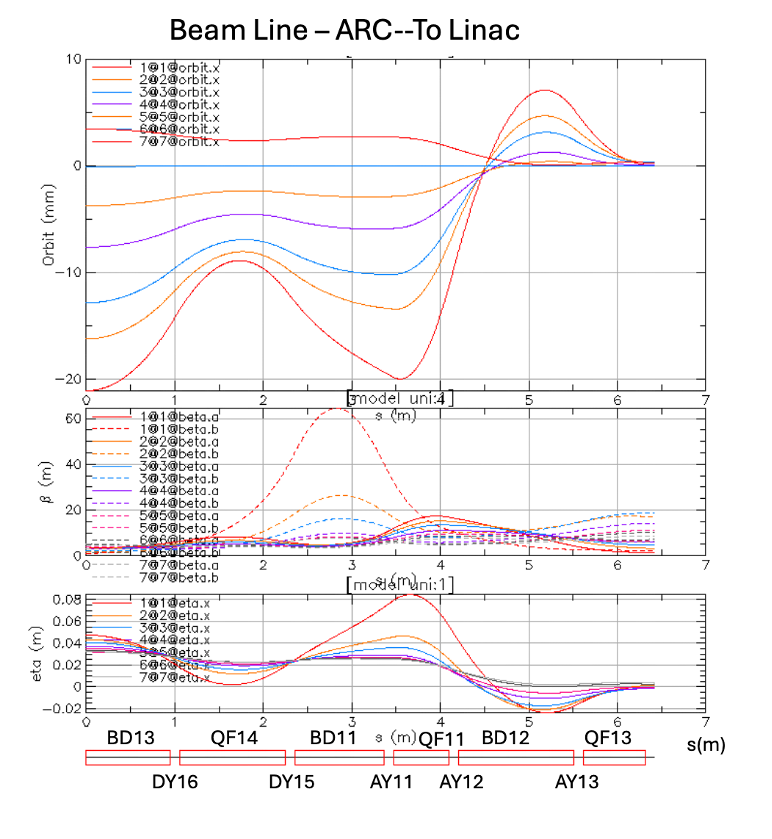}
    \caption{Beam Line matched FFA arc to the straight section with linac.}
    \label{fig:3.34}
\end{figure}

There are few approaches for the solution of this problem: The simplest
beam line solution is a very short beam line of only 6.6 meters long made
of six combined function magnets similar to the FFA arc magnets. It has
just six magnets of variable lengths, with magnetic dipole fields and
gradients within the permanent magnet limits set up for the magnets in the
FFA arcs. Their gradients are varied to establish perfect matching
conditions for the beam positions, dispersion functions, and zero slopes of
the betatron functions with appropriate maximum values. The beam line is
presented in Fig.~\ref{fig:3.34}.
 
Magnetic bending field and gradient values with the drift lengths are shown
in Table~\ref{tab:3.1}.
 
\begin{table}[htbp]
    \centering
    \caption{Magnetic field values.}
    \label{tab:3.1}
    \begin{tabular}{lr}
        \hline
        \textbf{Magnets and drifts} &
        \textbf{Magnetic field (T) and gradients (T/m)} \\
        \hline
        QF15\_ENT[B\_FIELD]      & $1.669$\,T     \\
        QF14\_ENT[B\_FIELD]      & $0.833$\,T     \\
        QF11[B\_FIELD]           & $0.429$\,T     \\
        QF13[B\_FIELD]           & $-1.038$\,T    \\
        BD15\_ENT[B\_FIELD]      & $-0.782$\,T    \\
        BD13\_ENT[B\_FIELD]      & $-1.798$\,T    \\
        BD11\_ENT[B\_FIELD]      & $0.222$\,T     \\
        BD12[B\_FIELD]           & $1.503$\,T     \\
        QF15\_ENT[B1\_GRADIENT]  & $26.216$\,T/m  \\
        QF14\_ENT[B1\_GRADIENT]  & $37.700$\,T/m  \\
        QF11[B1\_GRADIENT]       & $1.818$\,T/m   \\
        QF13[B1\_GRADIENT]       & $30.492$\,T/m  \\
        BD15\_ENT[B1\_GRADIENT]  & $-9.301$\,T/m  \\
        BD13\_ENT[B1\_GRADIENT]  & $-25.454$\,T/m \\
        BD11\_ENT[B1\_GRADIENT]  & $-12.695$\,T/m \\
        BD12[B1\_GRADIENT]       & $-30.703$\,T/m \\
        AY11[L]                  & $0.110$\,m     \\
        AY12[L]                  & $0.400$\,m     \\
        AY13[L]                  & $0.110$\,m     \\
        DY18[L]                  & $0.110$\,m     \\
        DY17[L]                  & $0.110$\,m     \\
        DY16[L]                  & $0.300$\,m     \\
        DY15[L]                  & $0.241$\,m     \\
        \hline
    \end{tabular}
\end{table}
 
These results followed previous work in matching arc-to-the
straights~\cite{trbojevic2021}.
 
The second approach for matching the arc-to-straight is the adiabatic
matching scheme. The bending angles of the consecutive periodic FFA FODO cells are gradually reduced following a third-order polynomial pattern, providing smooth control of the orbit and dispersion without perturbing the Twiss optical functions. The bending angle of the $i$'th cell is set to
\begin{equation}
\theta_i = A_i \, \theta_0,
\label{eq:bending}
\end{equation}
where $\theta_0$ is the bending angle of a regular arc FODO cell and $A_i$ is a scaling coefficient given by
\begin{equation}
A_i = 2\left(\frac{i}{C_n+1}\right)^3 - 3\left(\frac{i}{C_n+1}\right)^2 + 1,
\label{eq:scaling}
\end{equation}
with $C_n$ being the total number of FFA FODO cells used in the adiabatic section. In practice, the dipole field is first increased by about 3\% of its nominal value and then adiabatically reduced to zero to fit within the tunnel geometry. This approach suppresses the orbit and dispersion deviations, while keeping the betatron phase advance per cell nearly constant.

Residual orbit and dispersion offsets that remain after the adiabatic taper are further reduced using harmonic correction based on resonant excitation. In this scheme, a set of small corrector dipoles is used to selectively act on each of the different energy passes. The corrector strengths are varied along the beam line to provide a series of kicks, one per cell, synchronized with the betatron oscillations of a particular pass. The kicks accumulate coherently for the pass of choice, whereas they largely average out for the other passes. All of the passes can be controlled simultaneously by combining the kicker harmonics corresponding to the different-energy passes. All of the waveforms can be dialed into the same physical correctors. This approach can be applied in both transverse planes. The evolution of the particle coordinates under periodic kicks can be described as
\begin{equation}
\begin{pmatrix} x_N \\ x'_N \end{pmatrix}
= \sum_{n=1}^{N}
\begin{pmatrix}
A_{11} & A_{12} \\ A_{21} & A_{22}
\end{pmatrix}^{(N-n-1)}
\begin{pmatrix} 0 \\ \Delta x'(n) \end{pmatrix},
\label{eq:resonance_sum}
\end{equation}
where
\begin{equation}
\Delta x'(n) = x'_{\mathrm{corr}} \cos(2\pi\nu_{\mathrm{corr}}(n-1) + \phi_{\mathrm{corr}}),
\label{eq:kick}
\end{equation}
and $\nu_{\mathrm{corr}}$ represents the tune of the applied resonance. The single-cell transfer matrix coefficients are given by
\begin{equation}
\begin{aligned}
A_{11} &= \cos(2\pi\nu_x) + \alpha_x \sin(2\pi\nu_x), \\
A_{12} &= \beta_x \sin(2\pi\nu_x), \\
A_{21} &= -\frac{(1+\alpha_x^2)}{\beta_x} \sin(2\pi\nu_x), \\
A_{22} &= \cos(2\pi\nu_x) - \alpha_x \sin(2\pi\nu_x).
\end{aligned}
\label{eq:matrix}
\end{equation}
The resonance strength, proportional to $x'_{\mathrm{corr}}$, must remain much smaller than the tune separation between the passes to ensure that the excitation of one energy does not affect another. This method allows quasi-independent control of orbits and dispersions for all energies through a single beamline.

\begin{figure}[!htb]
  \centering
  \includegraphics[width=0.9\linewidth]{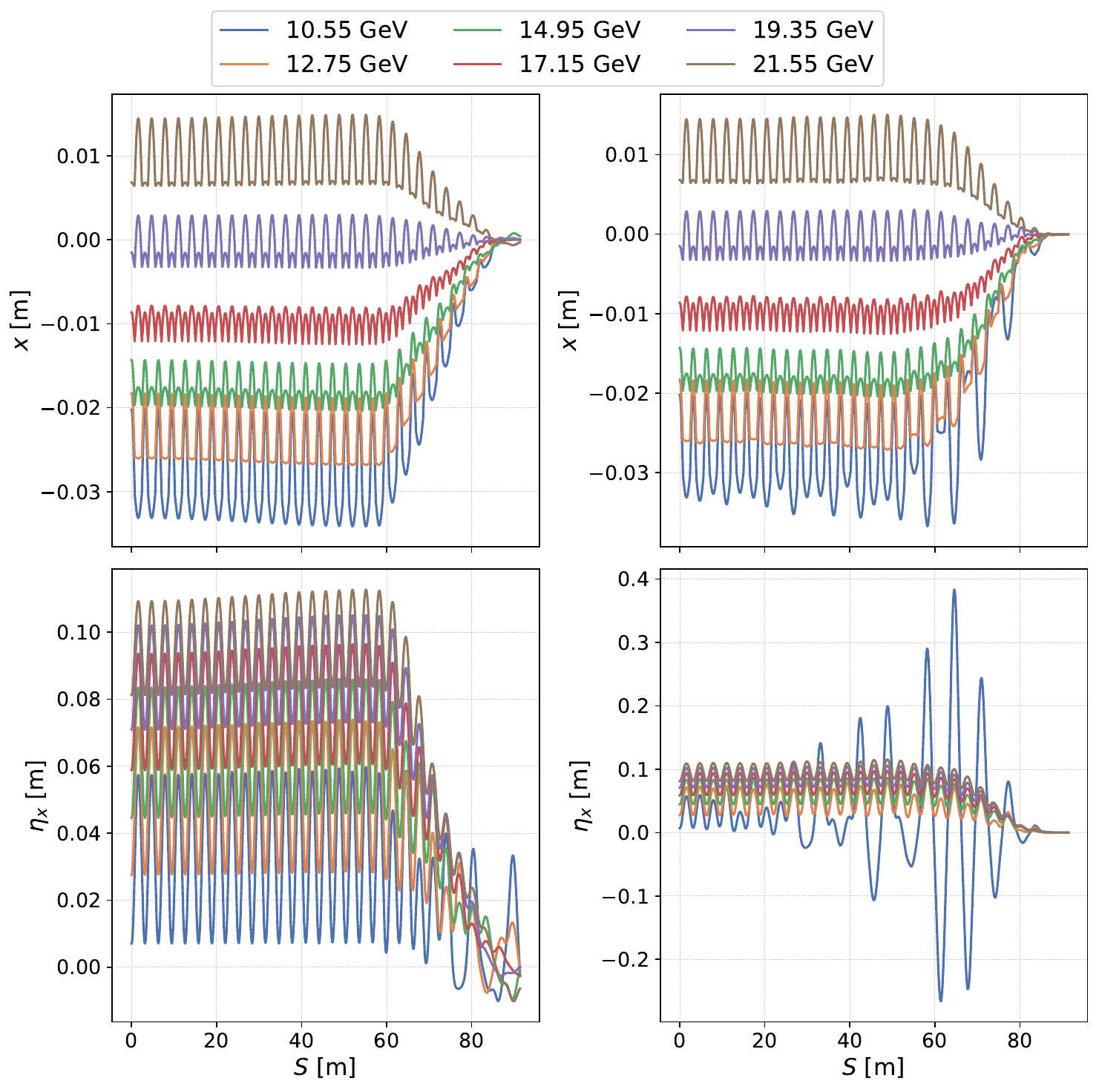}
  \caption{Comparison of beam dynamics in the north-west transition section with (left) and without (right) resonance correction in addition to the adiabatic correction which is applied to both. The top plots show the horizontal beam orbits, while the bottom plots show the corresponding horizontal dispersion functions.}
  \label{fig:orbit_dispersion}
\end{figure}

The CEBAF linac sections provide virtually no focusing to the high-energy passes. Therefore, to keep the beam envelope under control, its size must be significantly expanded in the long linac straights compared to the arcs. Thus, in terms of the Twiss parameters, matching to the linac requires rapid increase of the beam $\beta$ functions from an arc exit to the linac entrance with $\alpha$ functions kept reasonably small at the end of the matching section. 

\begin{figure}[!htb]
  \centering
  \includegraphics[width=0.9\linewidth]{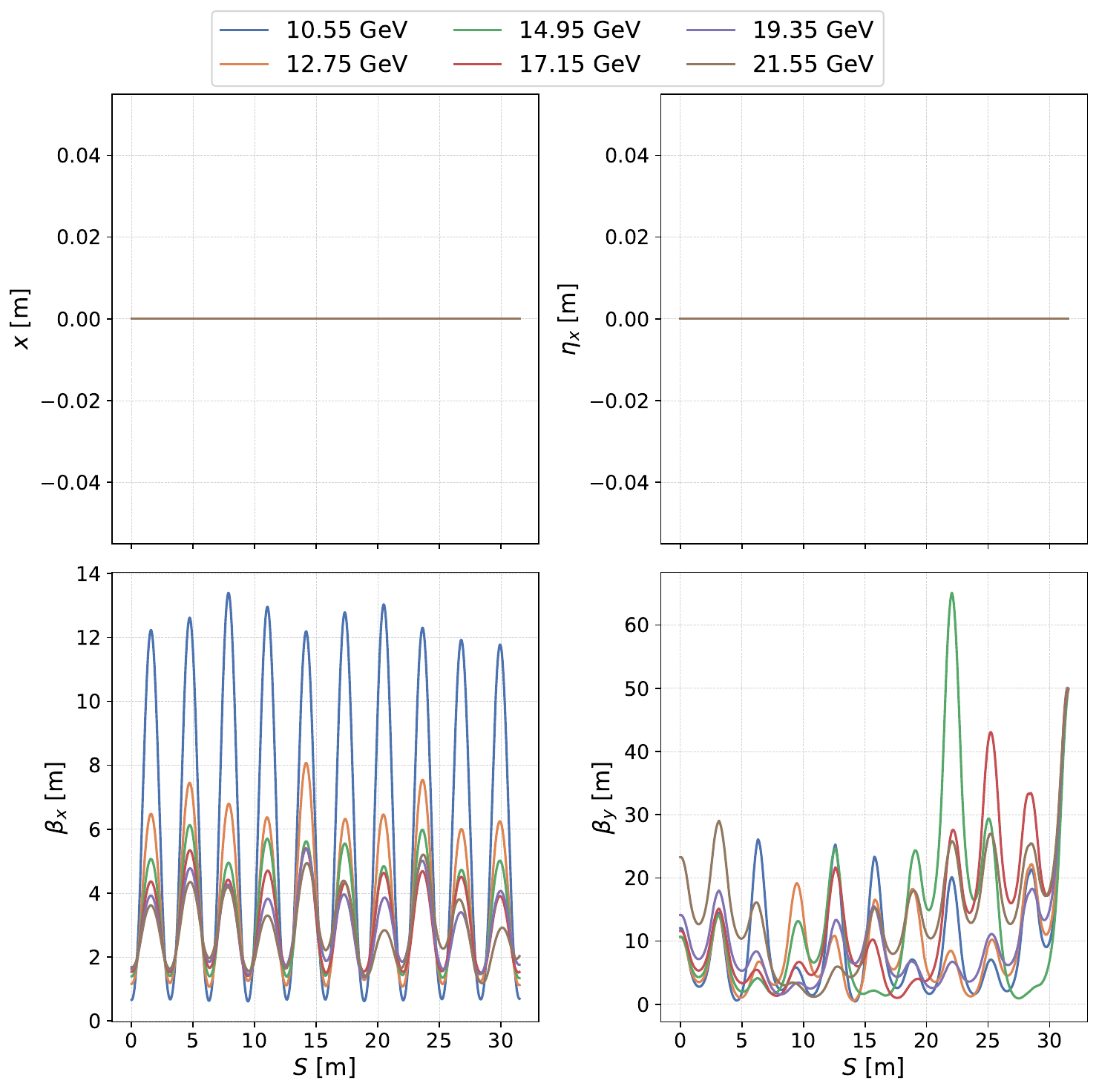}
  \caption{Twiss parameters of the north-west transition section with $\beta_y$ corrected for all high-energy passes. The top left plot shows the orbit and the top right plot shows the dispersion. The bottom left and right plots show the horizontal and vertical $\beta$ functions, respectively.}
  \label{fig:twiss_ga}
\end{figure}

It is challenging to achieve all of the matching requirements in the limited space available. We use a resonant approach to rapidly build up the $\beta$ function size in the underlying periodic FODO optics of the matching cell. A small quadrupole is added to each cell. The quadrupole strengths are modulated along the matching section to provide focusing kicks at twice the frequency of the betatron oscillations. The kicks then induce a parametric resonance that may lead to a rapid increase in the $\beta$ function. Due to a number of reasons such as the finite number of FODO cells in the matching section, non-linear effects, and finite widths of the resonance and betatron tunes, the resonant harmonics do not provide an entirely orthogonal control of the different passes, i.e., each resonance harmonic acts not only on its corresponding pass but influences other passes as well. Therefore, the optics of all passes are best optimized simultaneously. Figure~\ref{fig:twiss_ga} shows the result of such optimization, the resonant approach was used to expand the beams of all six passes in the vertical plane. Note that the resonance quads are placed at locations where $\beta_y$ is at maximum, while $\beta_x$ is at minimum. This results in the resonance quads having a much smaller effect in the horizontal plane. Another set of resonance quads can be used to induce a parametric resonance in the horizontal plane. This work is in progress. However, it is more challenging because, in contrast to $\beta_y$, the unperturbed $\beta_x$ of the underlying FODO lattice reaches minimum at the end of the matching section. This is not a favorable starting point for beginning increase of $\beta_x$.  

Given the high dimensionality of the variable space and the complexity of the constraints, a genetic algorithm (GA) was employed for optimization. The matching problem involves numerous coupled parameters, including magnet strengths, optics constraints, and targets for multiple energy passes, where conventional optimization methods often fail to converge. The GA efficiently searches large parameter spaces without relying on gradient information and can incorporate multiple objectives through its fitness evaluation. Using the \texttt{geneticalgorithm} library in Python, a viable solution was obtained that increased the vertical beta functions while maintaining acceptable horizontal betatron oscillations. The optimization used a population size of 10,000 over 100 generations and took advantage of the additional space available after the adiabatic matching section.

Achieving simultaneous optimization of both planes remains challenging because of the high number of constraints, non-orthogonality of the control knobs, and the aforementioned unfavorable situation for one of the two transverse planes. We are exploring several approaches to mitigate this problem including introduction of coupling and use of alternative options for the underlying optics, such as periodic doublet and triplet cells. 
\FloatBarrier